# IMPROVING THE PERFORMANCE OF OFDMA-BASED WI-FI NETWORK WITH HYBRID MEDIUM ACCESS CONTROL PROTOCOL DESIGN

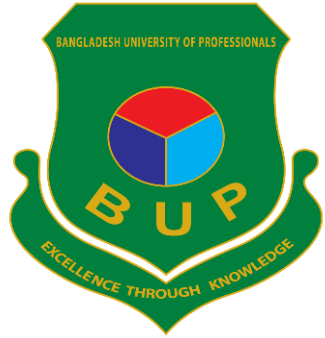

A Thesis Submitted to the Centre for Higher Studies and Research, Bangladesh University of Professionals for Partial Fulfillment of the Requirements for the Degree of Doctor of Philosophy

**By**

**Gazi Zahirul Islam**
PhD Researcher
Roll No. - P180043
Registration No. - 100401180043
Session: 2017-2018

**Under Supervision of**

**Mohammod Abul Kashem, PhD**
Professor
Department of Computer Science and Engineering
Faculty of Electrical and Electronic Engineering
Dhaka University of Engineering and Technology (DUET)

**December, 2021**

# DEDICATION

To my family, who inspired me throughout my studies and helped me in every sense to complete this thesis

# ACKNOWLEDGEMENTS

First and foremost, I praise and thank God, the Almighty, for His divine blessings throughout my PhD study to complete the research successfully. The feelings of accomplishment with satisfaction are an immense pleasure that could be felt in the heart but hardly be expressed in words. I am truly charmed to complete my PhD research work fulfilling the standards and reaching the obligatory benchmarks set by my supervisor and alma mater. This thesis would not have been possible without the help and support of many people. Herein, I would like to express sincere gratitude to all of them.

I would like to express deep and sincere gratitude to my PhD supervisor, Dr. Mohammod Abul Kashem, Professor, Department of Computer Science and Engineering, Dhaka University of Engineering and Technology (DUET) for his invaluable guidance throughout this study. His dynamism, vision, sincerity, and motivation have deeply inspired me. Without his relentless effort and support, it would not be possible for me to complete the work. It is a great privilege and honour to work and study under his guidance.

I would like to express profound gratitude to Brigadier General (Retd.) Syed Mofazzel Mawla, Dean, Centre for Higher Studies and Research (CHSR), Bangladesh University of Professionals (BUP) for his continuous guidance, invaluable suggestion, and firm support. I want to thank the members of the doctoral committee for their prudent evaluations and strong recommendations. Especially, I would like to mention the name of the seminar panel members, Dr. Md. Obaidur Rahman, Professor, Department of Computer Science and Engineering, Dhaka University of Engineering and Technology (DUET), and Dr. Md. Fokhray Hossain, Professor, Department of Computer Science and Engineering, Daffodil International University for their unconditional help and guidance. I am very grateful for their prudent remarks, helpful criticism, strong encouragement, and sincere appreciation that have helped me to produce a standard dissertation.

I would like to mention the Green Networking Research Group (GNR) led by Prof. Dr. Md. Abdur Razzaque. I am greatly benefitted from participating in many seminars and research activities arranged by the GNR. I would especially like to thank Dr. Fernaz Narin Nur (a member of GNR and my former colleague) for the insightful comments and suggestions that I had received. I also acknowledge the help and support received from Dr. A. B. M. Alim Al Islam, Professor, Department of Computer Science and Engineering, Bangladesh University of Engineering and Technology (BUET). I got opportunities to develop skills and expertise in many areas of wireless communications from the Professor.

I am very fortunate to pursue PhD at Bangladesh University of Professionals (BUP) under the Faculty of Science and Technology. I am indebted to the Centre for Higher Studies and Research (CHSR) for providing all supports and guiding me from the very beginning of the admission. I also pay sincere thanks to my fellow researchers at BUP, with whom I completed many projects and activities during the study. I gratefully acknowledge their support and cooperation throughout the journey of PhD. I would also like to thank the officials and staff of the BUP for providing valuable resources and information timely. Their continuous support and help enable me to conduct the mammoth research work smoothly.

I am incredibly grateful to my family for their love, prayers, caring, and sacrifices throughout my life. I am always indebted to my parents for giving birth to me in the first place and fulfilling all my needs. Last but not least, I would like to thank my dear wife for her admirable patience during the study and for helping me in every sense to complete the dissertation.

I would like to complete the acknowledgments by extending my gratitude to everyone else who, directly or not, consciously or not, has contributed to this thesis.

# DECLARATION

I hereby declare that the research work "Improving the Performance of OFDMA-based Wi-Fi Network with Hybrid Medium Access Control Protocol Design" has been carried out under the Faculty of Science and Technology, Bangladesh University of Professionals in fulfillment of the requirement for the degree of Doctor of Philosophy. I have composed this thesis based on original research findings from computer simulations acquired by me along with references from published literature. This has not been submitted in part or full to any other institution for any other degree. **I also certify that there is no plagiarized content in this thesis.**

03 December 2021	Gazi Zahirul Islam
Roll No.: P180043
Registration No.: 100401180043
Session: 2017-2018
Faculty of Science and Technology
Bangladesh University of Professionals

# CERTIFICATE OF THE SUPERVISOR

This is to certify that Mr. Gazi Zahirul Islam carried out his PhD study under my guidelines and supervision, and hence prepared the thesis entitled "Improving the Performance of OFDMA-based Wi-Fi Network with Hybrid Medium Access Control Protocol Design". So far as I am aware, the researcher duly acknowledged the other researchers' materials and sources used in this work. Further, the thesis was not submitted to any other universities or institutions for any other degree or diplomas.

It is thus recommended that the thesis be submitted to the Faculty of Science and Technology, Bangladesh University of Professionals, in fulfillment of the requirements for the award of the degree of Doctor of Philosophy. **I also certify that there is no plagiarized content in this thesis.**

04 December 2021                Professor Dr. Mohammod Abul Kashem
                                                           Department of Computer Science & Engineering
                                                           Dhaka University of Engineering and Technology
                                                           Gazipur-1707, Bangladesh

# ABSTRACT

Nowadays, the IEEE 802.11, i.e., Wi-Fi has emerged as a prevailing technology for broadband wireless networking. To meet the tremendous rise of demand for future generation wireless LANs, a robust and efficient MAC protocol is required for the Wi-Fi network. The acute rise of demand for high-speed wireless LANs has driven intensive research to improve the performance of MAC protocols leveraging a variety of MAC techniques. However, traditional MAC mechanisms are not suitable for next-generation communications due to some inherent constraints. In this regard, OFDMA technology could be adopted to design an efficient MAC protocol for the Wi-Fi network. The OFDMA technology supports multi-channel communications and hence ensures simultaneous communications between two or more users.

The purpose of this research is to provide a high-speed network for Wi-Fi users. The thesis presents three MAC protocols, namely, HTFA (High Throughput and Fair Access), ERA (Efficient Resource Allocation), and PRS (Proportional Resource Scheduling), by employing the OFDMA technology. The novel protocols improve Wi-Fi communication using the latest IEEE 802.11ax standard, i.e., Wi-Fi 6. In particular, the protocols improve several performance parameters of the MAC protocol, such as increasing the throughput, goodput, fairness index, and reducing the packet retransmissions, collisions, etc. The researcher designs the models and illustrates the working principles for the intended Wi-Fi protocols. The performance of the protocols is evaluated through comprehensive computer simulations using a powerful network simulator, i.e., NS-3. The saturation throughput is also measured effectively by constructing the analytical models. Simulation results validate that the new protocols are far better than the existing protocols.

The protocols designed in this thesis are compliant with the latest IEEE 802.11ax standard that promises to enhance the throughput (simply data rate) at least four times per user and support ten times users. Thus, the new protocols can ensure uninterrupted and smooth communication in highly dense environments (e.g., stations, airports, markets) where a large number of Wi-Fi users may need to send a massive volume of data concurrently. The thesis contains a lot of resources such as the state of the art of MAC protocols, analysis of contemporary protocols and their performance matrix; architecture of Wi-Fi system, OFDMA constraints and regulations; framework of protocols; relevant data, theory, and methods; analytical models; experimental results and analyses; etc. that would be the valuable resources to the future researchers for the research on the Wi-Fi network.

***Keywords:*** *Wi-Fi, Performance Parameters, MAC Protocol, OFDMA, IEEE 802.11ax, Resource Unit*

# TABLE OF CONTENTS

# LIST OF TABLES

# LIST OF FIGURES

# LIST OF ALGORITHMS

# LIST OF ACRONYMS AND ABBREVIATIONS

| Acronyms/Abbreviation | Expressions |
|---|---|
| ACK | Acknowledgment |
| AODV | Ad hoc On-Demand Distance Vector |
| AP | Access Point |
| BA | Block Acknowledgement |
| BEB | Binary Exponential Backoff |
| BSR | Buffer Status Report |
| BSS | Basic Service Set |
| CCC | Common Control Channel |
| CFP | Contention Free Period |
| CP | Contention Period |
| CRI | Channel Reservation Information |
| CSMA/CA | Carrier Sense Multiple Access / Collision Avoidance |
| CSMA/CD | Carrier Sense Multiple Access / Collision Detection |
| CTS | Clear to Send |
| CW | Contention Window |
| DCF | Distributed Coordination Function |
| DIFS | DCF Inter-Frame Space |
| DL | Downlink |
| DS | Distribution System |
| DSC | Dynamic Sensitivity Control |
| DSSS | Direct Sequence Spread Spectrum |
| EDCA | Enhanced Distributed Channel Access |
| EHT | Extremely High Throughput |
| EIFS | Extended Inter-Frame Space |
| eNB | Evolved Node B |
| ERA | Efficient Resource Allocation |
| ESS | Extended Service Set |
| ETSI | European Telecommunications Standards Institute |
| FHSS | Frequency Hopping Spread Spectrum |

| HCCA | HCF Controlled Channel Access |
| --- | --- |
| HCF | Hybrid Coordination Function |
| HIPERLAN | Higher Performance Radio LANs |
| HR-DSSS | High-Rate Direct Sequence Spread Spectrum |
| HTFA | High Throughput and Fair Access |
| IDE | Integrated Development Environment |
| IBSS | Independent Basic Service Set |
| IEEE | Institute of Electrical and Electronics Engineers |
| IFS | Inter-Frame Space |
| IoT | Internet of Things |
| LAN | Local Area Network |
| LRUs | Left Resource Units |
| LTE | Long Term Evolution |
| MAC | Medium Access Control |
| MCS | Modulation and Coding Scheme |
| MIMO | Multiple Input and Multiple Output |
| MPDU | MAC Protocol Data Unit |
| MR | Max Rate |
| MSDU | MAC Service Data Unit |
| MU MIMO | Multi-User MIMO |
| MUTAX | Minimizing Upload Time in 11AX |
| NAV | Network Allocation Vector |
| NIC | Network Interface Card |
| NS | Network Simulator |
| OFDM | Orthogonal Frequency Division Multiplexing |
| OFDMA | Orthogonal Frequency Division Multiple Access |
| OLSR | Optimized Link State Routing |
| OSI | Open Systems Interconnection |
| PC | Point Coordinator |
| PCF | Point Coordination Function |
| PF | Proportional Fair |
| PHY | Physical layer |
| PIFS | PCF Inter-Frame Space |

| PL | Path Loss |
|---|---|
| PLCP | Physical Layer Convergence Protocol |
| PMD | Physical Medium Dependent |
| PPDU | PLCP Protocol Data Unit |
| PRS | Proportional Resource Scheduling |
| QAM | Quadrature Amplitude Modulation |
| QoS | Quality of Service |
| RA | Random Access |
| RRUs | Right Resource Units |
| RSSI | Received Signal Strength Indicator |
| RTS | Request to Send |
| RU | Resource Unit |
| SA | Scheduled Access |
| SIFS | Short Inter-Frame Space |
| SR | Spatial Reuse |
| SRTF | Shortest Remaining Time First |
| SRU | Smallest Resource Unit |
| STA | Station |
| TF | Trigger Frame |
| TGax | Task Group ax |
| TWT | Target Wake Time |
| UORA | Uplink OFDMA-based Random Access |
| UL | Uplink |
| WEP | Wired Equivalent Privacy |
| Wi-Fi | Wireless Fidelity |
| Wi-Max | Worldwide Interoperability for Microwave Access |
| WLAN | Wireless Local Area Network |

# LIST OF NOTATIONS

| Notation | Meaning |
|---|---|
| $B$ | Bandwidth in Mbps |
| $T$ | i) System throughput and ii) Time duration of a TF cycle iii) Number of SRUs for RA-zone for initial scheduling |
| $t_i$ | Throughput of station i |
| $N$ | Number of stations ii) Distance power loss coefficient |
| $M$ | i) Number of sub-channels and ii) Number of resource units |
| $F$ | Max–min fairness |
| $l_i$ | Load of station i |
| $W$ | Contention window size |
| $W_{max}$ | Maximum contention window |
| $W_{min}$ | Minimum contention window |
| $\alpha$ | Number of backoff stages |
| $\delta$ | Propagation delay |
| $\tau$ | Probability of transmission |
| $R$ | i) Number of time slots ii) Radius of the basic service set |
| $E[p]$ | Average packet payload size |
| $T_{total}$ | Total average time |
| $T_{slot}$ | Time of a single slot |
| $p_k$ | Probability of successful access |
| $P_{jsuc}$ | Probability of successful transmission in the $j^{th}$ sub-channel |
| $P_{jcol}$ | Probability of collision in the $j^{th}$ sub-channel |
| $P_s(i)$ | Probability of successful transmission for i sub-channels |
| $P_c(i)$ | Probability of collision for i sub-channels |
| $E_s$ | Average number of sub-channels for successful transmission |
| $S$ | i) Saturation throughput ii) Number of SRUs for SA-zone for initial scheduling |
| $t_i/l_i$ | Normalized throughput |
| $L$ | Number of levels in a resource unit |
| RU $(l, i)$ | Resource unit with index $i$ and level value $l$ |
| $T_H$ | Duration of the header field transmission |
| $T_{TF}$ | Duration of sending the trigger frame to the stations |
| $T_P$ | Duration of sending the user data |
| $T_{ACK}$ | Duration of sending the acknowledgement frame |
| $PL$ | Path-loss |
| $f_c$ | Carrier frequency |
| $d$ | Distance between the transmitter and receiver |
| $A_{AP}$ | Number of antennas of the access point |
| $A_{STA}$ | Number of antennas of the stations |
| $p_i$ | Load of the ax STA i |
| $P$ | Total number of ax STAs |
| $L1$ | Total load of ax STAs |
| $q_i$ | Load of the non-ax STA i |
| $Q$ | Total number of non-ax STAs |
| $L2$ | Total load of non-ax STAs |
| $L3$ | Total load of all STAs in the network |
| $r_i$ | Number of SRUs for STA i |
| $U$ | Number of SRUs for SA-zone for revised scheduling |
| $V$ | Number of SRUs for RA-zone for revised scheduling |
| $P_f(n)$ | Floor loss penetration factor |
| $n$ | Number of floors between the transmitter and receiver |
| $f(x)$ | Function for the payload size |
| $Pmax$ | Maximum payload size |
| $Pmin$ | Minimum payload size |
| $s(t)$ | Stochastic process for backoff stage |
| $b(t)$ | Stochastic process for backoff time counter |
| $P_b$ | Probability the channel is busy |
| $P_S$ | Probability that a successful transmission occurs in a slot time |
| $T_S$ | Average successful transmission time of the channel |
| $T_C$ | Average collision time of the channel |
| σ | Duration of an empty slot time |
| $J(x)$ | Function of Jain's fairness index |

| $k$ | Number of users that equally shares the resource |
|---|---|
| $K$ | Number of attempts to transmit in BEB algorithm |
| $T_B$ | Backoff time in BEB algorithm |

# CHAPTER ONE

# Chapter One

# Introduction

## 1.1 Preliminary

The Wi-Fi network, also known as the Wireless Local Area Network (WLAN), is one of the major components of wireless communications. The emergence of wireless LANs brings the conveniences of user mobility and flexible network deployment in the local area communication. There are many areas of wireless communications such as WiMax, mobile or cellular, Wi-Fi, Bluetooth, etc. While the cellular network covers a large distance up to several kilometers, a Bluetooth network covers only a short distance up to several meters. However, Wi-Fi is a medium-range communication network that lies between the cellular network and the Bluetooth network. The range of Wi-Fi signals is up to 100 meters outdoors and up to 50 meters indoors.

The purpose of this research is to provide a high-speed network for Wi-Fi users. The performance of the Wi-Fi network mainly depends upon the performance of the MAC (Medium Access Control) protocol. However, the MAC layer of the Wi-Fi, i.e., IEEE 802.11, merely changed for more than a decade before the advent of the OFDMA (Orthogonal Frequency Division Multiple Access) technology. The latest standard of Wi-Fi, also known as Wi-Fi 6 (i.e., IEEE 802.11ax), adopts the OFDMA technology and promises to deliver high-speed communications in the Wi-Fi network even in congested areas.

In this thesis, three OFDMA-based MAC protocols are designed to enhance the throughput (i.e., data rate) in the Wi-Fi network. The protocols also increase the goodput and fairness index as well as reduce frame collisions and retransmissions in the network. Being OFDMA-based protocols, the new protocols are also highly appropriate for heterogeneous applications. The major contributions of the protocols are shown in Table 7.1 in Chapter 7. The significance and outcome of this research are summarized in Section 7.8.

## 1.2 Background of the Study

### 1.2.1 Why OFDMA-based Wi-Fi

Nowadays, Wi-Fi has emerged as a prevailing technology for broadband wireless networking. Along with lots of emerging applications and services over the Wi-Fi network, the demands for a high-speed and high-capacity Wireless LAN (WLAN) have been growing fast. The tremendous rise of demand for a faster and robust WLAN has led to intensive research to provide a high-speed wireless LAN leveraging a variety of Medium Access Control (MAC) techniques.

The highest body responsible for defining the communication standard is the Institute of Electrical and Electronics Engineers, known as the IEEE. The IEEE has defined the latest standard for the WLAN, which is known as IEEE 802.11ax or Wi-Fi 6. According to the specifications of the IEEE 802.11ax standard, the Wi-Fi 6 would gain at least four times enhancement in data throughput per user as well as should support ten times users (Standards IEEE 2019). The latest standard would provide good service even in highly dense areas such as stadiums, markets, stations, etc.

While the manufacturers and consumers rivet their eyes on Wi-Fi 6, the IEEE 802.11 working group turns to the next-generation communication standard, IEEE 802.11be or Wi-Fi 7. Wi-Fi 7 is supposed to provide an Extremely High Throughput (EHT) to fulfill the requirement of recent applications such as the 4k/8k video, augmented and virtual reality, online gaming, etc. The main technological difference between the recent Wi-Fi (Wi-Fi 6 and Wi-Fi 7) and the legacy Wi-Fi (Wi-Fi 1 to Wi-Fi 5) is the adoption of the OFDMA technology. Thus, a high-efficient MAC protocol is required for the future OFDMA-based wireless LANs that comply with the OFDMA specifications.

### 1.2.2 MAC Protocol and OFDMA Technology

In terms of computing, a communication protocol refers to the set of rules that computers have to follow to communicate with each other. Similarly, the MAC protocol defines the set of rules to communicate in the Wi-Fi network. The efficiency of the MAC protocol plays a significant role in enhancing the throughput. The throughput is defined as the successful message delivery over the network that simply refers to the network's speed. The details of the MAC protocol are discussed throughout Chapter 2.

One of the promising access methods for the MAC protocol is the OFDMA (Orthogonal Frequency Division Multiple Access) technology. The OFDMA technology is originated from the OFDM (Orthogonal Frequency Division Multiplexing) modulation techniques (National Instruments 2017). Inheriting the superiority of the OFDM, OFDMA-adopted MAC mechanisms could further raise the performance of WLAN by widening multi-station diversity. Because of some unprecedented advantages of OFDMA, some wireless systems, including WiMAX, are leveraging it from the very beginning. Section 3.4 and Section 3.5 of Chapter 3 describe how IEEE 802.11 network can utilize the OFDMA technology.

## 1.3 Problem Statement

The demand for faster, high-capacity, and efficient wireless LANs has been rising dramatically day by day. Nowadays, previous wireless LAN standards (Wi-Fi 1 to Wi-Fi 5) could not meet the user's demand satisfactorily. In this regard, the IEEE recently standardized IEEE 802.11ax (Wi-Fi 6) and also proposed another standard IEEE 802.11be (Wi-Fi 7) for the future wireless LAN. According to the requirements of the IEEE 802.11ax, wireless LAN should gain at least four times enhancement in data throughput per station (Standards IEEE 2019) as well as should support highly dense environments such as stadiums, markets, airports, etc. However, conventional MAC protocols for Wi-Fi could not meet the demand of the latest IEEE 802.11ax standard.

To fulfill the demand for high-speed wireless LANs, a new innovative protocol is required with sophisticated technological features. In this regard, OFDMA technology, which is adopted by Wi-Fi 6 and Wi-Fi 7, could be utilized to design an efficient and robust MAC protocol. Moreover, the OFDMA technology supports multi-channel communications and hence ensures simultaneous transmissions between two or more users.

In this paper, three MAC protocols are designed for the Wi-Fi network leveraging the OFDMA technology. The protocols are capable of enlarging the throughput (simply speed) of the wireless LAN significantly. The novel MAC mechanisms can also increase fairness in accessing the communication media as well as reduce frame collisions and retransmissions in the network.

## 1.4 Rationale of the Study

The main reason for choosing the study is to provide the users with high-speed wireless data in a WLAN environment. A high-capacity WLAN is necessary to support many users in dense areas such as the stadium, market, stations, etc. A robust WLNA is also of utmost importance to meet the growing needs of data-hungry applications such as real-time computer rendering modeling, augmented and virtual reality, cloud gaming, etc. Again, IEEE recent WLAN standards motivate and guide the researchers for designing OFDMA-based robust MAC protocol. Thus, the investigator conducts an exhaustive study on the MAC protocols and presents several innovative protocols for the OFDMA-based Wireless LAN.

To provide the high-speed network, at first, the researcher designs an efficient hybrid MAC protocol in this paper. The protocol is named High Throughput and Fair Access (HTFA) that adopts the latest OFDMA technology. Since WLANs have already employed the OFDM for the modulation, the OFDMA technology is highly recommended for the next-generation high-speed WLANs. OFDMA-adopted hybrid MAC protocols that inherit the properties of both random-access and reservation protocols are also suitable to support diverse applications (discussed in Section 2.6.1). Because of the heterogeneous applications, in the long run, the WLAN would deploy the hybrid MAC protocol supported by the OFDMA.

The efficiency of the OFDMA-based MAC protocols depends to a great extent on how the access point allocates the channel resources (e.g., resource units) to the intended users or terminals. These types of problems are commonly known as resource allocations or scheduling problems. Thus, nowadays, the study of the resource allocations approach for the MAC protocols is highly appropriate for the research on the IEEE 802.11ax. In this regard, the main challenge is to design the scheduling for the uplink (UL) path rather than the downlink (DL). Because, in the UL path, all sending STAs must be synchronized for the OFDMA transmissions (details in Section 2.6.2). After the HTFA protocol, this author designs two uplink scheduling protocols, namely, Efficient Resource Allocation (ERA) and Proportional Resource Scheduling (PRS), which promise to provide a high-throughput to the Wi-Fi network as well as reduce the retransmissions of the packets.

## 1.5 Research Questions

The main goal of this research is to improve Wi-Fi communication, specifically the performance parameters of the MAC protocol for OFDMA-based IEEE 802.11 networks (Wi-Fi 6 and Wi-Fi 7). To reach the goal, the researcher wishes to design new MAC protocols by adopting the OFDMA technology. In this paper, the researcher designs three innovative protocols, namely, HTFA, ERA, and PRS to enhance the throughput (i.e., data rate) of the network significantly. Thus, the first research question of this paper is:

> Question 1: How could an OFDMA-based MAC protocol enhance the throughput of the Wi-Fi network significantly?

In a typical Wi-Fi network, all stations or terminals usually do not receive equal time to transfer their packets or data. This limitation must be admitted due to the random features of the MAC protocols. The new protocols address this issue. The protocols increase fairness by a well-designed model. Hence, the researcher set the second research question as:

> Question 2: How to design a MAC protocol to ensure fair access of the stations to the communication media of the Wi-Fi network?

Reduction of the frame collisions and hence the retransmissions is another indicator of an efficient MAC protocol. Because the reduction of the number of retransmissions of frames increases the goodput of the network remarkably. All the protocols address the phenomena efficiently. Thus, the researcher sets the third research question as follow:

> Question 3: How to implement a MAC protocol effectively for reducing the number of collisions and retransmissions in the Wi-Fi network?

IEEE 802.11 networks (Wi-Fi 1 to Wi-Fi 5) predominantly use the random access mechanism for accessing the wireless channels. Recent IEEE standards (Wi-Fi 6 and Wi-Fi 7) adopt the OFDMA technology that provides the scheduled access mechanism along with the previous random access method. Since both mechanisms have some advantages, the researcher is interested to design a hybrid protocol that efficiently utilizes the random and scheduled access methods. Therefore, the fourth research question of this thesis is:

> Question 4: How to design a hybrid MAC protocol that implements random and scheduled access mechanisms for efficient transmissions?

The answers to the above questions lie in Chapter 4, i.e., Theoretical/Conceptual Framework, where the author designs the frameworks of all protocols in light of existing theories of the OFDMA technology and IEEE 802.11 recent standards.

## 1.6 Research Objectives

Following the problem statements and research questions, the author specifies the general objective and some specific objectives for the intended research that are outlined in the following subsections.

### 1.6.1 General Objective

The main purpose of this research is to provide a high-speed wireless LAN. As such, the researcher designs three OFDMA-based MAC protocols by adopting hybrid mechanisms for the latest IEEE 802.11 networks (i.e., Wi-Fi 6 and beyond). Hence, the general objective of this research is:

> To improve Wi-Fi communication for OFDMA-based IEEE 802.11 networks.

### 1.6.2 Specific Objectives

The novel protocols improve the Wi-Fi network's performance by increasing the throughput, goodput, fairness index, and reducing the packet retransmissions, collisions, etc. The specific objectives of the research are mentioned below:

1. To enhance the throughput of the OFDMA-based MAC protocol for the Wi-Fi network.
2. To increase fair access of the stations to the communication media in the Wi-Fi network.
3. To reduce the probability of frame collisions and retransmissions in the Wi-Fi network.
4. To increase the goodput and hence enhance the quality of user experience for the Wi-Fi users.
5. To implement an efficient hybrid protocol utilizing the random and scheduled access mechanisms

It is noted that the third specific objective also contributes to achieving the first specific objective. Because reduction of frame collisions also reduces packet loss in the network, which ultimately increases the throughput of the wireless LAN. Again, the first specific objective (i.e., throughput enhancement) also contributes to achieving a high goodput.

## 1.7 Limitations of the Study

There are many contributions to this research, along with a few limitations of the new protocols. Still, the investigator strongly believes there would be a tremendous tradeoff of the protocols. The limitations are summarized below,

The researcher's unique contribution to designing the HTFA protocol (i.e., first protocol of this thesis) is its innovative sub-channel distribution procedure leveraging the OFDMA technology (described in Section 4.2.2). In this protocol, it is assumed that all sub-channels created by the OFDMA are of equal length. For instance, if the total channel bandwidth is 100 kHz, then by the assumption, it is allowed to create five sub-channels, each of those having a bandwidth of 20 kHz or four sub-channels each of those having a bandwidth of 25 kHz. The author considers the sub-channels for the HTFA protocol are of equal length to avoid sheer complexity in designing the protocol.

The HTFA protocol is designed before the standardization of the IEEE 802.11ax (Standards IEEE 2019). As such, there are not enough guidelines provided by the IEEE about the length of the sub-channels, sub-channel formation rules, scheduling mechanisms, etc. However, this limitation is removed in the ERA protocol (i.e., second protocol), where the different lengths of RUs (Sub-channels) are available to assign to the STAs.

Resource scheduling in IEEE 802.11ax can be done by two methods, namely, random access and scheduled access. The ERA protocol is designed using the scheduled access mechanism where there is no scope for random transmissions. Due to the absence of random access feature, the ERA provides very high throughput and reduces retransmissions of packets. However, random access provision provides the facility of immediate data transmissions by the STAs whose BSR (Buffer Status Report) information is not available to the AP (details in Section 4.4). Thus to provide very high throughput, the random access feature is not included in the ERA protocol. At last, the researcher designs the PRS protocol, which is more superior and robust than the previous two protocols. The PRS eliminates the ERA protocol's limitations by providing both the random and scheduled access mechanisms for channel access.

## 1.8 Methodology and Performance Parameters

### 1.8.1 Methodology

In this dissertation, three MAC protocols are designed for the OFDMA-based Wi-Fi network. The models for the intended protocols are delineated in Chapter 4. Then the researcher performs mathematical analyses in Chapter 5 and computer simulations in Chapter 6 to measure the efficiency of the protocols. The analytical models are constructed for mathematical analyses, and the Network Simulator-3 is used to conduct the simulations. The investigator has to deal with a large volume of data and mathematical procedures for the validation of the protocol. Thus, the research follows a *quantitative approach* in this research to implement new protocols. A lot of tools, techniques, software, hardware, etc. are used to implement and validate the protocol. The details of the research methodology are described in Chapter 5.

### 1.8.2 Performance Parameters

Conduction of purely quantitative and technical research requires measuring many performance parameters during the investigation. A brief overview of several performance parameters of the current research is given below:

i). Saturation Throughput: It is the throughput measured in the saturation stage of the wireless system. In the saturation stage, every station (STA) always has some packets available for sending. The throughput is measured in Bytes, Kilobytes, Megabytes, etc.

ii). Collision Probability: Collision probability is measured in a scale range from 0 to 1. 0 indicates no collision at all, and 1 indicates 100% collision of data frames. The more collision in the system, the more packet losses in the system.

iii). System Throughput ($T$): System throughput is the summation of throughput of all STAs in the wireless LAN and is defined by the following equation:

$$T = \sum_{i=1}^{N} t_i \qquad (1.1)$$

where $t_i$ denotes the achieved throughput of the $i^{th}$ station.

iv). Goodput: Goodput is the measurement of useful data that reflects the real user experience in the network. Goodput is not the same as the throughput because goodput does not count the undesirable data that is arised from the retransmissions, protocol overheads, etc.

v). Contention Window Size ($W$): It is simply an integer number that represents the number of empty slots. The Binary Exponential Backoff (BEB) algorithm is used to determine the backoff time for the MAC protocols.

vi). Backoff Slot Time: Backoff slot time corresponds to the time slot used for the backoff purpose. The duration of a backoff slot is measured in the microsecond.

vii). Bandwidth: It is the data transfer capability of the wireless channel and is measured in bit/sec, Kbit/sec, Mbit/sec, etc. The bandwidth also represents the range of frequencies within a given band which is measured in hertz.

viii). Max–min fairness: Max–min fairness ($F$) is used to find the fairness of a protocol and is represented by the following equation:

$$F = \max \frac{t_i}{l_i} - \min \frac{t_i}{l_i} \qquad (1.2)$$

where,

$t_i$ denotes the achieved throughput of the $i^{th}$ terminal,

$l_i$ denotes the traffic load of the $i^{th}$ terminal,

$t_i/l_i$ denotes normalized throughput.

ix). Jain's fairness index: Jain's fairness index ($J$) is also used to evaluate the fairness of the wireless protocols. It can be calculated by using the following equation:

$$J(x) = \frac{(k \cdot \frac{M}{k})^2}{N \cdot k \cdot (\frac{M}{k})^2} = \frac{k}{N} \tag{1.3}$$

where,

$M$ denotes the number of resource units

$N$ denotes the number of users

$k$ denotes the number of users that equally shares the resource

## 1.9 Chapter Outlines

This section provides an overview of all chapters of this thesis. The thesis consists of 8 chapters which are discussed below:

***Chapter 1:*** The first chapter introduces the readers to wireless communications at the beginning in Section 1.1. Then the background study of the current research is discussed in Section 1.2. After that, the problem statement, rationale of the study, research questions, research objectives, limitations of the study are described in subsequent sections. Section 1.8 gives an overview of methodology and performance parameters which will be further illustrated throughout Chapter 5, i.e., Research Methodology. Section 1.9 (i.e., this section) provides the outlines of all of the chapters in this paper. Finally, Section 1.10 concludes Chapter 1.

***Chapter 2:*** Chapter 2 provides a literature review of the current research. Section 2.2 shows the performance matrix of some predominant Wi-Fi standards. Then the functions of the MAC layer and various timing intervals are described in Section 2.3. Section 2.4 delineates the MAC frame format. Section 2.5 explains the basic and advanced access mechanisms of the MAC. The author discusses MAC protocol classes and their features elaborately in Section 2.6. The author also goes through a lot of promising protocols designed by other researchers and summarizes their advantages and limitations. Section 2.6 also provides the background of the scheduling protocols. At last, the author summarizes the research gaps in Section 2.7.

*Chapter 3:* Chapter 3, i.e., Wi-Fi Architecture, starts with the discussion of the IEEE 802.11 regarding the OSI (Open Systems Interconnection) reference model. Then the architecture of the Wi-Fi network, in particular, the Basic Service Set (BSS) and the Extended Service Set (ESS), are delineated with comprehensive figures. Section 3.3 gives an overview of the physical layer of the IEEE 802.11, and Section 3.4 relates to the IEEE 802.11ax and OFDMA technology. OFDMA specifications and constraints are detailed in Section 3.5. In the end, the chapter describes various services (i.e., station services and distribution services) provided by Wi-Fi.

*Chapter 4:* Chapter 4, i.e., Theoretical / Conceptual Framework, establishes the foundation of this research. The researcher innovates three Wi-Fi protocols in this thesis. Section 4.2 describes the framework for the HTFA protocol ($1^{st}$ protocol) along with the relevant theory. In the same way, Section 4.3 and Section 4.4 illustrate the other two protocols, namely, ERA (second protocol) and PRS (third protocol), respectively. The models and working procedures of the protocols are also explained elaborately. The main novel feature of the HTFA is its unique principles for sub-channel distribution which are described in Subsection 4.2.2. The novelty of ERA is its innovative algorithm for scheduled access which is described in Subsection 4.3.3. The main contribution of the PRS protocol is designing two scheduling algorithms which are described in Subsection 4.4.2.

*Chapter 5:* This chapter illustrates the methodology employed for this research. Section 5.2 designs the target network models for the new protocols. Section 5.3 describes the simulation parameters and samples for the experiments. Section 5.4 describes the necessary instruments and software to establish the simulation environment for Network Simulator-3. Section 5.5 illustrates the procedures to measure the efficiency of the protocols. Here, the researcher employs two procedures, i.e., analytical model and computer simulation. Finally, Section 5.6 illustrates the analytical models using suitable figures for the three protocols.

*Chapter 6:* This chapter contains data analysis of the computer simulation regarding the intended protocols. In Section 6.2, the researcher evaluates the efficiency of the HTFA protocol. The researcher also compares the HTFA with some other contemporary protocols. Section 6.3 compares the ERA protocol with some other promising protocols implemented by other researchers. The section also describes the data of the performance comparison between different methods used for the MAC protocols. At last, Section 6.4 describes the simulation data of the PRS protocols. In this section, different access mechanisms are compared to each other as well as the comparison between different protocols is also illustrated.

*Chapter 7:* Chapter 7 discusses the significance and importance of the current research in-depth. In Section 7.2, the contributions of the HTFA protocol, such as throughput enhancement and fair access, are discussed scientifically in detail. This section also outlines the challenges and significance of the analytical model of the HTFA protocol. In Subsection 7.3, the investigator tries to persuade the importance of an optimal scheduler by conducting a simple simulation. Section 7.4 and Section 7.5 illustrate the contributions and significance of the ERA and PRS protocol, respectively. The utilities and impact of the Network Simulation-3 are described briefly in Section 7.6. The key contributions of the three protocols are shown in a table in Section 7.7. Finally, Section 7.8 summarizes the outcome and importance of this thesis.

*Chapter 8:* The last chapter, i.e., Conclusions and Future Works, comprises three sections. Section 8.1 emphasizes the necessity of a high-speed Wi-Fi network for future generation communications and discusses how this work would contribute to achieving that. Section 8.2 summarizes the works that have been completed for this research. In Section 8.3, the researcher comes up with some suggestions for future study, which are very much appropriate to extend the current research further.

## 1.10 Conclusion

The scientific research follows a set of steps sequentially from the topic selection to the conclusions. This chapter is organized in such a way so that it covers most of the steps of this research. Some of the steps, such as the problem statement, research questions, and objectives, are explained in detail in this chapter. Some other steps, such as literature review, research methods, data collection and analysis, are introduced a little bit in this chapter and will be elaborated on in the rest of the chapters.

In the preliminary section, the readers are introduced to the area of wireless communications. This section also summarizes the research work. Section 1.2 discusses the background of this research. Section 1.3 states the problem statement in a short context, and Section 1.4 explains the rationale of the current study. Section 1.5 formulates the research questions with explanations. The research objectives are outlined in Section 1.6 that contains the general objective and five specific objectives. The researcher mentions a few limitations of the work, which are discussed in Section 1.7. Section 1.8 provides an overview of the research methodology that will also be illustrated in Chapter 5. Section 1.8 also explains several performance parameters for the evaluation of the MAC protocols. Finally, Section 1.9 outlines each of the chapters of this paper in a short context.

# CHAPTER TWO

# Chapter 2

# Literature Review

## 2.1 Introduction

The author has searched many resources for this research from primary, secondary, and tertiary sources. All sorts of protocols for wireless LANs have been studied and perceived their superiority and limitations over each other. The researcher also goes through the details of Wi-Fi standards, MAC layer, access mechanisms, etc. The outcome of the literature search is discussed in this chapter.

Section 2.2 introduces the readers to different Wi-Fi standards released so far by the IEEE. Section 2.3 illustrates the functions of the MAC and various timing intervals. The details of the IEEE 802.11 MAC frame format are shown in Section 2.4. Section 2.5 describes the basic access mechanism and advanced access mechanism of the MAC. In Section 2.6, the paper discusses many promising protocols with their pros and cons proposed by other researchers. In Section 2.7, the investigator outlines the research gaps.

## 2.2 Wi-Fi Standards

In 1997, the IEEE formed a group with several members to oversee wireless LAN development. The group created the first Wi-Fi specifications termed the 802.11. Later IEEE released many improved versions such as 802.11a, 802.11b, 802.11n, 802.11ac, etc. The recent release is 802.11ax, which is also known as Wi-Fi 6. Table 2.1 shows the performance matrix of some predominant standards released from 1997 to date.

## 2.3 IEEE 802.11 MAC

The IEEE 802.11 Medium Access Control (MAC) has to ensure reliable data delivery over the wireless media, which is subject to unreliability due to noise, interference, and other propagational effects. The MAC also provides advanced LAN services that are equal to or beyond the 802.3 wired LAN.

Table 2.1: Performance matrix of some predominant Wi-Fi standards

| IEEE Standard | 802.11a | 802.11b | 802.11g | 802.11n | 802.11ac | 802.11ax |
|---|---|---|---|---|---|---|
| Year Adopted | 1999 | 1999 | 2003 | 2009 | 2014 | 2019 |
| Frequency | 5 GHz | 2.4 GHz | 2.4 GHz | 2.4/5 GHz | 5 GHz | 2.4/5 GHz |
| Maximum Data Rate | 54 Mbps | 11 Mbps | 54 Mbps | 600 Mbps | 1 Gbps | 10-12 Gbps |
| Typical Range (Indoors) | 100 feet | 100 feet | 125 feet | 225 feet | 90 feet | 230 feet |
| Typical Range (Outdoors) | 400 feet | 450 feet | 450 feet | 825 feet | 1000 feet | 1000 feet |

The primary function of the MAC is to ensure reliable data delivery to the STAs in the Wi-Fi network. The basic data transfer mechanism involves the two-frame exchange scheme. To further increase the reliability of successful transmission, a four-frame exchange scheme may be used. These two schemes are discussed elaborately in the following Subsection 2.3.1.

The second function of the MAC is to ensure fair access of the STA to the shared wireless medium. The IEEE 802.11 MAC proposes two different methods to access wireless media. One is known as the Distributed Coordination Function (DCF), and another one is known as the Point Coordination Function (PCF). The details of the DCF and PCF operations can be found in the Standards IEEE (2019) and O'Hara & Petrick (2005). The basic access mechanism is DCF which is predominantly used for MAC, and the PCF is used for centrally controlled access. The DCF and PCF operations are elaborated in Section 2.4.

Another important function of the MAC is to protect the user data while traversing through the wireless medium. In this regard, the IEEE 802.11 employs a very efficient algorithm which is known as the WEP (Wired Equivalent Privacy) algorithm. The Wi-Fi network is predominantly using the WEP algorithm for data encryption over the wireless medium.

### 2.3.1 Reliable Data Delivery

Like other wireless networks, the IEEE 802.11 network is also subject to considerable uncertainty of successful data transmissions due to the interference, noise, and other propagational effects in wireless communications. Even employing the error-correction scheme, some of the MAC frames may not be sent to the receiver successfully. The uncertainty of the wireless medium may result in the loss of a significant number of MAC frames. To deal with the situation, IEEE 802.11 MAC employs a frame exchange protocol. In this way, when the receiving STA receives the data frame, then it sends back an acknowledgement (ACK) frame to the sending STA. If the sending STA does not receive the ACK frame within the time threshold, then it has to resend the data frame. Not receiving the ACK frame implies two things: (i) the data frame sent from sending STA to receiving STA is damaged, or (ii) the ACK frame sent from receiving STA to sending STA is damaged.

The above method for achieving reliable data delivery is known as a two-frame exchange method. But this method cannot ensure a higher degree of reliability. As such, the four-frame exchange method can be used where an additional pair of frames is used. The frames are known as Request-to-Send (RTS) and Clear-to-Send (CTS). In this method, the sending STA at first sends an RTS frame to the receiving STA. The receiving STA then replies with a CTS frame. After receiving the CTS frame, the sending STA sends the data frame to the receiving STA. At last, receiving the data frame, the destination STA sends an ACK frame to the sending STA. The purpose of the RTS frame is to alert all wireless stations within the communication range of the sending STA and refrains from transmission in the medium. Similarly, the CTS frame refrains the STAs surrounding the receiving STA from transmissions.

### 2.3.2 Channel Access Timing

The MAC employs different types of timing intervals for channel access which are known as Inter-Frame Spaces (IFSs). These IFSs include slot time, Short IFS (SIFS),

DCF Inter-Frame Space (DIFS), PCF Inter-Frame Space (PIFS), and Extended Inter-Frame Space (EIFS) (Perahia & Stacey 2013). Figure 2.1 illustrates the different timing intervals used in the IEEE 802.11 MAC protocols. Both the DCF and PCF operations are implemented using these timing intervals.

**Slot time:** The slot time, also known as the backoff slot time, corresponds to the time slot used in the contention window for the backoff purpose. It is also used to find the duration of some other IFSs. The duration of other inter-frame spaces is SIFS plus an integer number of slot times.

**SIFS:** The Short Inter-Frame Space (SIFS) is primarily utilized for all immediate response-related activities such as the transmission of RTS, CTS, and ACK frames. The Short IFS is also utilized for the separation of individual data frames in a back-to-back data burst. SIFS interval is a function of several delays such as delay in decoding the PLCP (Physical Layer Convergence Protocol) preamble, the MAC processing delay, the transceiver turnaround time, etc.

**PIFS:** The PIFS interval is used while the WLAN is operating in PCF mode and ensures the next highest access priority. This interval is defined by the following equation:

$$PIFS = SIFS + Slot\ time \qquad (2.1)$$

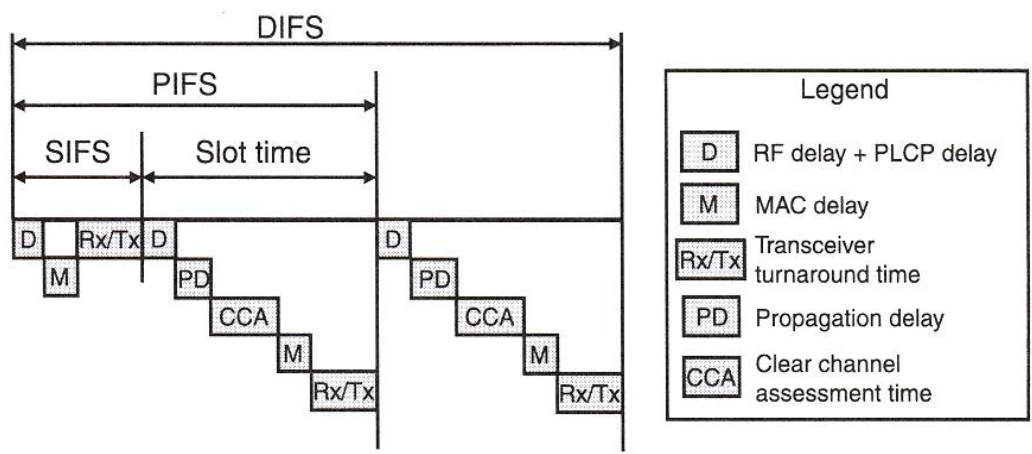

Figure 2.1: The Inter-Frame Space (IFS) definitions

**DIFS:** The DIFS interval is used while the wireless LAN is operating in the DCF mode and ensures a minimum time gap between two data packets. The PIFS interval is defined by the following equation:

$$DIFS = SIFS + 2 \text{ X Slot time} \tag{2.2}$$

**EIFS:** The EIFS is the longest interval used in WLAN to protect from hidden nodes. An STA utilizes EIFS interval instead of DIFS interval for deferral when a frame is detected but not successfully received. The EIFS is defined by the following equation:

$$EIFS = SIFS + DIFS + \text{Time to send ACK frame} \tag{2.3}$$

The IEEE 802.11 standard fixes the duration for the slot time and SIFS for a particular PHY. For example, the duration of slot time is 9 microseconds, and the duration of SIFS is 16 microseconds for 802.11n and 802.11ac (Perahia & Stacey 2013).

## 2.4 MAC Frame Format

The MAC layer receives the MSDUs (MAC Service Data Units) from higher layers in the protocol stack for reliable delivery to the equivalent layer to other STAs. Therefore, the MAC adds information to the MSDU in the form of headers and trailers to form the MAC Protocol Data Unit (MPDU). Then the MPDU is passed to the physical layer (PHY) for sending through the wireless medium to other STAs. To enhance the probability of successful delivery, the MAC may divide the MSDUs into several frames known as fragments.

The MAC Protocol Data Unit (MPDU) is simply known as the MAC frame, which is displayed in Figure 2.2 (a). The general format shown in Figure 2.2 is used for all data and control frames. However, all of the fields may not be required in all situations. The functions of each of the fields are explained below:

**Frame control:** The frame control field is a 2-byte (16 bits) long field that contains valuable information to control the MAC frames (Stallings 2004). There are 11 subfields (Figure 2.2 (b)) of the frame control field that is explained in the following:

Protocol version: This subfield is 2 bits in length that indicates the version of the IEEE 802.11 MAC protocol.

| Field Name | FC | D/I | Address | Address | Address | SC | Address | Frame body | CRC |
|---|---|---|---|---|---|---|---|---|---|
| Bytes | 2 | 2 | 6 | 6 | 6 | 2 | 6 | 0-2312 | 4 |

FC = Frame Control  
D/I = Duration/Connection ID  
SC = Sequence Control  
CRC = Cyclic Redundancy Check  

(a) MAC frame

| FC Subfield | Protocol Version | Type | Subtype | To DS | From DS | MF | RT | PM | MD | WEP | O |
|---|---|---|---|---|---|---|---|---|---|---|---|
| Bits | 2 | 2 | 4 | 1 | 1 | 1 | 1 | 1 | 1 | 1 | 1 |

DS = Distribution System  
MF = More Fragments  
RT = Retry  
PM = Power Management  
MD = More Data  
WEP = Wired Equivalent Privacy  
O = Order  

(b) Frame control field

Figure 2.2: IEEE 802.11 MAC frame format

Type: The MAC frames are mainly divided into three types i.e., data frame, control frame, and management frame. This 2-bit long subfield identifies the type of the current frame.

Subtype: Each of the control, management, and data frames has several subtypes. This 4-bit long subfield is used to choose the specific subtype frames. Appendix A shows all frame type and subtype combinations of the MAC frames.

To DS: If the frame is destined to the DS (Distribution System), then the 'To DS' bit is set to 1 otherwise 0.

From DS: If the MAC frame leaves the DS, then the bit is set to 1 otherwise 0.

More Fragments: The 1-bit subfield sets to 1 if more fragments are remaining to be sent. In the last fragment, the bit is set to 0.

Retry: The retry bit sets to 1 if the current frame is a retransmission of the previous frame. The bit sets to 0 for a frame to be transmitted the first time.

Power Management: This subfield is set to 1 if the transmitting STA is in the power management mode (sleep mode) to announce that it will not be available for further communication.

More Data: The 1-bit subfield is set to 1 to indicate that the sending STA has additional data to be sent over the medium to the destination STA.

WEP: This bit is set to 1 if the WEP (Wired Equivalent Privacy) protocol is used to encrypt the data.

Order: The order subfield is set to 1 to request an ordered service of the data. Thus, the receiving STA has to process the frames in order.

**Duration/ID:** This field is 2 bytes long. The field alternately contains the time duration information or identification number. If it is used as the time duration field, then the value shows the time in microseconds that the channel will be assigned for the transmission. If it is used for identification, then it contains the association or connection ID.

**Address:** 6 bytes are assigned for the address field. There may be up to four addresses, and each of those corresponds to the 48-bit MAC address. The addresses include the source, destination, transmitting STA, and receiving STA.

**Sequence control:** The 2-byte long sequence control field is divided into two subfields: fragment number and sequence number. The fragment number subfield is of 4 bits and is used for fragmentation and reassembly of the MAC frame. The sequence number subfield is of 12 bits and is used to keep the sequence of the MAC frames for a pair of transmitters and receivers.

**Frame body:** The frame body field contains the user's data, which is known as MSDU or MAC Service Data Unit. The length of the frame body varies and may be up to 2312 bytes.

**Cyclic Redundancy Check (CRC):** The 4-byte long CRC field is also known as the Frame Check Sequence (FCS) field. The 32-bit CRC is an error-detecting code used to validate the integrity of the MPDU that comprises the MAC header and the body.

## 2.5 Access Mechanisms

As discussed in Section 2.3, the IEEE 802.11 MAC proposes two different methods to access the wireless media, namely, Distributed Coordination Function (DCF) and Point Coordination Function (PCF). While DCF has been utilized predominantly as the basic access mechanism, PCF could barely attract the developers to be implemented. A variation of DCF is proposed later for the IEEE 802.11e to ensure prioritized QoS, which is known as Enhanced Distributed Channel Access (EDCA). Similarly, a variation of PCF is proposed for the IEEE 802.11e to support parameterized QoS, which is known as HCCA (HCF (i.e., Hybrid Coordination Function) Controlled Channel Access) (Perahia & Stacey 2013).

### 2.5.1 CSMA/CA with Binary Exponential Backoff

The DCF mechanism employs the famous CSMA/CA (Carrier Sense Multiple Access with Collision Avoidance) protocols with the BEB (Binary Exponential Backoff) algorithm. This basic mechanism of wireless LAN is somewhat similar to the IEEE 802.3 wired LAN with a few exceptions. The method used in the wired LAN is known as the Carrier Sense Multiple Access with Collision Detection (CSMA/CD). Both the WLAN and wired LAN employ the same distributed access mechanism known as the Carrier Sense Multiple Access (CSMA), while WLAN employs the collision avoidance technique and wired LAN employs the collision detection technique. This is because the STAs in a wired LAN can detect the collision while the STAs in wireless LAN are not able to detect the collision.

The CSMA/CA mechanism of MAC acts like "listen before talk". Before transmission of any frame, an STA has to sense the medium to know there is an ongoing transmission in the medium. If no, then the listening STA can attempt to send its frame to the destination STA. If yes, then the listening STA must wait for a prescribed period determined by the BEB algorithm. The BEB algorithm selects a number randomly that represents a certain amount of time that the listening STA must wait. The random number generated by the BEB algorithm is uniformly distributed in a range known as the Contention Window (*CW*). Initially, the contention window has the minimum value ($CW_{min}$) and then effectively doubles the value for every unsuccessful transmission. If the contention window reaches the maximum value ($CW_{max}$), it remains at that value until it is reset. The contention window is reset to the minimum value after every successful frame transmission. The details of the CSMA/CA with BEB algorithm is illustrated in the flowchart of Figure 2.3, where $K$ denotes the number of attempts and $T_B$ denotes the backoff time.

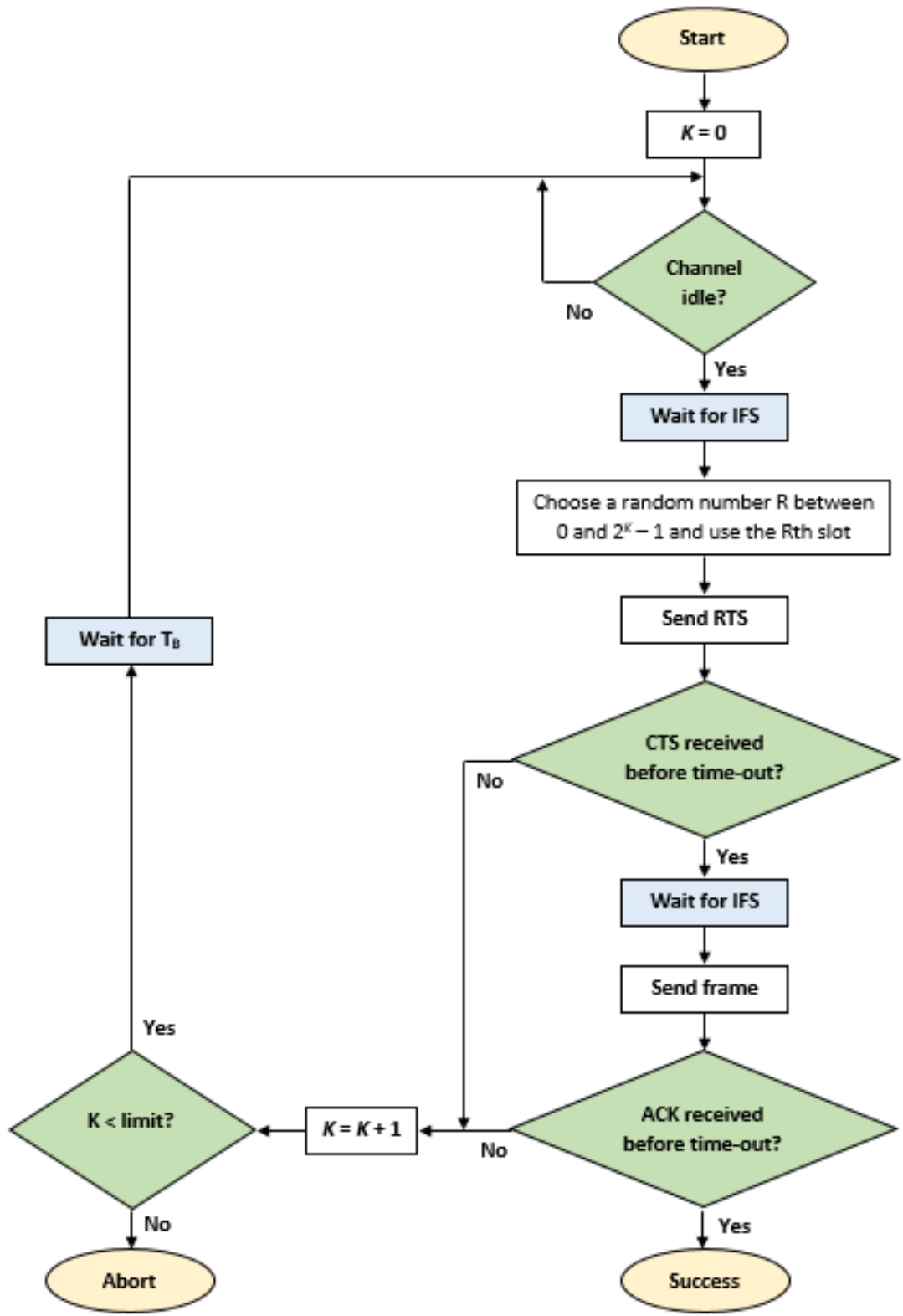

Figure 2.3: Flowchart of the CSMA/CA with BEB algorithm

**2.5.2 Advanced Access Mechanism**

The PCF is an advanced access mechanism which is an optional access mechanism of the IEEE standard. PCF is implemented on top of the DCF in an infrastructure BSS (i.e., not in independent or ad hoc BSS) that may be suitable for time-sensitive data (Forouzan 2013). PCF is a centrally controlled access mechanism that is controlled by the Point Coordinator (PC) that resides in the Access Point (AP). The PIFS (PCF IFS) interval (details in Subsection 2.3.2) is introduced to ensure higher priority for the PCF over the DCF. Since PCF has a higher priority than DCF, STAs using only DCF may not acquire the wireless channel to transmit. As such, a repetition interval consisting of the CFP (Contention Free Period) and CP (Contention Period) has been designed to support all STAs (Figure 2.4). The contention-free transmission during the CFP is based on the polling technique, which is also used in the Ethernet. During the CP, the network is controlled by the DCF mechanism, and during the CFP, the network is controlled by the PCF mechanism, which in turn is controlled by the PC.

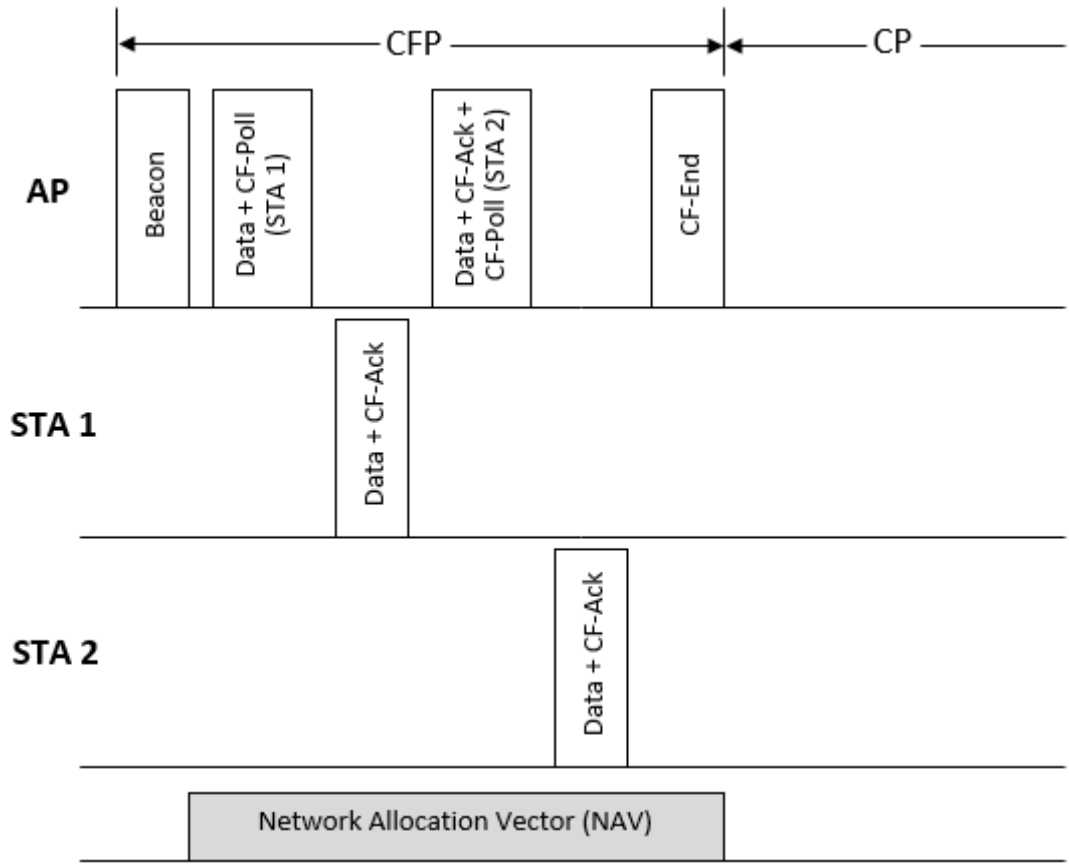

Figure 2.4: Data transfer during the PCF operation

Figure 2.4 illustrates data transmissions between the AP and two STAs using the PCF mechanism. The repetition interval during the PCF operation is initiated by the beacon frame sent by the AP. After receiving the beacon frame, the Network Allocation Vector (NAV) of all STAs in the BSS is updated according to the maximum expected duration of the CFP period (Perahia & Stacey 2013). The end of the CFP is indicated by the CF-End frame sent by the AP. At the end of the CFP, the AP resets the NAV of all STAs for contention-based access. The polling of the STAs is done by the PC or the AP by sending the CF-Poll frame. CF-Ack frame is an acknowledgement frame sent by the STAs or AP for the correct reception of the previous data. The CF-Poll and CF-Ack can be piggybacked on the data frame for increased efficiency.

## 2.6 Existing Promising Protocols

In this paper, three protocols, namely, HTFA, ERA, and PRS, are designed for upcoming wireless LANs. Hundreds of protocols have been studied carefully for the research, and the researcher is able to find their strength and weakness.

### 2.6.1 Related Pieces of Literature of HTFA Protocol

*Random Access Protocols*

Random access protocol allocates the communication channels to the participating stations randomly and need not reserve the resources before communication occurs. Random access protocols are very suitable for data networks such as the Internet. CSMA/CA, ALOHA, etc. are typical examples of the random-access protocol. Lots of works have been done on random access protocols. A few of those are discussed below:

Chi et al. (2016) implement a heuristic technique for network coding to improve the throughput of wireless LANs. Wireless radio inherently suffers from half-duplex limitations i.e., only one-way data transmission at a time (O'Hara & Petrick 2005). To overcome this constraint, Wang and Wang (2010) suggest utilizing two radios at a time where one radio will be dedicated to sensing the communication medium, and the other will be preserved for data transmission only. By the way, this innovative mechanism does not apply to the terminals, which have only one radio. Liao et al. (2015) propose a MAC protocol termed MAC (FD+) for full-duplex (FD) transmission to increase the throughput of Wi-Fi. The performance gain of MAC (FD+) is significantly higher than other contemporary protocols. However, the protocol is not evaluated in the non-saturation condition.

Choi and Park (2006) design an ALOHA-employed random-access protocol whose primary purpose is to mitigate time delay during medium access. Although it decreases time delay, the protocol suffers from high collision during frame transmission. Kwon et al. (2009) propose a multi-channel CSMA/CA protocol using a two-dimensional backoff counter. The performance analysis shows that the protocol performs better than a single-channel system. However, channel utilization of the protocol is not satisfactory as the STAs maintain only one backoff timer for all the sub-channels. To solve the problem, Wang and Wang (2010) uses multiple backoff timers for multiple sub-channels. Stojanova et al. (2021) find that optimal solution has a co-relation with special network configuration. In general, the larger channels are good for throughput maximization and smaller channels are good for fairness. The authors of Qu et al. (2015) propose an innovative protocol termed 'OMAX' that addresses the OFDMA synchronization difficulties and overhead problems. The OMAX utilizes the fast backoff for smooth synchronization and also reduces the protocol overhead using a new frame structure. However, the protocol ignores the OFDMA deterministic approach (i.e., scheduled access) and hence is deprived of utilizing the most robust feature of IEEE 802.11ax.

*Reservation Protocols*

Reservation protocols are appropriate for the connection-oriented application that must meet pre-defined traffic requirements. As such, these protocols reserve resources (e.g., sub-channels in OFDMA) before actual communication occurs. Thus, reservation protocols are more suitable than random access protocols in wireless communications to guarantee QoS. Polling, token bus, etc. are good representatives of the reservation protocol (Molisch 2010). However, reservation protocols are not appropriate for data networks such as the Internet. The Internet has to deal frequently with bursty traffic where data packets come in arbitrary fashion and patterns. Hence, it is not recommended to allocate some bandwidth or other resources priorly for bursty traffic as its real requirements and characteristics could not be predicted.

Yuan et al. (2011) and Lee (2018) devise two protocols for efficient resource utilization. Yuan et al. design the protocol 'CCRM' that sends the Channel Reservation Information (CRI) reliably. The CRI originated from an STA is relayed further by the cooperation of adjacent STAs in the network. Lee (2018) innovates a priority-based protocol that uses the Common Control Channel (CCC). The CCC serves the secondary STAs for sending the control packets to determine the priority for accessing primary channels.

*Hybrid Protocols*

Both the reservation and random-access protocols have their advantages and disadvantages. That is why the OFDMA-adopted WLAN would deploy a hybrid medium access control protocol i.e., a combination of random and reservation protocols. In wireless LAN, some applications are suited to random protocol while others to reservation protocol. In this regard, OFDMA-adopted wireless LANs come forward to boost diverse applications by leveraging the hybrid mechanism. Leveraging OFDMA technology, a large channel can be divided into many small sub-channels. Then an access point could provide some sub-channels random feature where multiple stations in a particular sub-channel randomly contend for channel access i.e., a random feature. On the other hand, an AP could also fix a sub-channel for a pair of sender and receiver, providing no channel access to other stations i.e., a reservation feature. Figure 4.1 in Section 4.2.1 illustrates the random and reservation feature of a MAC protocol in detail.

Since OFDMA, being a newer technology, there are only a few researches have been conducted for wireless LANs which adopt OFDMA technology. Some of the well-known articles are summarized below.

Bellalta (2016) the author reviews the expected future Wireless LAN scenarios and use cases that justify the push for a new PHY/MAC IEEE 802.11 amendment. Barrachina-Muñoz et al. (2019) propose a dynamic channel bonding technique that adapts the channel bandwidth on a per-packet transmission which in turn significantly outperforms traditional single-channel on average. Nguyen et al. (2016) innovate a hybrid protocol by coalescing the TDMA with CSMA-CA for efficient message broadcasting. It reduces the usage of control packets and hence enhances the performance on the control channel. Xuelin et al. (2015) design a protocol that combines the strength of the TDMA with the DCF (Distributed Coordination Function). It decreases protocol overheads and thereby increases the throughput remarkably. In the articles of Haile & Lim (2013) and Ferdous & Murshed (2010), the stations are classified into several groups, and stations in the common group use a common sub-channel for transmission. The RTS/CTS frames are used for assigning channel resources to the members of the group. In Xu et al. (2013), an intermittent carrier sense mechanism is introduced that permits a wireless station to access multiple sub-channels simultaneously.

**2.6.2 Background of ERA and PRS Protocol**

The efficiency of the OFDMA-based MAC protocols depends to a great extent on how the access point allocates the channel resources (i.e., RUs) to the users or stations. These types of problems are generally known as resource scheduling problems. Thus, the study of the resource scheduling algorithm is highly appropriate for further research on the IEEE 802.11ax MAC protocols. Perceiving the importance of resource scheduling, the author designs his second protocol i.e., ERA (Efficient Resource Allocation), and the third protocol i.e., PRS (Proportional Resource Scheduling), leveraging the OFDMA technology.

In this regard, it is worth mentioning that the main challenge is to design the scheduling for the uplink (UL) path rather than the downlink (DL). It is easier to design the scheduling for the downlink since only the access point would send data to the other stations. However, in the UL path, many stations would have to send data to the access point at a time. Therefore, the uplink scheduling in OFDMA is much more challenging than the downlink scheduling since the sending stations must be synchronized as well as the PHY preambles generated by the senders must be the same (Bankov et al. 2018).

The Wi-Fi 6 MAC multi-user OFDMA transmissions in the uplink path use two different mechanisms, namely, i) Random Access (RA) and ii) Scheduled Access (SA) (Bhattarai et al. 2019). The SA method eliminates contention from the STAs and thereby enhances the system throughput. The RA mechanism allows data transmission from the STAs whose BSR information is not available to the AP. It is noted both the ERA and PRS are designed for the uplink path for the IEEE 802.11ax network. The ERA utilizes only the SA mechanism (details in Section 4.3.1) for data transmission while PRS uses both the SA and RA mechanisms (details in Section 4.4.1).

In the past, a large number of scheduling algorithms have been carefully studied for cellular networks, like the LTE (Long Term Evolution), where OFDMA technology has already been adopted (Bankov et al. 2017). Now, the Wi-Fi researchers may hire some promising LTE schedulers and adapt to the wireless LAN. However, this task is a bit complex since OFDMA fundamentals in the wireless LAN are different from those in the cellular network. Moreover, the IEEE 802.11ax standard imposes some restrictions which must comply with to design the scheduler for the Wi-Fi. Thus, it is not guaranteed that an efficient LTE scheduler would remain the best for Wi-Fi as well. Thus, OFDMA adoption in the wireless LAN significantly changes the properties of classic OFDMA-adopted schedulers.

For instance, in the LTE networks, an STA or user can have more than one Resource Unit (RU) in the uplink or downlink transmissions. However, according to the IEEE 802.11ax specifications, an STA could not be assigned to more than one RU. IEEE 802.11ax standard also limits the maximal MCS (Modulation and Coding Scheme) that can be used in narrow resource units and the usage of MU MIMO (Multi-user Multiple Input and Multiple Output). Specifically, Multi-user MIMO is only available in the 106-tone and larger resource units, while the novel 1024 QAM (Quadrature Amplitude Modulation) can be used only in the 242-tone and larger resource units. These limitations further complicate the scheduling mechanism of the Wi-Fi network. In contrast to the LTE, in the Wi-Fi network, the rates in uplink resource units are non-additive, i.e., if an STA sends in twice wider resource unit, it would not guarantee that the STA delivers twice more data. Wilhelmi et al. (2021) outline the challenges and limitations of the IEEE 802.11ax Spatial Reuse (SR) operation. Therefore, to design an efficient scheduler for the Wi-Fi network, the Wi-Fi researchers would need to focus on all of the issues mentioned above.

**2.6.3 Promising Resource Scheduling Protocols**

The IEEE TGax (TASK GROUP AX 2013) keeps the provisions of using the random access along with the scheduled access for the 802.11ax network. In the 802.11ax, the random access mechanism is exclusively known as the UORA (Uplink OFDMA-based Random Access), which is described in Lanante et al. (2017) and Ghosh et al. (2015a). The UORA method inherits the legacy MAC (Medium Access Control) mechanism for the Wi-Fi 6 network. The authors of Ahn et al. (2016) and Avdotin et al. (2019) argue that UORA is not efficient in terms of throughput achievement. However, the UORA method can be efficiently used temporarily by the stations for sending the BSR. Because for using the deterministic method (i.e., SA method), an STA must send its BSR information to the AP. Khorov et al. (2016) adopt the EDCA (Enhanced Distributed Channel Access) method for the 802.11ax network. They use different values for EDCA parameters (e.g., minimum/maximum contention window size) for the fair and efficient distribution of channel resources. However, a rigorous investigation is required to validate the performance and flexibility of using such protocols as there are not sufficient works available yet.

Javed & Prakash (2014) develop a framework named 'CHAMELEON' to coexist the wireless technologies in an unlicensed band. The authors in (Bai et al. 2019) design an adaptive station grouping mechanism to overcome the dense network challenges in Wi-Fi 6 by utilizing a BSR-based Two-stage Mechanism. In (Sangdeh & Zeng 2021), the researchers have devised a new protocol for Wi-Fi 6 terms DeepMux, which is a deep-learning-based MU-MIMO-OFDMA transmission mechanism. In Chen et al. (2021), the authors integrate the MU transmissions with the Target Wake Time (TWT) technique. Simulation results show that their mechanism outperforms the existing techniques in several performance parameters. A.-K. Ajami et al. focus on single unlicensed frequency band transmissions, where the locations of APs, users, and LTE eNodeBs (eNBs) are modeled as three independent homogeneous Poisson point processes (Ajami et al. 2017) Their analysis quantifies both single-user and multi-user operation modes of the Wi-Fi 6.

OFDMA MU transmissions in the Wi-Fi 6 must be synchronized with the TF, and all STAs would get the same amount of time for data transmission. If the packet is too large to send within the allotted time, the packet can be fragmented as described in Ghosh et al. (2015b). Similarly, multiple short packets can be sent within a TF cycle. If there are no packets available for a TF cycle, then padding is the only option to fill up the blank space. To enhance the performance, Wang et al. (2015) permit the aggregation of packets from different access categories. A Multi-TID BA frame can be used to acknowledge a set of frames of different access categories. A Multi-STA BA frame is also be used to replace ACKs or BAs to several stations (Khorov et al. 2018). IEEE 802.11ax latest amendment allows the Block Acknowledgement (BA) (Merlin et al. 2015 and Kim et al. 2015) to acknowledge all STAs using a single frame.

The throughput of the Wi-Fi STAs can be achieved several times by using the multi-user MIMO (Multiple Input and Multiple Output) and spatial stream. It is possible to improve the spatial reuse in congested scenarios by tuning the carrier sensing mechanism using the Dynamic Sensitivity Control (DSC) (Thorpe & Murphy 2014). The authors in Afaqui et al. (2015) set different values for DSC parameters e.g., Received Signal Strength Indicator (RSSI) for the practical implementation of Wi-Fi protocol. Bellalta & Kosek-Szott (2019) illustrate the changes at PHY and MAC layer in the 802.11ax and measure the throughput for MU-MIMO transmissions effectively. The authors in Daldoul et al. (2020) compare uplink and downlink multi-user transmission for Wi-Fi 6 network and show that uplink OFDMA outperforms the downlink transmission when multiple STAs frequently send only a few MPDUs (MAC Protocol Data Units). Chadda et al. (2021) propose a fast, scalable, and fully graph-centric method for choosing a channel width and assignment for the access points for any IEEE 802.11 Wireless LAN. The method performs better than other mechanisms consisting of selecting the channel width regardless of the Wireless LAN topology by 15% in fairness and 20% in throughput.

The researcher of this thesis especially concentrates on some promising protocols for comparing and evaluating the ERA protocol. Those include Wang and Psounis (2018), Bankov et al. (2018), Bankov et al. (2017), etc. Wang and Psounis (2018) propose two algorithms, namely, Greedy and Recursive for scheduling the resource units. Although the throughput of the Greedy is not good enough, it executes faster than the Recursive Scheduling. The Recursive Scheduling splits the bandwidth into the resource units and then distributes those to the stations in a near-optimal fashion and hence provides a promising throughput. The IEEE 802.11ax standard imposes some constraints and restrictions for using OFDMA technology in the Wi-Fi network. Bankov et al. (2018) distinguish the OFDMA features for cellular and Wi-Fi networks. They modify some well-known LTE schedulers such as the PF and MR and adapt those to the Wi-Fi network. The researchers of Bankov et al. (2017) illustrate the OFDMA peculiarities for the Wi-Fi network and claim that classic cellular protocol would not suit the Wi-Fi network. The researchers also innovate a novel scheduler named MUTAX (Minimizing Upload Time in 11AX) that reduces the upload time and increases the goodput.

The author of this thesis has studied several cellular protocols such as the MR (Max Rate) discussed in Bankov et al. (2017), Khorov et al. (2018), Capozzi et al. (2013), etc., and PF (Proportional Fair) in Lee et al. (2009), Kwan et al. (2009), Li et al. (2010), etc. and compare those with the PRS protocol. The main objective of the MR protocol is to achieve the highest throughput for the network. Thus, the MR protocol maximizes the cumulative throughput $S = \sum_i S_i(t)$, where $S_i$ is the throughput of the station *i* at time *t*. Although MR protocol focuses on the throughput, it ignores the fairness of channel access. Hence, researchers come with the PF protocol to overcome the fairness problems. To increase fairness, PF ensures equal channel time to all stations in the steady-state (Kwan et al. 2009). By the way, to increase the fairness index, PF has sacrificed some throughput in contrast to the MR protocol. The PF maximizes the utility function, $U_{PF} = \sum_i \log R_i$, where $R_i$ is the service rate of station *i*. For maximizing the utility function, it is required to maximize $\sum_i \frac{d_i(t)}{R_i(t)}$, where $d_i(t)$ denotes the total data transmitted to station *i* at time *t*. Since the MR and PF are cellular protocols, those are modified and adapted to the 802.11ax architecture while conducting the simulations.

The PRS is also compared with the divide and conquer algorithm designed by Wang and Psounis (2018) that optimally assigns the RUs to the STAs. Unlike the cellular protocols, divide and conquer permits assigning more than one RUs to a single STA. The throughput gained by this algorithm is a tight upper bound of the optimal user schedule. Hence, the divide and conquer algorithm provides significantly higher throughput than other existing Wi-Fi-6 protocols. Wang and Psounis (2020) develop another promising algorithm named 'Recursive', which is also very appropriate for protocol comparison. They propose a recursive scheduling algorithm that splits the bandwidth into resource units and then schedules those to the STAs in a near-optimal fashion. The algorithm is very efficient in handling variable packet size and limited radio capabilities of the access point. The algorithm solves the scheduling and resource allocation (SRA) problem efficiently and dynamically adjusts the number of resource units. Both the 'Divide and Conquer' and 'Recursive' algorithms focus on the SRA problem and ensure good performance especially, in throughput and fairness.

## 2.7 Research Gaps

The Wi-Fi 6 standard focuses on high-speed and uninterrupted communications in the wireless LAN, even in congested environments. Thus, the most important task is to improve the throughput as the Wi-Fi standard promises to ensure four times enhancement in throughput per user. However, no MAC protocols to date meet the expectations (especially throughput) of the standard yet. Traditional protocols discussed in the previous section are far below the latest standard. The main reason is that those protocols have not adopted OFDMA technology and are thereby not able to provide high-speed wireless LAN. Another reason is receiving limited frequency bandwidth for data transfer. As such, future research should comply with the newest standard that permits utilizing larger channel bandwidth (e.g., 160 MHz bandwidth allowed in IEEE 802.11ax) along with the OFDMA technology.

As OFDMA is a newer technology in Wi-Fi, only a handful of works have been done to utilize it. In the previous section, the researcher also investigates some latest IEEE 802.11ax protocols that employ the OFDMA. These protocols are capable of providing more throughput than conventional protocols. However, the total throughput achievable by those protocols is not yet good enough to fulfill the goal of the IEEE 802.11ax standard. Moreover, all protocols have not addressed some other important issues of the wireless LAN precisely, such as reducing collisions and retransmissions, fair access to the medium, etc. Those issues are also fundamental because the performance of the Wi-Fi protocol largely depends on those characteristics.

Perceiving the above research gaps, constraints of existing protocols, and excellent opportunity of OFDMA technology, the researcher is motivated to design several efficient and robust MAC protocols for the OFDMA-based Wi-Fi network. In this paper, three novel Wi-Fi protocols are designed which are capable of enlarging the throughput of the wireless LAN significantly. The novel protocols also reduce frame collisions and retransmissions and increase fair access to the communication medium.

## 2.8 Conclusion

This chapter gives the state of the art of the current research. The chapter covers most of the important topics relating to the MAC protocols, such as the performance matrix of different Wi-Fi, MAC frame format, basic access mechanism, etc. Some of the related issues such as the IEEE 802.11 PHY and network architecture will be discussed in Chapter 3 i.e., Wi-Fi Architecture. Section 2.6 describes many existing protocols along with their advantages and limitations. The background study of the new protocols is also provided in Section 2.6. Based on the analysis of different protocols provided in Section 2.6, the researcher summarizes the research gaps in Section 2.7.

# CHAPTER THREE

# Chapter 3

# Wi-Fi Architecture

## 3.1 Introduction

Different organizations around the world have attempted to develop the standard of wireless LANs. In 1990, the highest body of engineering and technology, known as the Institute of Electrical and Electronics Engineers or the IEEE, formed a group of experts to deliver a standard for the WLANs to be operated at 1 and 2 Mbps. Another committee was formed in 1992 by the European Telecommunications Standards Institute (ETSI). The committee also attempted to develop a standard for the higher performance radio LANs known as the HIPERLAN to be operated in the 20 Mbps range. In recent times, WLAN standards also targeted specialized home applications integrating the concept of the Internet of Things (IoT). In 1997, the IEEE created the first Wi-Fi standard termed IEEE 802.11, which was the basis for the later updated versions. The specification of the IEEE 802.11 standard was remarkably influenced by the WLAN devices available at that time. The IEEE 802.11 standard development committee had taken a relatively long time to complete the standard due to several promising proposals coming from numerous product vendors. Nevertheless, it was the most popular standard at that time. After that, several updated versions were released by the IEEE, which are described in the previous Section 2.2 i.e., Wi-Fi Standards.

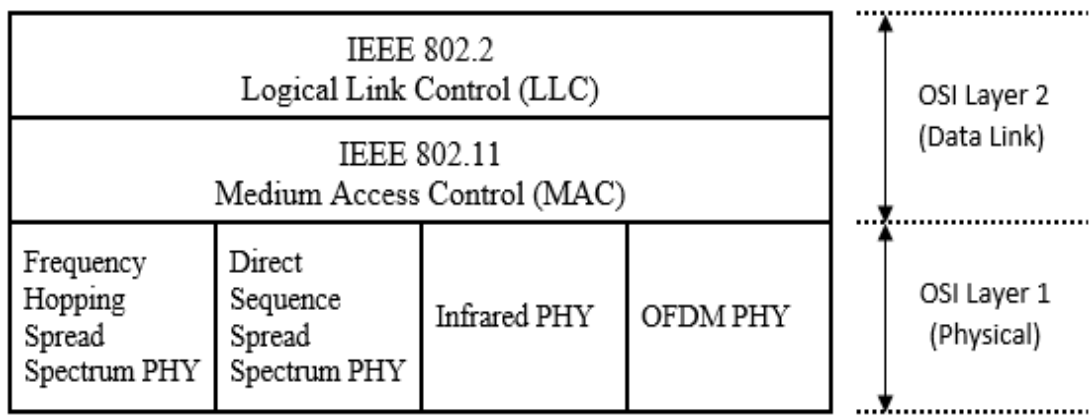

Figure 3.1: IEEE 802.11 standards mapped to the OSI reference model

The IEEE 802.11 standard specifies the rules of wireless connectivity for the wireless STAs in a geographically limited area known as the Wireless Local Area Network or WLAN. As previous IEEE 802.x standards such as the IEEE 802.3 i.e., Ethernet and the 802.5 i.e., Token Ring, the 802.11 WLAN standard specifies both the physical (PHY) layer and medium access control (MAC) layer as shown in Figure 3.1. In addition, the IEEE 802.11 MAC layer also supports some functions that are related to the upper layer protocols. These functions include mobility management, fragmentation, error recovery, power conservation, etc. (Bing 2000). These extended functionalities permit the MAC layer to abstract the unique features of the PHY layer from the upper layers in the wireless LAN. Figure 3.1 relates the IEEE 802.11 MAC and PHY in the OSI (Open Systems Interconnection) reference model that comprises seven layers. The MAC and PHY lie in layer 2 and layer 1, respectively in the OSI model.

The IEEE 802.3 i.e., Ethernet standard was created for the Local Area Network (LAN) that is connected with physical wires. Similarly, the IEEE 802.11 standard is created for wireless LAN or Wi-Fi. That's why the IEEE 802.11 standard is very similar to IEEE 802.3 standard in most respects. Specifically, the IEEE 802.11 standard emphasizes the following issues:

- Compliance of an 802.11 device with other similar devices and integration with wired LAN
- Mobility management of an 802.11 device in multiple wireless LANs that may be overlapped or separate.
- Specification of MAC level access control and data delivery services
- Defining physical layer signaling techniques and interfaces
- Ensure the privacy and security of the users' data in the wireless media.

## 3.2 IEEE 802.11 Network Architecture

The architecture of the IEEE 802.11 wireless LAN is designed in such a way so that most of the decision-making capability is assigned to the Access Point (AP). The architecture is very fault-tolerant for WLAN devices, and it removes any possible bottlenecks. This architecture is very flexible and supports a diverse class of networks e.g., transient, semi-permanent, or permanent. It also incorporates the deep power-saving mode that prolongs the battery life of a station without disconnecting from the Wi-Fi network (O'Hara & Petrick 2005). The Wi-Fi network may have several components such as the Station (STA), the Access Point (AP), the Distribution System (DS), the Basic service set (BSS), and the Extended Service Set (ESS). An overview of different components is given below:

*Station (STA)*

The station (STA) is the most elementary component of the Wi-Fi network. The STA must contain the functions of the IEEE 802.11 protocol and a connection to the wireless medium. Typically, the MAC and PHY functions of the IEEE 802.11 are integrated into the software and hardware of the Network Interface Card (NIC) (Intelligraphics.com 2007).

An STA may be a smartphone, laptop, desktop, or even an AP. An STA may be mobile, portable, or stationary, and it must support the four IEEE 802.11 station services (discussed in Subsection 3.5.1).

*Access Point (AP)*

An Access Point (AP) connects a wireless station to another station or a different network. It is a mandatory component for the Infrastructure Basic Service Set that will be discussed soon. Typically, the functionalities of the access point are implemented in a wireless router.

*Distribution System (DS)*

The Distribution System (DS) works like a backbone through which one AP can communicate with another AP. The distribution system may be a switch, a wired network, or a wireless network. The IEEE 802.11 does not impose any restrictions on implementing the distribution system. The distribution system provides five services which will be described in Subsection 3.5.2.

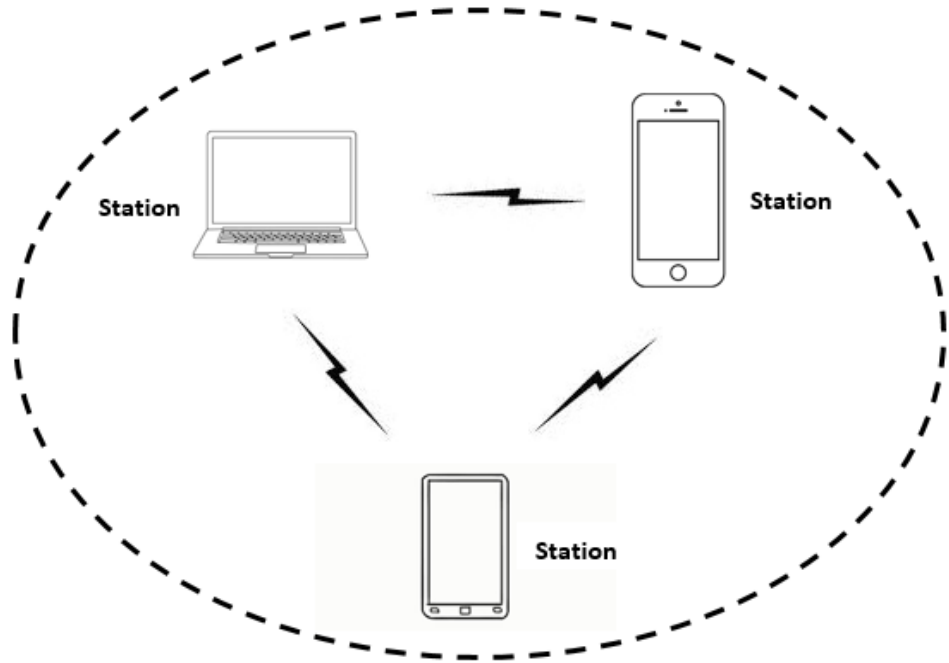

Figure 3.2: Independent basic service set

*Basic Service Set (BSS)*

The IEEE 802.11 Wi-Fi architecture is built around a basic service set (BSS), which is a set of STAs that communicate with each other. Each BSS has an identification known as BSSID that usually corresponds to the MAC address of the NIC (Bing 2000). The small geographical region within which the STAs of a BSS can communicate with one another is known as the basic service area.

### 3.2.1 Independent Basic Service Set

In an Independent Basic Service Set (IBSS), all mobile STAs communicate directly with one another. The Independent Basic Service Set (IBSS) is also known as the *Ad-hoc* network since the STAs in this topology communicate with each other in a peer-to-peer fashion. Figure 3.2 shows an IBSS where three stations are communicating directly with one another without the help of an AP.

In an IBSS, every STA might not be able to communicate with all other STAs due to the range limitations. No relay functions or AP exist for data forwarding in the IBSS. Therefore, all STAs must be within the distance threshold to communicate.

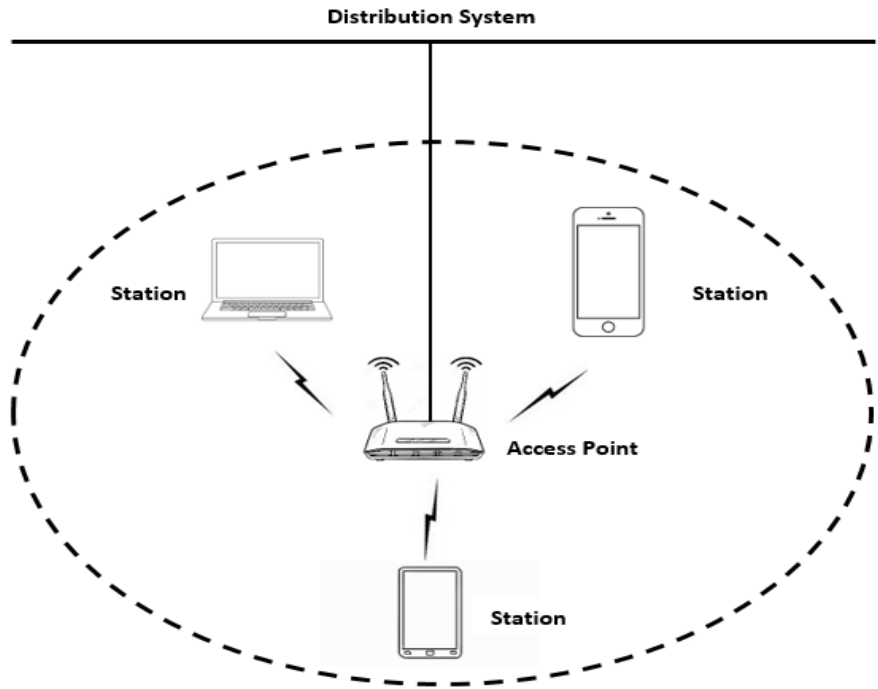

Figure 3.3: An infrastructure basic service set

### 3.2.2 Infrastructure Basic Service Set

In the Infrastructure Basic Service Set, the STAs communicate with each other through the AP, and they no longer communicate directly with one other. Thus, if one wireless STA in the BSS wants to communicate with another wireless STA, the communication is sent first to the access point and then from the access point to the other wireless STA. The access point provides the connection to the distribution system or the wired LAN (if any) and also does the local relay function for the BSS. Figure 3.3 shows an Infrastructure BSS where three STAs are communicating with each other through the AP.

### 3.2.3 Extended Service Set

Several infrastructure BSS may group to form an Extended Service Set (ESS) as shown in Figure 3.4. The 802.11 standard extends the range of mobility of the wireless LAN utilizing the concept of the ESS (O'Hara & Petrick 2005). Every infrastructure BSS has an AP, and the APs of different infrastructure BSS can communicate with each other through the DS.

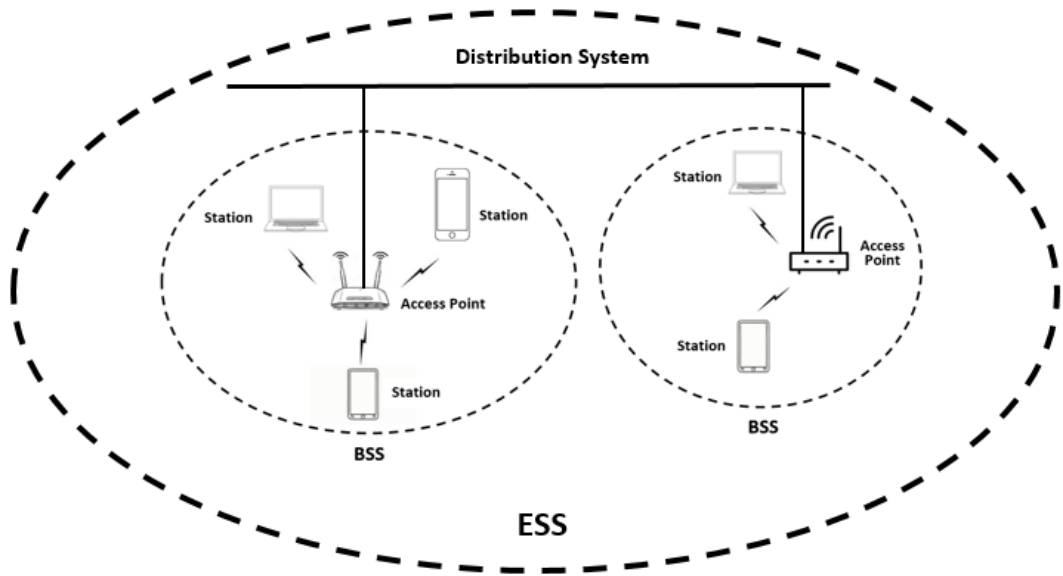

Figure 3.4: An extended service set

The distribution system may be considered as a special purpose box which implementation technique is out of concern of the IEEE 802.11. The DS acts as the backbone for communications in the Wi-Fi network, which may be built by a wired network or a wireless network. When the DS receives a frame from an AP, then the DS decides whether the frame be relayed to the same BSS or sent to another AP or even sent to a wired network. These mechanisms are implemented in the MAC frame format and discussed in Section 2.3.3.

## 3.3 IEEE 802.11 PHY

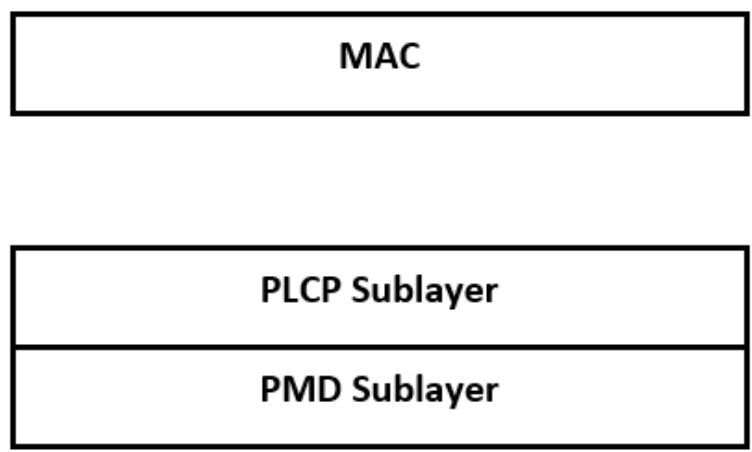

Figure 3.5: IEEE 802.11 PHY sublayers along with the MAC

Table 3.1: IEEE 802.11 PHY specifications

| Standard | Modulation Technique | Band | Data Rate (Mbps) |
|---|---|---|---|
| 802.11 | FHSS, DSSS | 2.4 GHz | 1 and 2 |
| 802.11a | OFDM | 5 GHz | 6 to 54 |
| 802.11b | DSSS, HR-DSSS | 2.4 GHz | up to 11 |
| 802.11g | OFDM | 2.4 GHz | up to 54 |
| 802.11n | OFDM | 2.4 GHz, 5 GHz | 600 |
| 802.11ac | OFDM | 5 GHz | 1.3 Gbps |
| 802.11ax | OFDM | 2.4 GHz, 5 GHz | 10-12 Gbps |

The physical layer of the IEEE 802.11 has two sublayers which are known as the PLCP (Physical Layer Convergence Protocol) and the PMD (Physical Medium Dependent). Figure 3.5 shows the PHY sublayers along with the MAC. The interaction between the MAC and the PHY is done by the PLCP sublayer. The PLCP maps the MAC Protocol Data Unit (MPDU) into the PLCP Protocol Data Unit (PPDU). This helps the PMD sublayer to send and receive data over the wireless media efficiently. The PPDU contains the PLCP preamble, signal part, and the data (Wireless Communication 2020). The PLCP preamble helps in synchronizing the incoming transmissions (Gast 2005). The requirements of the preamble may depend on the modulation techniques. The IEEE 802.11 PHY uses different modulation techniques such as Frequency Hopping Spread Spectrum (FHSS), Direct Sequence Spread Spectrum (DSSS), High-Rate DSSS (HR-DSSS), and Orthogonal Frequency Division Multiplexing (OFDM). Table 3.1 summarizes IEEE 802.11 PHY specifications.

## 3.4 IEEE 802.11ax and OFDMA Technology

The IEEE 802.11 committee proclaims that the new standard (i.e., IEEE 802.11ax) would achieve at least four times data per station and ensure uninterrupted communication in the dense areas as well. The physical data rate of the Wi-Fi network has been significantly boosted by utilizing increased bandwidths and leveraging sophisticated technologies such as the MIMO (Qu et al. 2015). However, the MAC layer of Wi-Fi has merely changed for the last 16 years. Since its birth, the IEEE 802.11 employed DCF (Distributed Coordination Function) mechanism for the MAC (Akyildiz & Wang 2009). According to the DCF mechanism, only one STA or user can access the wireless channel at a time (Standards IEEE 2015). The DCF rules adopted in the IEEE 802.11 is appropriate to sparsely dense system, while in the highly congested locations the MAC proficiency of the DCF will be extremely poor due to the provision of single-user accessibility (Standards IEEE 2013). To overcome the problems, a novel multi-user MAC mechanism is required instead of the legacy single-user MAC (Standards IEEE 2014).

The latest Wi-Fi 6 borrows several features from the cellular technologies especially, from the LTE to serve many users simultaneously in the same channel by employing the OFDMA technology (Islam & Kashem 2019). Apart from the adoption of the OFDM modulation, the Wi-Fi 6 allocates a group of non-overlapping subcarriers to the stations efficiently and ensures multi-user communications in the same channel. The Wi-Fi channel (i.e., total available bandwidth) is partitioned into several smaller sub-channels with a specified number of orthogonal subcarriers (National Instruments 2017). The Wi-Fi 6 terms the smallest sub-channel as the Resource Unit (RU), which contains at least 26 subcarriers (i.e., tones).

The access point observes the traffic requirement of the users then allocates the channels to stations. It may allocate the whole channel to a single station, or it may split the channel into several sub-channels to support many stations concurrently (Figure 3.6). In the dense areas where a lot of stations would normally compete for channel access inefficiently, the OFDMA technology can now serve them simultaneously with a smaller but dedicated sub-channel. Therefore, the average throughput per station improves significantly.

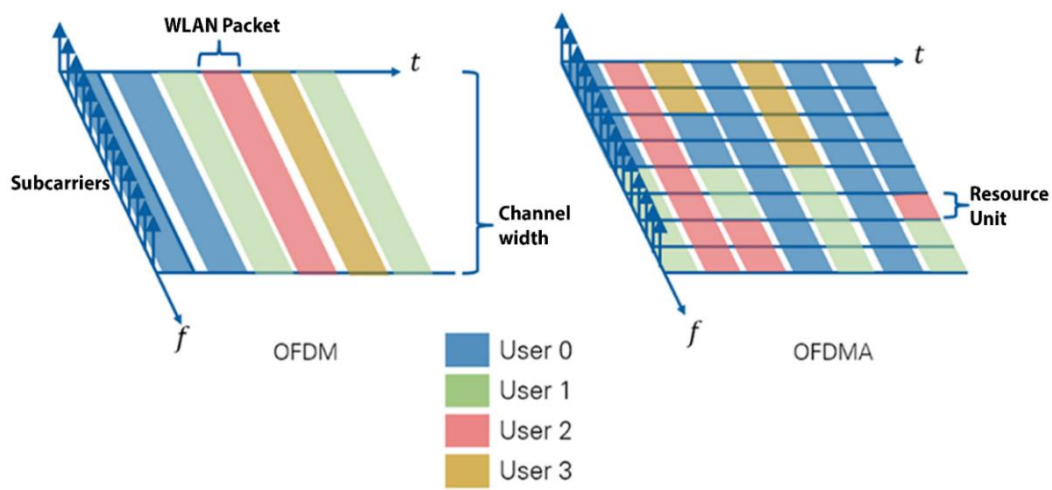

Figure 3.6: Contrast between OFDM and OFDMA

## 3.5 OFDMA Specifications and Constraints

The IEEE 802.11ax can operate both in 2.4 GHz and 5 GHz frequency bands and supports different channel bandwidths such as 20 MHz, 40 MHz, 80 MHz, and 160 MHz (Standards IEEE 2019). The whole bandwidth of a channel is divided into several sets of orthogonal sub-carriers, which are called the Resource Units i.e., RUs. Then the RUs are distributed to stations whose BSR information is available to the AP. Thus, utilizing different resource units, different STAs can communicate with the access point concurrently without suffering from co-channel interference and contentions.

Although Wi-Fi 6 adopts the OFDMA technology from the cellular network, the rules of usage of OFDMA technology in the Wi-Fi network are somewhat different. Section 3.5.1 provides the specification of OFDMA resource units, and Section 3.5.2 discusses the constraints and regulations for the utilization of OFDMA in the Wi-Fi network. The protocols designed in this paper follow the OFDMA specifications and constraints which are illustrated in this section.

### 3.5.1 Specification of Resource Units

Like the LTE network, the whole bandwidth of a Wi-Fi channel can be partitioned into several sets of orthogonal sub-carriers, which are called resource units. According to the OFDMA specification for IEEE 802.1ax, a resource unit must contain at least 26 tones i.e., sub-carriers, which the researcher terms Smallest Resource Unit (SRU). Thus, a larger channel (e.g., 80 MHz channel) may have more RUs than a smaller channel (e.g., 20 MHz). Based on the size, an RU may contain 26, 52, 106, 242, 484, 996, or $2 \times 996$ sub-carriers. Figure 3.7 portrays the resource unit formation in a 40 MHz channel. For demonstration purposes, the SRUs are numbered on top of the figure from left to right. There are 18 SRUs in a 40 MHz channel, where each of the RUs contains 26 tones. Similarly, the 20 MHz, 80 MHz, and 160 MHz channels can be split into at most 9, 37, and 74 SRUs, respectively.

Figure 3.7: OFDMA resource unit formation in a 40 MHz channel

Figure 3.7 portrays how a 40 MHz Wi-Fi channel is partitioned in terms of RUs (Islam & Kashem 2021). Eliminating the 1st level (i.e., bottom level in Figure 3.7) and then vertically splitting the channel into two halves then each half would be symmetrical to each other. After splitting, each half would represent the resource unit formation in a 20 MHz Wi-Fi channel. By the way, it is also possible to create larger RUs by merging in place of the division rule. For instance, merging two 40 MHz channels vertically and adding a 996-tone RU as the bottom level would show the arrangement of resource units in an 80 MHz Wi-Fi channel. The same method is also applicable to create a 160 MHz OFDMA channel from two 80 MHz channels.

### 3.5.2 Constraints and Regulations

In IEEE 802.11ax, adjacent smaller RUs can combine to form a larger RU, or a larger RU can be split into several smaller RUs. However, the merging and division of RUs are not arbitrarily, as portrayed in Figure 3.7. For example, SRU 15 and SRU 16 can be merged together to form a larger RU of 52 tones (lighter-gray colored). However, the combinations such as (SRU 14 and SRU 15) or (SRU 16 and SRU 17) are not valid for merging (underlined in the figure). Again, a larger RU having 242 tones may split into three smaller RUs as two 106-tone RUs and one 26-tone RU (lighter-gold colored).

The access point distributes the RUs to the STAs according to their traffic demands and availability of RUs. The rules for the distributions of RUs to the STA in the Wi-Fi are not the same as the LTE network. For instance, in the Wi-Fi 6 network, an RU cannot be assigned to more than one STA. Again, RUs which have more than 106 tones can only be assigned for the MU-MIMO transmissions, and RUs that have more than 242 tones can only be used for the novel 1024 QAM (Quadrature Amplitude Modulation) (Standards IEEE 2019 and Naik et al. 2018).

All resource units, irrespective of the size, get the same amount of time for data transmission (Islam & Kashem 2020). As such, STAs having larger RU (e.g., 242 tones) can send more data than those having smaller RU (e.g., 52 tones). For example, an STA having an RU of 106 tones can send approximately 4X data of an STA having an RU of 26 tones. In OFDMA, multi-user transmissions can be achieved by splitting the whole channel into several smaller RUs. All of the STAs in the network must be assigned the RUs for transmission then STAs start transmission simultaneously using their RUs. The Trigger Frame (TF) sent by the AP helps the STAs to be synchronized with the system.

## 3.6 IEEE 802.11 Services

IEEE 802.11 WLAN specifies nine distinct services to be provided by the stations (STAs) and Distribution System (DS) together (Stallings 2004). Among the nine services, four are provided by the STA, and five are provided by the DS. The services provided by the STAs are (i) authentication (ii) de-authentication (iii) privacy and (iv) data delivery. The services provided by the DS are (i) association, (ii) disassociation (iii) reassociation (iv) distribution, and (v) integration. Table 3.2 summarizes the nine services.

### 3.6.1 Station Services

The IEEE 802.11 WLAN provides four station services as stated above. These services are very similar in most respect to the wired network, like IEEE 802.3 Ethernet. Each of the station services is described below:

Table 3.2: IEEE 802.11 services

| Service Name | Service Provider | Purpose |
| --- | --- | --- |
| Authentication | STA | LAN access and security |
| De-authentication | STA | LAN access and security |
| Privacy | STA | LAN access and security |
| Data delivery | STA | MSDU delivery |
| Association | DS | MSDU delivery |
| Disassociation | DS | MSDU delivery |
| Reassociation | DS | MSDU delivery |
| Distribution | DS | MSDU delivery |
| Integration | DS | MSDU delivery |

*Authentication*

The 802.11 defines the authentication services to identify valid stations to grant access to the network. Due to limited physical security in the wireless LAN, 802.11 put effort into ensuring access control as much as the Wired LAN.

The authentication service provides the mechanisms for an STA to identify another STA. No STA is allowed to use the wireless LAN without proving their identity in the prescribed method. Thus, each and every STA in the WLAN must perform an authentication procedure for data delivery in the network.

The IEEE 802.11 specifies two different authentication services, namely, a) Open system authentication b) Shared key authentication. These are described below:

*a) Open system authentication:* This is the default authentication method. This method uses the two-step procedure and is very simple to implement. At first, the STA that wants to authenticate transmits a special management frame known as the authentication management frame to the receiving STA. The receiving STA examines the frame which contains the identity of the sender. Then, the receiver sends back a frame to declare whether it recognizes or not.

*b) Shared key authentication:* This method is widely used in today's Wi-Fi network. The secret information is shared by the STAs in the network through a secure encrypted channel. An STA can recognize another STA by matching the secret information shared before in the network. The encryption of the channel can be done using the well-known WEP algorithm (Stallings 2004).

*De-authentication*

De-authentication service is used to remove a previously authorized STA from the Wi-Fi network. Once an STA is de-authenticated, then the STA cannot access the network anymore. When an STA has used the network and completed its task, then it can notify the associated AP to remove it from the network. An AP can also forcibly de-authenticate an STA.

*Privacy*

The protection of user data is of utmost importance and challenging in wireless communication. In the wired network, the data can be protected from intruders by keeping the physical medium in a secure place. However, in wireless communication, it is very challenging as there is no existence of a physical medium. The purpose of the IEEE 802.11 privacy service is to protect user data at a level equivalent to a wired network. The IEEE 802.11 privacy service concerns the protection of the user data while traversing in the wireless medium only. In this regard, the WEP algorithm is the first choice to encrypt user data.

*Data Delivery*

Unlike the wired network, the wireless network is subject to considerable unreliability for successful data delivery. The purpose of the data delivery service provided by IEEE 802.11 is the same as IEEE 802.3 Ethernet. However, ensuring successful data delivery in the IEEE 802.11 network is more challenging than the IEEE 802.3 network due to the interference, noise, and other propagational effects of wireless communications. The IEEE 802.11 ensures reliable delivery of the MAC frames from one STA to other STAs with minimal duplication and reordering.

### 3.6.2 Distribution Services

The distribution services are provided by the AP across the distribution system (DS). A brief explanation of the five distribution services is given below:

*Association*

The IEEE 802.11 association service ensures the logical connection between the STA and the AP. Each wireless STA must be associated with the nearest AP to transmit data through the AP onto the DS. This connection is necessary for the AP to know where and how to deliver the data to the STA through the DS. Getting the knowledge of association, the AP can allocate sufficient resources for the STA. Typically, the STA invokes this service only once, when it enters into Basic Service Set (BSS) the first time.

*Disassociation*

The disassociation service can be used by the STA or the AP to terminate the logical connection between them. If an STA does not require the service of an AP anymore, then it can invoke the disassociation service. Similarly, An AP can also forcibly disassociate an STA or several STAs for a variety of reasons such as resource constraints, shutting down, etc. Neither the AP nor the STA can refuse the termination of the association. When an STA is disassociated from the AP, then the STS has to initiate another new association service to communicate further with the AP.

*Reassociation*

The reassociation service facilitates the STAs moving from one BSS to another BSS through the ESS. The STAs use the reassociation service repeatedly while those move throughout the extended service set, lose connections from the previous AP, and establish a connection with the new AP. This service helps the new AP to get the information of the previous associated AP of the STA.

*Distribution*

The access point utilizes the distribution service to determine where and how to deliver the MAC frames it receives from an STA. The access point uses the IEEE 802.11 distribution service to determine whether the received frame should be sent back to the current BSS or forwarded to the DS depending on the location of the destination STA. In this regard, three association-related services (i.e., association, disassociation, and re-association) provide the required information to operate.

*Integration*

The integration service provides a connection to the IEEE 802.11 WLAN with other wired or wireless LANs. The integration service translates the IEEE 802.11 MAC frames to other frames that the data may traverse in a different network and vice-versa (O'Hara & Petrick 2005). An abstract architectural concept known as the portal performs the integration service. The functionalities of the portal i.e., the integration service may be implemented by the AP or other similar components.

## 3.7 Conclusion

This chapter complements the previous chapter i.e., Literature Review. This chapter focuses on the IEEE 802.11 PHY, architecture, and services. Section 3.1 illustrates how IEEE 802.11 maps to the OSI reference model. This section also summarizes the functions of the IEEE 802.11 standard. Section 3.2 describes the network architecture of the IEEE 802.11, where several components such as the station, access point, distribution system along with the Basic Service Set (BSS) and Extended Service Set (ESS) are discussed elaborately.

An overview of IEEE 802.11 PHY is provided in Section 3.3. Section 3.4 relates the latest IEEE 802.11ax standard with OFDMA Technology. Details of the OFDMA specifications and constraints are discussed in Section 3.5. The IEEE 802.11 services are of two types, namely, station service and distribution service. The station services are authentication, de-authentication, privacy, and data delivery which are discussed in Subsection 3.6.1. The distribution services include association, reassociation, disassociation, integration, and distribution, which are discussed in Subsection 3.6.2.

# CHAPTER FOUR

# Chapter 4

# Theoretical / Conceptual Framework

## 4.1 Introduction

In this paper, the researcher designs three protocols, namely, HTFA (High Throughput and Fair Access), ERA (Efficient Resource Allocation), and PRS (Proportional Resource Scheduling) to enhance the performance of the Wi-Fi network. This chapter illustrates the frameworks and working principles of the protocols. The details of the HTFA, ERA, and PRS protocols are provided in Section 4.2, Section 4.3, and Section 4.4, respectively. At last, Section 4.5 concludes the chapter.

## 4.2 Framework for HTFA Protocol

The HTFA protocol adopts a two-dimensional time-frequency access model leveraging the OFDMA technology, which is described in Subsection 4.2.1. Subsection 4.2.2 describes the working principles of the protocol with the help of several comprehensive diagrams. Subsection 4.2.3 delineates the access mechanism of the HTFA protocol.

### 4.2.1 Two-Dimensional Time-Frequency Model

The HTFA protocol adopts a two-dimensional time-frequency access model. In a traditional single-channel WLAN, only one STA can acquire the channel at a time, and the remaining STAs must wait to transmit their data until the wireless medium is free (Bianchi 2000). By the grace of OFDMA technology, a large channel can be divided into many smaller sub-channels. Now, more than one STA can transmit data simultaneously located in different sub-channels. Thus, in an OFDMA multi-channel wireless LAN, stations could contend for resources in the time and frequency domain concurrently, as shown in Figure 4.1 (Islam & Kashem 2018). In the time domain, a station can gain the time slices of a sub-channel, and in the frequency domain, a station has the opportunity to choose one or more sub-channels (Tinnirello et al. 2010 and Rahman et al. 2010).

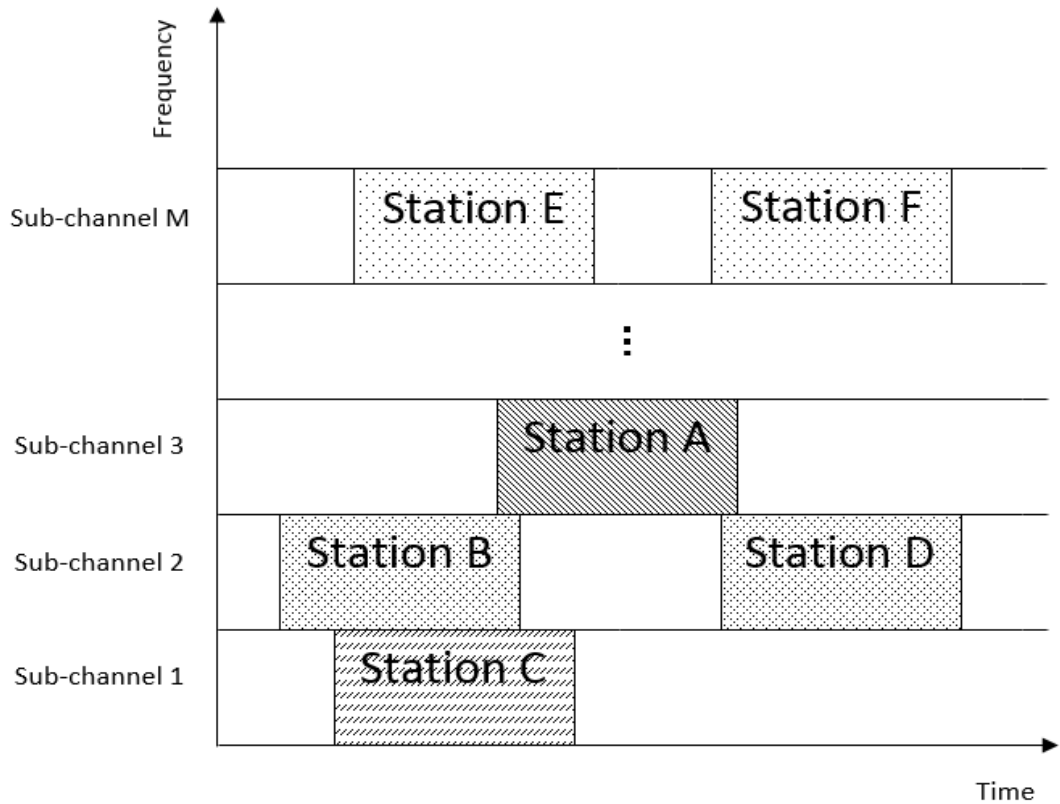

Figure 4.1: A 2-dimensional time-frequency access model

In Figure 4.1, a large OFDMA channel is partitioned into *M* smaller equal sub-channels in the frequency domain. In sub-channel 1 and 3, there is only one station. Thus, sub-channel 1 and 3 may be dedicated to their respective stations like a reservation protocol. In sub-channel 2 and *M*, there are two stations. Now, the stations have to contend to gain access to the channel in the time domain due to the existence of more than one station in the sub-channel. Thus, sub-channel 2 and *M* exhibit the features of a random-access protocol. In essence, Figure 4.1 represents a hybrid protocol since some of the sub-channels act like a reservation protocol, and some others act like random access protocol.

**4.2.2 Working Principles and Sub-channel Distribution**

The main distinguishing feature of the HTFA protocol is its unique algorithm for distributing the STAs (i.e., nodes/terminals/users) to the sub-channels in the Wi-Fi network. The author named the protocol HTFA (High Throughput and Fair Access) as the protocol promises to deliver high throughput of data along with fair access to the medium.

The protocol employs a 2-step algorithm i.e., Algorithm 1. Initially, step 1 will be applied to distributing the stations. After applying step 1, step 2 will be applicable. Step 2 describes how the stations in the sub-channels access the medium.

**Algorithm 1: STA Distribution and Channel Access**

*Step 1:* The stations are approximately evenly distributed to the sub-channels in the network.

*Step 2:* If a sub-channel contains only one station, then the sub-channel be dedicated to that station. But if a sub-channel contains more than one station, then the stations in that sub-channel contend randomly to access the medium.

In step 1, it is said "approximately evenly distributed"; since the STAs may not always exactly equally distributed to the sub-channels. However, the number of STAs distributed to each of the sub-channels would be differed by at most one. For example, there are 7 STAs and 2 sub-channels in the network. Definitely, one of the sub-channels gets 4 STAs, and another sub-channel gets 3 STAs.

The creation of the sub-channels as well as the distribution of the STAs to the sub-channels is done by the AP. There are five different scenarios or cases are considered for distributing the STAs among the sub-channels. The cases are:

Case 1: Number of the sub-channels is equal to the number of STAs ($M = N$)

Case 2: Number of the sub-channels is larger than the number of STAs ($M > N$)

Case 3: Some STAs leave the Wi-Fi network after a while

Case 4: Number of the STAs is larger than the number of sub-channels ($N > M$)

Case 5: Some STAs join the Wi-Fi network after a while

A brief description of each of the cases is given in the following:

Case 1: Number of the sub-channels is equal to the number of STAs ($M = N$): Suppose, initially there are three STAs and three sub-channels in the network. The three STAs are STA A, STA B, and STA C; and the three sub-channels are Sub-channel 1, Sub-channel 2, and Sub-channel 3. Since in case 1, $M = N$ and the STAs are evenly distributed to the sub-channels (i.e., Step 1); each STA will acquire exactly one sub-channel (Figure 4.2 (a)). According to step 1, it is not possible that one STA acquires 2 sub-channels, and the remaining two STAs acquire only 1 sub-channel. Thus, HTFA can ensure fair access by the STAs to the communication medium that is absent in the articles (Wang & Wang 2010) and (Xu et al. 2013). The protocols of the mentioned articles will be compared with the HTFA, which will be discussed in Subsection 6.3.1.

| Sub-channel 1 |
| :---: |
| Station B |
| Sub-channel 2 |
| Station A |
| Sub-channel 3 |
| Station C |
| (a) |

| Sub-channel 1 |
| :---: |
| Station B |
| Sub-channel 2 |
| Station A |
| Sub-channel 3 |
| Station B |
| (b) |

| Sub-channel 1 | |
| :---: | :---: |
| Station B | Station C |
| Sub-channel 2 | |
| Station A | |
| Sub-channel 3 | |
| Station C | |
| (c) | |

Figure 4.2: Sub-channel distribution (a) Case 1 (b) Case 2 (c) Case 3

Case 2: Number of the sub-channels is larger than the number of STAs ($M > N$): Now suppose; initially, there are two STAs and three sub-channels in the network. The two STAs are STA A and STA B; and the three sub-channels are Sub-channel 1, Sub-channel 2, and Sub-channel 3. Since in case 2, $M > N$ and the sub-channels are approximately evenly distributed, one STA receives 2 sub-channels, and another STA receives 1 sub-channel. For example, STA A acquires 1 sub-channel (sub-channel 2), and STA B acquires 2 sub-channels (Sub-channel 1 and Sub-channel 3), as shown in Figure 4.2 (b). Alternatively, it is also possible that STA A acquires 2 sub-channels and STA B acquires 1 sub-channel.

Case 3: Some STAs leave the Wi-Fi network after a while: Suppose, initially Sub-channel 1, Sub-channel 2, and Sub-channel 3 are acquired by STA B, STA A, and STA C, respectively. After elapsing some time, STA B may need not the network anymore and releases its sub-channel, i.e., Sub-channel 1. So, one of the two remaining STAs in the network (C or A) can gain B's sub-channel. For instance, in Figure 4.2 (c), STA C is acquiring the sub-channel (Sub-channel 1) of STA B.

Case 4: Number of the STAs is larger than the number of sub-channels ($N > M$): Now, suppose there are four STAs and three sub-channels in the Wi-Fi network. The four STAs are STA A, STA B, STA C, and STA D; and three sub-channels are Sub-channel 1, Sub-channel 2, and Sub-channel 3 as before. Since, in this case, the number of STAs is larger than the number of sub-channels, 2 sub-channels will be dedicated to two STAs, and 1 sub-channel will be shared by the remaining two STAs. For example, STA B and STA C get Sub-channel 1 and Sub-channel 3, respectively. So, the other two stations STA A and STA D have to share the remaining sub-channel (Sub-channel 2), as shown in Figure 4.3. Now that Sub-channel 2 is shared by multiple stations (STA A and STA D), the stations have to compete for channel access randomly in accordance with the DCF (Distributed Coordination Function) rules.

Figure 4.3 illustrates the random-access mechanism, where STA A and STA D are contending for channel-access in Sub-channel 2. Initially, STA D and STA A produce the random backoff values 7 and 5, respectively. As STA D produces a larger number than STA A, STA A acquires the sub-channel first for data transmission when its backoff number reaches 0. After elapsing one more slot, STA D's backoff number also reaches 0, and thereby STA D gains the sub-channel. In the second round, STA D and STA A produce new backoff numbers 2 and 5, respectively, and the same mechanism continues according to the legacy DCF rules.

Case 5: Some STAs join the Wi-Fi network after a while: How do the future STAs be included in the network depends on the current configuration of the network. The current configuration should be one of the following:

(i) $M > N$ (Number of the sub-channels is larger than the number of STAs)
(ii) $N \geq M$ (Number of the STAs is larger or equal to the number of sub-channels)

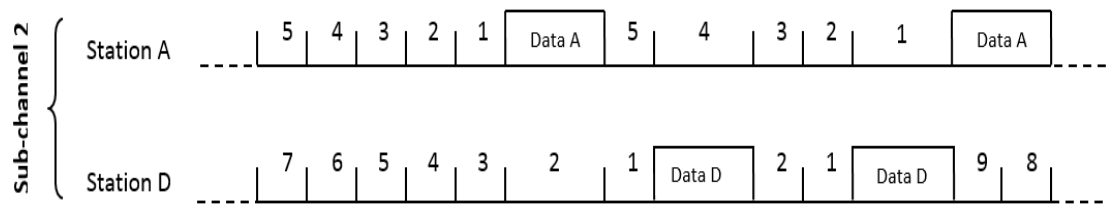

Figure 4.3: Contention of the STA A and STA D in Sub-channel 2

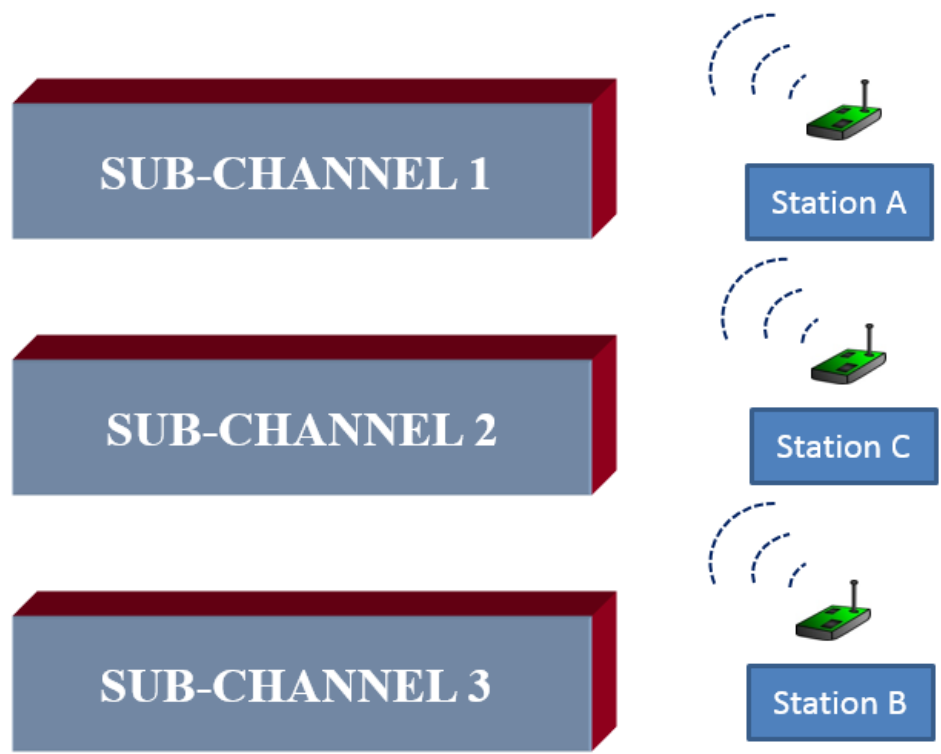

Figure 4.4: Network configuration for 3 sub-channels and 3 stations

If $M > N$, then the new STA acquires a sub-channel from the old STA, which has the maximum number of sub-channels. The new STA may need to wait for some time so that the old STA can send its ongoing packet. If $N \geq M$, then the new STA joins into the sub-channel, which has the minimum number of STAs.

The sub-channel distribution procedure described here is very intuitive. For further clarification of the procedure, let's see an exemplary scenario. Suppose at the beginning an AP is having no STAs and three sub-channels, namely, SUB-CHANNEL 1, SUB-CHANNEL 2, and SUB-CHANNEL 3 in the Wi-Fi network. After a while, STA A enters into the network, and it acquires all three sub-channels. Then STA B joins the network, and it seizes one of the sub-channels (e.g., SUB-CHANNEL 3) from STA A. At this stage, STA A has two sub-channels i.e., SUB-CHANNEL 1 and SUB-CHANNEL 2; and STA B has one sub-channel i.e., SUB-CHANNEL 3. Again, after some time, STA C joins into the network, and it seizes one of the sub-channels (e.g., SUB-CHANNEL 2) from STA A. Now, each of the three STAs possesses exactly one sub-channel, as shown in Figure 4.4.

Suppose, after some time, another station, i.e., STA D, also wants to be a member of the network. Now one of the three sub-channels must be shared with STA D. As such, the AP keeps STA D with STA A into the SUB-CHANNEL 1. So, STA D and STA A have to compete for acquiring SUB-CHANNEL 1 by generating random backoff numbers. Similarly, STA E arrives, and the AP assigns it to the sub-channel, which has the minimum number of STAs. Now SUB-CHANNEL 2 and SUB-CHANNEL 3 have the minimum number of STAs (i.e., 1 STA each) and suppose STA E gets SUB-CHANNEL 2. Until now, the network configurations are as shown in Figure 4.5, where STA A and STA D share SUB-CHANNEL 1; STA C and STA E share SUB-CHANNEL 2; and STA B utilizes SUB-CHANNEL 3 alone.

Now suppose STA E leaves the network. The leaving of STA E does not affect the configuration of the network as the sub-channels are approximately evenly distributed to the STAs. Again, after some time, STA C also leaves the network. Now the number of STAs is not approximately evenly distributed since SUB-CHANNEL 1 has two STAs (Station A and Station D), and SUB-CHANNEL 2 has no STA at all. Therefore, one STA of SUB-CHANNEL 1 has to migrate to SUB-CHANNEL 2. In the end, each sub-channel acquires exactly one STA, and sub-channels are evenly distributed to the STAs.

### 4.2.3 Access Mechanism

Every STA in the network must be associated with the AP for data transmission. The AP assigns sub-channels to the STAs according to the working principles discussed in Subsection 4.2.2. The access mechanism for a particular sub-channel depends on the number of available STAs in that sub-channel. There would be two cases:

i.        Multiple STAs in the sub-channel
ii.       Single STA in the sub-channel

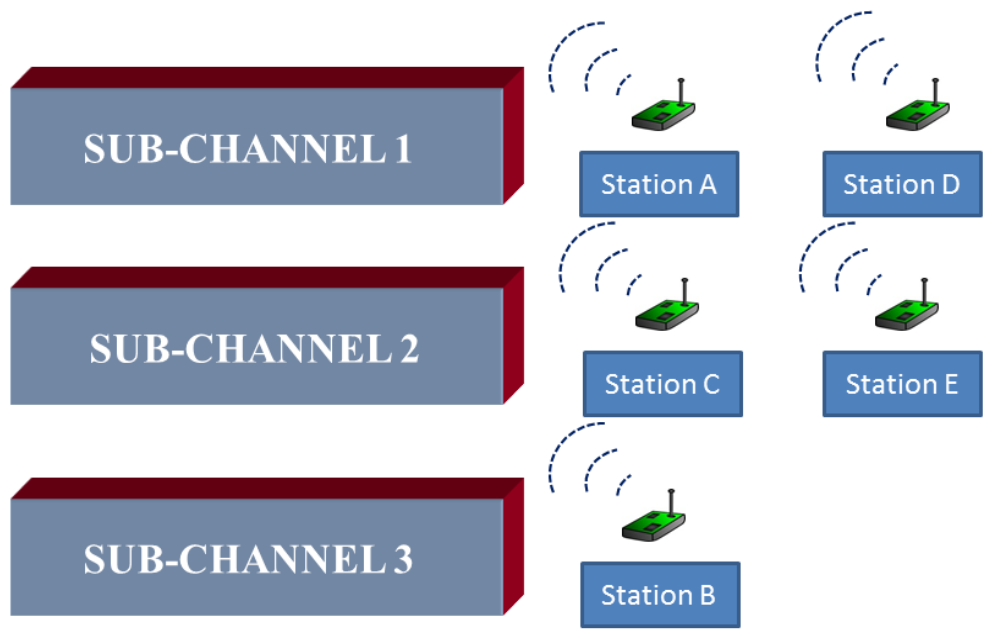

Figure 4.5: Network configuration for 3 sub-channels and 5 stations

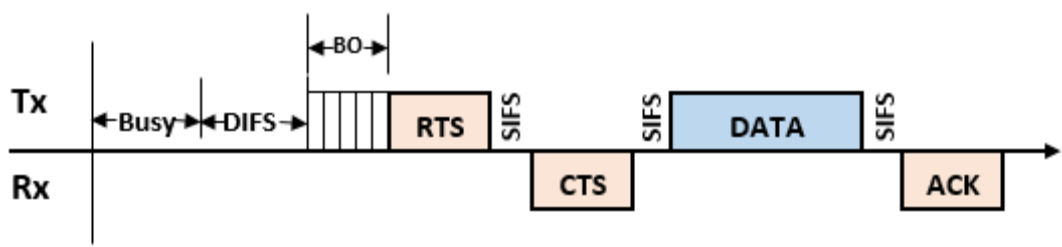

Figure 4.6: Four-way handshaking for multiple STAs in a sub-channel

i. Multiple STAs in the sub-channel: In this case, a four-way handshaking method (RTS/CTS, DATA/ACK) be employed (Islam & Kashem 2019) as shown in Figure 4.6. This method utilizes different control frames (discussed in Subsection 2.3.1) such as the RTS (Request-to-Send) frame, CTS (Clear-to-Send) frame, DATA (Data) frame, ACK (Acknowledgement) frame, and several timing intervals (stated in Subsection 2.3.2) such as the DIFS (Distributed Inter-Frame Space), SIFS (Short Inter-Frame Space). After the DIFS interval, the sending STA enters into the backoff interval. When the backoff number reaches 0, the sending STA transmits an RTS frame to the receiving STA. After the SIFS interval, the receiving STA replies using the CTS frame. Upon receiving the CTS frame, the sending STA waits for the SIFS interval and then sends the DATA frame to the receiver. After receiving the DATA frame, the receiving STA waits for another SIFS interval and then replies using the ACK frame.

According to the DCF rule, for sending data, the STAs have to generate the random backoff number to minimize the probability of frame collisions. If two or more STAs produce the same backoff value, a collision must occur, and hence the data of the colliding STAs will be lost. This unsolicited incident is illustrated in Figure 4.7, where STA B and STA C produce the same backoff value of 7.

The MAC protocol employs the BEB algorithm to compute the backoff number. The backoff value is uniformly chosen in the range of [0, $W-1$], where $W$ denotes the contention window size. At the first attempt to transmit, $W$ is set to its minimum value ($W_{min}$). This value is doubled at each backoff stage up to the maximum value after each unsuccessful transmission. The maximum value computes as $W_{max} = 2^{\alpha}W_{min}$, where $\alpha$ denotes the number of backoff stages (Islam & Kashem 2019).

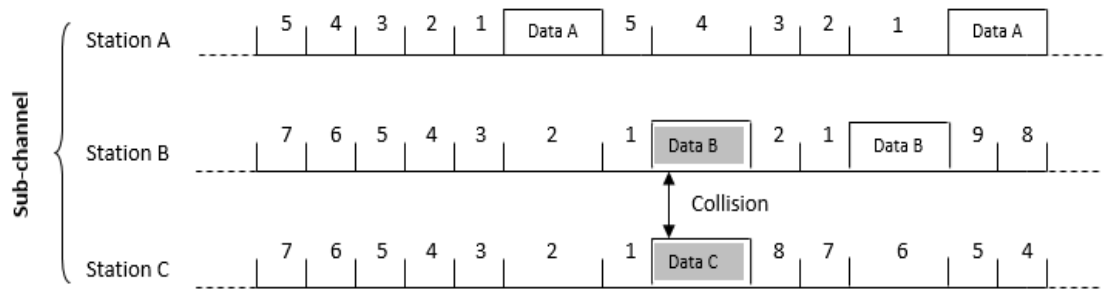

Figure 4.7: A Collision between Station B and Station C in the sub-channel

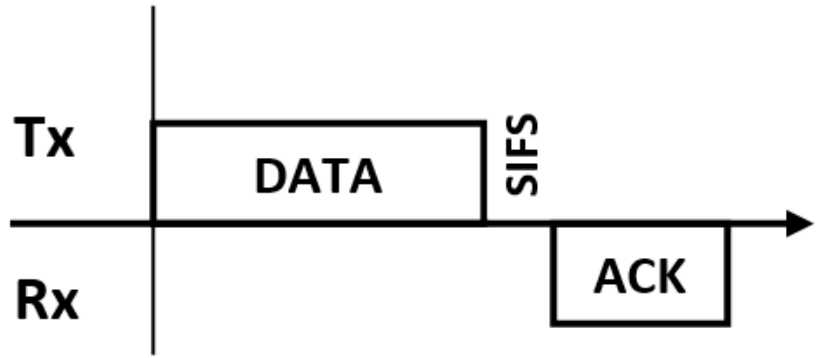

Figure 4.8: Two-way handshaking for a single STA in a sub-channel

ii. Single STA in the sub-channel: In this configuration, the two-way handshaking method (DATA/ACK) be used as shown in Figure 4.8. Here, the RTS/CTS frame pair of four-way handshaking method is not required since the sub-channel will not be shared with other STAs. So, the protocol need not consider the existence of any hidden STAs in the surrounding area.

The DIFS and backoff interval are also not required as there is no possibility of the collision of MAC frames. As a consequence, the overheads of the MAC protocol are reduced significantly, thereby increasing the throughput of the network. In this configuration, the sending STA sends the DATA to the receiver. The receiving STA waits for a SIFS interval after receiving the DATA, then sends an ACK frame to the sending STA.

## 4.3 Framework for ERA Protocol

The framework of the ERA (Efficient Resource Allocation) protocol is illustrated throughout this section. The ERA protocol is regulated by the OFDMA specifications and constraints described in Section 3.5. The model for the protocol is designed in Subsection 4.3.1, where resource units of the whole channel are arranged into different levels. Subsection 4.3.2 classifies the stations into different load groups using some equations. In Subsection 4.3.3, the researcher designs an algorithm for scheduling purposes.

| $l = 3$ | (3, 0) | (3, 1) | (3, 2) | (3, 3) | (3, 4) | (3, 5) | (3, 6) | (3, 7) |
|---|---|---|---|---|---|---|---|---|
| $l = 2$ | (2, 0) | | (2, 1) | | (2, 2) | | (2, 3) | |
| $l = 1$ | (1, 0) | | | | (1, 1) | | | |
| $l = 0$ | (0, 0) | | | | | | | |

Figure 4.9: Levelling and indexing the RUs in a 20 MHz channel

### 4.3.1 Model of the ERA Protocol

The bandwidth of the OFDMA channels can be assigned as Resource Units (RUs) to different stations. The RUs can be arranged into at most $L$ levels, where the number of levels ($L$) depends upon the size of the assigned bandwidth. In Figure 4.9, there are 4 levels (i.e., $L = 4$) in the 20 MHz channel, whereas, in the 40 MHz channel, there are 5 levels (i.e., $L = 5$). The number of levels increases according to the channel bandwidth and vice versa. Each level is denoted by assigning a level number ($l$) from the bottom (by assigning $l = 0$) to up, gradually increasing the value $l$. Thus, in Figure 4.9, $l = 0$ represents the largest RU level (at the bottom), and $l = 3$ represents the smallest RUs level (on the top) in a 20 MHz channel.

For simplicity, the researcher considers each of the RUs (having more than 26 tones) can be divided into two equal smaller RUs. Each of the RUs can be represented as RU ($l$, $i$), where $i$ denotes the index of an RU at level $l$. The whole bandwidth can be split into $2^l$ uniform (equal-size) RUs at level $l$ ($l \in \{0, 1, ..., L-1\}$), labeled as 0, 1, 2, ..., $2^l$-1. Each resource unit RU ($l$, $i$) with $l < L - 1$ can be divided into two resource units RU ($l + 1$, $2i$) and RU ($l + 1$, $2i + 1$). Following the above discussions, the lowest bandwidth (20 MHz) has four levels ($L = 4$), while the largest bandwidth (160 MHz) has 7 levels ($L = 7$).

### 4.3.2 Load Measurement

To send data in the network, the stations send their BSR information to the access point. The BSR contains the load information of a station. After receiving the BSRs from the stations, the access point categorizes those according to their loads as a) Low-Load (LL) b) Medium-Load (ML), and c) High-load (HL). The access point sets a reasonable/practical value for the LL parameter observing the load of the participating stations. Observing all available loads of the stations, the access point regulates the LL parameter so that most of the stations in the network fall in the LL group. Then, ML and HL parameters are calculated according to the following equations:

$$ML = 2 * LL. \tag{4.1}$$

$$HL = 2 * ML = 4 * LL. \tag{4.2}$$

Now, all stations in the basic service set would belong to a load group according to the load range mentioned in the equations (4.3) to (4.5):

$$0 < LL\ STAs \leq (LL + ML) / 2. \tag{4.3}$$

$$(LL + ML) / 2 < ML\ STAs \leq (HL + ML) / 2. \tag{4.4}$$

$$(HL + ML) / 2 < HL\ STAs. \tag{4.5}$$

Suppose a basic service set contains five stations, namely A, B, C, D, and E, which have loads of 4, 1.5, 2.7, 30, and 3 Mbps, respectively. If the access point sets the parameter LL = 2, then according to the above equations, the stations would be categorized as:

LL STAs: B, C, E; ML STAs: A; HL STAs: D

Since the wider RUs can provide substantially more data than the smaller RUs, the researcher allocates comparatively wider RUs to the HL stations. Thus, by Algorithm 2, which is designed in the following subsection, the ERA protocol ensures the larger RUs for the HL stations, intermediate RUs for the ML stations, and smaller RUs for the LL stations.

### 4.3.3 Resource Allocation

For simplicity and better efficiency, the researcher arranges the RUs at most tree levels irrespective of the channel bandwidth for the ERA protocol (Islam & Kashem 2020). Thus, all available channel bandwidths (i.e., 20, 40, 80, and 160 MHz) for the IEEE 802.11ax can utilize only three levels. The designer does not use the bottom level (i.e., $l = 0$) that provides only one resource unit, namely, RU (0, 0), which comprises the total channel bandwidth. Since RU (0, 0) can be assigned to only one of the stations, other stations in the network cannot access the channel and hence the protocol would be deprived of the utility of the OFDMA multi-channel mechanism. As the ERA protocol wants to serve more than one station simultaneously, the algorithm attempts to distribute RUs from the next three levels where the level number, $l$ = 1, 2, and 3.

Algorithm 2 is designed in the thesis as in Islam & Kashem (2020) to allocate the resource units to the intending stations in the basic service set. According to the algorithm, a wider resource unit such as RU (1, 0) is assigned to an HL STA (i.e., Step 03); a medium resource unit e.g., RU (2, 2) is assigned to an ML STA (i.e., Step 15), and two smaller resource units, namely, RU (3, 6), and RU (3, 7) is assigned to two LL STAs (i.e., Step 20). However, if there is no HL STA in the network at the moment then, the RU (1, 0) is divided into two as RU (2, 0) and (2, 1) (i.e., Step 05). Then, the RU (2, 0) is supposed to be assigned to an ML STA, and RU (2, 1) is further split into two shorter resource units as RU (3, 2) and RU (3, 3), which to be assigned to two LL STAs. Similarly, if there is no ML STA during the distribution, then its resource elements are to be divided into two smaller RUs that would be allocated to the LL STAs.

The ERA protocol is designed to allocate the resource units according to the applied loads of the participating stations. As such, Algorithm 2 ignores the channel condition on the resource units while distributing those. The difference in throughput for channel condition is little comparing the size of the resource units. For example, assigning RU (1, 1) instead of RU (1, 0) to an HL STA would not bring any remarkable difference in the throughput. Thus Algorithm 2, alternatively, can choose RU (1, 1) for an HL STA; RU (2, 1) for an ML STA; RU (3, 0) and RU (3, 1) for two LL STAs. Nonetheless, a good RU selection based on the channel condition is of particular interest in wireless communications where all resource units are of equal length.

Algorithm 2 repeatedly executes for each flow of the data. The algorithm supports at most one HL STA, one ML STA, and two LL STAs simultaneously in each flow. If the number of participating STAs (i.e., which submitted BSR earlier) is more than the capacity for a flow, then the remaining STAs would get a chance to transmit in the immediate next flow after completing this flow. To main the priority among the STAs, the AP creates the FCFS (First Come First Served) queue based on the BSR submissions of the STAs. Since there are three classes of STAs (i.e., HL, ML, and LL), the AP can create three FCFS queues for the three classes.

**Algorithm 2: RU Assignment**

01: //Assignment of larger RU to LL STAs

02: **IF** HL STA available THEN

03:     Assign RU (1, 0) to an HL STA

04: **ELSE**

05:     Split RU (1, 0) into RU (2, 0) and RU (2, 1)

06:     Split RU (2, 1) into RU (3, 2) and RU (3, 3) and assign to two LL STAs

07:     **IF** ML STA available THEN

08:         Assign RU (2, 0) to an ML STA

09:     **ELSE**

10:         Split RU (2, 0) into RU (3, 0) and RU (3, 1) and assign to two LL STAs

11:     **END IF**

12: **END IF**

13: //Assignment of intermediate RU to ML STAs

14: **IF** ML STA available THEN

15:     Assign RU (2, 2) to an ML STA

16: **ELSE**

17:     Split RU (2, 2) into RU (3, 4) and RU (3, 5) and assign to two LL STAs

18: **END IF**

19: //Assignment of smaller RUs to LL STAs

20: Assign RU (3, 6) and RU (3, 7) to two LL STAs

According to the Wi-Fi 6 standard, in this system, the access point schedules the STAs and regulates the transmission parameters for UL and DL paths. The AP sends the Trigger Frame (TF) to the STAs to propagate the scheduling and transmission information. The access point grants channel access, distributes resources, controls transmission parameters, and synchronizes the sending STAs utilizing the TF. The transmission parameters (Ali et al. 2018) comprise the Modulation and Coding Schemes (MCSs), transmission power, transmission duration, etc.

## 4.4 Framework for PRS Protocol

In this section, the framework and working procedure of the PRS (Proportional Resource Scheduling) protocol (Islam & Kashem 2021) is described. The system model (details in Section 5.2.3) for the PRS protocol is designed in light of IEEE 802.11ax and the earlier standards as well. As the PRS is designed for the future WLAN, it is also flexible to deploy in different scenarios where the previous standards may not. Thus, it helps future Wi-Fi networks be more flexible and robust to implement the SA and RA mechanisms concurrently. Like the ERA protocol, PRS is also regulated by the OFDMA specifications and constraints delineated in Section 3.5. The model for the PRS protocol is designed in Subsection 4.4.1. The PRS utilizes two algorithms for resource scheduling which are illustrated in Subsection 4.4.2. Subsection 4.4.3 describes the access mechanisms for the PRS protocol.

### 4.4.1 Model of the PRS Protocol

The working principle of Proportional Resource Scheduling, i.e., PRS protocol is delineated in Figure 4.10. The figure represents a window of Wi-Fi channel resources in a 40 MHz channel. The sliding bar of the window divides the whole bandwidth in terms of RUs into two portions. The RUs that reside at the left side of the bar are known as the LRUs (Left Resource Units) and those that reside at the right side are known as RRUs (Right Resource Units). The LRUs would be assigned to the Scheduled Access (SA) STAs, and RRUs would be assigned to the Random Access (RA) STAs. According to Wi-Fi 6, a resource unit must contain at least 26 sub-carriers (i.e., tones) which PRS terms as Smallest Resource Unit (SRU).

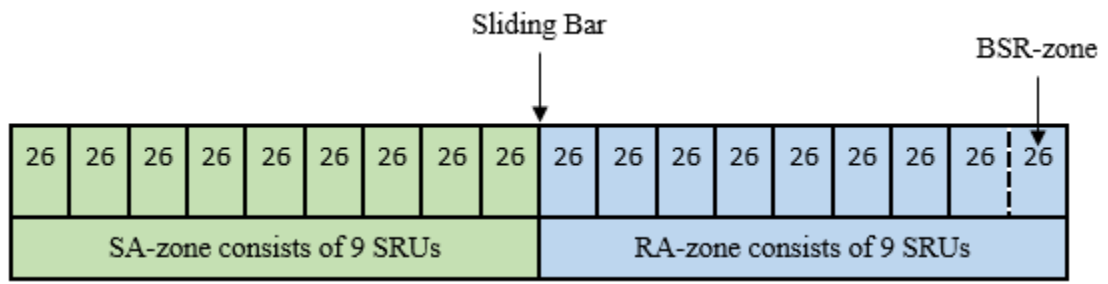

Figure 4.10: Dividing the RUs where SA-zone = RA-zone

The proportion of LRUs and RRUs is determined by the AP based on the total load of the candidate STAs (SA and RA stations). According to the PRS protocol, in general, SA STAs utilize the scheduled access mechanism for payload transmission, and RA STAs use the inherent random access mechanism. Thus LRU-zone (green colored) is dedicated to the SA STAs only. Similarly, the RRU-zone (blue colored) would be shared primarily by the RA STAs. A few SA STAs (removed from LRU-zone) may also share the RRU-zone, which will be discussed during revised scheduling. The right-most SRU that contains 26 tones is termed as BSR-zone, a sub-zone of RRU-zone. In the PRS protocol, the BSR-zone is of fixed size (1 SRU) and must be preserved. It can be used for sending the BSR (Buffer Status Report) or other control information even if no other RUs available for random access and can be used by both SA and RA STAs.

The PRS is a hybrid MAC protocol since it utilizes both the scheduled access and random access mechanism. It is noted for acquiring scheduled access, an STA must send its BSR to the access point. The BSR contains the load information of an STA, and AP needs this information for resource scheduling for that STA. Thus, for gaining the scheduled access, an SA STA has to send its BSR to the AP through the random access mechanism. That's why the PRS protocol reserves the BSR-zone.

### 4.4.2 Distribution of Resource Units

The Proportional Resource Scheduling protocol allocates RUs proportionally to LRU-zone (SA-zone) and RRU-zone (RA-zone) based on the total load of SA and RA STAs. For example, if the total load of SA STAs and RA STAs is equal, then the LRU-zone (for SA STAs) and RRU-zone (for RA STAs) are also equal or nearly equal. This example applies to the previous Figure 4.10, where both green-zone and blue-zone comprise 50% of the total resources. For simplicity of the design, RRU-zone also contains the BSR-zone which size (i.e., only 1 SRU) is very small comparing the whole channel. During initial scheduling, due to the presence of BSR-zone, RA STAs get a slightly reduced space than what should be by the exact proportional distribution. However, during the revised scheduling, this will be compensated.

Again, the proportional distribution also depends on the minimum RU size. Since the minimum RU size is 26 tones, the PRS divides the LRU-zone and RRU-zone on the boundary of the RUs (i.e., not in the middle of a RU). To under the phenomena, a scheduling problem is considered, which is demonstrated below.

***Scheduling Problem Example:*** Suppose there are 5 SA STAs (e.g., A, B, C, D, E) and 3 RA STAs (e.g., X, Y, Z) in an infrastructure basic service set. The load of the STAs A, B, C, D, E are 3.1, 2.2, 2.9, 1.3, 0.7 MB, respectively, and the load of the STAs X, Y, Z are 3.4, 1.2, 2.1 MB, respectively.

To solve the scheduling problems, the PRS protocol first applies the initial scheduling algorithm, i.e., Algorithm 3, then applies the revised scheduling algorithm, i.e., Algorithm 4. The revised scheduling algorithm works on the output of the initial scheduling algorithm and gives the final scheduling results (Islam & Kashem 2021). Table 4.1 shows the meaning of all notations used in the algorithms.

**Algorithm 3: Initial Scheduling**

01: // Load calculation

02: Find the total load of SA STAs, $L1 = \sum_{i=1}^{P} p_i$

03: Find the total load of RA STAs, $L2 = \sum_{i=1}^{Q} q_i$

04: Total load of all STAs, $L3 = L1 + L2$

05: // Finding the number of SRUs for the zones

06: Number of SRUs for SA-zone, $S = \lfloor (L1/L3)M \rfloor$

07: Number of SRUs for RA-zone, $T = \lceil (L2/L3)M \rceil$

Table 4.1: Notation glossary for the PRS algorithms

| Notation | Description |
|---|---|
| $p_i$ | Load of the SA STA i |
| $P$ | Total number of SA STAs |
| $L1$ | Total load of SA STAs |
| $q_i$ | Load of the RA STA i |
| $Q$ | Total number of RA STAs |
| $L2$ | Total load of RA STAs |
| $L3$ | Total load of all STAs in the network |
| $M$ | Total number of SRUs in the channel |
| $S$ | Number of SRUs for SA-zone for initial scheduling |
| $T$ | Number of SRUs for RA-zone for initial scheduling |
| $r_i$ | Number of SRUs for STA i |
| $U$ | Number of SRUs for SA-zone for revised scheduling |
| $V$ | Number of SRUs for RA-zone for revised scheduling |

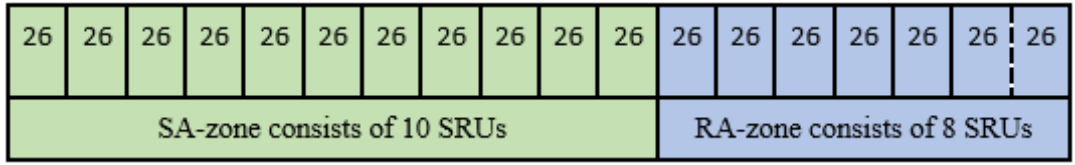

Figure 4.11: Distribution of SRUs after initial scheduling

*Apply Initial Scheduling Algorithm*

According to Algorithm 3, the total load of SA STAs ($L1$) i.e., Step 02:

$$L1 = \sum_{i=1}^{P} p_i \tag{4.6}$$

Thus, load of the SA STAs = (3.1 + 2.2 + 2.9 + 1.3 + 0.7) MB = 10.2 MB

Similarly, total load of RA STAs, ($L2$) i.e., Step 03:

$$L2 = \sum_{i=1}^{Q} q_i \tag{4.7}$$

Hence, load of the RA STAs = (3.4 + 1.2 + 2.1) MB = 6.7 MB

Total load of all STAs (in Step 04), $L3 = L1 + L2$ = (10.2 + 6.7) = 16.9 MB

Number of SRUs for SA-zone, ($S$) i.e., Step 06:

$$S = \lfloor (L1/L3)M \rfloor \tag{4.8}$$

In a 40 MHz channel, there are 18 SRUs (i.e., $M = 18$), as shown in Figure 4.10.

Thus, the total number of SRUs for SA-zone = $\lfloor 10.2/16.9 \times 18 \rfloor = \lfloor 10.86 \rfloor = 10$

Number of SRUs for RA-zone, ($T$) i.e., Step 07:

$$T = \lceil (L2/L3)M \rceil \tag{4.9}$$

Therefore, the total number of SRUs for RA-zone = $\lceil 6.7/16.9 \times 18 \rceil = \lceil 7.14 \rceil = 8$

In the above calculation, the algorithm takes the floor value for the SA method and the ceiling value for the RA method to compensate RA-zone, which contains the BSR-zone. Now, the resource window looks like as in Figure 4.11, where SA-zone contains 10 SRUs and RA-zone contains 8 SRUs. However, this proportion is subject to change after applying the revised scheduling algorithm.

**Algorithm 4: Revised Scheduling**

01: // Finding the number of SRUs for each of the SA STAs

02: Number of SRUs for STA i, $r_i = \lfloor (p_i/L1)S \rfloor$

03: // Finding the number of SRUs for the zones

04: Number of SRUs for SA-zone, $U = \sum_{i=1}^{P} r_i$

05: Number of SRUs for RA-zone, $V = M - U$

*Apply Revised Scheduling Algorithm*

Again, each of the SA STAs gets SRUs proportionally to their available loads. Hence, 10 SRUs getting from initial scheduling to be distributed as follows by the revised scheduling algorithm,

Number of SRUs for STA i, $(r_i)$, i.e., Step 02:

$$r_i = \lfloor (p_i/L1)S \rfloor \qquad (4.10)$$

Therefore,

STA A gets $\lfloor 3.1/10.2 \times 10 \rfloor = \lfloor 3.03 \rfloor = 3$ SRUs;

STA B gets $\lfloor 2.2/10.2 \times 10 \rfloor = \lfloor 2.16 \rfloor = 2$ SRUs;

STA C gets $\lfloor 2.9/10.2 \times 10 \rfloor = \lfloor 2.84 \rfloor = 2$ SRUs;

STA D gets $\lfloor 1.3/10.2 \times 10 \rfloor = \lfloor 1.27 \rfloor = 1$ SRU;

STA E gets $\lfloor 0.7/10.2 \times 10 \rfloor = \lfloor 0.69 \rfloor = 0$ SRU;

Since Algorithm 4 takes the floor value of the integer, the total number of SRUs for all SA STAs is now 8. There is no SRU for STA E and hence STA E is to be moved from SA-zone to RA-zone for random access. Since the initial distribution has a total of 10 SRUs and revised distribution requires 8 out of 10 SRUs, the residual (10-8) = 2 SRUs are to be moved from the green-zone (SA-zone) to the blue-zone (RA-zone). It compensates RA-zone to mitigate the following:

  i)    Compensate for the BSR-zone
  ii)   Compensate for the extra STAs that come from SA-zone

According to the revised and final distribution, the scheduling window looks like as in Figure 4.12, where SA-zone contains 8 SRUs and RA-zone contains 18-8 = 10 SRUs. On the top row of Figure 4.12, the boundary of SRUs in the RA-zone is also removed since all the resources in the RA-zone (blue-zone) work as a single random access channel. The RA-zone to be shared by the RA STAs X, Y, Z, and SA STA E. Recall that 1 SRU must be preserved for the BSR-zone, even if RA-zone gets 0 SRU after revised scheduling.

| A | | | B | | C | | D | X, Y, Z, E | | | | | | | | | |
|---|---|---|---|---|---|---|---|---|---|---|---|---|---|---|---|---|---|
| 26 | 26 | 26 | 26 | 26 | 26 | 26 | 26 | 26 | 26 | 26 | 26 | 26 | 26 | 26 | 26 | 26 | 26 |
| 1, 2 | | | | | 6, 7 | | | | | | | | | | | | |
| 1 | 2 | 3 | 4 | 5 | 6 | 7 | 8 | 9 | 10 | 11 | 12 | 13 | 14 | 15 | 16 | 17 | 18 |
| SA-zone consists of 8 SRUs | | | | | | | | RA-zone consists of 10 SRUs | | | | | | | | | |

Figure 4.12: Final distribution of SRUs after revised scheduling

The AP approximates the number of RUs for the SA-zone and RA-zone applying the algorithms and finally distributes the resources according to Figure 4.12. The LRUs that resided in the SA-zone would be distributed to the SA stations in terms of RUs leveraging the OFDMA technology. On the other hand, the RRUs that resided in the RA-zone would be distributed to the SA and RA stations in terms of bandwidth. It is noted that RA stations are unable to receive RUs, and the AP does not offer those RUs. As mentioned in the previous paragraph, the whole bandwidth of the RA-zone would work as a single random access channel, and the stations in that zone would compete for the channel access according to the legacy RA mechanism. After the final scheduling (See Figure 4.12), the capacity of the RA-zone may not be a multiple of 20 MHz, which is not necessary for future WLAN as assumed in the network model in Section 5.2.3.

It is also noted that the AP computes the resources in terms of RUs as dictates the algorithms. The AP can perform computation reliably without creating real RUs since the number of SRUs is fixed for a particular bandwidth. According to the IEEE 802.11ax, the number of SRUs in 20, 40, 80, and 160 MHz channels are 9, 18, 37, and 74, respectively. The AP creates the RUs only for the SA stations that belong to the SA-zone after finishing complete computations (i.e., after revised scheduling), as shown in Figure 4.12. So, the SRUs drawn in Figure 4.10 and Figure 4.11 have no real existence, and the paper has shown that to visualize how the algorithms work.

### 4.4.3 Access Mechanism

The PRS protocol is a hybrid one that utilizes both the SA mechanism and RA mechanism to access the wireless channel. In PRS, the candidates for the SA mechanism are those which get resource units in the SA-zone (green zone), and for the RA mechanism are those which get bandwidth in the RA-zone (blue zone), after the final scheduling. It is also observed that after revised scheduling, a few SA STAs whose load is comparatively lower may migrate to the RA-zone and hence use the RA mechanism to access the channel.

In the PRS protocol, the access point schedules the channel resources and regulates downlink and uplink transmission parameters. The access point grants channel access, distribute resources to the eligible STAs, regulates transmission parameters, and synchronizes STAs for uplink OFDMA transmission. The transmission parameters include transmission power, transmission duration, modulation and coding schemes, MIMO, etc.

*SA mechanism*

The IEEE 802.11ax innovates the Trigger Frame (TF) concept to serve the OFDMA transmissions (Avdotin et al. 2019). The AP sends the TF to all stations in the Wi-Fi network, which contains transmission information, scheduling information, and other parameters of the wireless channel. All participating STAs in the SA-zone synchronize using the TF and start transmissions at the same time, as shown in Figure 5.4 in Section 5.6.3. Each of the RUs that acts as a sub-channel will get the same amount of time for uplink transmission. However, their data transfer rates depend on the size of the resource unit. Stations having larger RU can send more data than those having smaller RU.

For the scheduling problem, Algorithm 4 finds the number SRUs for each of the SA stations. The PRS protocol assigns the SRUs to the stations on the FCFS (First Come First Served) basis from left to right, as shown in Figure 4.12. The left-most SRU numbered 1, and the right-most SRU numbered 18 (i.e., 2nd row from the bottom of Figure 4.12). STA A gets SRU 1, 2, 3; STA B gets SRU 4, 5; STA C gets SRU 6, 7; and STA D gets the SRU 8. According to IEEE 802.11ax, some adjacent smaller RUs can combine together to form a large RU. Details of the merging procedure are portrayed in Figure 3.7 of Section 3.5. Abide by the rules, STA A can merge SRUs 1 and 2 to form a larger resource unit RU (1, 2). Similarly, STA C can merge SRUs 6 and 7 to form the RU (6, 7). No more merging is possible for scheduling problem example in accordance with the configuration of Figure 4.12. The merging can increase the throughput of the channels. The merging reduces the number of individual RUs (i.e., sub-channels), and wider channels (e.g., 80 or 160 MHz) may have more merging configurations due to the presence of more SRUs.

*RA mechanism*

Random access mechanism is the legacy technique for accessing the channel that the Wi-Fi stations predominantly used before adopting the OFDMA technology in the latest Wi-Fi 6 and Wi-Fi 7 standards. The PRS keeps provision of using both the RA and SA mechanisms as Wi-Fi 6 standard. In the problem scheduling example, STAs X, Y, Z, and E use the RA mechanism shown in Figure 4.12. According to the novel protocol, all computed RUs in the RA-zone are merged to form a single random access channel after revised scheduling. As mentioned earlier, all SA STAs must use the RA mechanism momentarily to send their BSR to the AP. Thus RA-zone always reserves an SRU to support the STAs for sending the BSR. The RA mechanism is governed by the renowned Binary Exponential Backoff (BEB) algorithm. The two-frame (DATA/ACK) or four-frame (RTS/CTS, DATA/ACK) exchange methods can be used for the data transmission for the RA mechanism. To ensure a higher degree of reliability and deal with the hidden node problem (Forouzan 2013 and Perahia & Stacey 2013), the four-frame exchange method is preferable. Details of two-frame (See Figure 4.8) and four-frame (See Figure 4.6) exchange methods are illustrated in Section 4.2.3.

In the RA mechanism, every STA has to generate a random backoff counter using the BEB algorithm. In the random access method, only one STA can send data at a time. If more than one STA send data at the same time, then a collision must occur between the data frames, and therefore sending STA has to resend the data. The BEB algorithm reduces the probability of frame collision of the random access protocol by generating the random backoff counter. The algorithm calculates the backoff value in a range [0, $W-1$], where $W$ denotes the contention window size. Initially, the contention window size is set to a minimum value ($W_{min}$). After each of the failed transmissions, the contention window size is doubled, and it may reach up to the maximum value, $W_{max} = 2^{\alpha}W_{min}$, where $\alpha$ denotes the number of backoff stages.

## 4.5 Conclusion

The core contributions of this research are illustrated in this chapter. To implement a new hybrid MAC protocol, i.e., HTFA, the author adopts a 2-dimensional time-frequency model based on the existing theories (stated in Subsection 4.2.1). The two-dimensional time-frequency access facility for the WLAN and other cellular protocols is an excellent outcome of the OFDMA technology. The researcher describes the concept of the new MAC protocol leveraging the two-dimensional time-frequency model in Subsection 4.2.2. The working principles and sub-channel distribution procedure are the unique contributions of the HTFA protocol, which are delineated with several diagrams. Subsection 4.2.3 explains the access mechanism of the HTFA adhering to the basic rules of the MAC protocols.

The author designs another two Wi-Fi protocols, namely, ERA, and PRS that are elaborated throughout Section 4.3 and Section 4.4, respectively. The PRS protocol adopts both the RA and SA mechanisms for accessing the wireless channels, while ERA utilizes only the RA mechanism. All of the protocols presented in this thesis paper adopt the OFDMA technology that provides the opportunity for simultaneous data transmissions by multiple STAs.

# CHAPTER FIVE

# Chapter 5

# Research Methodology

## 5.1 Introduction

In this chapter, the details of the research methodology of the current research will be discussed elaborately. The author designs the models for the intended Wi-Fi protocols by adopting the OFDMA technology that is illustrated throughout the previous Chapter 4. Experiments and data analyses will be thoroughly examined in Chapter 6, i.e., Data Analysis for evaluation of the protocols. Thus, this research follows the *quantitative approach* to evaluate the protocols.

At first, the chapter describes the system and target network for the protocols in Section 5.2. Simulation parameters and samples are discussed briefly in Section 5.3. Then the list of the instruments is described in Section 5.4. Section 5.5 overviews the procedures to be adopted for this protocol. Section 5.6 constructs the analytical models for the mathematical analysis. The researcher adopts the Markov Chain model to perform the statistical analysis of the HTFA protocol and also explains the analytical model of the ERA and PRS protocol. The researcher effectively estimates the saturation throughput of all three protocols, which are explained in Subsection 5.6.1, Subsection 5.6.2, and Subsection 5.6.3, respectively. At last, Section 5.7 concludes this chapter.

## 5.2 System Design

The HTFA protocol adopts a two-dimensional time-frequency access model then specifies the methods for accessing the medium. The outline of the model, relevant theory along with literature are introduced and discussed in Section 4.2 of Chapter 4. The ERA protocol adopts a model that helps to identify and distribute the OFDMA RUs to the STAs efficiently. The details of the ERA model, along with a scheduling algorithm, are illustrated in Section 4.3 of Chapter 4. The last protocol, i.e., PRS develops a model for both RA and SA mechanisms of the Wireless LAN. The model and working procedure of this protocol are elaborated throughout Section 4.4.

Figure 5.1 shows a system model diagram to highlight the target network topology for all three protocols, namely, HTFA, ERA, and PRS. Subsection 5.2.1, Subsection 5.2.2, and Subsection 5.2.3 delineate the system model for the HTFA, ERA, and PRS, respectively. It is assumed that every STA keeps a timer for synchronization with other STAs in the system. For proper synchronization, the AP provides the reference time information to the STAs at a regular interval in accordance with the time synchronization function suggested by the IEEE 802.11 standard (Standards IEEE 2019). Imperfect time synchronization produces clock offset among the participating STAs in the network. In the OFDMA-adopted wireless LAN, orthogonality among the sub-channels is of utmost importance, and it could not be guaranteed if the clock offset surpassed the pre-defined threshold value (Kwon et al. 2009). As a consequence, it is assumed that clock synchronization would be maintained efficiently to confine the maximum clock offset within the threshold.

### 5.2.1 Target Network Model for HTFA

To measure the efficiency of the PRS protocol, the researcher considers an OFDMA-based wireless LAN where the total available bandwidth is denoted by $B$. There are $M$ sub-channels, $N$ STAs, and one AP in the system, as shown in Figure 5.1. The bandwidth is equally distributed to $M$ sub-channels. Therefore, each of the sub-channels receives a bandwidth of $B/M$. If each of the sub-channels has only one STA, then the STAs can communicate with the access point without any interruption. However, if a sub-channel can have multiple STAs, then the STAs in that sub-channel must compete randomly to access the medium. The working principle of the system is regulated by the 2-step algorithm designed in Subsection 4.2.2.

### 5.2.2 Target Network Model for ERA

The researcher considers an OFDMA-adopted wireless LAN that operates in the infrastructure BSS as portrayed in Figure 5.1. In the network, there are $N$ STAs that connect to one AP, and STAs are communicating through the AP. The system can operate in both of the Wi-Fi frequency bands (2.4 GHz, 5 GHz) using the bandwidth (20/40/80/160 MHz) that is specified for the 802.11ax by the IEEE standard (Standards IEEE 2019). The whole bandwidth of a channel is partitioned into several resource units by the AP, then the AP assigns the resource units to the STAs according to their available load (details in Section 4.3). Thus, employing different resource units, different STAs communicate through the AP simultaneously without suffering from channel contentions and co-channel interference.

### 5.2.3 Target Network Model for PRS

An OFDMA-based infrastructure BSS is considered for the PRS protocol, as like in Figure 5.1. There are $N$ STAs in the network where STAs communicate with each other through the AP. The PRS protocol is designed for the OFDMA-based future Wi-Fi network and supports both the SA and RA mechanisms simultaneously for channel access. As the IEEE 802.11ax standard, PRS can operate in 2.4 GHz and 5 GHz frequency bands and supports different channel bandwidths, 20, 40, 80, and 160 MHz [1]. The stations in the PRS are also able to operate in any frequency below the assigned channel bandwidth. In future WLAN, the stations would support both SA and RA mechanisms known as SA STAs and RA STAs, respectively. Therefore, the stations can seek any services (SA or RA), whatever they prefer. Unlike legacy RA stations, the future RA stations are capable of providing feedback to the AP.

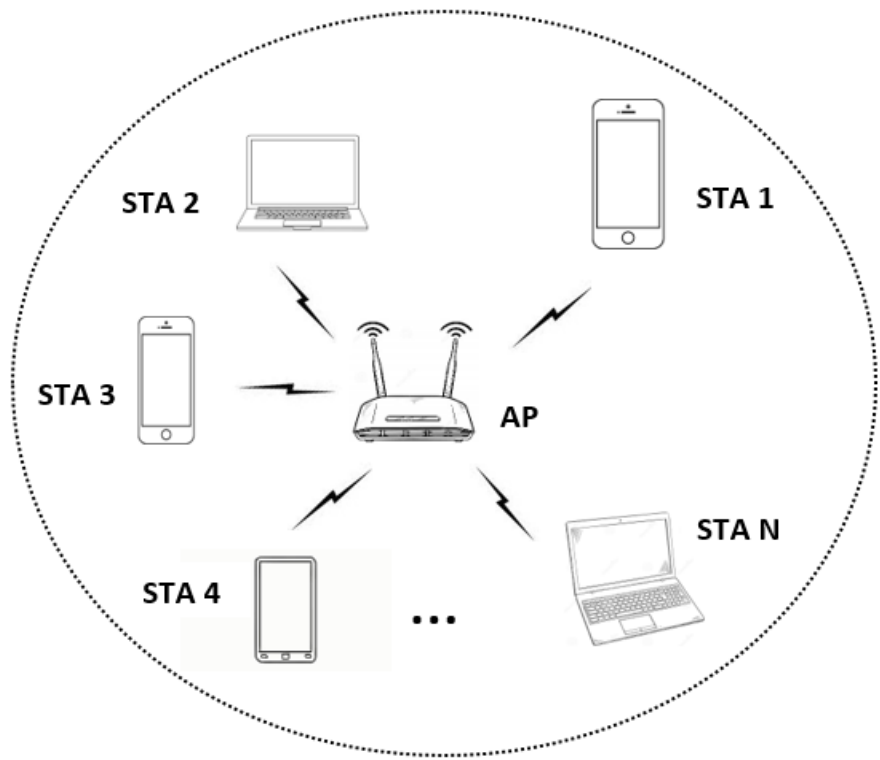

Figure 5.1: A Wi-Fi network with N STAs and one AP

Initially, the AP computes the entire channel bandwidth in terms of the smallest resource units. The calculated resource units are divided into two portions: one for the SA mechanism and another for the RA mechanism (details in Section 4.4). The SA mechanism is to be used only by the SA STAs whose BSR information is available to the AP. The RA mechanism is to be used by all RA STAs and a few low-load SA STAs (if any). The RUs computed for the RA mechanism by the AP will be merged to form a single random access channel. The tasks of all scheduling and decisions are made by the AP.

## 5.3 Simulation Parameters and Samples

All of the samples for computer simulation are generated by NS-3 software. The samples are produced to measure the efficiency, precisely the performance parameters of the new protocols. The experiments are repeatedly conducted by several computers having higher processing capabilities. The investigator generates a large number of samples for evaluation and comparison of the protocols through exhaustive programming using C++ and Python.

### 5.3.1 Parameters for HTFA

Some important samples for protocol evaluation are packet transmission time, backoff slot duration, DIFS duration, number of stations ($N$), number of backoff stages ($\alpha$), minimum contention window size ($W_{min}$). Some of the samples generated for protocol comparison are slot time, packet length, total channel bandwidth, number of stations, number of sub-channels, etc. The important performance parameters measured for the HTFA protocol are saturated throughput, collision probability, normalized throughput, max-min fairness, etc. Table 6.1 and Table 6.2 in Chapter 6 summarize the simulation parameters and sample size of the HTFA protocol.

### 5.3.2 Parameters for ERA

The most notable samples generated for the ERA protocol are the number of STAs ($N$), number of antennas of the AP ($A_{AP}$), number of antennas of the STAs ($A_{STA}$), transmit power, packet size, packet generation delay, etc. The performance parameters measured during the simulations are average throughput, average upload time, average goodput, and the number of retransmissions. Table 6.4 provides a glimpse of simulation parameters and sample size of the ERA protocol.

### 5.3.3 Parameters for PRS

The investigator also produces many samples for the PRS protocol, such as the number of STAs ($N$), distribution of STAs for RA and SA mechanisms, transmit power, number of antennas of the AP ($A_{AP}$), number of antennas of the STAs ($A_{STA}$), etc. Two performance parameters of the RA and SA mechanisms of the PRS protocol, namely, average throughput and the number of retransmissions, are measured efficiently for a different distribution of stations. The performance parameters of the PRS, such as the average throughput, fairness index, and average goodput, are also compared with some prominent protocols. Table 6.5 summarizes the simulation parameters and sample size of the PRS protocol.

## 5.4 Instruments

### 5.4.1 Hardware Requirement

A mid-range computer such as a desktop computer or laptop is the minimum hardware requirement to conduct the simulation. However, a high-configured computer such as the workstation is preferred for such experiments since some of the simulations take a lot of time to execute. Moreover, a high-configured machine is appropriate to install a diverse class of software to conduct the simulations for such research. The researcher conducted simulations several times for this research using different mid-range computers like desktops and laptops.

### 5.4.2 Software Requirement

A variety of software has been installed to perform the simulation and data analysis. A few of those are listed below:

Operating System: The latest Linux variant 'Ubuntu' (Ubuntu 2018) is one of the best platforms to install NS-3 and conduct simulation. All of the application software is installed on this operating system.

Network Simulator: Computer simulations were conducted using the well-known network simulator 'NS-3' as mentioned previously. NS-3 is the most powerful discrete-event network simulator developed for computer network researchers. More details of NS-3's utilities and impact will be discussed in Section 7.6.

Programming: Extensive programming knowledge is required both in C++ and Python to code for NS-3. All of the simulation codes developed for the experiments are appended at Appendices B, D, and F.

Data Presentation: OriginPro package was used for graphing and analyzing the data that are obtained from NS-3. All simulation data and corresponding graphs are shown in Appendices C, E, and G.

Eclipse: It is an integrated development environment (IDE) developed by Java. Although it is developed by Java, it can be used for customizing C++ and Python code.

Netanim Animator: Although this program is not essential for getting simulation data, using this animator, researchers can visualize the simulation events.

## 5.5 Procedures

As mentioned earlier, the investigator measured the efficiency of all protocols through mathematical analysis and computer simulation. Subsection 5.5.1 gives an outline of mathematical evaluation that will be further elaborated in Section 5.6. Subsection 5.5.2 gives an overview of all simulations that will be analyzed in detail in Chapter 6.

### 5.5.1 Mathematical Evaluation

A prudent mathematical analysis could convince the computer network researchers as well as enhance the credibility of the innovation. A statistical model is constructed to analyze the performance of the HTFA protocol. The researcher adopts the Markov Chain model to evaluate the protocol statistically. The researcher evaluates the throughput of the HTFA protocol utilizing the analytical model developed by G.

(Bianchi 2000). The details of the statistical analysis of the HTFA will be discussed in Section 5.6.1, and the challenges of deriving the model to be discussed in Section 7.2.3. Similarly, an analytical model is developed for the ERA protocol, which will be described in Subsection 5.6.2. The model also utilizes a path-loss model, which is represented by equation (5.10). The third protocol for this research, i.e., PRS leverages both the random and scheduled access mechanisms for channel access. The analytical models for both of the mechanisms are constructed in Subsection 5.6.3.

**5.5.2 Computer Simulation**

The researcher conducted a lot of laboratory experiments to measure the efficiency of the new protocols. The investigator conducts computer simulation with the most powerful network simulator ever designed named Network Simulator-3 or simply NS-3 (Network Simulator 2017). The following experiments have been performed through computer simulation for the evaluation of the protocols.

*Experiments for the HTFA protocol:*

Experiment 1: Saturated throughput with respect to $W_{min}$.

Experiment 2: Collision probability with respect to $W_{min}$.

Experiment 3: Relation between the throughput and average number of active STAs.

Experiment 4: Saturated throughput with respect to the backoff slot duration.

Experiment 5: Normalized throughput of the STAs with different traffic loads.

Experiment 6: Total system throughput and the max-min fairness.

*Experiments for the ERA protocol:*

Experiment 1: Comparison of the throughput between competing protocols.

Experiment 2: Impact of the number of stations on the throughput.

Experiment 3: Impact of the number of antennas of the AP on the throughput.

Experiment 4: Upload time vs number of stations.

Experiment 5: Goodput vs number of stations.

Experiment 6: Number of retransmissions vs number of stations.

*Experiments for the PRS protocol:*

Experiment 1: Throughput of the PRS protocol for different distribution of stations

Experiment 2: Number of retransmissions of the PRS protocol for different distribution of stations

Experiment 3: Throughput comparison between different protocols

Experiment 4: Fairness comparison between different protocols

Experiment 5: Impact of the number of stations on the throughput

Experiment 6: Impact of the number of antennas on the throughput

Experiment 7: Goodput vs number of stations

## 5.6 Analytical Models

### 5.6.1 Analytical Model for HTFA

Let's consider a Basic Service Set (BSS) having an Access Point (AP) placed at the center of BSS. The Basic Service Set has $N$ stations (STAs) and $M$ sub-channels. Also, assume that the number of STAs is extremely large than the number of sub-channels, i.e., $N \gg M$. The concept of the saturated condition is the same as in (Bianchi 2000) when WLAN carries the maximum load. It is assumed that every station always has some packets available for sending. This implies the input queue of each station in the wireless LAN is always non-empty in the saturation stage.

Like Bianchi (2000), several authors find the saturation throughput of traditional single-channel wireless LAN using the Markov chain model. However, in this thesis, the researcher attempts to find the saturation throughput of the multi-channel wireless

LAN. It is not feasible extending the single-channel Markov chain model for the OFDMA-employed multi-channel model. The equations in Bianchi (2000) reveal that the researcher examined the behaviors of a single STA in the wireless LAN and acquired the transmission probability ($\tau$) of a packet in an arbitrarily chosen time slot. As such, the author of this paper finds the probability of successful data transmissions from the side of the OFDMA sub-channels.

A time-frequency block is defined in Figure 5.2, where the time slots in the sub-channels are used for resource utilization. The number of time slots *(R)* for the successful transmission of the average packet payload *(E[p])* is $R = T_{total}/T_{slot}$, where $T_{total}$ represents the total average time, and $T_{slot}$ represents a single time slot. When an STA acquires an OFDMA sub-channel randomly in a time slot, then the successful transmission means that only this STA accesses this particular sub-channel. The probability of successful access in one sub-channel at each slot time is $p_k = 1/R$ due to the equal probability of all time slots under the assumption of the saturation conditions.

When an STA successfully transfers data in a sub-channel, for example, in the $j^{th}$ sub-channel in a time slot, then the successful transmission probability is defined by $P_{jsuc}$. Thus,

$$P_{jsuc} = \sum_{k=0}^{R-1} p_k \binom{N}{1}\tau(1-\tau)^{N-1}. \tag{5.1}$$

Under the assumption of the saturation stage, the probability of collision $P_{jcol}$ in the $j^{th}$ sub-channel must satisfy the following equation.

$$P_{jcol} = 1 - P_{jsuc}. \tag{5.2}$$

Since each STA could gain only one sub-channel at any time slot for sending the data in the saturation stage, the probability depends between $P_{1suc}$ and $P_{2suc}$. So, the probability of successful transmissions of *1$^{st}$* and *2$^{nd}$* sub-channels are as follows:

$$P(1suc, 2suc) = P_{1suc} P_{2suc|1suc} \tag{5.3}$$

$$= P_{1suc} \sum_{k=0}^{R-1} p_k \binom{N-1}{1}\tau(1-\tau)^{N-2}.$$

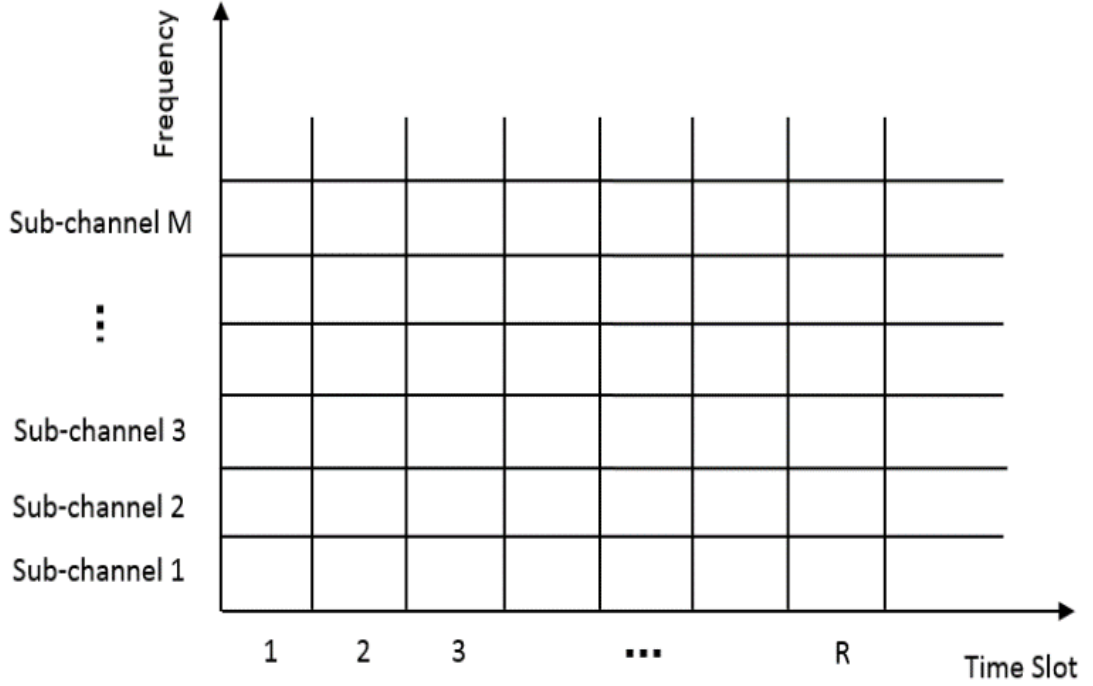

Figure 5.2: The time-frequency block

$P_s(i)$ and $P_c(i)$ denote the probability of the successful transmissions and collisions, respectively, considering the number of sub-channels in the OFDAM system is $i$.

Obviously, $P_s(1) = P_{1suc}$. In general, the probability of the successful transmission of $i$ sub-channels can be computed as below,

$$P_s(i) = \binom{M}{i} P(1suc, 2suc, \ldots, isuc)\, P\big((i+1)col, (i+2)col, \ldots, M\, col\big), \qquad (5.4)$$

We can get $P\big((i+1)col, (i+2)col, \ldots, M\, col\big)$ following the procedure of equation (5.3).

As previously assumed $N \gg M$; $P_{1suc}$ and $P_{2suc}$ are independent and also equal to each other. Thus, $P(1suc, 2suc)$ could be rewritten as below,

$$P(1suc, 2suc) = P_{1suc}\, P_{2suc} = P_{1suc}^2.$$

Simplifying equation (5.4) yields,

$$P_s(i) = \binom{M}{i} P(1suc, 2suc, \ldots, isuc)\, P\big((i+1)col, (i+2)col, \ldots, M\, col\big), \qquad (5.5)$$

$$= \binom{M}{i} P_{1suc}^{i} (1 - P_{1suc})^{M-i}, \tag{5.6}$$

The $P_s(i)$ can be used to find the average number of OFDMA sub-channels ($E_s$) that provide successful transmission at any single time slot as follow,

$$E_s = \sum_{i=0}^{M-1} i\, P_s(i). \tag{5.7}$$

Finally, the saturation throughput ($S$) is obtained as follow:

$$S = \frac{E[p].E_s}{T_{slot}}. \tag{5.8}$$

### 5.6.2 Analytical Model for ERA

The ERA protocol is developed to allocate the Resource Units (RUs) employing the SA mechanism for uplink transmission. In this subsection, the author designs an analytical model for the ERA where only the 802.11ax stations are eligible for receiving the SA RUs from the AP. There is no capture effect or collisions in the channel caused by the hidden nodes (Forouzan 2013 and Perahia & Stacey 2013), since all participating stations already have sent their BSR (Buffer Status Report) to the AP for gaining the scheduled access.

For the analytical model, a Wi-Fi 6 infrastructure network is considered that comprises one AP and $N$ STAs. It is assumed that the network is in the saturation stage, where all stations always have some packets to send. To start the uplink transmission, the access point sends the Trigger Frame (TF) to the stations. The TF contains necessary scheduling information, which is received from Algorithm 2. The TF also contains transmission information (e.g., transmission power, MCSs, transmission duration, etc.) and helps the stations to be synchronized with the system.

The duration of the TF cycle ($T$) is shown in Figure 5.3 that is calculated as follows,

$$T = T_H + (T_{TF} + SIFS + \delta) + (T_P + SIFS + \delta) + (T_{ACK} + SIFS + \delta). \tag{5.9}$$

where $T_H, T_{TF}, T_P,$ and $T_{ACK}$ represent the duration of the header field, trigger frame, payload, and acknowledgement frame transmission time, respectively. $\delta$ is the propagation delay, and $SIFS$ is Short Inter-frame Space duration.

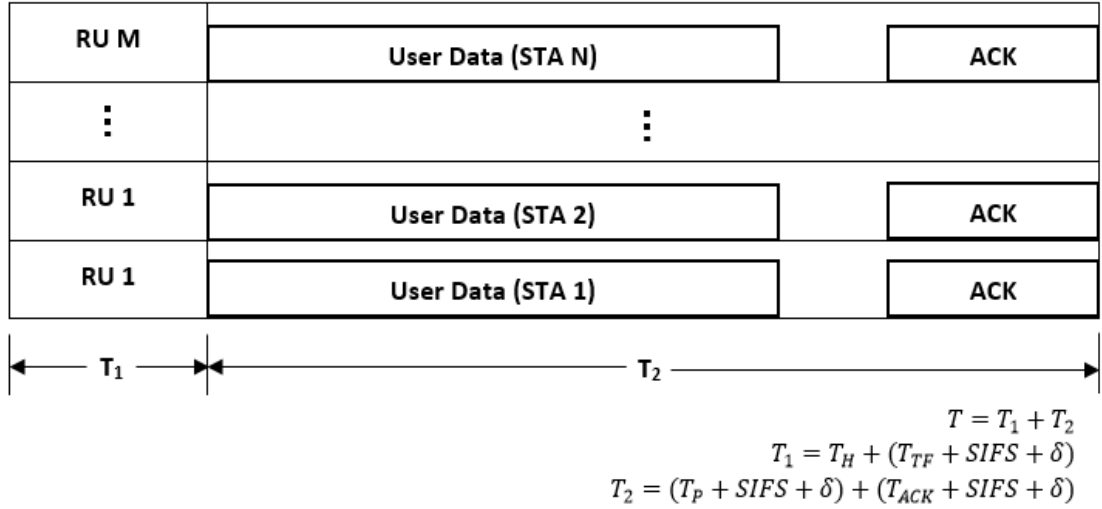

$$T = T_1 + T_2$$
$$T_1 = T_H + (T_{TF} + SIFS + \delta)$$
$$T_2 = (T_P + SIFS + \delta) + (T_{ACK} + SIFS + \delta)$$

Figure 5.3: Scheduled access in uplink transmissions leveraging OFDMA

The path-loss model is considered as described in Meinila et al. (2009), and the Path Loss ($PL$) is calculated by the following equation,

$$PL = A\, log_{10}(d[m]) + B + C\, log_{10}(f_c\,[GHz]/5.0) + X. \qquad (5.10)$$

where $d$ denotes the distance between the station and the access point; A, B, C, and X are the path-loss parameters which are described in Meinila et al. (2009).

Finally, the Saturation Throughput ($S$) is measured using the following equation,

$$S = \frac{M\, E[P]}{T}. \qquad (5.11)$$

where $M$ denotes the number of resource units, $T$ denotes the duration of the TF cycle represented by equation (5.9), and $E[P]$ is the average payload size in bits that can be estimated by the integral formula in equation (5.12),

$$E[P] = \frac{1}{Pmax - Pmin} \int_{Pmin}^{Pmax} f(x)\, dx. \qquad (5.12)$$

where $f(x)$ represents the function for the packet's payload size, $Pmax$ and $Pmin$ denote the maximum and minimum payload size, respectively.

### 5.6.3 Analytical Model for PRS

Being a hybrid protocol, the PRS protocol adopts both the SA mechanism and RA mechanism concurrently for accessing the Wi-Fi channels. Therefore, separate analytical models are required to evaluate the performance of this protocol. The investigator considers an infrastructure network that contains both SA and RA devices. There are only one access point and $N$ stations in the network that communicate through the access point. The network is in the saturation stage, where every STAs always has some packets for transmission.

Before protocol analysis, the investigator models the path loss for the PRS protocol in the indoor environment. The free space path loss ($PL_{FS}$) is defined (in dB) by the following equation, where $d$ denotes the distance between the transmitter and receiver in meter, and $f$ denotes the carrier frequency in hertz.

$$PL_{FS}(d) = 20 \, log_{10}(d) + 20 \, log_{10}(f) - 147.55. \tag{5.13}$$

IEEE 801.11n (Standards IEEE 2011 and Erceg et al. 2004) follows the following equation (5.14) for the path loss for indoors, which is also applicable for the IEEE 802.11ax (Standards IEEE 2021 and Liu et al. 2014). Erceg et al. (2004) and Rhodes (2003) define the indoor propagation model where path loss is having a slope of 2 up to a breakpoint distance ($d_{BP}$) and a slope of 3.5 after the breakpoint distance. Thus, the path loss for indoors ($PL$) as follows,

$$PL(d) = PL_{FS}(d), if \ d \leq \ d_{BP}$$

$$PL(d) = PL_{FS}(d) + 35 \, log_{10}(d/d_{BP}), if \ d > d_{BP}. \tag{5.14}$$

IEEE 802.11ax (Liu et al. 2014) specifies that extra floor penetration loss ($PEL_{floor}$) and wall penetration loss ($PEL_{wall}$) should be added to the path loss in the equation in (5.14). Thus, the overall indoor path loss ($PL_{overall}$) is represented by equation (5.15) as follows,

$$PL_{overall} = \ PL(d) + PEL_{floor} + PEL_{wall}. \tag{5.15}$$

*Analysis of SA Mechanism*

As discussed in the model of the PRS protocol in Section 4.4.1, the scheduled access method is applicable for only the SA-zone (green-zone) stations which contains only the SA devices. The SA mechanism leverages the OFDMA technology and hence supports multi-channel communication. Each individual RUs acts as a sub-channel, and STAs send data using their acquired RUs according to the PRS protocol. Figure 5.4 shows the uplink OFDMA transmissions using the SA method where for example, STA A gets RU 1, STA B gets RU 2, and so on. All RUs get the same amount of time to send the payload. All STAs start transmission at the same time after receiving the TF from the AP. The TF contains all necessary information, such as the scheduling and transmission information, to facilitate the OFDMA transmission (Islam & Kashem 2021). The duration of the TF cycle (T) is calculated as below (See Figure 5.4),

$$T = T_1 + T_2 = T_H + (T_{TF} + SIFS + \delta) + (T_P + SIFS + \delta) + (T_{ACK} + SIFS + \delta). \tag{5.16}$$

where $T_H, T_{TF}, T_P,$ and $T_{ACK}$ represent the duration of the header field, trigger frame, payload, and acknowledgement frame transmission time, respectively. $\delta$ is the propagation delay, and $SIFS$ is Short Inter-frame Space duration.

Finally, the saturation throughput can be measured using the following equation,

$$S = \frac{M\,E[P]}{T}. \tag{5.17}$$

where $M$ denotes the number of RUs for SA mechanism, and $E[P]$ is the average payload size in bits that can be estimated by the following integral formula,

$$E[P] = \frac{1}{Pmax-Pmin} \int_{Pmin}^{Pmax} f(x)\,dx \tag{5.18}$$

where $f(x)$ represents the function of the packet payload size, $Pmax$ and $Pmin$ denote the maximum and minimum payload size, respectively.

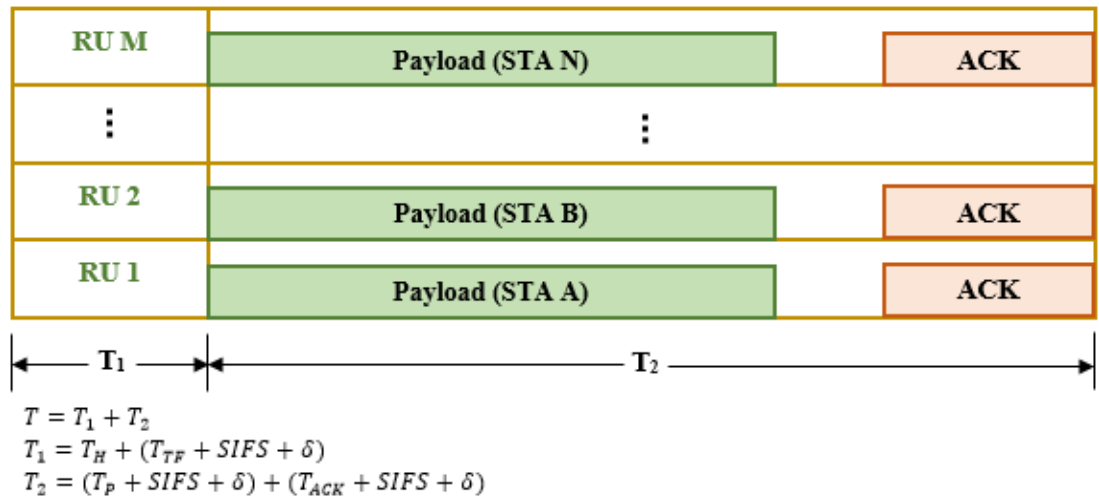

Figure 5.4: OFDMA transmissions using the SA mechanism

*Analysis of RA Mechanism*

The random access mechanism follows the DCF (Distributed Coordination Function) method that is standardized for the legacy Wi-Fi network. For efficient random access, DCF adopts the CSMA/CA (Carrier Sense Multiple Access / Collision Avoidance) mechanisms along with the BEB algorithm. The BEB algorithm restricts the maximum contention window size as $W_{max} = 2^{\alpha}W_{min}$ (details in Section 4.4.3). This assumption is to be kept for the Markov chain analysis for the RA mechanism. The candidate STAs for the RA mechanism are those which reside in the RA-zone (blue-zone) after applying the revised scheduling algorithm (i.e., Algorithm 4).

A lot of pieces of literature such as Tinnirello et al. (2010), Bianchi (2000), Ziouva & Antonakopoulos (2002), etc. are available that measure the performance of the random access MAC protocol. Based on the literature, the Markov chain model is adopted for the RA mechanism of the PRS protocol. The transition diagram for the contention window size is shown in Figure 5.5, where $p$ denotes the conditional collision probability of a packet in the RA-zone. The transition diagram is drawn using the Markov chain model for the 802.11 RA operation.

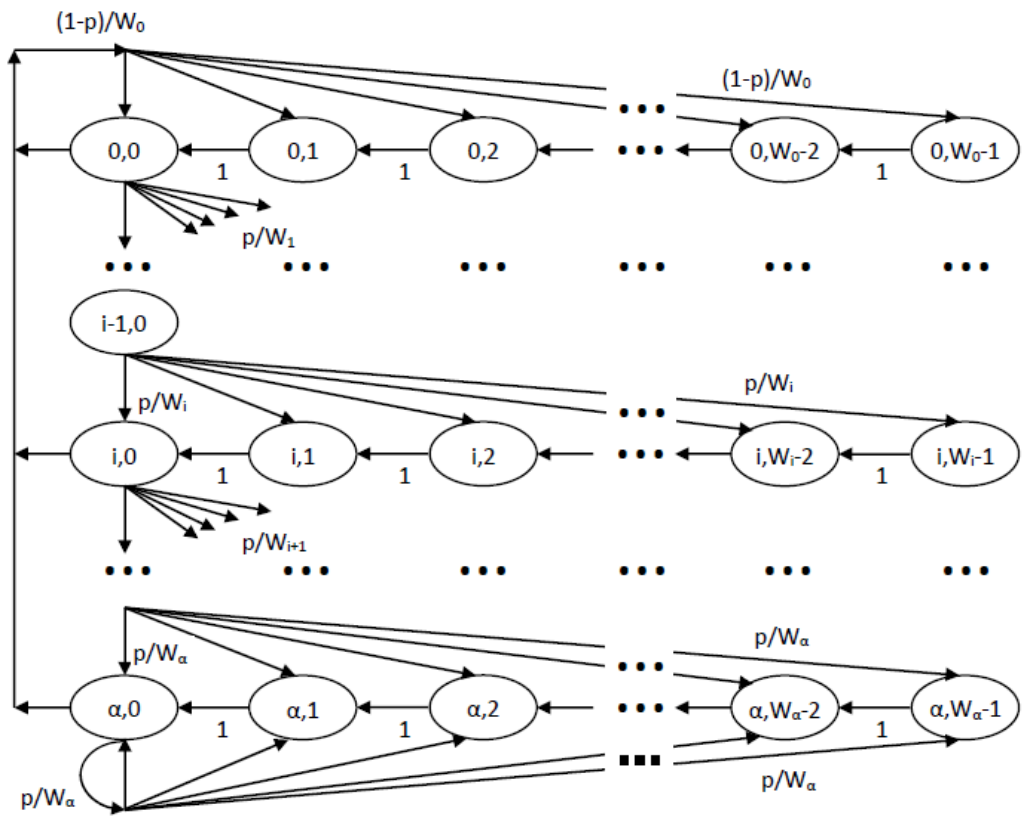

Figure 5.5: State transition diagram by the Markov chain model

Let $s(t)$ and $b(t)$ be the stochastic processes that represent the backoff stage and backoff time counter, respectively, for a given STA. In the Markov chain, the non-null one-step transition probabilities are represented by the equation (5.19). In equation (5.19), short notation is used as $P\{i_1, k_1 \mid i_0, k_0\} = P\{s(t+1) = i_1, b(t+1) = k_1 \mid s(t) = i_0, b(t) = k_0\}$.

$$\begin{cases} P\{i, k \mid i, k+1\} = 1 \text{ where } k \in [0, W_i - 2], i \in [0, \alpha] \\ P\{0, k \mid i, 0\} = (1-p)/W_0 \text{ where } k \in [0, W_0 - 1], i \in [0, \alpha] \\ P\{i, k \mid i-1, 0\} = p/W_i \text{ where } k \in [0, W_i - 1], i \in [1, \alpha] \\ P\{\alpha, k \mid \alpha, 0\} = p/W_\alpha \text{ where } k \in [0, W_\alpha - 1]. \end{cases}$$

(5.19)

Let $P_b$ denotes the probability the channel is busy, and $P_S$ denotes the probability that a successful transmission occurs in a slot time. The $P_b$ and $P_S$ can be expressed by the equations (5.20) and (5.21) respectively as follows,

$$P_b = 1 - (1-\tau)^N \tag{5.20}$$

$$P_S = n\tau(1-\tau)^{N-1} \tag{5.21}$$

where $\tau$ denotes the probability that an STA transmits in randomly chosen slot time, and $N$ is the number STAs for the RA mechanism. Then the normalized throughput can be measured as follow,

$$S = \frac{P_S E(P)}{(1-P_b)\sigma + P_S T_S + (P_b - P_S)T_C} \tag{5.22}$$

where,

$T_S$ denotes the average successful transmission time of the channel,

$T_C$ denotes the average collision time of the channel,

$\sigma$ denotes the duration of an empty slot time and

$E(P)$ denotes the average packet payload size, which can be expressed by the same integral formula represented by equation (5.18) in the SA mechanism.

## 5.7 Conclusion

This chapter explains the research methodology employed for this work. Section 5.1 introduces the chapter with a brief overview of the research. A standard target network model of Wi-Fi is portrayed in Section 5.2. The working principles of this target network lie in the models of the protocols, which are illustrated throughout the previous Chapter 4. The analytical models and simulation parameters of the protocols depend on this target network.

Section 5.3 of this chapter gives a glimpse of the simulation parameters and samples for the experiments. All of the samples for computer simulation are generated by well-known NS-3 software. The hardware and software required for the simulation are described in Section 5.4. Two types of procedure, i.e., 'Mathematical Evaluation' and 'Computer Simulation' are used to measure the efficiency of the protocols. Subsection 5.5.1 introduces the mathematical model, and Subsection 5.5.2 lists the experiments for this work. The analytical models of all novel protocols are explained mathematically using suitable figures and equations throughput Section 5.6.

# CHAPTER SIX

# Chapter 6

# Data Analysis

## 6.1 Introduction

In this paper, three OFDMA-based IEEE 802.11 protocols are innovated to provide a high-speed Wi-Fi network. The concept of the protocols, along with the related theories, are detailed in Chapter 4. To validate the protocols, it is necessary to measure the efficiency of the new protocols using state-of-the-art methods. The researcher used two such methods (i.e., mathematical analysis and computer simulation) to evaluate the protocols. The details of the methods are discussed in Chapter 5.

This chapter contains the data analysis of computer simulations for measuring the efficiency of the protocols. The simulation results of the new protocols, namely, HTFA, ERA, and PRS, are analyzed exhaustively with the help of graphs. Section 6.2, Section 6.3, and Section 6.4 describe the data of HTFA, ERA, and PRS protocols, respectively. All simulations are conducted using the simulator NS-3 (Network Simulator 2017).

## 6.2 Analysis of HTFA Protocol

In this section, at first, HTFA (High Throughput and Fair Access) protocol is evaluated for different simulation parameters. Then HTFA is compared with several prominent protocols designed by other researchers.

### 6.2.1 HTFA Evaluation

The researcher conducted several experiments to measure several performance parameters such as the saturation throughput, collision probability, etc. of the HTFA protocol. The experimental parameters for experiments 1 to 4 are listed in Table 6.1. For all of the experiments, the transmission failures resulting from the channel errors are not considered.

Table 6.1: Simulation parameters for HTFA protocol evaluation

| Parameters | Value |
|---|---|
| Packet transmission time | 2.5 ms |
| DIFS duration | 110 μs |
| Backoff slot duration | 50 μs |
| Number of stations ($N$) | 10 |
| Number of backoff stages ($\alpha$) | 6 |
| Minimum contention window ($W_{min}$) | 32 |

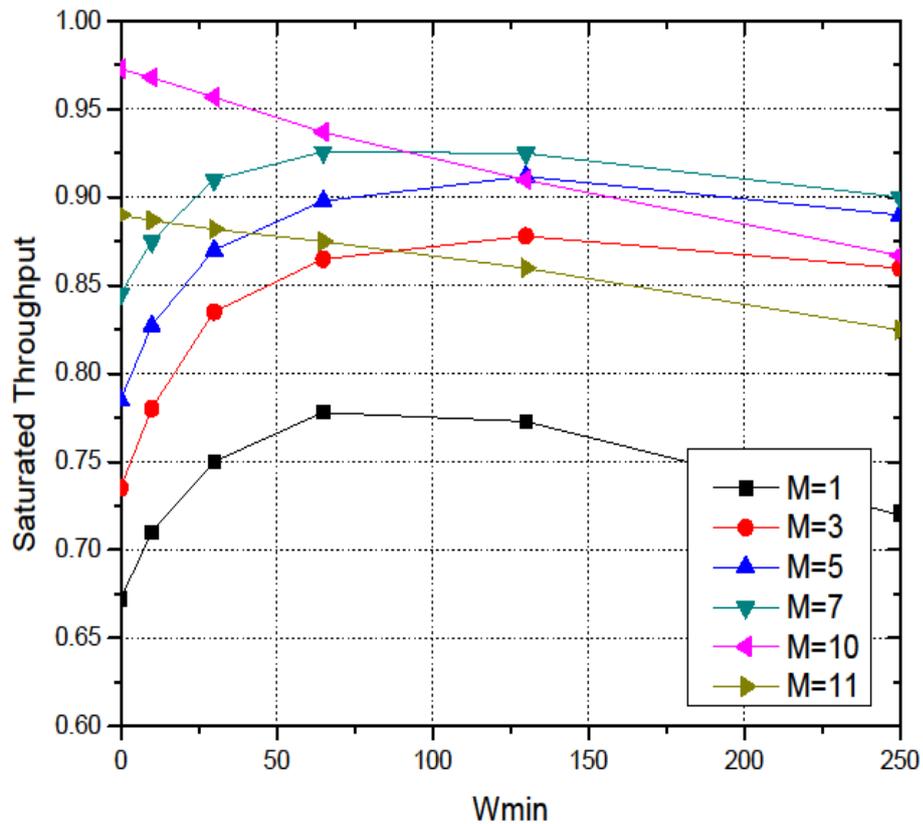

Figure 6.1: Saturated throughput with respect to $W_{min}$

**Experiment 1:**

Figure 6.1 shows the relationship between the saturation throughput and minimum contention window size ($W_{min}$) for a different number of sub-channels ($M$) of the HTFA protocol. In this simulation, 10 STAs are used for a different number of sub-channels such as $M = 1, 3, 5, 7, 10,$ and 11. The diagram shows that the maximum saturated throughput of the system is increasing gradually until the number of sub-channels increases to the number of STAs. The saturated throughput reduces when the number of sub-channels surpasses the number of STAs. This happens due to the redundant sub-channels that are not utilized exhaustively and also the presence of the channelization overhead. The diagram shows the irregular behavior for the higher number of sub-channels, such as for $M = 10$ and 11. Because if the number of the sub-channels is equal to or greater than the number of STAs ($M \geq N$), increasing the contention window size also enlarges the idle time while hardly contributing to reducing the collision probability. This incident suggests that when M approaches N, the protocol should employ a small number of backoff slots for the resolution of collisions as the probability of collision is very low for a large number of sub-channels.

**Experiment 2:**

Figure 6.2 shows the relationship between the collision probability and the minimum contention window size ($W_{min}$) for the different number of OFDMA sub-channels. The graph shows that the probability of collision reduces as $W_{min}$ increases and vice-versa. The graph also reveals that the collision probability decreases if the number of sub-channels increases for a particular contention window size. This happens because the more sub-channels in the system, the fewer contentions of the STAs for acquiring the medium. Thus, it is noticed that for the higher number of sub-channels, the probability of collision is zero because the number of STAs (i.e., $N = 10$) is less than or equal to the number of sub-channels (i.e., $M = 10$ and 11).

**Experiment 3:**

The throughput of the HTFA protocol with respect to the average number of active STAs under non-saturated load is shown in Figure 6.3. The diagram shows that the throughput enhances until the number of active STAs ($N$) surpasses the number of sub-channels ($M$) and reduces gradually beyond that because collisions occur more frequently among the STAs. As throughput loss due to the redundant sub-channels is considerably larger than that caused by the frame collisions, it is desirable to keep the number of OFDMA sub-channels slightly smaller than the average number of active STAs.

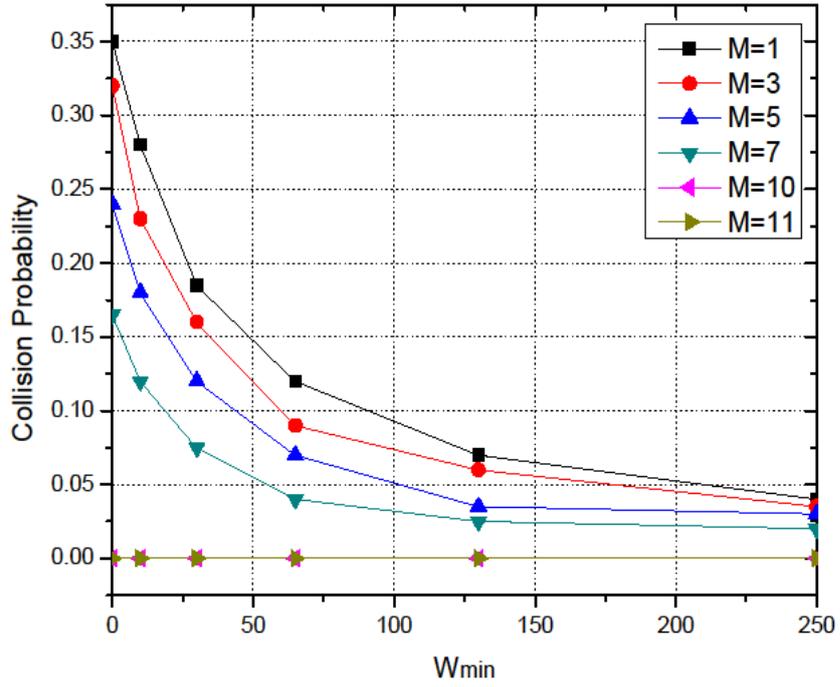

Figure 6.2: Collision probability with respect to $W_{min}$

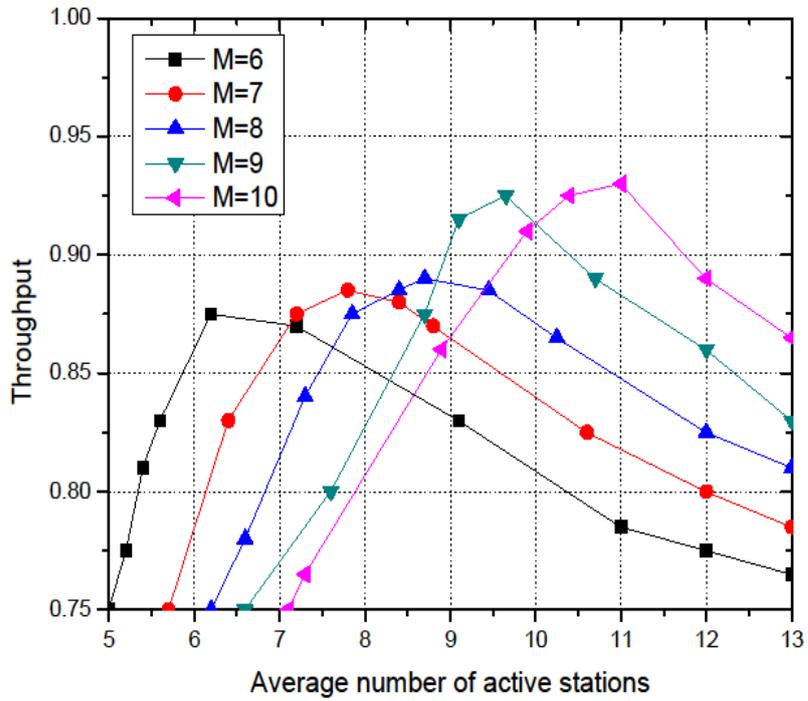

Figure 6.3: Relation between the throughput and average number of active STAs

Based on the above findings, it is advised to adopt a system where the number of sub-channels would be fixed adaptively according to the number of active STAs in the network. Thus, an adaptive control mechanism could be incorporated in such a way that the access point first calculates the number of active STAs in the system, then determines the number of optimum sub-channels from the calculation, and finally announces the result to the participating STAs through the control channel using the beacon messages.

**Experiment 4:**

The researcher further investigates the impact of increasing the backoff slot duration on the performance of the new OFDMA-adopted multi-channel hybrid system. So, he examines the saturation throughput of the wireless LAN with respect to the different backoff slot sizes. Figure 6.4 shows the result for a system of 10 STAs where the slot size ranges from 20-120 μs. As expected, the saturated throughput decreases as the backoff slot size incremented and vice-versa. Still, the OFDMA-adopted system provides more throughput than the conventional single-channel system up to a certain value of the backoff slot duration, for example, up to about 82 microseconds for *M*=3. It is also noted, if the number of OFDMA sub-channels increases, the outperforming range of the backoff slot size (i.e., slot duration) also rises.

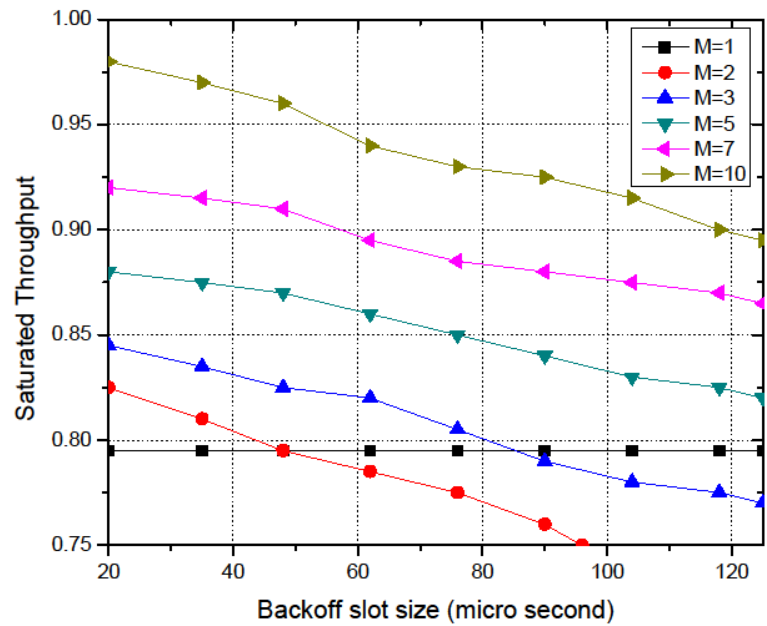

Figure 6.4: Saturated throughput with respect to the backoff slot duration

### 6.2.2 Protocol Comparison

Let $l_i$ denotes the traffic load, and $t_i$ denotes the achieved throughput of the $i^{th}$ STA. Obviously, $t_i \leq l_i$. As the traffic loads of different STAs may vary remarkably, the researcher wishes to find the normalized throughput denoted by $t_i/l_i$ for a fair comparison of the protocols. Here, two different metrics are used to compare the performance of the HTFA protocol with some other promising protocols. These two metrics, i.e., the system throughput ($T$), and the max-min fairness ($F$), are represented by the following equations:

$$T = \sum_{i=1}^{N} t_i \qquad (6.1)$$

$$F = \max \frac{t_i}{l_i} - \min \frac{t_i}{l_i} \qquad (6.2)$$

The simulation parameters for protocol comparison are shown in Table 6.2. It is assumed that the wireless STAs create packets in accordance with the Poisson distribution, and different wireless STAs may have various traffic loads. The investigator compares the new HTFA protocol with the CM-CSMA/CA protocol designed by Wang & Wang (2010) and SRMC-CSMA/CA protocol designed by Xu et al. (2013) in the following scenario.

**Experiment 5:**

The investigator conducts simulations for three STAs with loads 12, 18, and 24 Mbit/s, respectively, which are known as low-load STA, medium-load STA, and high-load STA, respectively. The total bandwidth of the channel is 54 Mbit/s, and it is divided into 3 smaller sub-channels. The sub-channels are of equal length and hence each of the sub-channels receives a bandwidth of 54/3 = 18 Mbit/s. In this experiment, the normalized throughput of the STAs having different loads is measured for selected 3 protocols and is displayed in the bar chart of Figure 6.5. It is observed that the normalized throughput of the CM-CSMA/CA protocol decreases sharply with the increase in the traffic load.

Table 6.2: Simulation parameters for protocol comparison

| Parameters | Value |
|---|---|
| Packet length | 1500 bytes |
| Number of stations ($N$) | 3 |
| Number of sub-channels ($M$) | 3 |
| Total channel bandwidth ($B$) | 54 Mbit/sec |
| Minimum contention window ($CW_{min}$) | 32 |
| Maximum contention window ($CW_{max}$) | 1024 |
| Backoff Slot time | 10 µs |

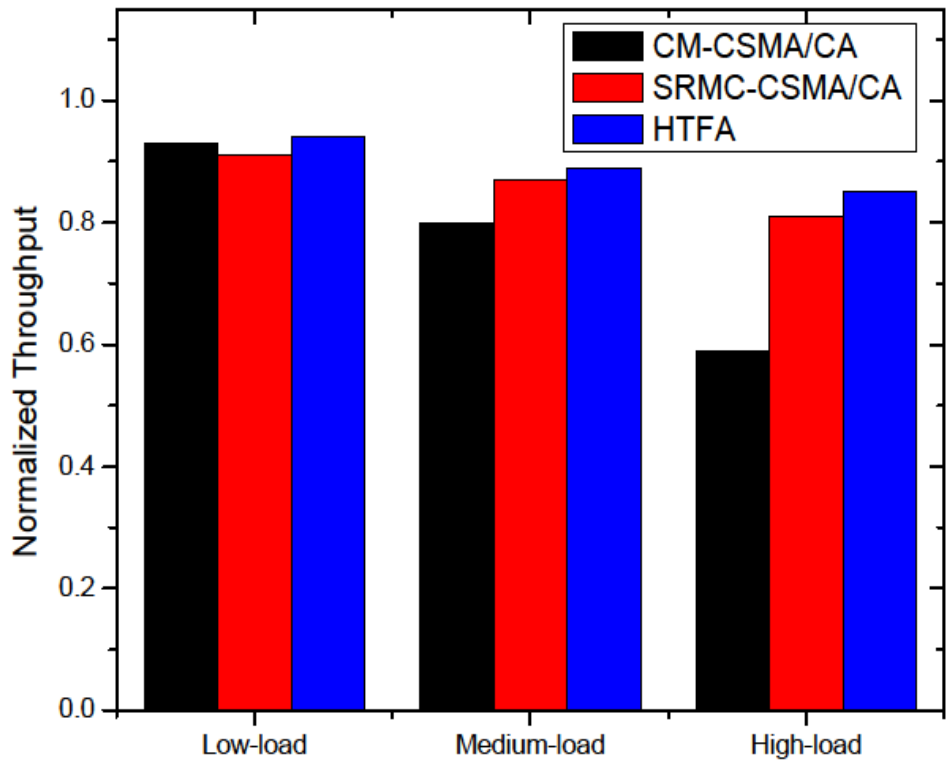

Figure 6.5: Normalized throughput of the STAs with different traffic loads

Table 6.3: Total system throughput and max-min fairness

| Performance Metric | CM-CSMA/CA | SRMC-CSMA/CA | HTFA |
|---|---|---|---|
| System throughput ($T$) | 41.20 Mbps | 47.6 Mbps | 49.3 Mbps |
| Max–min fairness ($F$) | 0.30 | 0.07 | 0.05 |

According to the CM-CSMA/CA model, an STA could access only one sub-channel in most of the cases, and thereby the normalized throughput of the high-load STA does not exceed 0.60. However, the HTFA and SRMC-CSMA/CA protocols permit an STA to access many sub-channels simultaneously. So, the normalized throughput of the high-load STA of the HTFA and SRMC-CSMA/CA protocol is not restricted by the bandwidth of an individual OFDMA sub-channel. It is observed that the normalized throughput of all three types of load (low, medium, and high) is above 0.81 of the HTFA and SRMC-CSMA/CA protocol. However, the HTFA protocol performs slightly better than the SRMC-CSMA/CA protocol due to the fewer contentions and fewer collisions ensured by the HTFA's mechanisms.

**Experiment 6:**

The researcher measures the overall system throughput and the max-min fairness of the intended three protocols, and the outcome is shown in Table 6.3. The STAs leveraging the HTFA mechanisms need not compete for the sub-channel acquisition rather than the sub-channels are dedicated to the STAs if the number of participating STAs ($N$) in the system is smaller or equal to the number of sub-channels ($M$). So, when, $N \leq M$, the STAs monopolize the sub-channels as each of the STAs gets at least one sub-channel. Thus, in this case, the random backoff slot is not required, which reduces the delay. As a result, the HTFA protocol delivers higher throughput than the CM-CSMA/CA and SRMC-CSMA/CA protocol. As the HTFA protocol adopts a hybrid mechanism for distributing the sub-channels among the STAs, its max-min fairness is also promising than the CM-CSMA/CA and SRMC-CSMA/CA protocol.

## 6.3 Analysis of ERA Protocol

In this section, the efficiency of the Efficient Resource Allocation, i.e., ERA protocol is evaluated and compared with some promising protocols and methods. Again, the robust NS-3 simulator is chosen to conduct the experiments. The performance of the protocols is measured for the uplink OFDMA transmissions. An IEEE 802.11ax infrastructure basic service set is considered that operates in a 40 MHz channel at 5GHz band. The path-loss model for the ERA is represented by equation (5.10), where the channel is modeled according to the WINNER II model as described in Meinila et al. (2009). All terminals in the network use a power of 15 dBm for data transmission. The experiments are set up considering the analytical model depicted in Section 5.6.2. The simulation parameters are shown in Table 6.4. All experiments hold the same parameters unless otherwise specified.

### 6.3.1 Protocol Comparison

The investigator compares several performance parameters of the ERA protocol with that of Greedy designed by Wang and Psounis (2018), SRTF designed by Bankov et al. (2018), and MUTAX designed by Bankov et al. (2017). These protocols are very promising than other existing protocols. Details of these protocols are described in Section 2.6 in the 'Literature Review' chapter.

**Experiment 1:**

At first, the researcher measures the throughput of the competing protocols. Figure 6.6 shows the average throughput in Mbps of the four protocols. It is evident from the chart that the ERA ensures the highest throughput, i.e., 238 Mbps among the competitors. On the other hand, the throughput of the Greedy scheduler is the lowest, which is only around 196 Mbps. It happens due to the assignment of a smaller portion of the RUs to the STAs. The throughput of the SRTF algorithm is also not satisfactory due to the exhaustive service. By the way, the throughput of the MUTAX that employs a channel-splitting scheduler is significantly higher than the throughput of the Greedy and SRTF. The throughput of the ERA is maximum among all schedulers because of its dynamic and adaptive scheduling that is based on the load of the STAs (explained in Section 4.3).

Table 6.4: General simulation parameters for ERA

| Parameters | Value |
|---|---|
| Number of stations ($N$) | 25 |
| Radius of the basic service set ($r$) | 15 meters |
| Number of antennas of the AP ($A_{AP}$) | 4 |
| Number of antennas of the stations ($A_{STA}$) | 1 |
| Transmit power | 15 dBm |
| Frequencey band | 5 GHz |
| Channel bandwidth | 40 MHz |

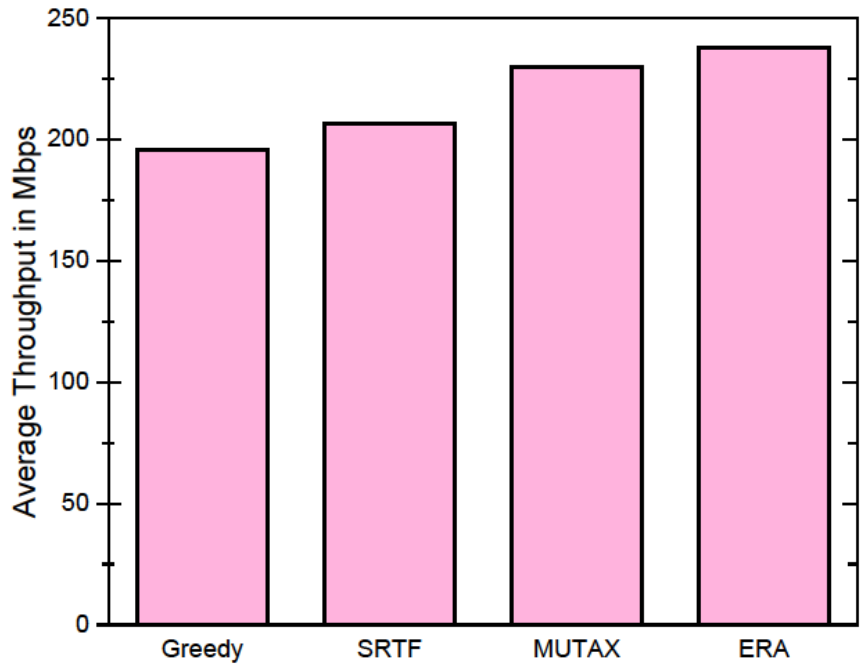

Figure 6.6: Comparison of the throughput between competing protocols

**Experiment 2:**

Figure 6.7 compares the average throughput of the protocols for several congestion environments, namely, low, medium, and high that contain 5, 20, and 60 stations, respectively. Again, the ERA performs the best, whereas the Greedy performs the worst due to the causes discussed in the previous experiment. It is noticed from the chart of Figure 6.7 that the average throughput increments slightly as congestion increases and vice versa. The reason is that a larger number of stations are accessing the Wi-Fi network in the same area (r = 15 meters). It indicates that more stations approach the access point, thereby reducing the communication distance, and hence the stations receive stronger signals. It is also observed that the MUTAX protocol comparatively performs best in high congestion than its medium and low congestion environments. For instance, from low to high congestions the throughput of the MUTAX increased by (238 – 225.5) Mbps = 12.5 Mbps (i.e., 5.54 %) whereas the throughput of the SRTF increased by (212 – 202) Mbps = 10 Mbps (i.e., 4.95 %).

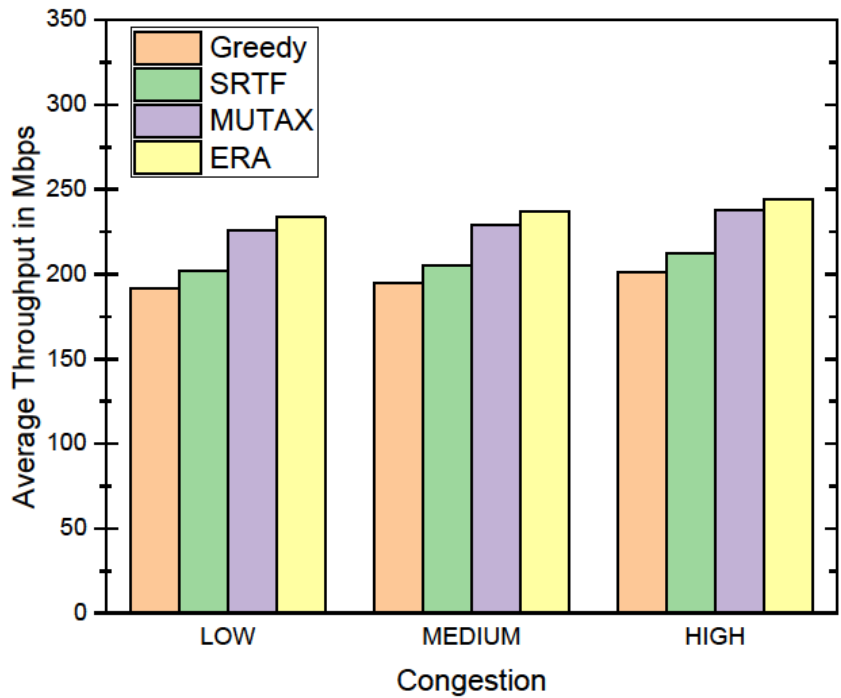

Figure 6.7: Impact of the number of stations on the throughput

**Experiment 3:**

Figure 6.8 displays the impact of the number of antennas of the access point ($A_{AP}$) on the average throughput. For this experiment, the investigator varies the number of antennas of the access point for observing the changes in the throughput of the four competing protocols. The number of antennas of the stations ($A_{STA}$) is always kept to only 1 for all experiments conducted for the ERA protocol as the stations (especially the smartphones) are low-power devices and are also small in size.

The bar chart of Figure 6.8 depicts that an increment in the values of $A_{AP}$ increments the average throughput as well. However, the changes in the throughput are not linear since the total transmit power is constant. For instance, an increment of the number of antennas of the access point from 2 to 4 improves the throughput by 25 Mbps of the ERA protocol while it is only around 16.5 Mbps for incrementing the number of antennas from 4 to 6 for the same protocol. The same explanation applies to the remaining competing schedulers.

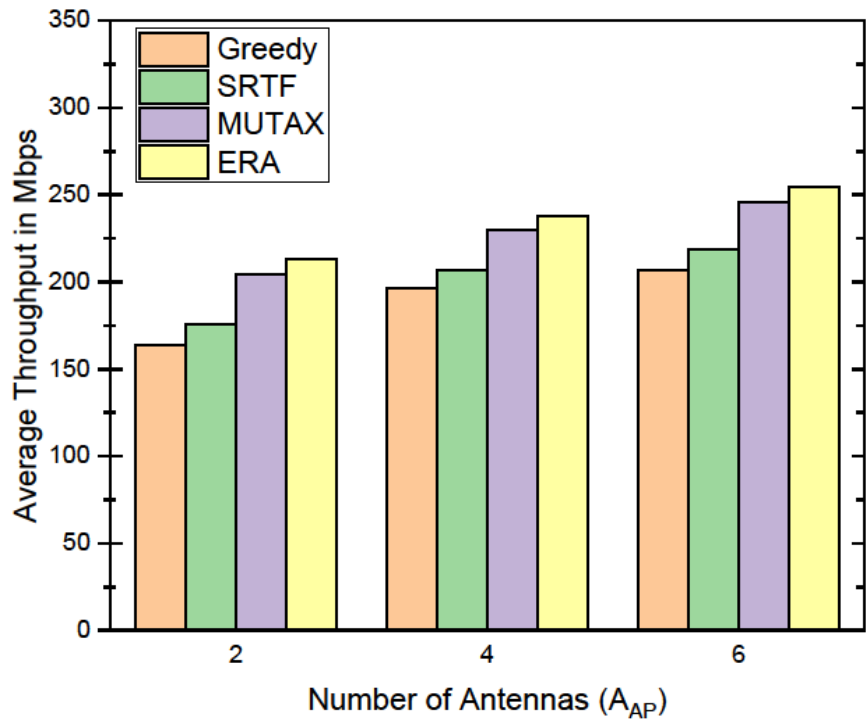

Figure 6.8: Impact of the number of antennas of the AP on the throughput

**Experiment 4:**

Another experiment is conducted to perceive the impact of the number of stations on the average upload time. For this simulation, the flow sizes are drawn from the truncated lognormal distribution, where packet sizes are distinguished as the minimum, average, and maximum that contain the data 1 KB, 500 KB, and 5 MB, respectively. After delivering a flow, the next flow is generated after a random delay drawn from the truncated exponential distribution with the minimum, average, and maximum values of 0.1, 0.3, and 0.6 seconds, respectively. Figure 6.9 portrays that the upload time increases exponentially with the increase of the number of stations and vice versa. For a fewer terminals, the difference in upload time between the algorithms is not significant. However, for a larger number of terminals, the difference in the upload time is remarkable. The MUTAX and SRTF algorithms are specially designed to minimize the upload time. Hence, these two protocols take lesser time (e.g., 0.60 s to 0.66 s for 60 stations) than other competitors (i.e., Greedy, and ERA).

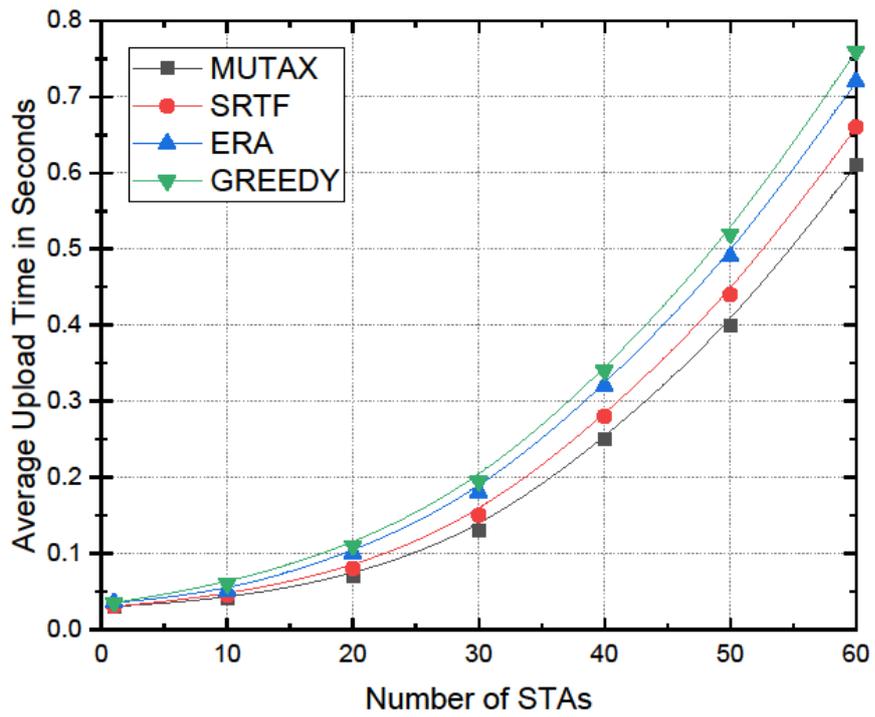

Figure 6.9: Upload time vs number of stations

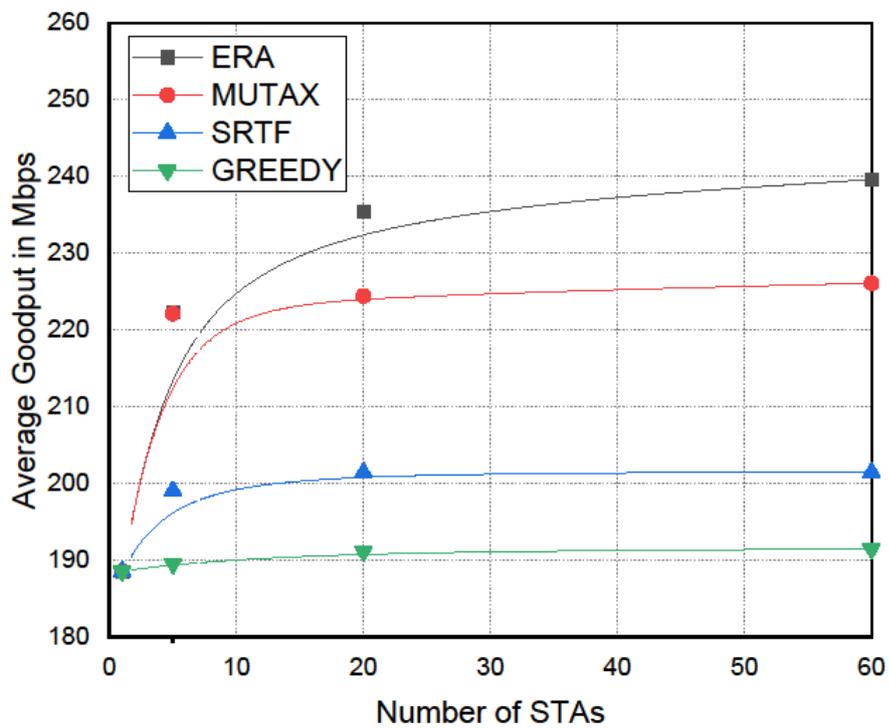

Figure 6.10: Goodput vs number of stations

**Experiment 5:**

The investigator conducts another experiment to analyze the goodput of the protocols for a different number of STAs, keeping the simulation parameters the same as the previous experiment. Figure 6.10 displays the average goodput of the competing schedulers for the number of STAs 1 to 60. As expected, the ERA shows the best performance among all participating protocols for its innovative algorithm that is designed solely for the ax STAs. For only one STA, all schedulers give almost the same goodput, which is around 188.5 Mbps. However, the difference in goodput becomes very high when a larger number of STAs join the network. For instance, when the number of STAs is 20, ERA ensures a goodput of 235.34 Mbps while Greedy gives around 191.1 Mbps only. Again, when the number of STAs is 60, ERA provides an extremely high goodput of around 240 Mbps while SRTF gives only 201.4 Mbps.

### 6.3.2 Comparison between Methods

The purpose of the last experiment is to examine the number of retransmissions for a different number of terminals of numerous methods. Lanante et al. (2017) discussed various generalized methods adopted by different MAC protocols.

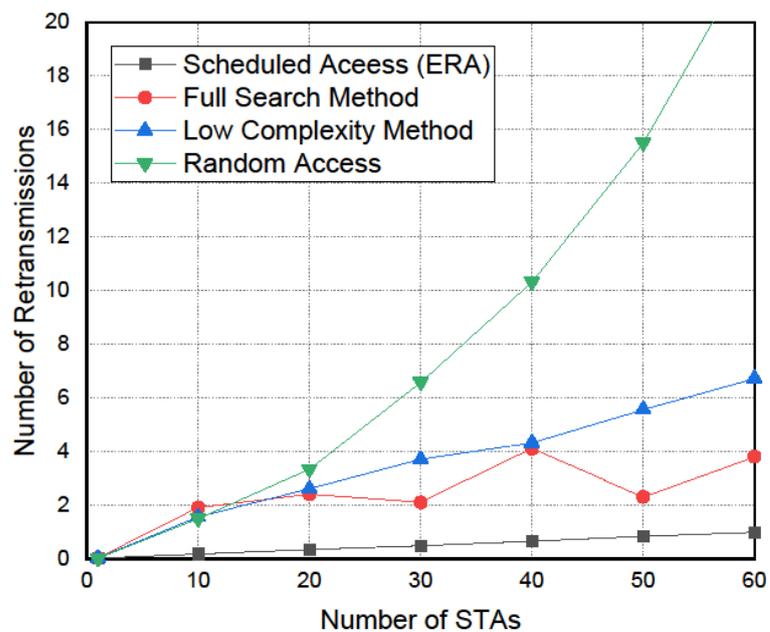

Figure 6.11: Number of retransmissions vs number of stations

**Experiment 6:**

The author categorizes the MAC protocols into four main types according to their design for the experimental purpose. Those are Scheduled Access, Random Access, Full Search, and Low Complexity. The novel ERA scheduler is a pure scheduled access protocol that keeps no provision for random access. It is evident from Figure 6.11 that the performance (in terms of packet retransmissions) of the random-access protocols is the worst among all methods. On the other hand, the scheduled access protocols such as the ERA displays the best performance.

The main reason for the retransmissions of packets is the presence of hidden nodes in the surrounding area. The hidden nodes degrade the performance of the random-access protocols severely as long as the number of stations increases. According to Figure 6.11, when there is only 1 station in the network, the number of retransmissions of all methods is almost 0 since there is no possibility of collisions between the packets. As the number of stations in the network is increasing gradually, the possibility of the presence of hidden stations in the surrounding areas is also increasing.

However, the hidden stations cannot impact the performance of the pure scheduled access protocols because there is no provision for random channel access here. That's why, even in a highly congested network (e.g., 60 stations), the number of retransmissions of the ERA protocol is below 1. On the other hand, for a similar scenario (e.g., 60 stations), the number of retransmissions of the random-access protocols is around 23. The only reason for the retransmissions of the ERA protocol is the transmission impairments. The transmission impairments are noise, distortion, attenuation, etc., which are the general propagation problems of the wireless channels and are negligible here. The number of retransmissions of full search and low complexity methods is not severe like the random-access mechanism, although the full search method depicts an unpredictable characteristic for different numbers of stations.

## 6.4 Analysis of PRS Protocol

A lot of simulations are conducted to measure the efficiency of the PRS protocol using the Network Simulator (NS-3). The investigator measures the performance for the uplink transmissions as PRS proposes the scheduling for the uplink purpose. The downlink scheduling in OFDMA is easier to implement (Islam & Kashem 2020 and Lee 2019) and does not be considered for the simulations. In the first two simulations (in Subsection 6.4.1), the researcher evaluates the PRS protocol for different proportions of SA and RA STAs, and in the rest of the simulations (in Subsection 6.4.2), the PRS is compared with other contemporary promising protocols. All simulations are conducted for the indoor network, and the network is in the saturation stage. The simulations follow the analytical models developed in Section 5.6.3. In general, all experiments are regulated by the simulation parameters shown in Table 6.5 unless otherwise specified.

### 6.4.1 Comparison between Access Mechanisms

At first, the throughput and number of retransmissions of the PRS protocol are measured for different distributions of STAs according to their access mechanisms. The distributions are classified as SA, RA, and Hybrid. In the SA distribution, all STAs belong to the SA-zone (green-zone) that utilizes the SA mechanism. Similarly, in the RA distribution, all STAs belong to the RA-zone (blue-zone) that utilizes the RA mechanism. In the Hybrid distribution, half of the STAs belong to the SA-zone (green-zone), and another half of the STAs belong to the RA-zone (blue-zone). According to the revised scheduling algorithm (i.e., Algorithm 4 in Section 4.2), although there may produce a variety of distributions of the STAs, only three distributions (i.e., SA, RA, Hybrid) are considered here for the simplicity and to observe the superiority of the SA mechanism over the RA mechanism in terms of throughput and retransmissions.

Table 6.5: Typical simulation parameters for PRS

| Parameters | Values |
| --- | --- |
| Number of stations (N) | 60, 35, 10 for high, medium, and low congestion, respectively |
| Radius of BSS (r) | 15 meters |
| Transmit power | 15 dBm |
| Frequency band | 5 GHz |
| Channel bandwidth | 40 MHz |
| Number of antennas of the stations ($A_{STA}$) | 1 |
| Number of antennas of the access point ($A_{AP}$) | 4 |

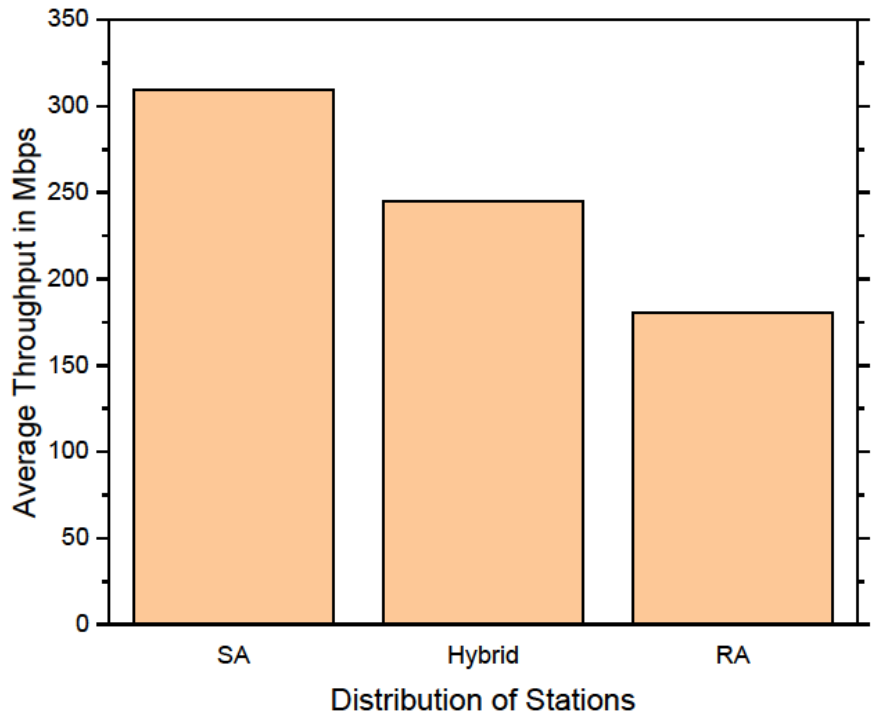

Figure 6.12: Throughput of the PRS protocol for different distribution of stations

Figure 6.12 shows the average throughput for the different distribution of stations. The simulation considers a dense environment (i.e., 60 STAs). Thus, in the SA distribution, all 60 STAs utilize the SA mechanism, in the RA distribution, all 60 STAs utilize the RA mechanism, and in the Hybrid distribution, half of the stations (30 STAs) uses the SA, and another half of the stations (30 STAs) uses the RA mechanism. It is observed that SA distribution provides the highest throughput, i.e., 309.2 Mbps, while RA distribution provides only around 180.6 Mbps. In the SA distribution, all 60 STAs are utilizing the scheduled access mechanism using the latest OFDMA technology. In the SA mechanism, there is no possibility of collisions since the absence of the hidden nodes and also no need for the backoff window like the RA mechanism. Thus, the SA distribution significantly outperforms the RA distribution in terms of throughput. The Hybrid distribution provides approximately 245 Mbps which is around the average of SA and RA distributions as half of the STAs use the SA method and another half use the RA method for the channel access.

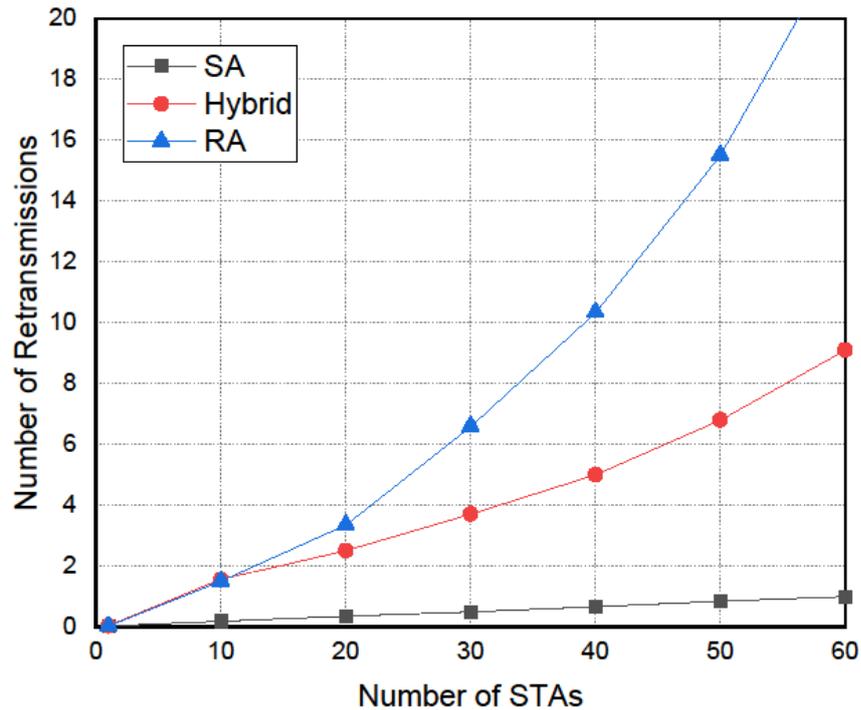

Figure 6.13: Number of retransmissions of the PRS protocol for different distribution of stations

Figure 6.13 shows the number of retransmissions for the different distribution of stations. As expected, SA distribution performs the best, and RA distribution shows the worst performance. The retransmission of packets reduces the goodput of the network, and it happens due to several reasons such as the collision of packets and transmission impairment. The causes of transmission impairment are noise, distortion, attenuation, etc., which are the general propagation problems of the wireless channels, and it may ignore because this impact is comparatively little than the collision problem. Again, the collision problem occurs to only random access mechanism due to two reasons. One is the presence of the hidden nodes, and another is the same backoff value produced by two RA STAs. It is observed in Figure 6.13, when the number of STAs is very low (e.g., 1 STA), the number of retransmissions is nearly zero for all of the distributions. As the number of STAs is increasing, the impact of RA distribution becomes severe while SA distribution shows excellent performance. Even in the dense scenario (i.e., 60 STAs), the number of retransmissions of SA distribution is below 1. As in the previous simulation, the impact of Hybrid distribution on the retransmission is in between the SA and RA distributions. The robustness of the OFDMA scheduled access over the legacy random access mechanism can be perceived by the above two simulations shown in Figure 6.12 and Figure 6.13.

**6.4.2 Protocol Comparison**

In the rest of the simulations (Figure 6.14 to Figure 6.18), the PRS protocol is compared with other promising protocols, namely, Divide and Conquer, Recursive, MR, and PF. As discussed earlier, MR and PF are basically the cellular protocols, which are adapted to the 802.11ax network. More about the competing protocols are discussed in Section 2.6.3. A medium congestion scenario (e.g., 35 STAs) is considered for the rest of the simulations in this article unless otherwise specified. During simulations, all protocols are executing only OFDMA transmissions, and all STAs are OFDMA-capable (i.e., SA STAs). Thus, during the comparison in the rest of the simulations, PRS leverages only the OFDMA SA mechanism.

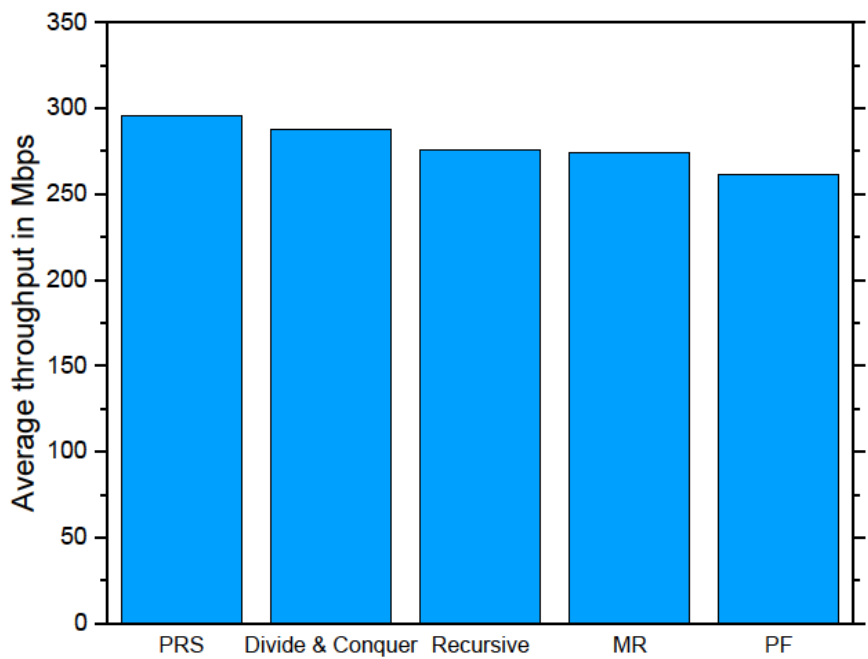

Figure 6.14: Throughput comparison between different protocols

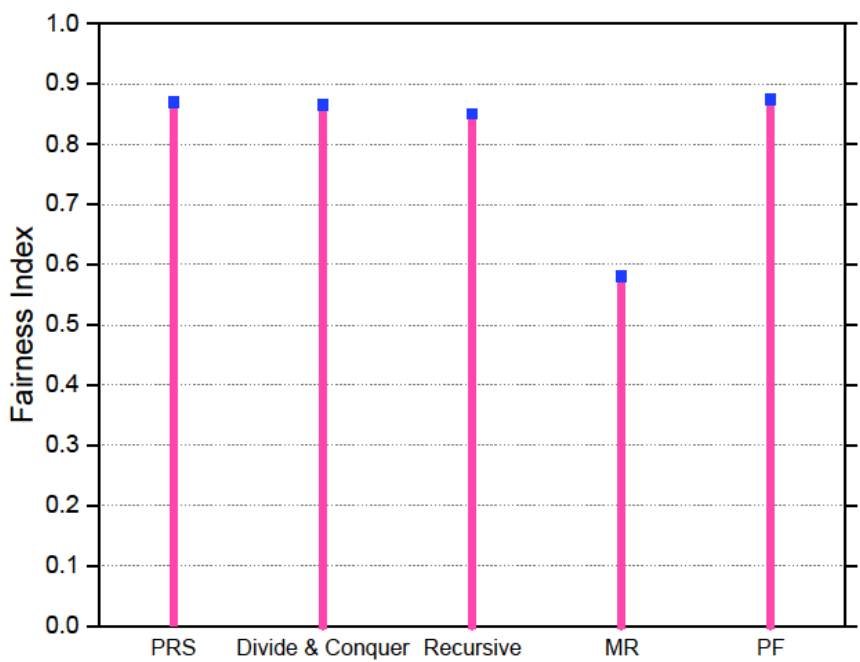

Figure 6.15: Fairness comparison between different protocols

The comparison of throughput between the PRS and other protocols is shown in Figure 6.14, where PRS provides the highest throughput, i.e., 295.7 Mbps. The PRS algorithm distributes the RUs proportionally to the SA stations and hence ensures efficient exploitation of RUs by the stations. The Divide and Conquer algorithm, which is closest to the PRS, also provides a very high throughput, i.e., 287.5 Mbps. The throughput gained by this algorithm is a tight upper bound for the optimal user schedule. Another ax scheduler, i.e., Recursive that schedules RUs to the STAs in a near-optimal fashion and adjusts the number of resource units dynamically, also provides a good throughput of 275.4 Mbps. The MR, which is very good in providing high throughput in the cellular network, is not the best for the Wi-Fi network, giving a throughput of around 273.8 Mbps. The PF, another cellular protocol designed for fair access to the medium, provides the lowest throughput among all competing protocols.

The fairness in accessing the Wi-Fi channel of the competing protocols is shown in Figure 6.15. Jain's fairness index is used for measuring the fairness of the protocols. The Proportional Fair, i.e., PF protocol shows the best performance in this regard. The PF, which is designed for the LTE network for increasing fairness, performs very well (i.e., index 0.874) in the Wi-Fi network as well. The PRS also ensures very good fairness (i.e., index 0.87) since it distributes RUs according to the available loads of the stations. The performance of Divide & Conquer and Recursive algorithms are also satisfactory since the algorithms solve the SRA problem optimally and hence ensure an index of 0.865 and 0.85, respectively. The MR performs the worst (i.e., index 0.58) as it ignores the fairness effect and focuses on the throughput in the cellular network.

Let's examine the overall performance of the protocols in Figure 6.14 and Figure 6.15. The overall performance of ax-based protocols (PRS, Divide and Conquer, Recursive) is better than the cellular protocols (PF, MR), although cellular protocols are adapted in Wi-Fi 6 environment during the simulation. These phenomena suggest that it is not good to hire the LTE protocols for Wi-Fi 6 network. The LTE network predominantly uses the OFDMA technology that IEEE 802.11ax has recently adopted. However, the OFDMA specifications are somewhat different for Wi-Fi 6 than those for cellular networks. That's why the overall performance of PF and MR is not satisfactory in the Wi-Fi network, which is validated by the last two simulations.

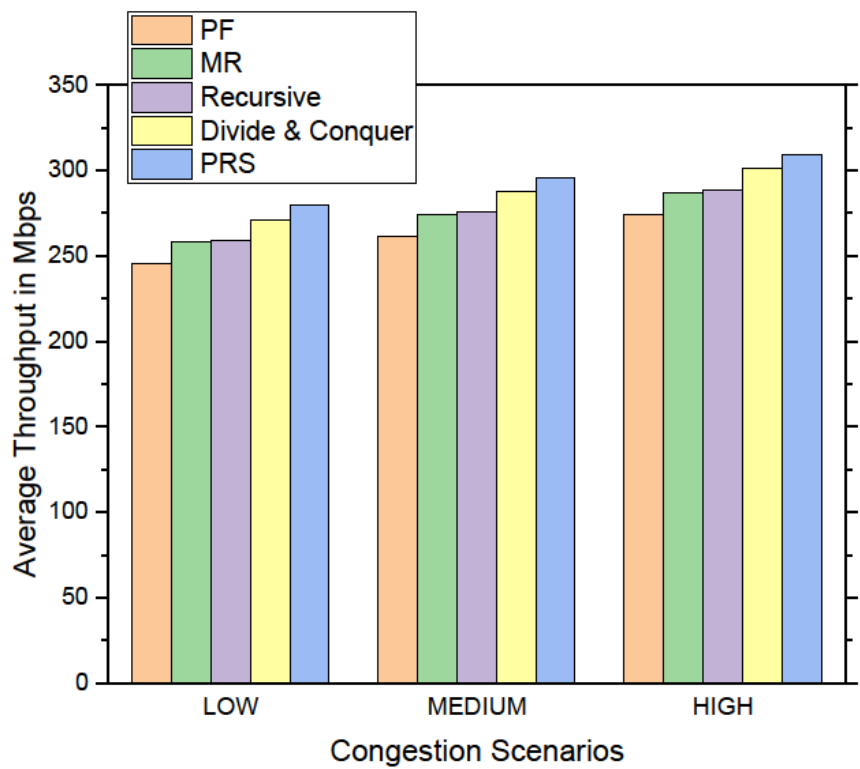

Figure 6.16: Impact of the number of stations on the throughput

Throughput for different distributions of stations of the competing protocols is compared in Figure 6.16. Three congestion scenarios, namely, Low (i.e., 10 STAs), Medium (i.e., 35 STAs), and High (i.e., 60 STAs), are considered here. The throughput of all protocols increases as the number of STAs increases and vice versa. When the number of STAs increases, more STAs reside in the same radius (r = 15 meters) and approach the access point. Thus, STAs get a stronger signal and increase the throughput thereby. For example, the PRS protocol provides a throughput of 279.5 Mbps, 295.7 Mbps, and 309.2 Mbps for Low, Medium, and High congestions, respectively. All protocols show similar performance (i.e., throughput enhancement) as the density of the STAs increases.

Figure 6.17 shows how the throughput varies according to the number of antennas of the access point ($A_{AP}$). The average throughput of the protocols is observed for the number of antennas 2, 4, and 6. The number of antennas of the STA ($A_{STA}$) has always been kept to the minimum value of 1 for all simulations conducted for PRS protocol since the STAs (especially the cell phones) are low-power devices and are small in size. It is evident from the chart that the throughput increases gradually with the number of antennas of the AP. However, this enhancement is not linear because the total transmit power is constant. For example, the throughput of MR protocol increases by 29.6 Mbps for incrementing the number of antennas from 2 to 4. On the other hand, this enhancement is only 22.1 Mbps for incrementing antennas from 4 to 6.

Figure 6.18 exhibits the goodput scenario of the participating protocols for a different number of STAs. The difference between the throughput and goodput is that goodput does not count the undesirable data (e.g., retransmissions, overhead, etc.). Unlike the random access, in OFDMA transmissions, the number of retransmissions is very low (illustrated in Figure 6.13); hence the goodput is almost the same as the throughput. The number of STAs is varied from 1 to 60 to observe the change in goodput provided by different protocols. As the number of STAs increases, the goodput also increases and vice versa. When the number of STAs is only one, the goodput of each of the protocols is almost the same, i.e., around 225 Mbps. As the number of STAs is increasing, the difference in goodput becomes significant. For example, when the number of STAs is 30, the difference between the goodput of PRS and PF protocols is (293.8 - 258.5) = 35.3 Mbps.

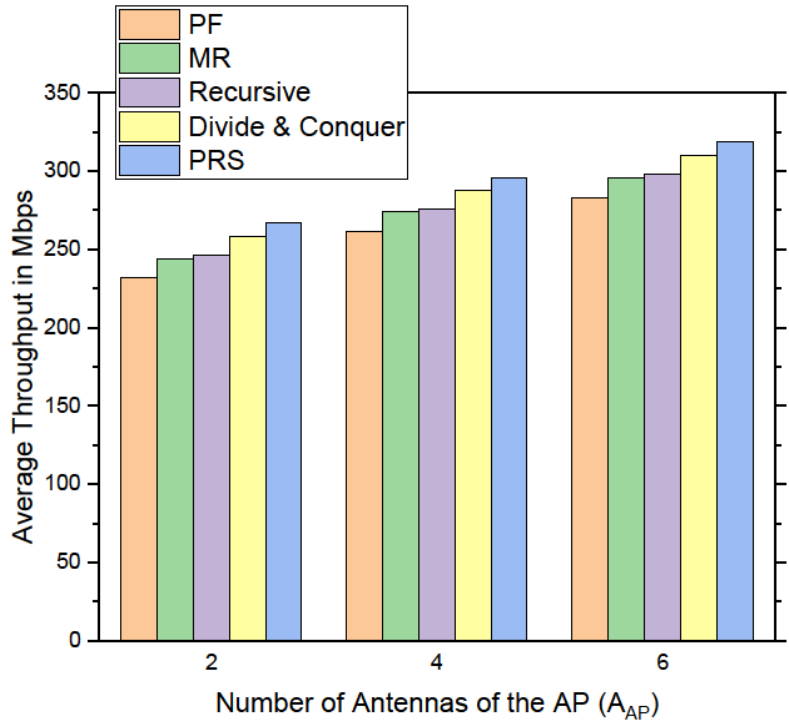

Figure 6.17: Impact of the number of antennas on the throughput

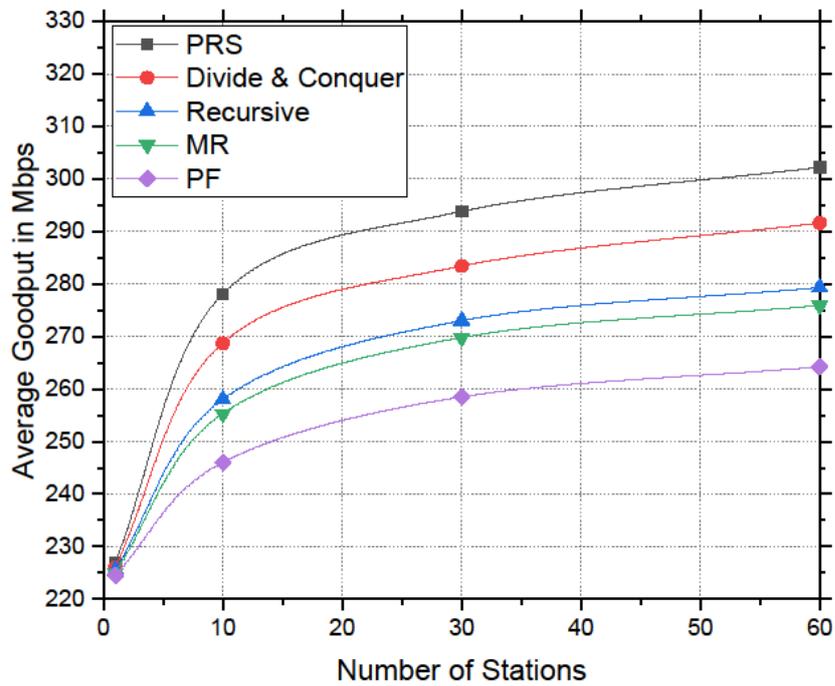

Figure 6.18: Goodput vs number of stations

## 6.5 Conclusion

This chapter contains data analysis of the computer simulations that validate the robustness of the protocols. The researcher conducts a good number of computer simulations for all three novel protocols presented in this thesis. Section 6.2 evaluates the HTFA protocol through several experiments such as the throughput vs. minimum contention window, collision probability vs. minimum contention window, throughput vs. average number of active stations, saturation throughput vs. backoff slot size, and so on. Section 6.2 also compares the HTFA with two other promising protocols, namely, CM-CSMA/CA designed by Wang & Wang (2010) and SRMC-CSMA/CA designed by Xu et al. (2013).

The experiments for the ERA protocol are described in Section 6.3. For evaluating the efficiency of the ERA protocol, several simulations are conducted, such as the throughput comparison between different competing protocols, the impact of the number of stations on the throughput, impact of the number of antennas of the access point on the throughput, upload time versus the number of stations, goodput versus the number of stations, number of retransmissions versus the number of stations.

Section 6.4 illustrates the results and experimental data of the PRS protocol. In this section, at first, the researcher measures the throughput and number of retransmissions of the PRS protocol for a different distribution of stations and proves that the SA mechanism is more efficient than the RA mechanism. Then the researcher compares the performance of the PRS with other promising protocols through several simulations, namely, throughput comparison between different protocols, fairness comparison between different protocols, the impact of the number of antennas on the throughput, goodput vs number of stations, etc. The data analyses in this chapter prove the robustness of the novel protocols. The contributions of the protocols are further described in the following Chapter 7.

# CHAPTER SEVEN

# Chapter 7

# Discussions and Significance of the Research

## 7.1 Introduction

In this chapter, the contributions and significant features of the protocols are discussed elaborately. The primary objective of all of the protocols is to enhance the throughput of the Wi-Fi network. Throughput enhancement of HTFA protocol is achieved by reducing the overheads, frame collisions, etc. which is described in Subsection 7.2.1. The HTFA protocol also increases fairness in accessing the communication medium, which is discussed in Subsection 7.2.2. These tremendous benefits have been achieved by the innovative design of the new MAC protocol. The impact and challenges of the analytical model of HTFA are discussed in Subsection 7.2.3.

The author designs two efficient schedulers, i.e., ERA and PRS, which are primarily designed for the IEEE 802.11ax and IEEE 802.11be network respectively. Section 7.3 discusses the importance of an optimal scheduler over a random scheduler. Section 7.4 explains the significance and contributions of the ERA protocol elaborately. Due to the innovative scheduling scheme of the ERA protocol, it effectively increases the throughput and reduces the retransmissions that are discussed in Section 7.4.

The PRS is the last protocol for this thesis which significant contributions are discussed elaborately in Section 7.5. Section 7.6 highlights the key features and importance of the NS-3 simulator. The key contributions of all three protocols are summarized in Table 7.1 in Section 7.7. This section also introduces the readers to several papers where the new protocols (i.e., HTFA, ERA, and PRS) of this researcher are published. Finally, Section 7.8 concludes this chapter by summarizing the contributions of the dissertation.

## 7.2 Importance and Challenges of HTFA Protocol

Leveraging the OFDMA technology, the researcher innovates a hybrid MAC mechanism for the HTFA protocol. The significant contributions of the protocol are discussed in the following subsections.

### 7.2.1 Throughput Enhancement

The Wi-Fi 6 standard, i.e., IEEE 802.11ax has a challenging ambition of enhancing the average throughput per station four times in highly congested areas such as stadiums, airports, railway stations, etc. The standard also directs to support ten times stations to provide uninterrupted Internet access to the users (Sin et al. 2015). In this regard, the HTFA protocol provides a prototype for a high-speed wireless LAN.

The HTFA protocol ensures higher throughput than some other promising MAC protocols, which is shown in Section 6.2.2 of Chapter 6. The main reason for increasing the throughput of the protocol is its innovative working principles and sub-channel distribution procedures that are discussed in Section 4.2.2 of Chapter 4. The researcher explained 5 cases to distribute the STAs among the sub-channels. The protocol directly benefitted by case 1 and case 2 for increasing the throughput. Again, the cases are listed below:

> Case 1: Number of sub-channels is equal to the number of STAs (*M=N*)
>
> Case 2: Number of sub-channels is larger than the number of STAs (*M>N*)

Let's consider two configurations in the Wi-Fi system. In the first configuration, there are 3 sub-channels, and 3 stations exist in the system. So, Case 1 is applicable for this configuration, and according to Case 1, each station would get exactly one sub-channel. In the second configuration, there are 5 sub-channels, and 3 stations exist in the system. So, Case 2 is applicable for this configuration, and according to Case 2, two stations would get 2 sub-channels each, and one station would get 1 sub-channel.

Thus, in Case 1 and Case 2, a station is getting at least one dedicated sub-channel for data transmission. In other words, a sub-channel need not serve more than one station when Case 1 or Case 2 is applicable. Since the sub-channels are dedicated to specific stations, random properties of the MAC protocol would not be utilized. Thus, the protocol would not work like a random-access protocol; rather, it would work as a reservation protocol. Thus, in these two cases, the stations monopolize the sub-channel, and no existence of the random backoff slot. In essence, the wireless stations in the HTFA protocol need not compete to acquire the sub-channels rather than sub-channels are dedicated to the stations if $N \leq M$. The absence of the random backoff slot definitely increases the throughput of the system due to the following reasons:

  i.  Less overhead

  ii.  Reduced delay

  iii.  Collision reduction

i. Less overhead: When, $N \leq M$ (Case 1 and Case 2), The sub-channels would be dedicated to the stations. Thus, only two-way handshaking is needed, as shown in Figure 4.8. Thus, the RTS frame, CTS frame, and contention window are irrelevant in those scenarios. Thus, overheads of the protocol are reducing significantly, and thereby, the assigned bandwidth would be more efficiently utilized for usable data.

ii. Reduced delay: The traditional random-access protocols have to suffer from delay, especially for the contention window. The large the contention window size, the more delay in the network. Since there is no requirement of the contention window here, the delay would be minimized here.

iii. Collision reduction: One of the serious problems of the random-access protocol is the frequent collision of frames. The more collision, the more packet loss in the system. When two or more stations send the data frames simultaneously, then there must occur a collision. The HTFA protocol ensures no frame collision when $N \leq M$. In those configurations, the probability of frame collision would be zero because a sub-channel need not contain more than one station. By the way, there is a probability of frame collision in some sub-channels if and only if $N > M$. However, total frame collision probability in the HTFA protocol is much smaller than any non-OFDMA protocol. Hence, the reduced frame collision would ensure more throughput of the Wi-Fi system.

### 7.2.2 Fair Access

The strength of the HTFA protocol lies in its working principles. The new OFDMA-based protocol also deploys a hybrid technique for distributing the sub-channels among the stations to access the medium. Thereby, HTFA provides not only high throughput but also improves fair access to the wireless medium by the stations. According to the working principles, the HTFA protocol works in two steps. In the first step, STAs are approximately evenly distributed to the sub-channels. As such, the number of STAs between the sub-channels would not differ more than one. The examples of the previous Subsection 7.2.1 are also applicable to understanding this phenomenon. Thus, in the first step, the protocol ensures a fair distribution of the sub-channels to the participating stations and thereby increases fair access to the wireless channels. In Section 6.2.2, it is shown that the HTFA protocol performs better than several promising protocols such as SRMC-CSMA/CA and CM-CSMA/CA introduced in (Xu et al. 2013) and (Wang & Wang 2010), respectively both in terms of throughput and fairness.

### 7.2.3 Analytical Model

It is urged to evaluate a new protocol by adopting an analytical model that eventually enhances the credibility and acceptability of the conducted research. Here, the researcher investigates the throughput of the HTFA protocol using the analytical model presented in Bianchi (2000). Bianchi formulates an ideal model for analyzing the saturated throughput of DCF, which is followed by some other researchers. He designed the model using a discrete-time Markov chain, where the backoff mechanism is regulated by the traditional single-channel CSMA/CA. Several other papers, including Tinnirello et al. (2010), which are considered extensions of Bianchi (2000), investigated the enhanced mathematical model for the actual backoff mechanisms by considering the existence of anomalous slots. Pioneer model in Bianchi (2000) and some of its extensions estimate saturation throughput of the single-channel terminal employing CSMA/CA mechanism. In this paper, the researcher evaluates the saturation throughput of multi-channel stations rather than the single-channel station, which are also regulated by the CSMA/CA protocol.

## 7.3 Importance of Optimal Scheduler

The performance of the OFDMA-adopted MAC protocol mainly relies on the design of the resource scheduling algorithm (Islam & Kashem 2021). The bandwidth of the whole channel can be distributed to the appropriate STAs by selecting a variety of RU combinations (See Figure 3.7). To observe the significance of an efficient resource scheduling algorithm, the researcher conducted a simple simulation for uplink OFDMA transmission. The simulation result is shown in Figure 7.1.

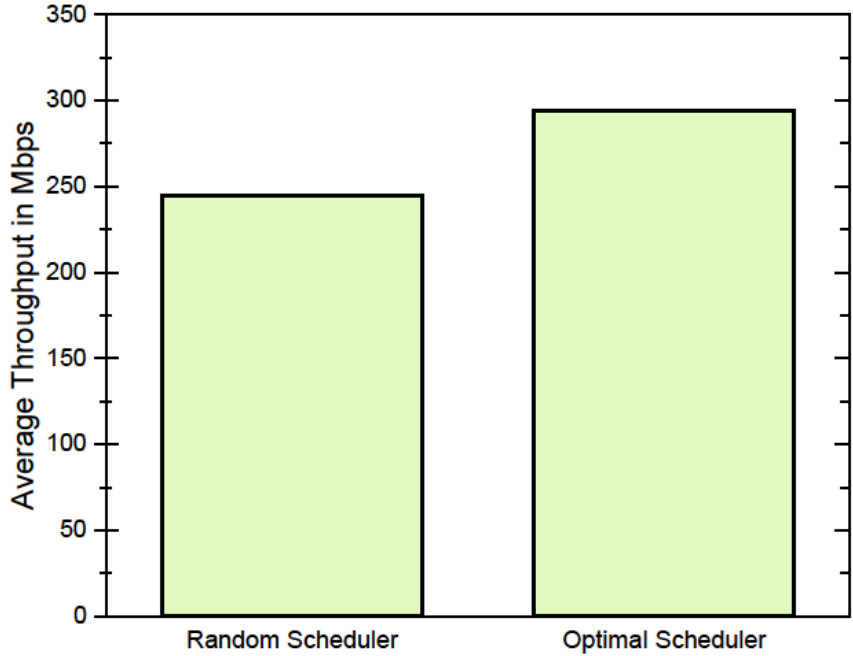

Figure 7.1: Comparison between a random scheduler and an optimal scheduler

The network contains 35 802.11ax STAs where the number of antennas of the AP and STAs is 4 and 1, respectively. The system operates in the 5 GHz band using a bandwidth of 40 MHz, and it uses the best Wi-Fi channel. The transmit power of the devices is 15 dBm, and the radius of the network is 15 meters. The figure shows the throughput of a random scheduler that arbitrarily selects the users without optimizing RUs and an optimal scheduler that selects RUs and users carefully. An optimal scheduler maximizes the throughput of the system as $max \sum_{i=1}^{N} t_i$, where $t_i$ is the throughput of the $i^{th}$ station, and $N$ is the total number of stations in the network. It is observed in Figure 7.1, the average throughput of the optimal scheduler (about 294 Mbps) is much higher than a random scheduler (about 245 Mbps).

Perceiving the significance of a robust resource scheduling protocol for the recent WLAN standards, the researcher designed the ERA (Efficient Resource Allocation) and PRS (Proportional Resource Scheduling) protocol. These protocols enhance the throughput and goodput of the network significantly as well as reduce the number of retransmissions of packets. The researcher conducted rigorous experiments using Network Simulator-3 to observe the performance of the novel ERA and PRS scheduler over several well-known schedulers. The details of the experiments and data are described in Chapter 6.

## 7.4 Contributions of ERA Protocol

### 7.4.1 Throughput Enhancement

The throughput of the Wi-Fi protocol largely depends on the efficient and wise scheduling of RUs to the STAs. This paper innovates the ERA protocol that promises to deliver a high throughput to the Wi-Fi network. The author compared the ERA protocol with some existing promising protocols, namely, Greedy (Wang and Psounis 2018), SRTF (Bankov et al. 2018), and MUTAX (Bankov et al. 2017). The simulation results in Section 6.3.1 proved that the performance of ERA in terms of throughput is the best among the competing protocols.

Examining Figure 6.6 in Chapter 6, it is observed that the throughput of the Greedy is the lowest among the four competing protocols because it assigns a smaller portion of the RUs to the STAs. The throughput of the SRTF scheduler is also not satisfactory due to the exhaustive service method of the protocol. However, the throughput of the MUTAX is promising due to the usage of a channel-splitting scheduler. The MUTAX applies an adaptive method to split the channel into the RUs. For each of the configurations of the RUs, MUTAX solves an optimization problem and hence gets the best scheduling. The throughput of the ERA protocol is the highest among all of the competing protocols because of its dynamic and adaptive scheduling approach. Based on the BSR, in every flow, ERA classified the STAs according to their loads. Then AP assigns larger RUs to the higher-load STAs and smaller RUs to the lower-load STAs and thus makes a balance between the needs and grants.

It is noted, the changes in throughput of the ERA protocol for different congestions are comparatively uniform than its competitors while always ensuring the highest throughput (refer to Figure 6.7). The phenomena suggest that the ERA protocol is suitable at homes and dense areas (e.g., stadium, station, etc.) equally, while the MUTAX is less suitable at homes than the congested areas. The throughput is also proportional to the transmit power. Due to the power constraint of the Wi-Fi devices, the researcher always keeps the transmit power the same for all simulations. Thus, doubling the number of antennas of the access point does not enhance the throughput by 2X (refer to Figure 6.8). Thus, the reason for increasing the throughput in Figure 6.8 is due to the incrementing of the number of antennas of the access point. Nonetheless, the throughput could be enhanced further by increasing the transmit power of the antennas.

### 7.4.2 Reduction of Retransmissions

Reduction of the number of retransmissions is one of the key factors to increase the goodput of the network. Goodput is the measurement of the useful data that reflects the real user experience in the network. Goodput is not the same as the throughput because goodput does not count the undesirable data that is arised from the retransmissions, protocol overheads, etc. The ERA protocol reduces the number of retransmissions of packets to nearly zero even in the congested scenarios, as proven in experiment 6 (refer to Figure 6.11). Thus, the ERA protocol enhances the goodput of the network remarkably (refer to Figure 6.10), thereby improving the quality of the user experience.

The main reason for packet retransmissions in the wireless LAN is the interruption of the hidden nodes during transmission. The ERA protocol is designed by the pure scheduled access mechanism where the STAs would not compete for channel access randomly. As such, ERA seldom requires retransmissions, while pure random access protocols need the highest number of retransmission among different classes of protocols, as shown in Figure 6.11.

## 7.5 Significance of PRS Protocol

### 7.5.1 A Framework for Concurrent SA and RA

Previously, wireless LAN predominantly uses the RA mechanism for channel access. After adopting the OFDMA technology (by Wi-Fi 6 and Wi-Fi 7), the SA mechanism can be utilized for data transmission. Since both mechanisms have some advantages, many researchers design MAC protocols that are based on either RA or SA techniques. However, these protocols have no provisions for simultaneous implementation of RA and SA mechanisms. The PRS protocol designed for the future wireless LANs (specifically Wi-Fi 7), implements RA and SA mechanisms concurrently and significantly improves network performance.

It is observed in Figure 4.10 of Chapter 4; the PRS divides the RUs into two: LRUs (Left Resource Units) and RRUs (Right Resource Units). The LRUs would be assigned to the Scheduled Access (SA) STAs and RRUs (i.e., equivalent bandwidth) be assigned to the Random Access (RA) STAs. The PRS algorithms ensure a fair distribution of resources to the SA and RA mechanisms according to the load of the SA and RA stations.

### 7.5.2 Performance Improvement

The performance of the OFDMA-based wireless LAN largely depends on the scheduling protocol. The researcher designs two algorithms for the resource scheduling of the PRS protocol. Algorithm 3 provides the initial scheduling information, which is received by Algorithm 4 as input. After performing revision, Algorithm 4 provides the final scheduling information to the access point. The PRS distributes the channel resources proportionally to the stations according to their available loads. Thus, it utilizes the resources efficiently and increases the throughput and fairness in accessing the channel.

The investigator constructs analytical models both for the SA and RA mechanisms and conducts extensive simulations to measure the efficiency of the PRS protocol. Section 5.6.3 describes the analytical models, and Section 6.4 shows the data analysis for the PRS protocol. The analyses validate the robustness of the protocol in terms of throughput, goodput, fairness, and retransmissions.

### 7.5.3 Superiority of Scheduled Access over Random Access

The new Wi-Fi supports both the Scheduled Access (SA) and Random Access (RA) methods for channel access. The SA method is a new technique for data transmission in the IEEE 802.11ax and IEEE802.11be standards, and this method does not apply to the previous standards, e.g., IEEE 802.11ac, IEEE 802.11n, etc. The main goal of incorporating the SA method is to improve the performance, especially the throughput of the wireless LAN. The PRS protocol supports both the SA and RA mechanisms to provide service to both SA and RA devices. To the best of author knowledge, this is the first scheduling protocol for Wi-Fi 7 that estimates the amount of RUs for both mechanisms proportionally according to the available loads. Besides, the PRS also calculates the number of RUs for each of the STAs of the SA mechanism. The process of scheduling is thoroughly investigated using a practical scheduling problem which is illustrated in Section 4.4.2. The author designs two exclusive scheduling algorithms (i.e., Algorithm 3 and Algorithm 4) for the PRS protocol to perform the scheduling. The main novelty of the PRS protocol lies in the design of the algorithms.

The investigator conducts two exclusive experiments (illustrated in Section 6.4.1) to contrast the performance between the SA and RA mechanisms. The first experiment (refer to Figure 6.12) compares the throughput, and the second experiment (refer to Figure 6.13) compares the number of retransmissions between the mechanisms. It is observed from Figure 6.12 that SA provides around 1.7 times throughput of the RA mechanism. Again Figure 6.13 shows that the number of retransmissions for 60 STAs is below 1 for the SA mechanism, while for the RA mechanism, it exceeds 20. From the experiments, it is perceived that the SA mechanism is more robust than the RA mechanism in terms of throughput, retransmissions, and similar performance parameters, e.g., goodput, probability of collisions.

## 7.6 Simulation Using NS-3

The researcher evaluates the performance of protocols through exhaustive computer simulations. For simulation, the most robust simulator to date, named 'Network Simulator-3' or simply NS-3, is used to get the data precisely. NS-3 is a discrete-event network simulator developed for computer network researchers. The Network Simulator-3 is a free software package that is licensed under the GNU GPLv2, and the simulator is publicly available for research, development, and use. The large majority of its users focuses on different wireless and wired networks simulations that involve models for Wi-Fi, 5G, or LTE for the physical and data link layers, as well as a variety of static or dynamic routing protocols such as the OLSR (Optimized Link State Routing) and AODV (Ad hoc On-Demand Distance Vector) for IP-based applications.

Implementation of NS-3 in the operating system is quite complex as lots of programs need to integrate with each other. The NS-3 provides the best result for lower stack protocols, and for upper stack protocol, its performance may decrease (Rampfl 2013). However, it is the first choice for the research community, both for the wired and wireless environment, to get the best result.

The investigator successfully conducted many experiments using NS-3 to find a lot of performance parameters of the HTFA, ERA, and PRS protocols. The details of the simulation results of the HTFA protocol can be found in Section 6.2. Section 6.3 and Section 6.4 described the simulation results of the ERA and PRS protocol, respectively. The aggregated outcome of the simulations proved the robustness of the newly designed protocols

## 7.7 Three Protocols at a Glance

The author designed three Wi-Fi protocols, namely, HTFA (High Throughput and Fair Access), ERA (Efficient Resource Allocation), and PRS (Proportional Resource Scheduling), in this thesis. The novel protocols primarily focus on throughput enhancement and some other parameters such as fairness, collisions, retransmissions, etc. The author drafted the HTFA protocol (Islam & Kashem 2018) for a conference proceeding which is later fully implemented and analyzed in the article Islam & Kashem (2019). The HTFA provides a prototype for the OFDMA-based hybrid protocol to future researchers.

Then, the researcher focused on the scheduling problems as the scheduling algorithm is one of the key players on which the performance of a Wi-Fi 6 protocol depends. Therefore, the researcher came with the ERA protocol, which is published in the article Islam & Kashem (2020). The ERA protocol designed a scheduling scheme that adopts the SA mechanism. Later, the author designed another scheduling protocol (i.e., PRS) in Islam & Kashem (2021), which is more promising than the previous two protocols. The PRS protocol utilized both the SA and RA mechanisms and designed two innovative algorithms for the scheduling purpose. Table 7.1 summarizes the major contributions of the three protocols designed for this dissertation.

Table 7.1: Key contributions of the three protocols

| Protocol Name | Significant Contributions |
|---|---|
| High Throughput and Fair Access (HTFA) | • The HTFA provides high throughput than the contemporary Wi-Fi protocols<br>• It also reduces frame collisions and increases fair access to the communication medium<br>• Provides a statistical analytical model for multi-channel MAC protocols using the Markov chain<br>• It designs a unique algorithm (i.e., Algorithm 1 in Section 4.2.2) for STA distribution and channel Access<br>• The protocol is developed before the standardization of the IEEE 802.11ax (Standards IEEE 2019) and hence provides a prototype for the OFDMA-based multi-channel protocols to the future researchers |
| Efficient Resource Allocation (ERA) | • The ERA provides very high throughput leveraging the OFDMA SA mechanism<br>• It also increases the goodput and reduces retransmissions in the network significantly, thereby improving the quality of user experience<br>• The ERA provides an analytical model for the OFDMA SA method<br>• The protocol employs an innovative algorithm (i.e., Algorithm 2 in Section 4.3.3) for the RU assignment<br>• It is believed that ERA is the first protocol for Wi-Fi 6 where the RUs are distributed to the STAs according to their available loads |
| Proportional Resource Scheduling (PRS) | • The PRS ensures extremely high throughput while operating in the SA mode<br>• The protocol provides a framework for simultaneous implementation of SA and RA mechanisms for future Wi-Fi<br>• In terms of novelty and robustness, the PRS is the best among the three protocols<br>• It also improves the goodput, fairness index and minimizes retransmissions of packets<br>• The PRS gives the analytical models both for the SA and RA mechanisms<br>• The protocol designs two outstanding and unique algorithms (Algorithm 3 and Algorithm 4 designed in Section 4.4.2) for performing resource scheduling |

This research aims to provide high-speed and smooth communications in wireless LAN for Wi-Fi users. Thus, the protocols improve several performance parameters especially, the throughput for the latest OFDMA-based IEEE 802.11 standards. The high-speed or maximum achievable throughput depends upon the available bandwidth. For example, if the PRS protocol provides a throughput of 309 Mbps using a 40 MHz channel, then it can provide around 4 times throughput using a 160 MHz channel. Again, the throughput of the Wi-Fi STAs can be raised several times by using the multi-user MIMO (Multiple Input and Multiple Output) and multiple spatial streams.

## 7.8 Conclusion

A high-speed Wi-Fi network is of utmost importance for future generation communications. This research would contribute a lot to achieving tremendous speed in the wireless local area network. The author innovates three novel Wi-Fi protocols in light of the IEEE 802.11ax standard to meet the demand for next-generation communication. The main contributions of this research are summarized as follows,

- Three novel protocols namely, HTFA, ERA, and PRS are designed for the OFDMA-adopted Wi-Fi network
- The protocols improve several performance parameters, especially the throughput, and hence promise to deliver a high-speed Wi-Fi network
- The protocols leverage the most promising OFDMA technology, and focus latest IEEE 802.11ax and IEEE 802.11be standards
- Along with the high throughput, the research also focuses on improving other performance parameters such as goodput, fairness, retransmissions, etc.
- Analytical models are constructed to validate the protocols
- Conducted extensive simulation to measure the efficiency of the protocols
- The dissertation contains a lot of resources such as the analysis of contemporary protocols; OFDMA constraints and regulations; framework of protocols; relevant data, theory, and methods; etc. that would be valuable resources for the research on the Wi-Fi network to future researchers.

# CHAPTER EIGHT

# Chapter 8

# Conclusions and Future Works

## 8.1 Conclusions

To meet the tremendous rise of demand for future generation wireless LANs, a robust and efficient Wi-Fi protocol is required. The acute rise of demand for high-speed WLANs has driven rigorous research to increase the throughput leveraging a variety of MAC techniques. The efficiency of the MAC layer plays a significant role in boost-up the throughput of the WLANs. However, the MAC layer of the WLAN has barely changed for the last 16 years. Although some IEEE 802.11n proposals claim to support up to 450 Mbps using three spatial streams (Perahia & Stacey 2013), the MAC layer restrains the throughput enhancement due to its overheads (Rahman et al. 2010). In this regard, the IEEE declares its latest WLAN standard, which is known as IEEE 802.11ax or Wi-Fi 6. According to the new standard, the wireless LAN should support four times data improvement and ten times users in the dense environment. Recently, IEEE proposes another new standard i.e., IEEE 802.11be to provide extremely high throughput and to support multifarious data-hungry applications. Both of the recent standards adopt the OFDMA technology for future wireless LANs. This paper designs three innovative protocols, namely, HTFA, ERA, and PRS for OFDMA-based high-speed Wi-Fi networks.

In this thesis, a hybrid MAC protocol named HTFA is designed leveraging OFDMA technology that could provide a high throughput of data as well as could ensure fair medium access by the stations. The unique properties of the protocol are its working principles and sub-channel distribution procedure among the participating stations. The simulation results validate the robustness of the novel mechanism in terms of throughput, collision reduction, and fair access. The protocol will gain these benefits at the expense of grown complexity in the channel distribution method. Still, the researcher is convinced that there would be a tremendous tradeoff of the new OFDMA-based MAC protocol. The throughput of the protocol has also been evaluated by statistical analysis.

Another contribution of this research is designing a scheduled access protocol named ERA. Besides very high throughput, the ERA reduces retransmissions of packets significantly. Low retransmissions provide very high goodput as well and hence ensure high quality of user experience. The researcher designs an innovative algorithm for distributing the resource units to the stations. The ERA algorithm ensures that stations would get resource units according to their needs. For that, the protocol estimates loads of all stations in every flow before the algorithm executes. To the best of researcher knowledge, the ERA is a unique protocol of its kind, where the RUs are allocated to the STAs according to the demands of the users.

The third protocol of this thesis is the PRS protocol that adopts both the scheduled and random access mechanisms. The researcher designs two algorithms for the resource scheduling of the PRS protocol. The first algorithm provides the initial scheduling information, which is received by the second algorithm as input. After performing revision, the second algorithm provides the final scheduling information to the access point. The PRS distributes the resources proportionally to the stations according to their available loads. Thus, it utilizes the resources efficiently and increases the throughput and fairness in accessing the channel. The investigator constructs analytical models both for the SA and RA mechanisms and conducts extensive simulations to measure the efficiency of the PRS protocol. The analyses validate the robustness of the protocol in terms of throughput, goodput, fairness, and retransmissions.

The investigator constructs analytical models and conducts rigorous computer simulations for all three protocols with the help of the robust NS-3 simulator. The researcher designs, implements, simulates, and analyzes the protocols rigorously. This work will contribute remarkably to delivering a high-speed Wi-Fi network for next-generation communications.

## 8.2 Summary of Research

The first and foremost goal of this research is to provide a high-speed Wi-Fi network for upcoming wireless LANs. In this regard, the author designs three novel protocols (i.e., HTFA, ERA, and PRS) for the wireless LAN. The protocols are designed to comply with the latest IEEE 802.11ax and IEEE 802.be standards. Thus, the new protocols can ensure uninterrupted and smooth communication in highly dense environments where a large number of users may need to send a huge volume of data simultaneously.

### 8.2.1 Summary of HTFA Protocol

The researcher employs the concept of a hybrid mechanism to implement the HTFA protocol. The salient feature of a hybrid protocol is that it is very efficient in supporting a diverse class of applications. The protocol implements the hybrid mechanism leveraging OFDMA technology. It is the OFDMA that facilitates the creation of multiple channels dividing a single large channel. The HTFA protocol regulates some sub-channels to work as a random-access protocol and some others to work as reservation or scheduled protocol. Hence, the new protocol would be more acceptable than the traditional random access and reservation protocols.

The MAC protocols, specifically the random access protocol, often suffer from frame collisions. This unwanted frame collision reduces the throughput of the network remarkably. The HTFA protocol reduces frame collision significantly by innovative design and hence enlarges the throughput of the network. It also increases fair access to the communication media by the stations. As such, wireless stations have the opportunity to gain access to the wireless channel more evenly.

It is also urged to construct a statistical model for evaluating the performance of a communication protocol scientifically. A prudent statistical analysis would convince the researchers about innovation. The author adopts the Markov Chain model to evaluate the new multi-channel MAC protocol mathematically. A good number of computer simulations were conducted successfully to measure the efficiency of the protocol. The simulation results reveal the potential of the protocol by measuring some crucial parameters of MAC protocols such as saturation throughput, collision probability, fairness, etc.

### 8.2.2 Summary of ERA Protocol

The ERA protocol employs the concept of the scheduled access mechanism to implement an efficient scheduler for the latest wireless LAN, i.e., Wi-Fi 6. The performance of the Wi-Fi 6 network largely depends on an efficient scheduler that distributes the RUs to the STAs wisely and prudently. The main challenge is to design the scheduling mechanism for the uplink (UL) path rather than the downlink (DL). The reason is that for the UL transmission, all sending STAs must be synchronized for the OFDMA transmission. The thesis designs a UL scheduler known as the ERA (Efficient Resource Allocation) that ensures a good throughput to the Wi-Fi network as well as reduces the retransmissions of the packets. The experiments and analytical model validate that the ERA protocol is very efficient and satisfies the requirements of the Wi-Fi 6 standard.

The main reason for providing the high-throughput by the ERA is its innovative design. The novelty of the ERA protocol is that it distinguishes the STAs according to their loads (i.e., data volume) and then distributes the resource units proportionately according to their load size. Due to the proportionate distribution of resource units by the AP, the assigned channel bandwidth would be used more efficiently and hence increase the throughput. The author designs Algorithm 2 (details in Section 4.3.3) for efficient and wise distribution of the RUs to the STAs. Another salient feature of the ERA is it reduces the retransmissions packets and increases the goodput of the network remarkably. In this protocol, the number of retransmissions is almost zero since hidden nodes could not interrupt the stations while transmitting data.

### 8.2.3 Summary of PRS Protocol

The Proportional Resource Scheduling i.e., PRS protocol is the most remarkable contribution of this thesis. It is believed that the PRS is the first scheduling protocol for Wi-Fi 7 that investigates both the SA and RA mechanisms together for scheduling and estimates the amount of resources for both mechanisms proportionally according to the available loads. Besides, the PRS also calculates the number of RUs for each of the STAs of the SA mechanism. The process of resource scheduling is thoroughly examined using an ideal problem in Section 4.4.2. The PRS employs two exclusive scheduling algorithms (i.e., Algorithm 3 and Algorithm 4) to perform the tasks of scheduling. The main novelty of the PRS protocol lies in the design of the algorithms.

The recent IEEE 802.11 standards keep provisions of the scheduled access leveraging the OFDMA along with the legacy random access for accessing the wireless medium while previous standards do not permit the scheduled access. The PRS is designed to support both the SA and RA devices and also provides a framework for simultaneous implementation of random access to scheduled access mechanisms. Lastly and most importantly, the PRS enhances the performance of the wireless LAN by increasing the throughput, goodput, and fairness index as well as reducing the retransmission of packets.

## 8.3 Future Works

To mitigate the increasing demands for a high-speed Wi-Fi network, IEEE 802 LAN/MAN Standard Committee has recently developed the IEEE 802.11ax standard (Standards IEEE 2019). The latest amendment of the Wi-Fi proposes a number of ways to enhance the performance of the wireless LAN. As stated earlier, the most promising feature of the IEEE 802.11ax is adopting the OFDMA technology. While the legacy Wi-Fi network employs only one random access channel, the Wi-Fi 6 network facilitates multi-channel communication leveraging the OFDMA technology. OFDMA technology derived from the OFDM modulation is highly suitable for the Wi-Fi 6 network to raise the performance further.

This thesis delivers several OFDMA-based protocols that improve the performance of the Wi-Fi 6 network. In addition to Wi-Fi 6, the thesis also focuses on Wi-Fi 7 (proposed by IEEE) while designing the last protocol i.e., PRS. The performance of the new protocols is analyzed and evaluated through analytical models and computer simulations. To further validate and observe the real behavior of the protocols, the researcher can attempt to set up a testbed. A testbed is a platform to conduct experiments for large development projects. Testbeds facilitate exhaustive, transparent, and replicable testing of scientific theories, computational tools, and new technologies. To set up a testbed, a lot of resources such as the specified hardware, software, skills, and expertise are required and thereby may require industry-academia collaboration. In this regard, the investigator can seek industry-academia collaboration for setting up a testbed.

To extend the research, the researcher may also investigate some other issues such as group scheduling (Er-Rahmadi et al. 2015) and Target Wake Time (TWT) mechanism (Nurchis & Bellalta 2019) to design a high-efficient scheduler. Group scheduling would reduce signaling overhead, thereby enhancing system throughput remarkably. The TWT mechanism incorporated in the IEEE 802.11ax amendment could play a tremendous role in lessening the adversary effects of the channel contentions as it provides a simple but efficient technique to schedule the OFDMA transmissions in time. The TWT mechanism can also provide a low-power consumption mode for the STAs having low traffics and periodic transmissions like the IoT (Internet of Things) applications.

As usual, the development of future wireless technologies often starts at around the same time as the latest standard begins to enter the market. While the users just rivet their eyes on Wi-Fi 6, the IEEE 802.11 working group is turning to the next-generation communication standard known as IEEE 802.11be or Wi-Fi 7 (Khorov et al. 2020). The development of the Wi-Fi 7 standard is going on, and it is expected the standard would be released by 2024. The new amendment also promises a revolution of unlicensed wireless connectivity (Garcia-Rodriguez et al. 2021). Wi-Fi 7 is supposed to provide an Extremely High Throughput (EHT) (Au 2019) to fulfill the requirement of recent applications such as the 4k/8k video, augmented and virtual reality, online gaming, etc. Due to some constraints and stringent requirements, the current Wi-Fi 6 is not capable of supporting these applications smoothly.

IEEE 802.11be would provide at least a 30 Gbps data rate roughly three times the maximum throughput of the 802.11ax. The standard will be operating in 2.4, 5, and 6 GHz frequency bands and its maximum channel bandwidth would be 320 MHz (i.e., twice 802.11ax). The main technical issues to be considered for the 802.11be are advanced PHY and MAC techniques, channel-sounding optimization, multi-link aggregation, supporting multiple RUs, 4096-QAM, cooperation between the APs, etc. (Deng et al. 2020). There are hundreds of open problems that can be addressed by the researchers who would like to contribute to the development of Wi-Fi 7.

# Appendix A

# Frame Type and Subtype Combination

Frame type and subtype combination of the MAC frame are shown in the following chart:

| Type Value | Type Description | Subtype Value | Subtype Description |
|---|---|---|---|
| 00 | Management | 0000 | Association request |
| 00 | Management | 0001 | Association response |
| 00 | Management | 0010 | Reassociation request |
| 00 | Management | 0011 | Reassociation response |
| 00 | Management | 0100 | Probe request |
| 00 | Management | 0101 | Probe response |
| 00 | Management | 1000 | Beacon |
| 00 | Management | 1001 | Announcement traffic indication message (ATIM) |
| 00 | Management | 1010 | Dissociation |
| 00 | Management | 1011 | Authentication |
| 00 | Management | 1100 | Deauthentication |
| 01 | Control | 1010 | Power save – poll |
| 01 | Control | 1011 | Request to send |
| 01 | Control | 1100 | Clear to send |
| 01 | Control | 1101 | Acknowledgement |
| 01 | Control | 1110 | Contention-free (CF)-end |
| 01 | Control | 1111 | CF-end + CF-ack |
| 10 | Data | 0000 | Data |
| 10 | Data | 0001 | Data + CF-Ack |
| 10 | Data | 0010 | Data + CF- Poll |
| 10 | Data | 0011 | Data + CF-Ack + CF-Poll |
| 10 | Data | 0100 | Null function (no data) |
| 10 | Data | 0101 | CF-Ack (no data) |
| 10 | Data | 0110 | CF-Poll (no data) |
| 10 | Data | 0111 | CF-Ack + CF-Poll (no data) |

# Appendix B

# NS-3 Code for HTFA Protocol

## Experiment 1: Saturated throughput with respect to Wmin

```
#include "ns3/stats-module.h"
#include "ns3/wifi-module.h"
#include "ns3/core-module.h"
#include "ns3/network-module.h"
#include "ns3/mobility-module.h"
#include "ns3/propagation-module.h"

using namespace ns3;
NS_LOG_COMPONENT_DEFINE ("MinstrelHtWifiManagerExample");
const int STA = 10;
double Throughput;
int SC_M; //Number of sub-channels
double g_intervalBytes = 0;
void PacketRx (Ptr<const Packet> pkt, const Address &addr)
{
  NS_LOG_DEBUG ("Received size " << pkt->GetSize ());
  g_intervalBytes += pkt->GetSize ();
}
void
RateChange (uint64_t newVal, Mac48Address dest)
{
  NS_LOG_DEBUG ("Change to " << newVal);
  g_intervalRate = newVal;
}
struct Step
{
  double stepSize;
  double stepTime;
};
  }
  std::string m_name;
  enum WifiPhyStandard m_standard;
  uint32_t m_width;
  bool m_sgi;
  double m_snrLow;
  double m_snrHigh;
  double m_xMin;
  double m_xMax;
  double m_yMax;
};
int main (int argc, char *argv[])
{
  std::vector <StandardInfo> standards;
  int steps;
  uint32_t rtsThreshold = 65535;
  uint32_t BE_MaxAmpduSize = 65535;
  double stepSize = 1; // dBm
  double stepTime = 1; // seconds
  uint32_t packetSize = 1024;  // bytes
```

```cpp
  int broadcast = 0;
  int ap1_x = 0;
  int ap1_y = 0;
  int sta1_x = 5;
  int sta1_y = 0;
  uint16_t nss = 1;
  bool shortGuardInterval = false;
  uint32_t channelWidth = 20;
  std::string standard ("802.11b");
  StandardInfo selectedStandard;
  std::string outfileName ("minstrel-ht-");
  CommandLine cmd;
  cmd.AddValue ("rtsThreshold", "RTS threshold", rtsThreshold);
  cmd.AddValue ("BE_MaxAmpduSize", "BE Max A-MPDU size", BE_MaxAmpduSize);
  cmd.AddValue ("stepSize", "Power between steps (dBm)", stepSize);
  cmd.AddValue ("stepTime", "Time on each step (seconds)", stepTime);
  cmd.AddValue ("broadcast", "Send broadcast instead of unicast", broadcast);
  cmd.AddValue ("channelWidth", "Set channel width (valid only for 802.11n, ac or ax)", channelWidth);
  cmd.AddValue ("shortGuard", "Set short guard interval (802.11n/ac)", shortGuardInterval);
  if (standard == "802.11b")
 {
    NS_ABORT_MSG_IF (channelWidth != 20 && channelWidth != 22, "Invalid channel width for standard " << standard);
    NS_ABORT_MSG_IF (nss != 1, "Invalid nss for standard " << standard);
   }
  else if (standard == "802.11a" || standard == "802.11g")
    {
      NS_ABORT_MSG_IF (channelWidth != 20, "Invalid channel width for standard " << standard);
      NS_ABORT_MSG_IF (nss != 1, "Invalid nss for standard " << standard);
    }
  else if (standard == "802.11n-5GHz" || standard == "802.11n-2.4GHz")
    {
      NS_ABORT_MSG_IF (channelWidth != 20 && channelWidth != 40, "Invalid channel width for standard " << standard);
      NS_ABORT_MSG_IF (nss == 0 || nss > 4, "Invalid nss " << nss << " for standard " << standard);
    }
  else if (standard == "802.11ax")
    {
      NS_ABORT_MSG_IF (channelWidth != 20 && channelWidth != 40 && channelWidth != 80 && channelWidth != 160, "Invalid channel width for standard " << standard);
      NS_ABORT_MSG_IF (nss == 0 || nss > 4, "Invalid nss " << nss << " for standard " << standard);
    }
  outfileName.append (standard);
  if (standard == "802.11n-5GHz" || standard == "802.11n-2.4GHz" || standard == "802.11ax")
    {
      std::ostringstream oss;
      std::string gi;
      if (shortGuardInterval)
        {
          gi = "SGI";
        }
      else
        {
```

```
          gi = "LGI";
        }
      oss << "-" << channelWidth << "MHz-" << gi << "-" <<nss << "SS";
      outfileName += oss.str ();
    }
  std::string tmp = outfileName + ".plt";
  std::ofstream outfile (tmp.c_str ());
  tmp = outfileName + ".eps";
  Gnuplot gnuplot = Gnuplot (tmp.c_str ());
  // The first number is channel width, second is minimum throughput, third is maximum
  // fourth and fifth provide xrange axis limits, and sixth the yaxis
  // maximum
  standards.push_back (StandardInfo ("802.11a", WIFI_PHY_STANDARD_80211a, 20, false, 3, 27, 0, 30, 60));
  standards.push_back (StandardInfo ("802.11b", WIFI_PHY_STANDARD_80211b, 22, false, -5, 11, -6, 15, 15));
  standards.push_back (StandardInfo ("802.11g", WIFI_PHY_STANDARD_80211g, 20, false, -5, 27, -6, 30, 80));
  standards.push_back (StandardInfo ("802.11n-5GHz",
WIFI_PHY_STANDARD_80211n_5GHZ, channelWidth, shortGuardInterval, 5, 30, 0, 35, 80));
  standards.push_back (StandardInfo ("802.11n-2.4GHz",
WIFI_PHY_STANDARD_80211n_2_4GHZ, channelWidth, shortGuardInterval, 5, 30, 0, 35, 80));
  standards.push_back (StandardInfo ("802.11ax", WIFI_PHY_STANDARD_80211ax,
channelWidth, shortGuardInterval, 5, 30, 0, 35, 80));
  standards.push_back (StandardInfo ("802.11-holland", WIFI_PHY_STANDARD_holland, 20, false, 3, 27, 0, 30, 60));
  standards.push_back (StandardInfo ("802.11-10MHz",
WIFI_PHY_STANDARD_80211_10MHZ, 10, false, 3, 27, 0, 30, 60));
  standards.push_back (StandardInfo ("802.11-5MHz",
WIFI_PHY_STANDARD_80211_5MHZ, 5, false, 3, 27, 0, 30, 60));
  // Set channel width
  // Adjust sub-channel width
  // Obtain pointer to the WifiPhy
  Ptr<NetDevice> ndClient = clientDevice.Get (0);
  Simulator::Schedule (Seconds (0.5 + stepTime), &ChangeSignalAndReportRate, rssLossModel, step, rssCurrent, rateDataset, actualDataset);
  PacketSocketHelper packetSocketHelper;
  packetSocketHelper.Install (serverNode);
  packetSocketHelper.Install (clientNode);
  PacketSocketAddress socketAddr;
  socketAddr.SetSingleDevice (serverDevice.Get (0)->GetIfIndex ());
  if (broadcast)
    {
      socketAddr.SetPhysicalAddress (serverDevice.Get (0)->GetBroadcast ());
    }
  else
    {
      socketAddr.SetPhysicalAddress (serverDevice.Get (0)->GetAddress ());
    }
  // Arbitrary protocol type.
  // Note: PacketSocket doesn't have any L4 multiplexing or demultiplexing
  //       The only mux/demux is based on the protocol field
  socketAddr.SetProtocol (1);
  Ptr<PacketSocketClient> client = CreateObject<PacketSocketClient> ();
  client->SetRemote (socketAddr);
  client->SetStartTime (Seconds (0.5));
```

```
  client->SetAttribute ("MaxPackets", UintegerValue (0));
  client->SetAttribute ("PacketSize", UintegerValue (packetSize));
  client->SetAttribute ("Interval", TimeValue (MicroSeconds (20)));
  clientNode->AddApplication (client);
  Ptr<PacketSocketServer> server = CreateObject<PacketSocketServer> ();
  server->SetLocal (socketAddr);
  server->TraceConnectWithoutContext ("Rx", MakeCallback (&PacketRx));
  serverNode->AddApplication (server);
  Simulator::Stop (Seconds ((steps + 1) * stepTime));
  Simulator::Run ();
  Simulator::Destroy ();
  gnuplot.AddDataset (rateDataset);
  gnuplot.AddDataset (actualDataset);
  std::ostringstream xMinStr, xMaxStr, yMaxStr;
  std::string xRangeStr ("set xrange [");
  xMinStr << selectedStandard.m_xMin;
  xRangeStr.append (xMinStr.str ());
  xRangeStr.append (":");
  xMaxStr << selectedStandard.m_xMax;
  xRangeStr.append (xMaxStr.str ());
  xRangeStr.append ("]");
  std::string yRangeStr ("set yrange [0:");
  yMaxStr << selectedStandard.m_yMax;
  yRangeStr.append (yMaxStr.str ());
  yRangeStr.append ("]");
  gnuplot.SetTerminal ("postscript eps color enh \"Times-BoldItalic\"");
  gnuplot.SetLegend ("Thorughput ", "Rate (Mb/s)");
  gnuplot.SetTitle (title);
  gnuplot.SetExtra  (xRangeStr);
  gnuplot.AppendExtra (yRangeStr);
  gnuplot.AppendExtra  ("set key reverse left Left");
  gnuplot.GenerateOutput (outfile);
  outfile.close ();
  return 0;
}
```

# Experiment 2: Collision probability with respect to Wmin

```
#include "ns3/core-module.h"
#include "ns3/network-module.h"
#include "ns3/applications-module.h"
#include "ns3/wifi-module.h"
#include "ns3/mobility-module.h"
#include "ns3/ipv4-global-routing-helper.h"
#include "ns3/internet-module.h"

// The simulation assumes 10 station in an infrastructure network:
//
//  STA    AP
//   *      *
//   |      |
//n1-n10   n11
//
NS_LOG_COMPONENT_DEFINE ("vht-wifi-network");
int main (int argc, char *argv[])
```

```
{
  double collision_prob;
  const int STA = 10;
  double simulationTime = 10; //seconds
  int CW; //Contention Window Size
  int SC_M; //Number of sub-channels
  double distance = 1.0; //meters
  CommandLine cmd;
  cmd.AddValue ("distance", "Distance in meters between the station and the access point", distance);
  cmd.AddValue ("simulationTime", "Simulation time in seconds", simulationTime);
  cmd.Parse (argc,argv);
  std::cout << "MCS value" << "\t\t" << "Channel width" << "\t\t" << "short GI" << "\t\t" << "Throughput" << '\n';
  for (int i = 0; i <= 9; i++) //MCS
    {
      for (int j = 20; j <= 160; ) //channel width
        {
          if (i == 9 && j == 20)
            {
              j *= 2;
              continue;
            }
          for (int k = 0; k < 2; k++) //GI: 0 and 1
            {
              uint32_t payloadSize; //1500 byte IP packet
              if (udp)
                {
                  payloadSize = 1472; //bytes
                }
              else
                {
                  payloadSize = 1448; //bytes
                  Config::SetDefault ("ns3::TcpSocket::SegmentSize", UintegerValue (payloadSize));
                }
              NodeContainer wifiStaNode;
              wifiStaNode.Create (1);
              NodeContainer wifiApNode;
              wifiApNode.Create (1);
UintegerValue (j));
              // mobility.
              MobilityHelper mobility;
              Ptr<ListPositionAllocator> positionAlloc = CreateObject<ListPositionAllocator> ();
              positionAlloc->Add (Vector (0.0, 0.0, 0.0));
              positionAlloc->Add (Vector (distance, 0.0, 0.0));
              mobility.SetPositionAllocator (positionAlloc);
              mobility.SetMobilityModel ("ns3::ConstantPositionMobilityModel");
              mobility.Install (wifiApNode);
              mobility.Install (wifiStaNode);
              /* Internet stack*/
              InternetStackHelper stack;
              stack.Install (wifiApNode);
              stack.Install (wifiStaNode);
              Ipv4AddressHelper address;
              address.SetBase ("192.168.1.0", "255.255.255.0");
              Ipv4InterfaceContainer staNodeInterface;
              Ipv4InterfaceContainer apNodeInterface;
```

```cpp
          staNodeInterface = address.Assign (staDevice);
          apNodeInterface = address.Assign (apDevice);
          /* Setting applications */
          ApplicationContainer serverApp, sinkApp;
          if (udp)
            {
              //UDP flow
              UdpServerHelper myServer (9);
              serverApp = myServer.Install (wifiStaNode.Get (0));
              serverApp.Start (Seconds (0.0));
              serverApp.Stop (Seconds (simulationTime + 1));
              UdpClientHelper myClient (staNodeInterface.GetAddress (0), 9);
              myClient.SetAttribute ("MaxPackets", UintegerValue (4294967295u));
              myClient.SetAttribute ("Interval", TimeValue (Time ("0.00001"))); //packets/s
              myClient.SetAttribute ("PacketSize", UintegerValue (payloadSize));
              ApplicationContainer clientApp = myClient.Install (wifiApNode.Get (0));
              clientApp.Start (Seconds (1.0));
              clientApp.Stop (Seconds (simulationTime + 1));
            }
          else
            {
              //TCP flow
              uint16_t port = 50000;
              Address apLocalAddress (InetSocketAddress (Ipv4Address::GetAny (), port));
              PacketSinkHelper packetSinkHelper ("ns3::TcpSocketFactory", apLocalAddress);
              sinkApp = packetSinkHelper.Install (wifiStaNode.Get (0));
              sinkApp.Start (Seconds (0.0));
              sinkApp.Stop (Seconds (simulationTime + 1));
              OnOffHelper onoff ("ns3::TcpSocketFactory",Ipv4Address::GetAny ());
              onoff.SetAttribute ("OnTime",  StringValue ("ns3::ConstantRandomVariable[Constant=1]"));
              onoff.SetAttribute ("OffTime", StringValue ("ns3::ConstantRandomVariable[Constant=0]"));
              onoff.SetAttribute ("PacketSize", UintegerValue (payloadSize));
              onoff.SetAttribute ("DataRate", DataRateValue (1000000000)); //bit/s
              ApplicationContainer apps;
              AddressValue remoteAddress (InetSocketAddress (staNodeInterface.GetAddress (0), port));
              onoff.SetAttribute ("Remote", remoteAddress);
              apps.Add (onoff.Install (wifiApNode.Get (0)));
              apps.Start (Seconds (1.0));
              apps.Stop (Seconds (simulationTime + 1));
            }
          Ipv4GlobalRoutingHelper::PopulateRoutingTables ();
          Simulator::Stop (Seconds (simulationTime + 1));
          Simulator::Run ();
          Simulator::Destroy ();
          double throughput = 0;
          if (udp)
            {
              //UDP
              uint32_t totalPacketsThrough = DynamicCast<UdpServer> (serverApp.Get (0))->GetReceived ();
              throughput = totalPacketsThrough * payloadSize * 8 / (simulationTime * 1000000.0); //Mbit/s
            }
          else
```

```
            {
              //TCP
              uint32_t totalPacketsThrough = DynamicCast<PacketSink> (sinkApp.Get (0))->GetTotalRx ();
              throughput = totalPacketsThrough * 8 / (simulationTime * 1000000.0); //Mbit/s
            }
            std::cout << i << "\t\t\t" << j << " MHz\t\t\t" << k << "\t\t\t" << throughput << " Mbit/s" << std::endl;
          }
        j *= 2;
        }
    }
  return 0;
}
```

# Experiment 3: Relation between the throughput and average number of active STAs

```
#include "ns3/core-module.h"
#include "ns3/wifi-net-device.h"
#include "ns3/yans-wifi-channel.h"
#include "ns3/yans-wifi-phy.h"
#include "ns3/propagation-delay-model.h"
#include "ns3/ptr.h"
#include "ns3/packet.h"
#include "ns3/simulator.h"
#include "ns3/nstime.h"
#include "ns3/command-line.h"
#include "ns3/wifi-tx-vector.h"

using namespace ns3;
class InterferenceExperiment
{
public:
  struct Input
  {
    Input ();
        double collision_prob;
        const int STA = 10;
        double simulationTime = 10; //seconds
        int SC_M; //Number of sub-channels
        Time interval;
    uint32_t packetSize;
    enum WifiPhyStandard standard;
    enum WifiPreamble preamble;
  };
  InterferenceExperiment ();
  void Run (struct InterferenceExperiment::Input input);
private:
  void SendA (void) const;
  void SendB (void) const;
  Ptr<YansWifiPhy> m_txA;
  Ptr<YansWifiPhy> m_txB;
  struct Input m_input;
};
```

```cpp
void
InterferenceExperiment::SendA (void) const
{
  Ptr<Packet> p = Create<Packet> (m_input.packetSizeA);
  WifiTxVector txVector;
  txVector.SetTxPowerLevel (m_input.txPowerLevelA);
  txVector.SetMode (WifiMode (m_input.txModeA));
  m_txA->SendPacket (p, txVector, m_input.preamble);
}
void
InterferenceExperiment::SendB (void) const
{
  Ptr<Packet> p = Create<Packet> (m_input.packetSizeB);
  WifiTxVector txVector;
  txVector.SetTxPowerLevel (m_input.txPowerLevelB);
  txVector.SetMode (WifiMode (m_input.txModeB));
  m_txB->SendPacket (p, txVector, m_input.preamble);
}
InterferenceExperiment::InterferenceExperiment ()
{
}
void
InterferenceExperiment::Run (struct InterferenceExperiment::Input input)
{
  m_input = input;
  double range = std::max (std::abs (input.xA), input.xB);
  Config::SetDefault ("ns3::RangePropagationLossModel::MaxRange", DoubleValue (range));
  Ptr<YansWifiChannel> channel = CreateObject<YansWifiChannel> ();
  channel->SetPropagationDelayModel (CreateObject<ConstantSpeedPropagationDelayModel> ());
  Ptr<RangePropagationLossModel> loss = CreateObject<RangePropagationLossModel> ();
  channel->SetPropagationLossModel (loss);
  Ptr<MobilityModel> posTxA = CreateObject<ConstantPositionMobilityModel> ();
  posTxA->SetPosition (Vector (input.xA, 0.0, 0.0));
  Ptr<MobilityModel> posTxB = CreateObject<ConstantPositionMobilityModel> ();
  posTxB->SetPosition (Vector (input.xB, 0.0, 0.0));
  Ptr<MobilityModel> posRx = CreateObject<ConstantPositionMobilityModel> ();
  posRx->SetPosition (Vector (0.0, 0.0, 0.0));
  m_txA = CreateObject<YansWifiPhy> ();
  m_txB = CreateObject<YansWifiPhy> ();
  Ptr<YansWifiPhy> rx = CreateObject<YansWifiPhy> ();
  Ptr<ErrorRateModel> error = CreateObject<YansErrorRateModel> ();
  m_txA->SetErrorRateModel (error);
  m_txB->SetErrorRateModel (error);
  rx->SetErrorRateModel (error);
  m_txA->SetChannel (channel);
  m_txB->SetChannel (channel);
  rx->SetChannel (channel);
  m_txA->SetMobility (posTxA);
  m_txB->SetMobility (posTxB);
  rx->SetMobility (posRx);
  m_txA->ConfigureStandard (input.standard);
  m_txB->ConfigureStandard (input.standard);
  rx->ConfigureStandard (input.standard);
  Simulator::Schedule (Seconds (0), &InterferenceExperiment::SendA, this);
  Simulator::Schedule (Seconds (0) + input.interval, &InterferenceExperiment::SendB, this);
  Simulator::Run ();
```

```
  Simulator::Destroy ();
}
int main (int argc, char *argv[])
{
  InterferenceExperiment::Input input;
  std::string str_standard = "WIFI_PHY_STANDARD_80211a";
  std::string str_preamble = "WIFI_PREAMBLE_LONG";
  double delay = 0; //microseconds
  CommandLine cmd;
  cmd.AddValue ("delay", "Delay in microseconds between frame transmission from sender A and frame transmission from sender B", delay);
  cmd.AddValue ("xA", "The position of transmitter A (< 0)", input.xA);
  cmd.AddValue ("xB", "The position of transmitter B (> 0)", input.xB);
  cmd.AddValue ("packetSizeA", "Packet size in bytes of transmitter A", input.packetSizeA);
  cmd.AddValue ("packetSizeB", "Packet size in bytes of transmitter B", input.packetSizeB);
  cmd.AddValue ("txPowerA", "TX power level of transmitter A", input.txPowerLevelA);
  cmd.AddValue ("txPowerB", "TX power level of transmitter B", input.txPowerLevelB);
  cmd.AddValue ("txModeA", "Wifi mode used for payload transmission of sender A", input.txModeA);
  cmd.AddValue ("txModeB", "Wifi mode used for payload transmission of sender B", input.txModeB);
  cmd.AddValue ("standard", "IEEE 802.11 flavor", str_standard);
  cmd.AddValue ("preamble", "Type of preamble", str_preamble);
  cmd.Parse (argc, argv);
  LogComponentEnable ("YansWifiPhy", LOG_LEVEL_ALL);
  LogComponentEnable ("InterferenceHelper", LOG_LEVEL_ALL);
  input.interval = MicroSeconds (delay);
    return 0;
   }
  InterferenceExperiment experiment;
  experiment.Run (input);
  return 0;
}
```

# Experiment 4: Saturated throughput with respect to the backoff slot duration

```
#include <cmath>
#include <sstream>
#include "ns3/core-module.h"
#include "ns3/network-module.h"
#include "ns3/wifi-module.h"
#include "ns3/stats-module.h"
#include "ns3/mobility-module.h"
#include "ns3/propagation-module.h"

using namespace ns3;
NS_LOG_COMPONENT_DEFINE ("IdealWifiManagerExample");
// 290K @ 20 MHz
const double NOISE_DBM_Hz = -174.0;
double noiseDbm = NOISE_DBM_Hz;
double g_intervalBytes = 0;
uint64_t g_intervalRate = 0;
void
PacketRx (Ptr<const Packet> pkt, const Address &addr)
```

```
{
  NS_LOG_DEBUG ("Received size " << pkt->GetSize ());
  g_intervalBytes += pkt->GetSize ();
}
void
RateChange (uint64_t oldVal, uint64_t newVal)
{
  NS_LOG_DEBUG ("Change from " << oldVal << " to " << newVal);
  g_intervalRate = newVal;
}
struct Step
{
  double stepSize;
  double stepTime;
};
struct StandardInfo
{
  StandardInfo ()
  {
    m_name = "none";
  }
  std::string m_name;
  enum WifiPhyStandard m_standard;
  uint32_t m_width;
  double m_xMin;
  double m_xMax;
  double m_yMax;
};
void
ChangeSignalAndReportRate (Ptr<FixedRssLossModel> rssModel, struct Step step, double rss,
Gnuplot2dDataset& rateDataset, Gnuplot2dDataset& actualDataset)
{
  NS_LOG_FUNCTION (rssModel << step.stepSize << step.stepTime << rss);
  double snr = rss - noiseDbm;
  rateDataset.Add (snr, g_intervalRate / 1000000.0);
  // Calculate received rate since last interval
  double currentRate = ((g_intervalBytes * 8)/step.stepTime) / 1e6; // Mb/s
  actualDataset.Add (snr, currentRate);
  rssModel->SetRss (rss - step.stepSize);
  NS_LOG_INFO ("At time " << Simulator::Now ().As (Time::S) << "; observed rate " <<
currentRate << "; setting new power to " << rss - step.stepSize);
  g_intervalBytes = 0;
  Simulator::Schedule (Seconds (step.stepTime), &ChangeSignalAndReportRate, rssModel, step,
(rss - step.stepSize), rateDataset, actualDataset);
}
int main (int argc, char *argv[])
{
  std::vector <StandardInfo> standards;
  uint32_t steps;
  uint32_t rtsThreshold = 999999;  // disabled even for large A-MPDU
  double stepSize = 1; // dBm
  double stepTime = 0.5; // seconds
  uint32_t packetSize = 1024;  // bytes
  int broadcast = 0;
  int ap1_x = 0;
  int ap1_y = 0;
  int sta1_x = 5;
```

```cpp
  int sta1_y = 0;
  uint16_t nss = 1;
  bool shortGuardInterval = false;
  uint32_t channelWidth = 20;
  std::string standard ("802.11b");
  StandardInfo selectedStandard;
  CommandLine cmd;
  cmd.AddValue ("rtsThreshold", "RTS threshold", rtsThreshold);
  cmd.AddValue ("stepSize", "Power between steps (dBm)", stepSize);
  cmd.AddValue ("stepTime", "Time on each step (seconds)", stepTime);
  cmd.AddValue ("broadcast", "Send broadcast instead of unicast", broadcast);
  if (standard == "802.11n-5GHz" || standard == "802.11n-2.4GHz" || standard == "802.11ac")
    {
      std::ostringstream oss;
      std::string gi;
      if (shortGuardInterval)
        {
          gi = "SGI";
        }
      else
        {
          gi = "LGI";
        }
      oss << "-" << channelWidth << "MHz-" << gi << "-" <<nss << "SS";
      plotName += oss.str ();
      dataName += oss.str ();
    }
  mobility.SetPositionAllocator (positionAlloc);
  mobility.SetMobilityModel ("ns3::ConstantPositionMobilityModel");
  mobility.Install (clientNode);
  mobility.Install (serverNode);
  Gnuplot2dDataset rateDataset (selectedStandard.m_name + std::string ("-rate selected"));
  Gnuplot2dDataset actualDataset (selectedStandard.m_name + std::string ("-observed"));
  struct Step step;
  step.stepSize = stepSize;
  step.stepTime = stepTime;

  // Only set the channel width and guard interval for HT and VHT modes
  if (selectedStandard.m_name == "802.11n-5GHz" ||
      selectedStandard.m_name == "802.11n-2.4GHz" ||
      selectedStandard.m_name == "802.11ac")
    {
      wifiPhyPtrClient->SetChannelWidth (selectedStandard.m_width);
      wifiPhyPtrServer->SetChannelWidth (selectedStandard.m_width);
      wifiPhyPtrClient->SetGuardInterval (shortGuardInterval);
      wifiPhyPtrServer->SetGuardInterval (shortGuardInterval);
    }
  NS_LOG_DEBUG ("Channel width " << wifiPhyPtrClient->GetChannelWidth () << " noiseDbm " << noiseDbm);
  NS_LOG_DEBUG ("NSS " << wifiPhyPtrClient->GetSupportedTxSpatialStreams ());
  // Configure signal and noise, and schedule first iteration
  noiseDbm += 10 * log10 (selectedStandard.m_width * 1000000);
  double rssCurrent = (selectedStandard.m_snrHigh + noiseDbm);
  rssLossModel->SetRss (rssCurrent);
  NS_LOG_INFO ("Setting initial Rss to " << rssCurrent);
  //Move the STA by stepsSize meters every stepTime seconds
```

```
  Simulator::Schedule (Seconds (0.5 + stepTime), &ChangeSignalAndReportRate, rssLossModel,
step, rssCurrent, rateDataset, actualDataset);
  PacketSocketHelper packetSocketHelper;
  packetSocketHelper.Install (serverNode);
  packetSocketHelper.Install (clientNode);

  PacketSocketAddress socketAddr;
  socketAddr.SetSingleDevice (serverDevice.Get (0)->GetIfIndex ());
  if (broadcast)
    {
      socketAddr.SetPhysicalAddress (serverDevice.Get (0)->GetBroadcast ());
    }
  else
    {
      socketAddr.SetPhysicalAddress (serverDevice.Get (0)->GetAddress ());
    }
  clientNode->AddApplication (client);
  Ptr<PacketSocketServer> server = CreateObject<PacketSocketServer> ();
  server->SetLocal (socketAddr);
  server->TraceConnectWithoutContext ("Rx", MakeCallback (&PacketRx));
  serverNode->AddApplication (server);
  Simulator::Stop (Seconds ((steps + 1) * stepTime));
  Simulator::Run ();
  Simulator::Destroy ();
  gnuplot.AddDataset (rateDataset);
  gnuplot.AddDataset (actualDataset);
  std::ostringstream xMinStr, xMaxStr, yMaxStr;
  std::string xRangeStr ("set xrange [");
  xMinStr << selectedStandard.m_xMin;
  xRangeStr.append (xMinStr.str ());
  xRangeStr.append (":");
  xMaxStr << selectedStandard.m_xMax;
  xRangeStr.append (xMaxStr.str ());
  xRangeStr.append ("]");
  std::string yRangeStr ("set yrange [0:");
  yMaxStr << selectedStandard.m_yMax;
  yRangeStr.append (yMaxStr.str ());
  yRangeStr.append ("]");
  std::ostringstream widthStrStr;
  std::ostringstream nssStrStr;
  std::string title ("Wi-Fi ideal rate control: ");
  title.append (standard);
  title.append (" channel width: ");
  widthStrStr << selectedStandard.m_width;
  title.append (widthStrStr.str ());
  title.append (" MHz nss: ");
  nssStrStr << nss;
  title.append (nssStrStr.str ());
  if (shortGuardInterval == true)
    {
      title.append (" shortGuard: true");
    }
  gnuplot.SetTerminal ("postscript eps color enh \"Times-BoldItalic\"");
  gnuplot.SetLegend ("SNR (dB)", "Rate (Mb/s)");
  gnuplot.SetTitle (title);
  gnuplot.SetExtra (xRangeStr);
  gnuplot.AppendExtra (yRangeStr);
```

```
  gnuplot.AppendExtra ("set key top left");
  gnuplot.GenerateOutput (outfile);
  outfile.close ();
  return 0;
}
```

# Experiment 5: Normalized throughput of the STAs with different traffic loads

```
#include <string.h>
#include "ns3/core-module.h"
#include "ns3/config-store-module.h"
#include "ns3/network-module.h"
#include "ns3/wifi-module.h"

using namespace ns3;
NS_LOG_COMPONENT_DEFINE ("WifiPhyConfigurationExample");
Ptr<YansWifiPhy>
GetYansWifiPhyPtr (const NetDeviceContainer &nc)
{
  Ptr<WifiNetDevice> wnd = nc.Get (0)->GetObject<WifiNetDevice> ();
  Ptr<WifiPhy> wp = wnd->GetPhy ();
  return wp->GetObject<YansWifiPhy> ();
}
void
PrintAttributesIfEnabled (bool enabled)
{
  if (enabled)
    {
      ConfigStore outputConfig;
      outputConfig.ConfigureAttributes ();
    }
}
int main (int argc, char *argv[])
{
  uint32_t testCase = 0;
  bool printAttributes = false;
  CommandLine cmd;
  cmd.AddValue ("testCase", "Test case", testCase);
  cmd.AddValue ("printAttributes", "If true, print out attributes", printAttributes);
  cmd.Parse (argc, argv);
  NodeContainer wifiStaNode;
  wifiStaNode.Create (1);
  NodeContainer wifiApNode;
  wifiApNode.Create (1);
  YansWifiChannelHelper channel = YansWifiChannelHelper::Default ();
  YansWifiPhyHelper phy = YansWifiPhyHelper::Default ();
  phy.SetChannel (channel.Create ());
  WifiHelper wifi;
  wifi.SetRemoteStationManager ("ns3::IdealWifiManager");
  // Configure and declare other generic components of this example
  Ssid ssid;
  ssid = Ssid ("wifi-phy-configuration");
  WifiMacHelper macSta;
  macSta.SetType ("ns3::StaWifiMac",
```

```
          "Ssid", SsidValue (ssid),
            "ActiveProbing", BooleanValue (false));
  WifiMacHelper macAp;
  macAp.SetType ("ns3::ApWifiMac",
           "Ssid", SsidValue (ssid),
             "BeaconInterval", TimeValue (MicroSeconds (102400)),
             "BeaconGeneration", BooleanValue (true));
  NetDeviceContainer staDevice;
  NetDeviceContainer apDevice;
  Ptr<YansWifiPhy> phySta;
  Config::SetDefault ("ns3::ConfigStore::Filename", StringValue ("output-attributes-" + std::to_string (testCase) + ".txt"));
  Config::SetDefault ("ns3::ConfigStore::FileFormat", StringValue ("RawText"));
  Config::SetDefault ("ns3::ConfigStore::Mode", StringValue ("Save"));
  switch (testCase)
  {
    case 0:
      // Default configuration, without WifiHelper::SetStandard or WifiHelper
      phySta = CreateObject<YansWifiPhy> ();
      // The default results in an invalid configuration of channel 0,
      // width 20, and frequency 0 MHz
      NS_ASSERT (phySta->GetChannelNumber () == 0);
      NS_ASSERT (phySta->GetChannelWidth () == 20);
      NS_ASSERT (phySta->GetFrequency () == 0);
      PrintAttributesIfEnabled (printAttributes);
      break;
    // The following cases test the setting of WifiPhyStandard alone;
    // i.e. without further channel number/width/frequency configuration
    case 1:
      // By default, WifiHelper will use WIFI_PHY_STANDARD_80211a
      staDevice = wifi.Install (phy, macSta, wifiStaNode.Get(0));
      apDevice = wifi.Install (phy, macAp, wifiApNode.Get(0));
      phySta = GetYansWifiPhyPtr (staDevice);
      // The research expects channel 36, width 20, frequency 5180
      NS_ASSERT (phySta->GetChannelNumber () == 36);
      NS_ASSERT (phySta->GetChannelWidth () == 20);
      NS_ASSERT (phySta->GetFrequency () == 5180);
      PrintAttributesIfEnabled (printAttributes);
      break;
    case 2:
      wifi.SetStandard (WIFI_PHY_STANDARD_80211b);
      staDevice = wifi.Install (phy, macSta, wifiStaNode.Get(0));
      apDevice = wifi.Install (phy, macAp, wifiApNode.Get(0));
      phySta = GetYansWifiPhyPtr (staDevice);
      //The research expects channel 1, width 22, frequency 2412
      NS_ASSERT (phySta->GetChannelNumber () == 1);
      NS_ASSERT (phySta->GetChannelWidth () == 22);
      NS_ASSERT (phySta->GetFrequency () == 2412);
      PrintAttributesIfEnabled (printAttributes);
      break;
    case 3:
      wifi.SetStandard (WIFI_PHY_STANDARD_80211g);
      staDevice = wifi.Install (phy, macSta, wifiStaNode.Get(0));
      apDevice = wifi.Install (phy, macAp, wifiApNode.Get(0));
      phySta = GetYansWifiPhyPtr (staDevice);
      // The research expects channel 1, width 20, frequency 2412
      NS_ASSERT (phySta->GetChannelNumber () == 1);
```

```
        NS_ASSERT (phySta->GetChannelWidth () == 20);
        NS_ASSERT (phySta->GetFrequency () == 2412);
        PrintAttributesIfEnabled (printAttributes);
        break;
      case 4:
        wifi.SetStandard (WIFI_PHY_STANDARD_80211n_5GHZ);
        staDevice = wifi.Install (phy, macSta, wifiStaNode.Get(0));
        apDevice = wifi.Install (phy, macAp, wifiApNode.Get(0));
        phySta = GetYansWifiPhyPtr (staDevice);
        // The research expects channel 36, width 20, frequency 5180
        NS_ASSERT (phySta->GetChannelNumber () == 36);
        NS_ASSERT (phySta->GetChannelWidth () == 20);
        NS_ASSERT (phySta->GetFrequency () == 5180);
        PrintAttributesIfEnabled (printAttributes);
        break;
      case 5:
        wifi.SetStandard (WIFI_PHY_STANDARD_80211n_2_4GHZ);
        staDevice = wifi.Install (phy, macSta, wifiStaNode.Get(0));
        apDevice = wifi.Install (phy, macAp, wifiApNode.Get(0));
        phySta = GetYansWifiPhyPtr (staDevice);
        // The research expects channel 1, width 20, frequency 2412
        NS_ASSERT (phySta->GetChannelNumber () == 1);
        NS_ASSERT (phySta->GetChannelWidth () == 20);
        NS_ASSERT (phySta->GetFrequency () == 2412);
        PrintAttributesIfEnabled (printAttributes);
        break;
      case 6:
        wifi.SetStandard (WIFI_PHY_STANDARD_80211ax);
        staDevice = wifi.Install (phy, macSta, wifiStaNode.Get(0));
        apDevice = wifi.Install (phy, macAp, wifiApNode.Get(0));
        phySta = GetYansWifiPhyPtr (staDevice);
        //The research expects channel 42, width 80, frequency 5210
        NS_ASSERT (phySta->GetChannelNumber () == 42);
        NS_ASSERT (phySta->GetChannelWidth () == 80);
        NS_ASSERT (phySta->GetFrequency () == 5210);
        PrintAttributesIfEnabled (printAttributes);
        break;
      case 7:
        wifi.SetStandard (WIFI_PHY_STANDARD_80211_10MHZ);
        staDevice = wifi.Install (phy, macSta, wifiStaNode.Get(0));
        apDevice = wifi.Install (phy, macAp, wifiApNode.Get(0));
        phySta = GetYansWifiPhyPtr (staDevice);
        //The research expects channel 172, width 10, frequency 5860
        NS_ASSERT (phySta->GetChannelNumber () == 172);
        NS_ASSERT (phySta->GetChannelWidth () == 10);
        NS_ASSERT (phySta->GetFrequency () == 5860);
        PrintAttributesIfEnabled (printAttributes);
        break;
      case 8:
        wifi.SetStandard (WIFI_PHY_STANDARD_80211_5MHZ);
        staDevice = wifi.Install (phy, macSta, wifiStaNode.Get(0));
        apDevice = wifi.Install (phy, macAp, wifiApNode.Get(0));
        phySta = GetYansWifiPhyPtr (staDevice);
        // The research expects channel 0, width 5, frequency 5860
        // Channel 0 because 5MHz channels are not officially defined
        NS_ASSERT (phySta->GetChannelNumber () == 0);
        NS_ASSERT (phySta->GetChannelWidth () == 5);
```

```cpp
      NS_ASSERT (phySta->GetFrequency () == 5860);
      PrintAttributesIfEnabled (printAttributes);
      break;
    case 9:
      wifi.SetStandard (WIFI_PHY_STANDARD_holland);
      staDevice = wifi.Install (phy, macSta, wifiStaNode.Get(0));
      apDevice = wifi.Install (phy, macAp, wifiApNode.Get(0));
      phySta = GetYansWifiPhyPtr (staDevice);
      // The research expects channel 36, width 20, frequency 5180
      NS_ASSERT (phySta->GetChannelNumber () == 36);
      NS_ASSERT (phySta->GetChannelWidth () == 20);
      NS_ASSERT (phySta->GetFrequency () == 5180);
      PrintAttributesIfEnabled (printAttributes);
      break;
    case 10:
      wifi.SetStandard (WIFI_PHY_STANDARD_80211n_5GHZ);
      phy.Set ("ChannelNumber", UintegerValue(44));
      staDevice = wifi.Install (phy, macSta, wifiStaNode.Get(0));
      apDevice = wifi.Install (phy, macAp, wifiApNode.Get(0));
      phySta = GetYansWifiPhyPtr (staDevice);
      // The research expects channel 44, width 20, frequency 5220
      NS_ASSERT (phySta->GetChannelNumber () == 44);
      NS_ASSERT (phySta->GetChannelWidth () == 20);
      NS_ASSERT (phySta->GetFrequency () == 5220);
      PrintAttributesIfEnabled (printAttributes);
      break;
    case 11:
      wifi.SetStandard (WIFI_PHY_STANDARD_80211n_5GHZ);
      phy.Set ("ChannelNumber", UintegerValue(44));
      staDevice = wifi.Install (phy, macSta, wifiStaNode.Get(0));
      apDevice = wifi.Install (phy, macAp, wifiApNode.Get(0));
      phySta = GetYansWifiPhyPtr (staDevice);
      // Post-install reconfiguration to channel number 40
      Config::Set ("/NodeList/1/DeviceList/*/$ns3::WifiNetDevice/Phy/$ns3::YansWifiPhy/ChannelNumber", UintegerValue(40));
      // The research expects channel 40, width 20, frequency 5200
      NS_ASSERT (phySta->GetChannelNumber () == 40);
      NS_ASSERT (phySta->GetChannelWidth () == 20);
      NS_ASSERT (phySta->GetFrequency () == 5200);
      PrintAttributesIfEnabled (printAttributes);
      break;
    case 12:
      wifi.SetStandard (WIFI_PHY_STANDARD_80211n_5GHZ);
      phy.Set ("ChannelNumber", UintegerValue (44));
      staDevice = wifi.Install (phy, macSta, wifiStaNode.Get(0));
      apDevice = wifi.Install (phy, macAp, wifiApNode.Get(0));
      phySta = GetYansWifiPhyPtr (staDevice);
      // Post-install reconfiguration to channel width 40 MHz
      Config::Set ("/NodeList/1/DeviceList/*/$ns3::WifiNetDevice/Phy/$ns3::YansWifiPhy/ChannelWidth", UintegerValue(40));
      return 0;
}
```

# Experiment 6: Total system throughput and the max–min fairness

```cpp
#include "ns3/wifi-net-device.h"
#include "ns3/yans-wifi-channel.h"
#include "ns3/yans-wifi-phy.h"
#include "ns3/ptr.h"
#include "ns3/mobility-model.h"
#include "ns3/vector.h"
#include "ns3/packet.h"
#include "ns3/simulator.h"
#include "ns3/nstime.h"
#include "ns3/command-line.h"
#include "ns3/flow-id-tag.h"
#include "ns3/wifi-tx-vector.h"

using namespace ns3;
class PsrExperiment
{
public:
  struct Input
  {
    Input ();
    double distance;
    std::string txMode;
    uint8_t txPowerLevel;
    uint32_t packetSize;
    uint32_t nPackets;
  };
  struct Output
  {
    uint32_t received;
  };
  PsrExperiment ();
  struct PsrExperiment::Output Run (struct PsrExperiment::Input input);
private:
  void Send (void);
  void Receive (Ptr<Packet> p, double snr, WifiTxVector txVector, enum WifiPreamble preamble);
  Ptr<WifiPhy> m_tx;
  struct Input m_input;
  struct Output m_output;
};
void
void
PsrExperiment::Receive (Ptr<Packet> p, double snr, WifiTxVector txVector, enum WifiPreamble preamble)
{
  m_output.received++;
}
PsrExperiment::PsrExperiment ()
{
}
PsrExperiment::Input::Input ()
  : distance (5.0),
    txMode ("OfdmRate6Mbps"),
    txPowerLevel (0),
    packetSize (2304),
```

```
    nPackets (400)
struct PsrExperiment::Output
class CollisionExperiment
{
public:
  struct Input
  {
    Input ();
    Time interval;
    double xA;
    double xB;
    std::string txModeA;
    std::string txModeB;
    uint32_t packetSizeA;
    uint32_t packetSizeB;
    uint32_t nPackets;
  };
  struct Output
  {
    uint32_t receivedA;
    uint32_t receivedB;
  };
  CollisionExperiment ();
  struct CollisionExperiment::Output Run (struct CollisionExperiment::Input input);
private:
  void SendA (void) const;
  void SendB (void) const;
  void Receive (Ptr<Packet> p, double snr, WifiTxVector txVector, enum WifiPreamble preamble);
  Ptr<WifiPhy> m_txA;
  Ptr<WifiPhy> m_txB;
  uint32_t m_flowIdA;
  uint32_t m_flowIdB;
  struct Input m_input;
  struct Output m_output;
};
void
CollisionExperiment::SendA (void) const
{
  Ptr<Packet> p = Create<Packet> (m_input.packetSizeA);
  p->AddByteTag (FlowIdTag (m_flowIdA));
  WifiTxVector txVector;
  txVector.SetTxPowerLevel (m_input.txPowerLevelA);
  txVector.SetMode (WifiMode (m_input.txModeA));
  m_txA->SendPacket (p, txVector, WIFI_PREAMBLE_LONG);
}
void
CollisionExperiment::SendB (void) const
{
  Ptr<Packet> p = Create<Packet> (m_input.packetSizeB);
  p->AddByteTag (FlowIdTag (m_flowIdB));
  WifiTxVector txVector;
  txVector.SetTxPowerLevel (m_input.txPowerLevelB);
  txVector.SetMode (WifiMode (m_input.txModeB));
  m_txB->SendPacket (p, txVector, WIFI_PREAMBLE_LONG);
}
void
```

```
CollisionExperiment::Receive (Ptr<Packet> p, double snr, WifiTxVector txVector, enum
WifiPreamble preamble)
{
  FlowIdTag tag;
  if (p->FindFirstMatchingByteTag (tag))
    {
      if (tag.GetFlowId () == m_flowIdA)
        {
          m_output.receivedA++;
        }
      else if (tag.GetFlowId () == m_flowIdB)
        {
          m_output.receivedB++;
        }
    }
}
CollisionExperiment::CollisionExperiment ()
{
}
CollisionExperiment::Input::Input ()
 : interval (MicroSeconds (0)),
   xA (-5),
   xB (5),
   txModeA ("OfdmRate6Mbps"),
   txModeB ("OfdmRate6Mbps"),
   txPowerLevelA (0),
   txPowerLevelB (0),
   packetSizeA (2304),
   packetSizeB (2304),
   nPackets (400)
{
}
struct CollisionExperiment::Output
CollisionExperiment::Run (struct CollisionExperiment::Input input)
{
  m_output.receivedA = 0;
  m_output.receivedB = 0;
  m_input = input;
  m_flowIdA = FlowIdTag::AllocateFlowId ();
  m_flowIdB = FlowIdTag::AllocateFlowId ();
  Ptr<YansWifiChannel> channel = CreateObject<YansWifiChannel> ();
  channel->SetPropagationDelayModel (CreateObject<ConstantSpeedPropagationDelayModel>
());
  Ptr<LogDistancePropagationLossModel> log =
CreateObject<LogDistancePropagationLossModel> ();
  channel->SetPropagationLossModel (log);
    Simulator::Schedule (Seconds (i), &CollisionExperiment::SendA, this);
  }
  for (uint32_t i = 0; i < m_input.nPackets; ++i)
    {
      Simulator::Schedule (Seconds (i) + m_input.interval, &CollisionExperiment::SendB, this);
    }
  m_txA = txA;
  m_txB = txB;
  Simulator::Run ();
  Simulator::Destroy ();
  return m_output;
```

```cpp
}
double CalcPsr (struct PsrExperiment::Output output, struct PsrExperiment::Input input)
{
  double psr = output.received;
  psr /= input.nPackets;
  return psr;
}
int main (int argc, char *argv[])
{
  if (argc <= 1)
    {
      std::cout << "Available experiments: "
                << "Psr "
                << "SizeVsRange "
                << "PsrVsDistance "
                << "PsrVsCollisionInterval "
                << std::endl;
      return 0;
    }
  std::string type = argv[1];
  argc--;
  argv[1] = argv[0];
  argv++;
  if (type == "Psr")
    {
      PrintPsr (argc, argv);
    }
  else if (type == "SizeVsRange")
    {
      PrintSizeVsRange (argc, argv);
    }
  else if (type == "PsrVsDistance")
    {
      PrintPsrVsDistance (argc, argv);
    }
  else if (type == "PsrVsCollisionInterval")
    {
      PrintPsrVsCollisionInterval (argc, argv);
    }
  else
    {
      std::cout << "Wrong arguments!" << std::endl;
    }
  return 0;
}
```

# Appendix C

# HTFA Protocol Data for Graphing

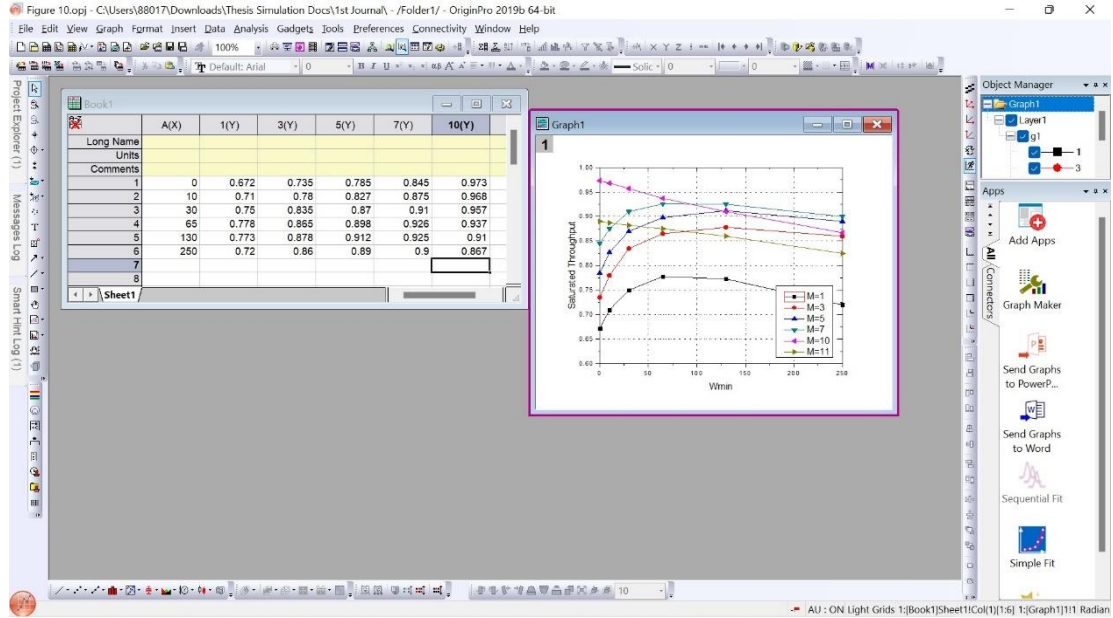

**Experiment 1: Saturated throughput with respect to $W_{min}$**

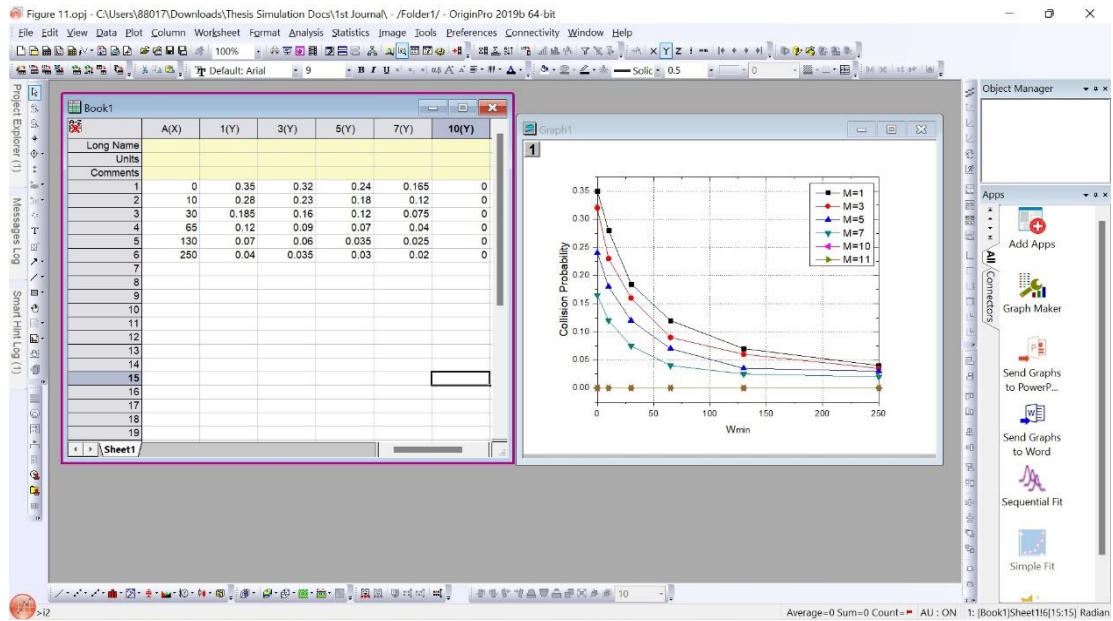

**Experiment 2: Collision probability with respect to $W_{min}$**

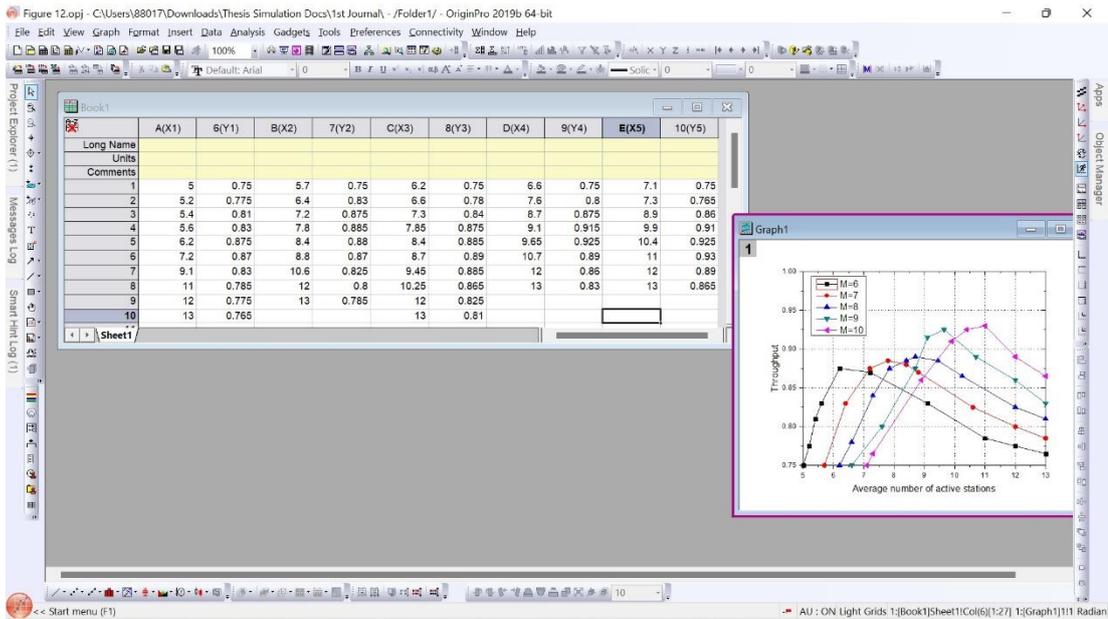

**Experiment 3: Relation between the throughput and average number of active STAs**

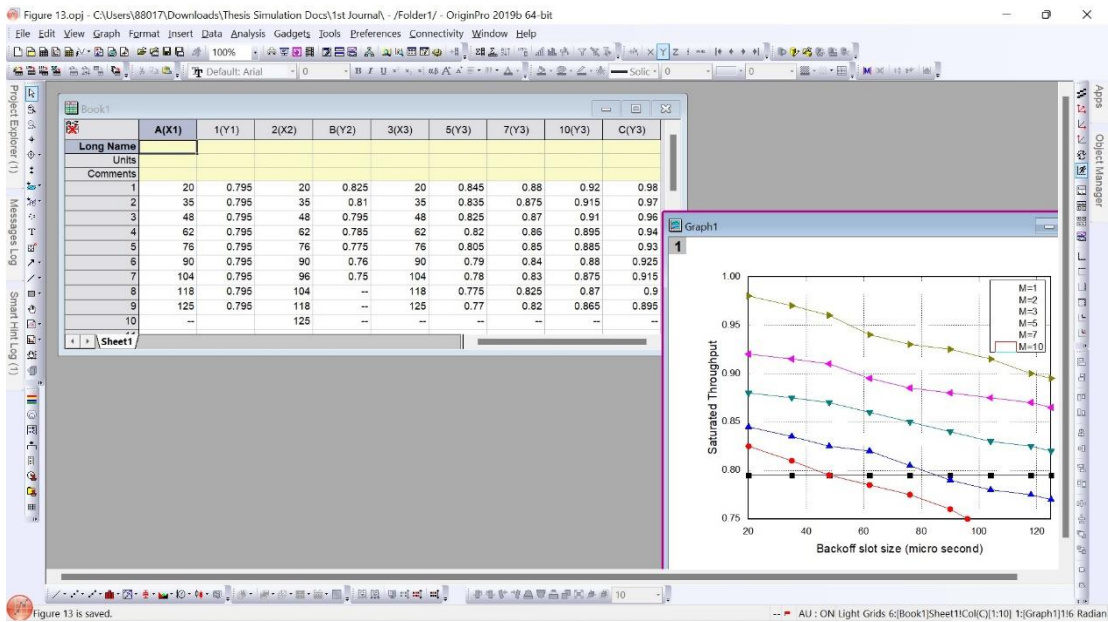

**Experiment 4: Saturated throughput with respect to the backoff slot duration**

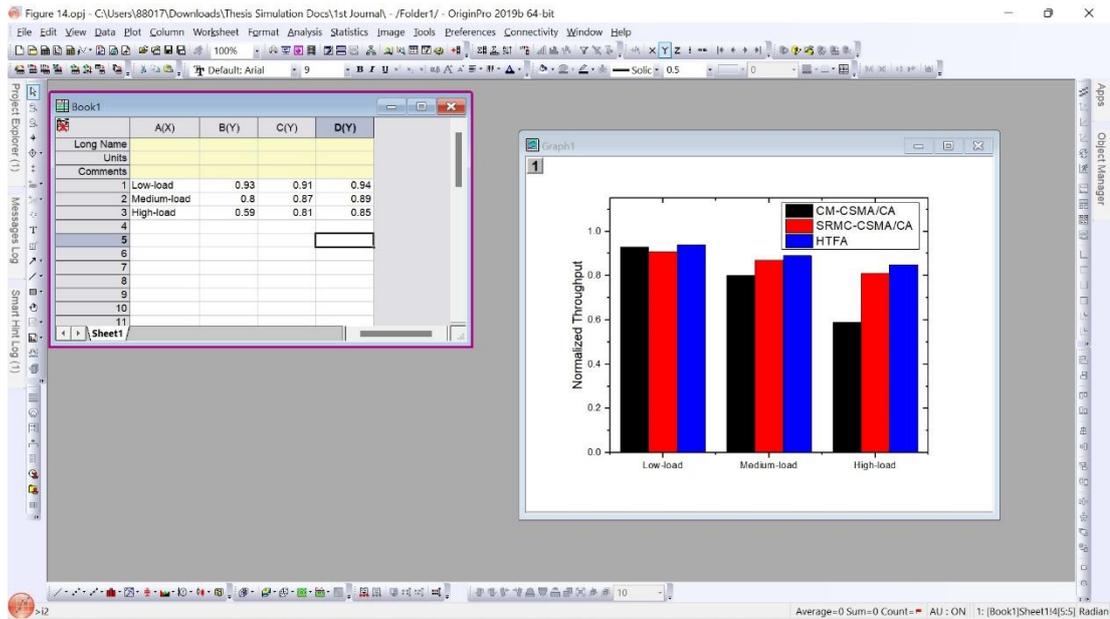

**Experiment 5: Normalized throughput of the STAs with different traffic loads**

TOTAL THROUGHPUT AND MAX-MIN FAIRNESS COMPARISON

| Metric | CM-CSMA/CA | SRMC-CSMA/CA | HTFA |
|---|---|---|---|
| $T$ (Mbit/sec) | 41.20 | 47.6 | 49.3 |
| $F$ | 0.31 | 0.07 | 0.05 |

**Experiment 6: Total system throughput and the max-min fairness**

# Appendix D

# NS-3 Code for ERA Protocol

## Experiment 1: Comparison of the throughput between different protocols

```
#include "ns3/mobility-module.h"
#include "ns3/core-module.h"
#include "ns3/network-module.h"
#include "ns3/applications-module.h"
#include "ns3/wifi-module.h"
#include "ns3/internet-module.h"

using namespace ns3;
//    STA      AP
//     *        *
//     |        |
//   n1-n25    n26
NS_LOG_COMPONENT_DEFINE ("ht-wifi-network");
int main (int argc, char *argv[])
{
  bool udp = true;
  double simulationTime = 10; //seconds
  double distance = 1.0; //meters
  double frequency = 5.0; //whether 2.4 or 5.0 GHz
  const int STA = 25;
  radius_meter = 15;
  antenna_AP = 4;
  antenna_STAs = 1;
  power_dBm = 15;
  band_GHz = 5;
  bandwidth_MHz = 40;
  CommandLine cmd;
  cmd.AddValue ("frequency", "Whether working in the 2.4 or 5.0 GHz band (other values gets rejected)", frequency);
  cmd.AddValue ("distance", "Distance in meters between the station and the access point", distance);
  cmd.AddValue ("simulationTime", "Simulation time in seconds", simulationTime);
  cmd.AddValue ("udp", "UDP if set to 1, TCP otherwise", udp);
  cmd.Parse (argc,argv);
  std::cout << "MCS value" << "\t\t" << "Channel width" << "\t\t" << "short GI" << "\t\t" << "Throughput" << '\n';
  for (int i = 0; i <= 7; i++) //MCS
    {
      for (int j = 20; j <= 40; ) //channel width
        {
          for (int k = 0; k < 2; k++) //GI: 0 and 1
            {
              uint32_t payloadSize; //1500 byte IP packet
              if (udp)
                {
                  payloadSize = 1472; //bytes
                }
```

```cpp
      else
        {
          payloadSize = 1448; //bytes
          Config::SetDefault ("ns3::TcpSocket::SegmentSize", UintegerValue (payloadSize));
        }
    NodeContainer wifiStaNode;
    wifiStaNode.Create (1);
    NodeContainer wifiApNode;
    wifiApNode.Create (1);
    YansWifiChannelHelper channel = YansWifiChannelHelper::Default ();
    YansWifiPhyHelper phy = YansWifiPhyHelper::Default ();
    phy.SetChannel (channel.Create ());
    // Set guard interval
    phy.Set ("ShortGuardEnabled", BooleanValue (k));
    WifiMacHelper mac;
    WifiHelper wifi;
    if (frequency == 5.0)
        {
  std::ofstream yansfile ("yans-frame-success-rate-n.plt");
  std::ofstream nistfile ("nist-frame-success-rate-n.plt");
  std::vector <std::string> modes;
  modes.push_back ("HtMcs0");
  modes.push_back ("HtMcs1");
  modes.push_back ("HtMcs2");
  modes.push_back ("HtMcs3");
  modes.push_back ("HtMcs4");
  modes.push_back ("HtMcs5");
  modes.push_back ("HtMcs6");
  modes.push_back ("HtMcs7");
  CommandLine cmd;
  cmd.AddValue ("FrameSize", "The frame size", FrameSize);
  cmd.Parse (argc, argv);
      std::ostringstream oss;
      oss << "HtMcs" << i;
      wifi.SetRemoteStationManager ("ns3::ConstantRateWifiManager","DataMode",
StringValue (oss.str ()),
                                    "ControlMode", StringValue (oss.str ()));

      Ssid ssid = Ssid ("ns3-80211n");
      mac.SetType ("ns3::StaWifiMac",
                   "Ssid", SsidValue (ssid));
      NetDeviceContainer staDevice;
      staDevice = wifi.Install (phy, mac, wifiStaNode);
      mac.SetType ("ns3::ApWifiMac",
                   "Ssid", SsidValue (ssid));
      NetDeviceContainer apDevice;
      apDevice = wifi.Install (phy, mac, wifiApNode);
      // Set channel width
      Config::Set ("/NodeList/*/DeviceList/*/$ns3::WifiNetDevice/Phy/ChannelWidth",
UintegerValue (j));
      // mobility.
      MobilityHelper mobility;
      Ptr<ListPositionAllocator> positionAlloc = CreateObject<ListPositionAllocator> ();
      positionAlloc->Add (Vector (0.0, 0.0, 0.0));
      positionAlloc->Add (Vector (distance, 0.0, 0.0));
      mobility.SetPositionAllocator (positionAlloc);
      mobility.SetMobilityModel ("ns3::ConstantPositionMobilityModel");
```

```cpp
      mobility.Install (wifiApNode);
      mobility.Install (wifiStaNode);
      /* Internet stack*/
      InternetStackHelper stack;
      stack.Install (wifiApNode);
      stack.Install (wifiStaNode);
      Ipv4AddressHelper address;

      address.SetBase ("192.168.1.0", "255.255.255.0");
      Ipv4InterfaceContainer staNodeInterface;
      Ipv4InterfaceContainer apNodeInterface;
      staNodeInterface = address.Assign (staDevice);
      apNodeInterface = address.Assign (apDevice);
        sinkApp.Start (Seconds (0.0));
        sinkApp.Stop (Seconds (simulationTime + 1));
        OnOffHelper onoff ("ns3::TcpSocketFactory",Ipv4Address::GetAny ());
        onoff.SetAttribute ("OnTime",  StringValue ("ns3::ConstantRandomVariable[Constant=1]"));
        onoff.SetAttribute ("OffTime", StringValue ("ns3::ConstantRandomVariable[Constant=0]"));
        onoff.SetAttribute ("PacketSize", UintegerValue (payloadSize));
        onoff.SetAttribute ("DataRate", DataRateValue (1000000000)); //bit/s
        ApplicationContainer apps;
        AddressValue remoteAddress (InetSocketAddress (staNodeInterface.GetAddress (0), port));
        onoff.SetAttribute ("Remote", remoteAddress);
        apps.Add (onoff.Install (wifiApNode.Get (0)));
        apps.Start (Seconds (1.0));
        apps.Stop (Seconds (simulationTime + 1));
      }

      double throughput = 0;
      if (udp)
        {
          //UDP
          uint32_t totalPacketsThrough = DynamicCast<UdpServer> (serverApp.Get (0))->GetReceived ();
          throughput = totalPacketsThrough * payloadSize * 8 / (simulationTime * 1000000.0); //Mbit/s
        }
      else
        {
          //TCP
          uint32_t totalPacketsThrough = DynamicCast<PacketSink> (sinkApp.Get (0))->GetTotalRx ();
          throughput = totalPacketsThrough * 8 / (simulationTime * 1000000.0); //Mbit/s
        }
      std::cout << i << "\t\t\t" << j << " MHz\t\t\t" << k << "\t\t\t" << throughput << " Mbit/s" << std::endl;
      }
    j *= 2;
    }
  }
 return 0;
}
```

# Experiment 2: Impact of the number of STAs on the throughput

```
#include "ns3/nist-error-rate-model.h"
#include "ns3/core-module.h"
#include "ns3/yans-error-rate-model.h"
#include "ns3/gnuplot.h"
#include <vector>
#include <cmath>

using namespace ns3;
int main (int argc, char *argv[])
{
 radius_meter = 15;
 antenna_AP = 4;
 antenna_STAs = 1;
 power_dBm = 15;
 band_GHz = 5;
 bandwidth_MHz = 40;

 uint32_t FrameSize = 2000;
 std::ofstream yansfile ("yans-frame-success-rate.plt");
 std::ofstream nistfile ("nist-frame-success-rate.plt");
 std::vector <std::string> modes;
 modes.push_back ("OfdmRate6Mbps");
 modes.push_back ("OfdmRate9Mbps");
 modes.push_back ("OfdmRate12Mbps");
 modes.push_back ("OfdmRate18Mbps");
 modes.push_back ("OfdmRate24Mbps");
 modes.push_back ("OfdmRate36Mbps");
 modes.push_back ("OfdmRate48Mbps");
 modes.push_back ("OfdmRate54Mbps");
 CommandLine cmd;
 cmd.AddValue ("FrameSize", "The frame size", FrameSize);
 cmd.Parse (argc, argv);
 Gnuplot yansplot = Gnuplot ("yans-frame-success-rate.eps");
 Gnuplot nistplot = Gnuplot ("nist-frame-success-rate.eps");
 Ptr <YansErrorRateModel> yans = CreateObject<YansErrorRateModel> ();
 Ptr <NistErrorRateModel> nist = CreateObject<NistErrorRateModel> ();
 WifiTxVector txVector;

    yansplot.AddDataset (yansdataset);
    nistplot.AddDataset (nistdataset);
   }
 // Define the APs
 NodeContainer wifiApNodes;
 wifiApNodes.Create (1);
 //Define the STAs
 NodeContainer wifiStaNodes;
 wifiStaNodes.Create (1);
 YansWifiPhyHelper wifiPhy = YansWifiPhyHelper::Default ();
 YansWifiChannelHelper wifiChannel = YansWifiChannelHelper::Default ();
 wifiPhy.SetChannel (wifiChannel.Create ());
 wifiPhy.Set("ShortGuardEnabled", BooleanValue(shortGuardInterval));
 NetDeviceContainer wifiApDevices;
 NetDeviceContainer wifiStaDevices;
 NetDeviceContainer wifiDevices;
            //UDP flow
```

```
            UdpServerHelper myServer (9);
            serverApp = myServer.Install (wifiStaNode.Get (0));
            serverApp.Start (Seconds (0.0));
            serverApp.Stop (Seconds (simulationTime + 1));
            UdpClientHelper myClient (staNodeInterface.GetAddress (0), 9);
            myClient.SetAttribute ("MaxPackets", UintegerValue (4294967295u));
            myClient.SetAttribute ("Interval", TimeValue (Time ("0.00001"))); //packets/s
            myClient.SetAttribute ("PacketSize", UintegerValue (payloadSize));
            ApplicationContainer clientApp = myClient.Install (wifiApNode.Get (0));
            clientApp.Start (Seconds (1.0));
            clientApp.Stop (Seconds (simulationTime + 1));
          }
        else
          {
            //TCP flow
            uint16_t port = 50000;
            Address apLocalAddress (InetSocketAddress (Ipv4Address::GetAny (), port));
            PacketSinkHelper packetSinkHelper ("ns3::TcpSocketFactory", apLocalAddress);
            sinkApp = packetSinkHelper.Install (wifiStaNode.Get (0));
            sinkApp.Start (Seconds (0.0));
            sinkApp.Stop (Seconds (simulationTime + 1));
            OnOffHelper onoff ("ns3::TcpSocketFactory",Ipv4Address::GetAny ());
            onoff.SetAttribute ("OnTime",  StringValue
("ns3::ConstantRandomVariable[Constant=1]"));

txVector, std::pow (10.0,snr / 10.0), FrameSize * 8);
      yansdataset.Add (snr, ps);
      if (ps < 0 || ps > 1)
        {
         //error
         return 0;
        }
      ps = nist->GetChunkSuccessRate (WifiMode (modes[i]), txVector, std::pow (10.0,snr /
10.0), FrameSize * 8);
      if (ps < 0 || ps > 1)
        {
         //error
         return 0;
        }
      nistdataset.Add (snr, ps);
     }
    yansplot.AddDataset (yansdataset);
    nistplot.AddDataset (nistdataset);
  }
  yansplot.SetTerminal ("postscript eps color enh \"Times-BoldItalic\"");
  yansplot.SetLegend ("SNR(dB)", "Frame Success Rate");
  yansplot.SetExtra  ("set xrange [-5:30]\n\
set yrange [0:1.2]\n\
set style line 1 linewidth 5\n\
set style line 2 linewidth 5\n\
set style line 3 linewidth 5\n\
set style line 4 linewidth 5\n\
set style line 5 linewidth 5\n\
set style line 6 linewidth 5\n\
set style line 7 linewidth 5\n\
set style line 8 linewidth 5\n\
```

```
set style increment user"
);
  yansplot.GenerateOutput (yansfile);
  yansfile.close ();
  nistplot.SetTerminal ("postscript eps color enh \"Times-BoldItalic\"");
  nistplot.SetLegend ("SNR(dB)", "Frame Success Rate");
  nistplot.SetExtra  ("set xrange [-5:30]\n\
  yansplot.GenerateOutput (yansfile);
  yansfile.close ();
  nistplot.SetTerminal ("postscript eps color enh \"Times-BoldItalic\"");
  nistplot.SetLegend ("SNR(dB)", "Frame Success Rate");
  nistplot.SetExtra  ("set xrange [-5:30]\n\
set yrange [0:1.2]\n\
set style line 1 linewidth 5\n\
set style line 2 linewidth 5\n\
set style line 3 linewidth 5\n\
set style line 4 linewidth 5\n\
set style line 5 linewidth 5\n\
set style line 6 linewidth 5\n\
set style line 7 linewidth 5\n\
set style line 8 linewidth 5\n\
set style increment user"
);
  nistplot.GenerateOutput (nistfile);
  nistfile.close ();
}
```

# Experiment 3: Impact of the number of antennas of the AP on the throughput

```
#include <sstream>
#include <fstream>
#include <math.h>

#include "ns3/core-module.h"
#include "ns3/network-module.h"
#include "ns3/internet-module.h"
#include "ns3/mobility-module.h"
#include "ns3/wifi-module.h"
#include "ns3/applications-module.h"
#include "ns3/stats-module.h"
#include "ns3/flow-monitor-module.h"

using namespace ns3;
using namespace std;
NS_LOG_COMPONENT_DEFINE ("RateAdaptationDistance");
class NodeStatistics
{
public:
  NodeStatistics (NetDeviceContainer aps, NetDeviceContainer stas);
  void CheckStatistics (double time);
  void RxCallback (std::string path, Ptr<const Packet> packet, const Address &from);
  void SetPosition (Ptr<Node> node, Vector position);
  void AdvancePosition (Ptr<Node> node, int stepsSize, int stepsTime);
  Vector GetPosition (Ptr<Node> node);
```

```cpp
  Gnuplot2dDataset GetDatafile ();
private:
  uint32_t m_bytesTotal;
  Gnuplot2dDataset m_output;
};
NodeStatistics::NodeStatistics (NetDeviceContainer aps, NetDeviceContainer stas)
{
  m_bytesTotal = 0;
}
void
NodeStatistics::RxCallback (std::string path, Ptr<const Packet> packet, const Address &from)
{
  m_bytesTotal += packet->GetSize ();
}
void
NodeStatistics::CheckStatistics (double time)
void
NodeStatistics::SetPosition (Ptr<Node> node, Vector position)
{
  Ptr<MobilityModel> mobility = node->GetObject<MobilityModel> ();
  mobility->SetPosition (position);
}
Vector
NodeStatistics::GetPosition (Ptr<Node> node)
{
  Ptr<MobilityModel> mobility = node->GetObject<MobilityModel> ();
  return mobility->GetPosition ();
}

int main (int argc, char *argv[])
{
  uint32_t rtsThreshold = 65535;
  std::string staManager = "ns3::MinstrelHtWifiManager";
  std::string apManager = "ns3::MinstrelHtWifiManager";
  std::string standard = "802.11n-5GHz";
  std::string outputFileName = "minstrelHT";
  uint32_t BE_MaxAmpduSize = 65535;
  bool shortGuardInterval = false;
  uint32_t chWidth = 20;
  int ap1_x = 0;
  int ap1_y = 0;
  int sta1_x = 5;
  int sta1_y = 0;
  int steps = 100;
  int stepsSize = 1;
  int stepsTime = 1;

  const int STA = 25;
  radius_meter = 15;
  antenna_AP;
  antenna_STAs = 1;
  power_dBm = 15;
  band_GHz = 5;
  bandwidth_MHz = 40;

  CommandLine cmd;
  cmd.AddValue ("staManager", "PRC Manager of the STA", staManager);
```

D-7

```
  cmd.AddValue ("apManager", "PRC Manager of the AP", apManager);
  cmd.AddValue ("standard", "Wifi Phy Standard", standard);
  cmd.AddValue ("shortGuardInterval", "Enable Short Guard Interval in all stations", shortGuardInterval);
  cmd.AddValue ("channelWidth", "Channel width of all the stations", chWidth);
  cmd.AddValue ("rtsThreshold", "RTS threshold", rtsThreshold);
  cmd.AddValue ("BE_MaxAmpduSize", "BE Mac A-MPDU size", BE_MaxAmpduSize);
  cmd.AddValue ("outputFileName", "Output filename", outputFileName);
  cmd.AddValue ("steps", "How many different distances to try", steps);
  cmd.AddValue ("stepsTime", "Time on each step", stepsTime);
  cmd.AddValue ("stepsSize", "Distance between steps", stepsSize);
  cmd.AddValue ("AP1_x", "Position of AP1 in x coordinate", ap1_x);
  cmd.AddValue ("AP1_y", "Position of AP1 in y coordinate", ap1_y);
  cmd.AddValue ("STA1_x", "Position of STA1 in x coordinate", sta1_x);
  cmd.AddValue ("STA1_y", "Position of STA1 in y coordinate", sta1_y);
  cmd.Parse (argc, argv);

  int simuTime = steps * stepsTime;

     //Configure the STA node
     wifi.SetRemoteStationManager (staManager, "RtsCtsThreshold", UintegerValue (rtsThreshold));

     Ssid ssid = Ssid ("AP");
     wifiMac.SetType ("ns3::StaWifiMac",
                "Ssid", SsidValue (ssid));
     wifiStaDevices.Add (wifi.Install (wifiPhy, wifiMac, wifiStaNodes.Get (0)));

     //Configure the AP node
     wifi.SetRemoteStationManager (apManager, "RtsCtsThreshold", UintegerValue (rtsThreshold));

     ssid = Ssid ("AP");
     wifiMac.SetType ("ns3::ApWifiMac",
                "Ssid", SsidValue (ssid));
     wifiApDevices.Add (wifi.Install (wifiPhy, wifiMac, wifiApNodes.Get (0)));
   }

     Ssid ssid = Ssid ("AP");
     wifiMac.SetType ("ns3::StaWifiMac",
                "Ssid", SsidValue (ssid),
                "BE_MaxAmpduSize", UintegerValue (BE_MaxAmpduSize));
     wifiStaDevices.Add (wifi.Install (wifiPhy, wifiMac, wifiStaNodes.Get (0)));
     //Configure the AP node
     wifi.SetRemoteStationManager (apManager, "RtsCtsThreshold", UintegerValue (rtsThreshold));

     ssid = Ssid ("AP");
     wifiMac.SetType ("ns3::ApWifiMac",
                "Ssid", SsidValue (ssid),
                "BE_MaxAmpduSize", UintegerValue (BE_MaxAmpduSize));
     wifiApDevices.Add (wifi.Install (wifiPhy, wifiMac, wifiApNodes.Get (0)));
   }
  else if (standard == "802.11ac")
   {
     wifi.SetStandard (WIFI_PHY_STANDARD_80211ac);
     VhtWifiMacHelper wifiMac = VhtWifiMacHelper::Default ();
```

```cpp
      //Configure the STA node
      wifi.SetRemoteStationManager (staManager, "RtsCtsThreshold", UintegerValue (rtsThreshold));

      Ssid ssid = Ssid ("AP");
      wifiMac.SetType ("ns3::StaWifiMac",
                "Ssid", SsidValue (ssid),
                "BE_MaxAmpduSize", UintegerValue (BE_MaxAmpduSize));
      wifiStaDevices.Add (wifi.Install (wifiPhy, wifiMac, wifiStaNodes.Get (0)));

      //Configure the AP node
      wifi.SetRemoteStationManager (apManager, "RtsCtsThreshold", UintegerValue (rtsThreshold));
      ssid = Ssid ("AP");
      wifiMac.SetType ("ns3::ApWifiMac",
                "Ssid", SsidValue (ssid),
                "BE_MaxAmpduSize", UintegerValue (BE_MaxAmpduSize));
      wifiApDevices.Add (wifi.Install (wifiPhy, wifiMac, wifiApNodes.Get (0)));
    }
  wifiDevices.Add (wifiStaDevices);
  wifiDevices.Add (wifiApDevices);
  NodeStatistics atpCounter = NodeStatistics (wifiApDevices, wifiStaDevices);
  //Move the STA by stepsSize meters every stepsTime seconds
  Simulator::Schedule (Seconds (0.5 + stepsTime), &NodeStatistics::AdvancePosition,
&atpCounter, wifiStaNodes.Get (0), stepsSize, stepsTime);

  //Configure the IP stack
  InternetStackHelper stack;
  stack.Install (wifiApNodes);
  stack.Install (wifiStaNodes);
  Ipv4AddressHelper address;
  address.SetBase ("10.1.1.0", "255.255.255.0");
  Ipv4InterfaceContainer i = address.Assign (wifiDevices);
  Ipv4Address sinkAddress = i.GetAddress (0);
  uint16_t port = 9;

  apps_sink.Start (Seconds (0.5));
  apps_sink.Stop (Seconds (simuTime));
  //Register packet receptions to calculate throughput
  Config::Connect ("/NodeList/1/ApplicationList/*/$ns3::PacketSink/Rx",
              MakeCallback (&NodeStatistics::RxCallback, &atpCounter));
  //Callbacks to print every change of rate
  Config::Connect ("/NodeList/0/DeviceList/*/$ns3::WifiNetDevice/RemoteStationManager/$" + apManager + "/RateChange",
              MakeCallback (RateCallback));
  Simulator::Stop (Seconds (simuTime));
  Simulator::Run ();
  std::ofstream outfile (("throughput-" + outputFileName + ".plt").c_str ());
  Gnuplot gnuplot = Gnuplot (("throughput-" + outputFileName + ".eps").c_str (), "Throughput");
  gnuplot.SetTerminal ("post eps color enhanced");
  gnuplot.SetLegend ("Time (seconds)", "Throughput (Mb/s)");
  gnuplot.SetTitle ("Throughput (AP to STA) vs time");
  gnuplot.AddDataset (atpCounter.GetDatafile ());
  gnuplot.GenerateOutput (outfile);
  Simulator::Destroy ();
  return 0;
}
```

# Experiment 4: Upload time vs number of STAs

```cpp
#include "ns3/core-module.h"
#include "ns3/network-module.h"
#include "ns3/applications-module.h"
#include "ns3/mobility-module.h"
#include "ns3/config-store-module.h"
#include "ns3/wifi-module.h"
#include "ns3/athstats-helper.h"
#include <iostream>

using namespace ns3;
static bool g_verbose = true;

void
DevTxTrace (std::string context, Ptr<const Packet> p)
{
  if (g_verbose)
    {
      std::cout << " TX p: " << *p << std::endl;
    }
}
void
DevRxTrace (std::string context, Ptr<const Packet> p)
{
  if (g_verbose)
    {
      std::cout << " RX p: " << *p << std::endl;
    }
}
void
PhyRxOkTrace (std::string context, Ptr<const Packet> packet, double snr, WifiMode mode,
enum WifiPreamble preamble)
{
  if (g_verbose)
    {
      std::cout << "PHYRXOK mode=" << mode << " snr=" << snr << " " << *packet << std::endl;
    }
}
void
PhyRxErrorTrace (std::string context, Ptr<const Packet> packet, double snr)
{
  if (g_verbose)
    {
      std::cout << "PHYRXERROR snr=" << snr << " " << *packet << std::endl;
    }
}
void
PhyTxTrace (std::string context, Ptr<const Packet> packet, WifiMode mode, WifiPreamble
preamble, uint8_t txPower)
{
  if (g_verbose)
    {
      std::cout << "PHYTX mode=" << mode << " " << *packet << std::endl;
    }
}
```

```cpp
void
PhyStateTrace (std::string context, Time start, Time duration, enum WifiPhy::State state)
{
  if (g_verbose)
    {
      std::cout << " state=" << state << " start=" << start << " duration=" << duration << std::endl;
    }
}
  //Configure the CBR generator
  PacketSinkHelper sink ("ns3::UdpSocketFactory", InetSocketAddress (sinkAddress, port));
  ApplicationContainer apps_sink = sink.Install (wifiStaNodes.Get (0));

  OnOffHelper onoff ("ns3::UdpSocketFactory", InetSocketAddress (sinkAddress, port));
  onoff.SetConstantRate (DataRate ("200Mb/s"), 1420);
  onoff.SetAttribute ("StartTime", TimeValue (Seconds (0.5)));
  onoff.SetAttribute ("StopTime", TimeValue (Seconds (simuTime)));
  ApplicationContainer apps_source = onoff.Install (wifiApNodes.Get (0));
 // Only set the channel width and guard interval for HT and VHT modes
  if (selectedStandard.m_name == "802.11n-5GHz" ||
      selectedStandard.m_name == "802.11n-2.4GHz" ||
      selectedStandard.m_name == "802.11ac")
    {
      wifiPhyPtrClient->SetChannelWidth (selectedStandard.m_width);
      wifiPhyPtrServer->SetChannelWidth (selectedStandard.m_width);
      wifiPhyPtrClient->SetGuardInterval (shortGuardInterval);
      wifiPhyPtrServer->SetGuardInterval (shortGuardInterval);
    }
static void
SetPosition (Ptr<Node> node, Vector position)
{
  Ptr<MobilityModel> mobility = node->GetObject<MobilityModel> ();
  mobility->SetPosition (position);
}
static Vector
GetPosition (Ptr<Node> node)
{
  Ptr<MobilityModel> mobility = node->GetObject<MobilityModel> ();
  return mobility->GetPosition ();
}
static void
AdvancePosition (Ptr<Node> node)
{
  Vector pos = GetPosition (node);
  pos.x += 5.0;
  if (pos.x >= 210.0)
    {
      return;
    }
  SetPosition (node, pos);

  if (g_verbose)
    {
      //std::cout << "x="<<pos.x << std::endl;
    }
  Simulator::Schedule (Seconds (1.0), &AdvancePosition, node);
}
```

```cpp
int main (int argc, char *argv[])
{
  int STA;
  radius_meter = 15;
  antenna_AP = 4;
  antenna_STAs = 1;
  power_dBm = 15;
  band_GHz = 5;
  bandwidth_MHz = 40;

  CommandLine cmd;
  cmd.AddValue ("verbose", "Print trace information if true", g_verbose);
  cmd.Parse (argc, argv);
  stas.Create (2);
  ap.Create (1);
  // give packet socket powers to nodes.
  packetSocket.Install (stas);
  packetSocket.Install (ap);
  WifiMacHelper wifiMac;
  YansWifiPhyHelper wifiPhy = YansWifiPhyHelper::Default ();
  YansWifiChannelHelper wifiChannel = YansWifiChannelHelper::Default ();
  wifiPhy.SetChannel (wifiChannel.Create ());
  Ssid ssid = Ssid ("wifi-default");
  wifi.SetRemoteStationManager ("ns3::ArfWifiManager");
  // setup stas.
  wifiMac.SetType ("ns3::StaWifiMac",
          "Ssid", SsidValue (ssid));
  staDevs = wifi.Install (wifiPhy, wifiMac, stas);
  // setup ap.
  wifiMac.SetType ("ns3::ApWifiMac",
          "Ssid", SsidValue (ssid));
  wifi.Install (wifiPhy, wifiMac, ap);
  YansWifiPhyHelper wifiPhy =  YansWifiPhyHelper::Default ();
  // This is one parameter that matters when using FixedRssLossModel
  // set it to zero; otherwise, gain will be added
  wifiPhy.Set ("RxGain", DoubleValue (0) );
  // ns-3 supports RadioTap and Prism tracing extensions for 802.11b
  wifiPhy.SetPcapDataLinkType (YansWifiPhyHelper::DLT_IEEE802_11_RADIO);

  YansWifiChannelHelper wifiChannel;
  wifiChannel.SetPropagationDelay ("ns3::ConstantSpeedPropagationDelayModel");
  // The below FixedRssLossModel will cause the rss to be fixed regardless
  // of the distance between the two stations, and the transmit power
  wifiChannel.AddPropagationLoss ("ns3::FixedRssLossModel","Rss",DoubleValue (rss));
  wifiPhy.SetChannel (wifiChannel.Create ());

  AthstatsHelper athstats;
  athstats.EnableAthstats ("athstats-sta", stas);
  athstats.EnableAthstats ("athstats-ap", ap);
  Simulator::Run ();
  Simulator::Destroy ();
  return 0;
}
```

# Experiment 5: Goodput vs number of STAs

```cpp
#include "ns3/core-module.h"
#include "ns3/network-module.h"
#include "ns3/mobility-module.h"
#include "ns3/config-store-module.h"
#include "ns3/wifi-module.h"
#include "ns3/internet-module.h"
#include <iostream>
#include <fstream>
#include <vector>
#include <string>

using namespace ns3;
NS_LOG_COMPONENT_DEFINE ("WifiSimpleInfra");
void ReceivePacket (Ptr<Socket> socket)
{
  while (socket->Recv ())
    {
      NS_LOG_UNCOND ("Received one packet!");
    }
}
static void GenerateTraffic (Ptr<Socket> socket, uint32_t pktSize,
                             uint32_t pktCount, Time pktInterval )
{
  if (pktCount > 0)
    {
      socket->Send (Create<Packet> (pktSize));
      Simulator::Schedule (pktInterval, &GenerateTraffic,
                           socket, pktSize,pktCount-1, pktInterval);
    }
  else
    {
      socket->Close ();
    }
}

int main (int argc, char *argv[])
{
  int STA;
  radius_meter = 15;
  antenna_AP = 4;
  antenna_STAs = 1;
  power_dBm = 15;
  band_GHz = 5;
  bandwidth_MHz = 40;

  std::string phyMode ("DsssRate1Mbps");
  double rss = -80;  // -dBm
  uint32_t packetSize = 1000; // bytes
  uint32_t numPackets = 1;
  double interval = 1.0; // seconds
  bool verbose = false;

  CommandLine cmd;
  cmd.AddValue ("phyMode", "Wifi Phy mode", phyMode);
  cmd.AddValue ("rss", "received signal strength", rss);
```

```
  cmd.AddValue ("packetSize", "size of application packet sent", packetSize);
  cmd.AddValue ("numPackets", "number of packets generated", numPackets);
  cmd.AddValue ("interval", "interval (seconds) between packets", interval);
  cmd.AddValue ("verbose", "turn on all WifiNetDevice log components", verbose);
  cmd.Parse (argc, argv);
  // Convert to time object
  Time interPacketInterval = Seconds (interval);
  // disable fragmentation for frames below 2200 bytes
  Config::SetDefault ("ns3::WifiRemoteStationManager::FragmentationThreshold", StringValue ("2200"));
  // turn off RTS/CTS for frames below 2200 bytes
  Config::SetDefault ("ns3::WifiRemoteStationManager::RtsCtsThreshold", StringValue ("2200"));
  // Fix non-unicast data rate to be the same as that of unicast
  Config::SetDefault ("ns3::WifiRemoteStationManager::NonUnicastMode",
                     StringValue (phyMode));

  NodeContainer c;
  c.Create (2);
  macAp.SetType ("ns3::ApWifiMac",
           "Ssid", SsidValue (ssid),
           "BeaconInterval", TimeValue (MicroSeconds (102400)),
           "BeaconGeneration", BooleanValue (true));
  NetDeviceContainer staDevice;
  NetDeviceContainer apDevice;
  Ptr<YansWifiPhy> phySta;
  Config::SetDefault ("ns3::ConfigStore::Filename", StringValue ("output-attributes-" + std::to_string (testCase) + ".txt"));
  Config::SetDefault ("ns3::ConfigStore::FileFormat", StringValue ("RawText"));
  Config::SetDefault ("ns3::ConfigStore::Mode", StringValue ("Save"));

  // The below set of helpers will help us to put together the wifi NICs the research wants
  // WifiHelper wifi;
  if (verbose)
    {
      wifi.EnableLogComponents ();  // Turn on all Wifi logging
    }
  wifi.SetStandard (WIFI_PHY_STANDARD_80211b);
  // Add a mac and disable rate control
  WifiMacHelper wifiMac;
  wifi.SetRemoteStationManager ("ns3::ConstantRateWifiManager",
                                "DataMode",StringValue (phyMode),
                                "ControlMode",StringValue (phyMode));
  // mobility.
  mobility.Install (stas);
  mobility.Install (ap);
      spectrumPhy.SetChannel (spectrumChannel);
      spectrumPhy.SetErrorRateModel (errorModelType);
      spectrumPhy.Set ("Frequency", UintegerValue (5180));
      spectrumPhy.Set ("TxPowerStart", DoubleValue (1)); // dBm  (1.26 mW)
      spectrumPhy.Set ("TxPowerEnd", DoubleValue (1));

      if (i <= 7)
        {
          spectrumPhy.Set ("ShortGuardEnabled", BooleanValue (false));
          spectrumPhy.Set ("ChannelWidth", UintegerValue (20));
        }
```

```cpp
      else if (i > 7 && i <= 15)
        {
          spectrumPhy.Set ("ShortGuardEnabled", BooleanValue (true));
          spectrumPhy.Set ("ChannelWidth", UintegerValue (20));
        }
      else if (i > 15 && i <= 23)
        {
          spectrumPhy.Set ("ShortGuardEnabled", BooleanValue (false));
          spectrumPhy.Set ("ChannelWidth", UintegerValue (40));
        }
      else
        {
          spectrumPhy.Set ("ShortGuardEnabled", BooleanValue (true));
          spectrumPhy.Set ("ChannelWidth", UintegerValue (40));
        }
    }
  else
        NS_FATAL_ERROR ("Unsupported WiFi type " << wifiType);

  OnOffHelper onoff ("ns3::PacketSocketFactory", Address (socket));
  onoff.SetConstantRate (DataRate ("500kb/s"));
  ApplicationContainer apps = onoff.Install (stas.Get (0));
  apps.Start (Seconds (0.5));
  apps.Stop (Seconds (43.0));
  // Setup the rest of the mac
  Ssid ssid = Ssid ("wifi-default");
  // setup sta.
  wifiMac.SetType ("ns3::StaWifiMac",
            "Ssid", SsidValue (ssid));
  NetDeviceContainer staDevice = wifi.Install (wifiPhy, wifiMac, c.Get (0));
  NetDeviceContainer devices = staDevice;
  // setup ap.
  wifiMac.SetType ("ns3::ApWifiMac",
            "Ssid", SsidValue (ssid));
  NetDeviceContainer apDevice = wifi.Install (wifiPhy, wifiMac, c.Get (1));
  devices.Add (apDevice);
  // Note that with FixedRssLossModel, the positions below are not
  // used for received signal strength.
  MobilityHelper mobility;
  Ptr<ListPositionAllocator> positionAlloc = CreateObject<ListPositionAllocator> ();
  positionAlloc->Add (Vector (0.0, 0.0, 0.0));
  positionAlloc->Add (Vector (5.0, 0.0, 0.0));
  mobility.SetPositionAllocator (positionAlloc);
  mobility.SetMobilityModel ("ns3::ConstantPositionMobilityModel");
  mobility.Install (c);
  InternetStackHelper internet;
  internet.Install (c);
  // Output what the research is doing
  NS_LOG_UNCOND ("Testing " << numPackets  << " packets sent with receiver rss " << rss );
  Simulator::ScheduleWithContext (source->GetNode ()->GetId (),
                    Seconds (1.0), &GenerateTraffic,
                    source, packetSize, numPackets, interPacketInterval);
  Simulator::Stop (Seconds (30.0));
  Simulator::Run ();
  Simulator::Destroy ();
  return 0;
}
```

# Experiment 6: Number of retransmissions vs number of STAs

```cpp
#include "ns3/network-module.h"
#include "ns3/wifi-module.h"
#include <sstream>
#include <iomanip>
#include "ns3/core-module.h"
#include "ns3/config-store-module.h"
#include "ns3/mobility-module.h"
#include "ns3/spectrum-module.h"
#include "ns3/internet-module.h"

// Network topology:
//   Wi-Fi 192.168.1.0
//
//    STA           AP
//     *            *
//     |            |
//   n1-n60        n61
//
using namespace ns3;

// Global variables for use in callbacks.
double g_signalDbmAvg;
double g_noiseDbmAvg;
uint32_t g_samples;
uint16_t g_channelNumber;
uint32_t g_rate;

void MonitorSniffRx (Ptr<const Packet> packet, uint16_t channelFreqMhz,
            uint16_t channelNumber, uint32_t rate,
            WifiPreamble preamble, WifiTxVector txVector,
            struct mpduInfo aMpdu, struct signalNoiseDbm signalNoise)

    WifiHelper wifi;
    wifi.SetStandard (WIFI_PHY_STANDARD_80211n_5GHZ);
    WifiMacHelper mac;
    Ssid ssid = Ssid ("ns380211n");
  Simulator::Schedule (Seconds (1.0), &AdvancePosition, ap.Get (0));
  PacketSocketAddress socket;
  socket.SetSingleDevice (staDevs.Get (0)->GetIfIndex ());
  socket.SetPhysicalAddress (staDevs.Get (1)->GetAddress ());
  socket.SetProtocol (1);

{
  g_samples++;
  g_signalDbmAvg += ((signalNoise.signal - g_signalDbmAvg) / g_samples);
  g_noiseDbmAvg += ((signalNoise.noise - g_noiseDbmAvg) / g_samples);
  g_rate = rate;
  g_channelNumber = channelNumber;
}

NS_LOG_COMPONENT_DEFINE ("WifiSpectrumPerExample");
int main (int argc, char *argv[])
{
 int STA;
 radius_meter = 15;
```

```
  antenna_AP = 4;
  antenna_STAs = 1;
  power_dBm = 15;
  band_GHz = 5;
  bandwidth_MHz = 40;

  bool udp = true;
  double distance = 50;
  double simulationTime = 10; //seconds
  uint16_t index = 256;
  std::string wifiType = "ns3::SpectrumWifiPhy";
  std::string errorModelType = "ns3::NistErrorRateModel";
  bool enablePcap = false;
  const uint32_t tcpPacketSize = 1448;
  CommandLine cmd;
  cmd.AddValue ("simulationTime", "Simulation time in seconds", simulationTime);
  cmd.AddValue ("udp", "UDP if set to 1, TCP otherwise", udp);
  cmd.AddValue ("distance", "meters separation between nodes", distance);
  cmd.AddValue ("index", "restrict index to single value between 0 and 31", index);
  cmd.AddValue ("wifiType", "select ns3::SpectrumWifiPhy or ns3::YansWifiPhy", wifiType);
  cmd.AddValue ("errorModelType", "select ns3::NistErrorRateModel or
ns3::YansErrorRateModel", errorModelType);
  cmd.AddValue ("enablePcap", "enable pcap output", enablePcap);
  cmd.Parse (argc,argv);

  uint16_t startIndex = 0;
  uint16_t stopIndex = 31;
  if (index < 32)
    {
      startIndex = index;
      stopIndex = index;
    }
  std::cout << "wifiType: " << wifiType << " distance: " << distance << "m; sent: 1000 TxPower: 1 dBm (1.3 mW)" << std::endl;
  std::cout << std::setw (5) << "index" <<
    std::setw (6) << "MCS" <<
    std::setw (12) << "Rate (Mb/s)" <<
    std::setw (12) << "Tput (Mb/s)" <<
    std::setw (10) << "Received " <<
    std::setw (12) << "Signal (dBm)" <<
    std::setw (12) << "Noise (dBm)" <<
    std::setw (10) << "SNR (dB)" <<
    std::endl;
  for (uint16_t i = startIndex; i <= stopIndex; i++)
    {
      uint32_t payloadSize;
      if (udp)
        {
          payloadSize = 972; // 1000 bytes IPv4
        }
      else
        {
          payloadSize = 1448; // 1500 bytes IPv6
          Config::SetDefault ("ns3::TcpSocket::SegmentSize", UintegerValue (payloadSize));
        }

      NodeContainer wifiStaNode;
```

```
  wifiStaNode.Create (1);
NodeContainer wifiApNode;
wifiApNode.Create (1);
YansWifiPhyHelper phy = YansWifiPhyHelper::Default ();
SpectrumWifiPhyHelper spectrumPhy = SpectrumWifiPhyHelper::Default ();
if (wifiType == "ns3::YansWifiPhy")
  {
    YansWifiChannelHelper channel;
    channel.AddPropagationLoss ("ns3::FriisPropagationLossModel");
    channel.SetPropagationDelay ("ns3::ConstantSpeedPropagationDelayModel");
    phy.SetChannel (channel.Create ());
    phy.Set ("TxPowerStart", DoubleValue (1)); // dBm (1.26 mW)
    phy.Set ("TxPowerEnd", DoubleValue (1));
    if (i <= 7)
      {
        phy.Set ("ShortGuardEnabled", BooleanValue (false));
        phy.Set ("ChannelWidth", UintegerValue (20));
      }
    else if (i > 7 && i <= 15)
      {
        phy.Set ("ShortGuardEnabled", BooleanValue (true));
        phy.Set ("ChannelWidth", UintegerValue (20));
      }
    else if (i > 15 && i <= 23)
      {
        phy.Set ("ShortGuardEnabled", BooleanValue (false));
        phy.Set ("ChannelWidth", UintegerValue (40));
      }
    else
      {
        phy.Set ("ShortGuardEnabled", BooleanValue (true));
        phy.Set ("ChannelWidth", UintegerValue (40));
      }
  }
else if (wifiType == "ns3::SpectrumWifiPhy")
  {
    Config::SetDefault ("ns3::WifiPhy::CcaMode1Threshold", DoubleValue (-62.0));

    Ptr<MultiModelSpectrumChannel> spectrumChannel
      = CreateObject<MultiModelSpectrumChannel> ();
    Ptr<FriisPropagationLossModel> lossModel
      = CreateObject<FriisPropagationLossModel> ();
    spectrumChannel->AddPropagationLossModel (lossModel);
    spectrumChannel->SetPropagationDelayModel (delayModel);
// mobility.
MobilityHelper mobility;
Ptr<ListPositionAllocator> positionAlloc = CreateObject<ListPositionAllocator> ();

Ipv4AddressHelper address;
address.SetBase ("192.168.1.0", "255.255.255.0");
Ipv4InterfaceContainer staNodeInterface;
Ipv4InterfaceContainer apNodeInterface;
staNodeInterface = address.Assign (staDevice);
apNodeInterface = address.Assign (apDevice);
/* Setting applications */
ApplicationContainer serverApp, sinkApp;
if (udp)
```

```
      {
RetransmissionExperiment::Run (struct CollisionExperiment::Input input)
{
  m_output.receivedA = 0;
  m_output.receivedB = 0;
  m_input = input;
  m_flowIdA = FlowIdTag::AllocateFlowId ();
  m_flowIdB = FlowIdTag::AllocateFlowId ();
  Ptr<YansWifiChannel> channel = CreateObject<YansWifiChannel> ();
  channel->SetPropagationDelayModel (CreateObject<ConstantSpeedPropagationDelayModel> ());
  Ptr<LogDistancePropagationLossModel> log = CreateObject<LogDistancePropagationLossModel> ();
  channel->SetPropagationLossModel (log);
      Simulator::Schedule (Seconds (i), &CollisionExperiment::SendA, this);
    }
  for (uint32_t i = 0; i < m_input.nPackets; ++i)
    {
      Simulator::Schedule (Seconds (i) + m_input.interval, &CollisionExperiment::SendB, this);
    }
      OnOffHelper onoff ("ns3::TcpSocketFactory",Ipv4Address::GetAny ());
      onoff.SetAttribute ("OnTime",  StringValue ("ns3::ConstantRandomVariable[Constant=1]"));
      onoff.SetAttribute ("OffTime", StringValue ("ns3::ConstantRandomVariable[Constant=0]"));
      onoff.SetAttribute ("PacketSize", UintegerValue (payloadSize));
      onoff.SetAttribute ("DataRate", DataRateValue (1000000000)); //bit/s
      ApplicationContainer apps;
      AddressValue remoteAddress (InetSocketAddress (staNodeInterface.GetAddress (0), port));
      onoff.SetAttribute ("Remote", remoteAddress);
      apps.Add (onoff.Install (wifiApNode.Get (0)));
      apps.Start (Seconds (1.0));
      apps.Stop (Seconds (simulationTime + 1));
    }

    Config::ConnectWithoutContext ("/NodeList/0/DeviceList/*/Phy/MonitorSnifferRx", MakeCallback (&MonitorSniffRx));
        //TCP
      uint32_t totalBytesRx = DynamicCast<PacketSink> (sinkApp.Get (0))->GetTotalRx ();
      totalPacketsThrough = totalBytesRx / tcpPacketSize;
      throughput = totalBytesRx * 8 / (simulationTime * 1000000.0); //Mbit/s
    }
    std::cout << std::setw (5) << i <<
      std::setw (6) << (i % 8) <<
      std::setw (10) << datarate <<
      std::setw (12) << throughput <<
      std::setw (8) << totalPacketsThrough;
  return 0;
}
```

# Appendix E

# ERA Protocol Data for Graphing

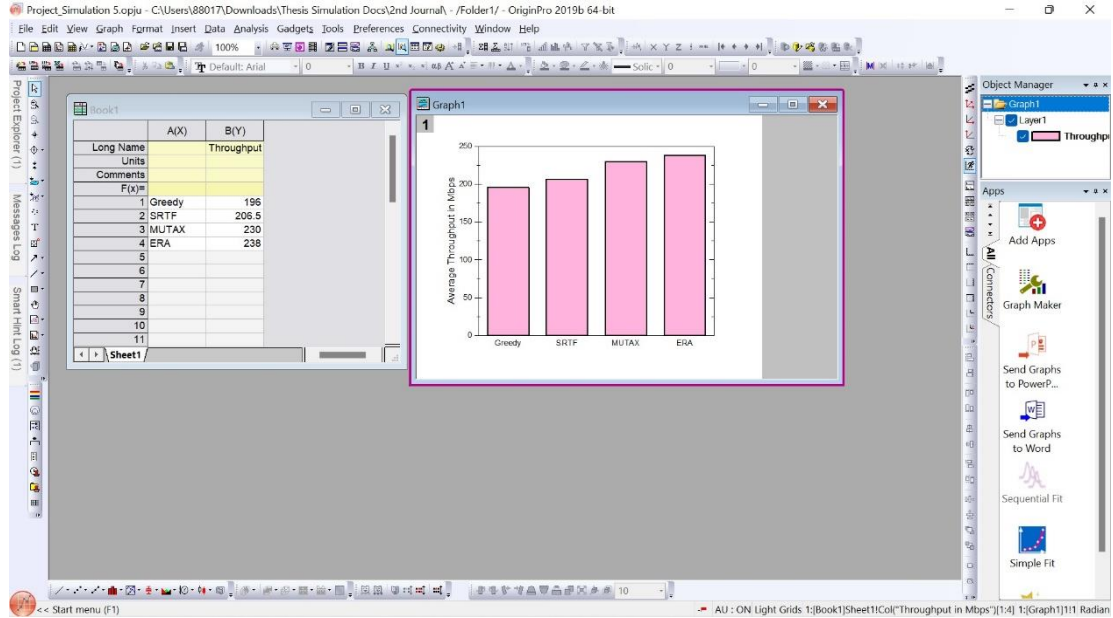

Experiment 1: Comparison of the throughput between different protocols

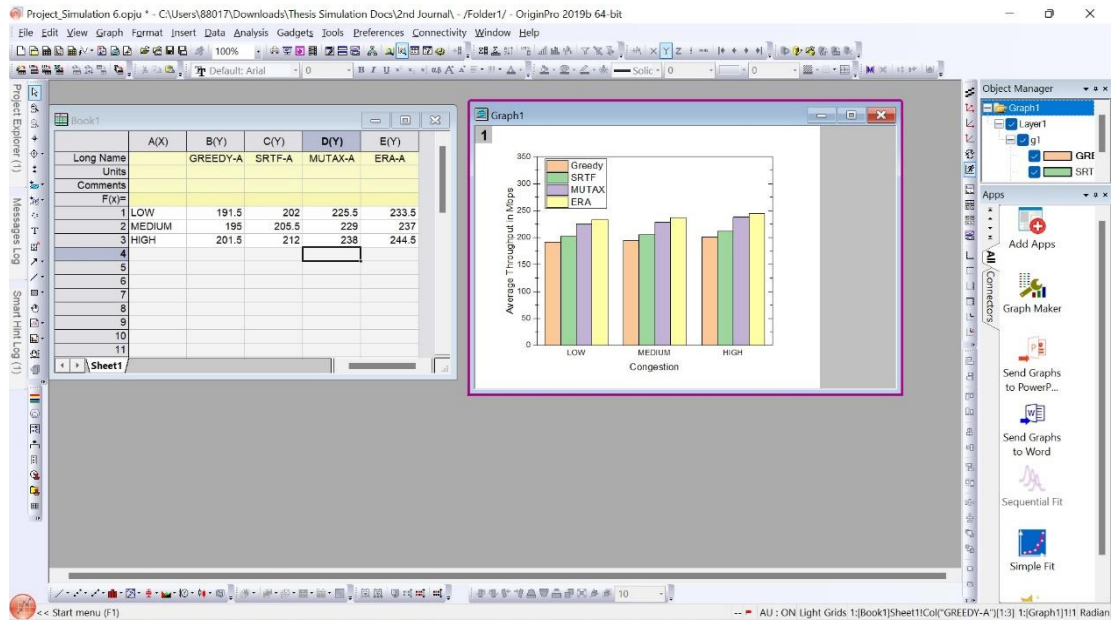

Experiment 2: Impact of the number of stations on the throughput

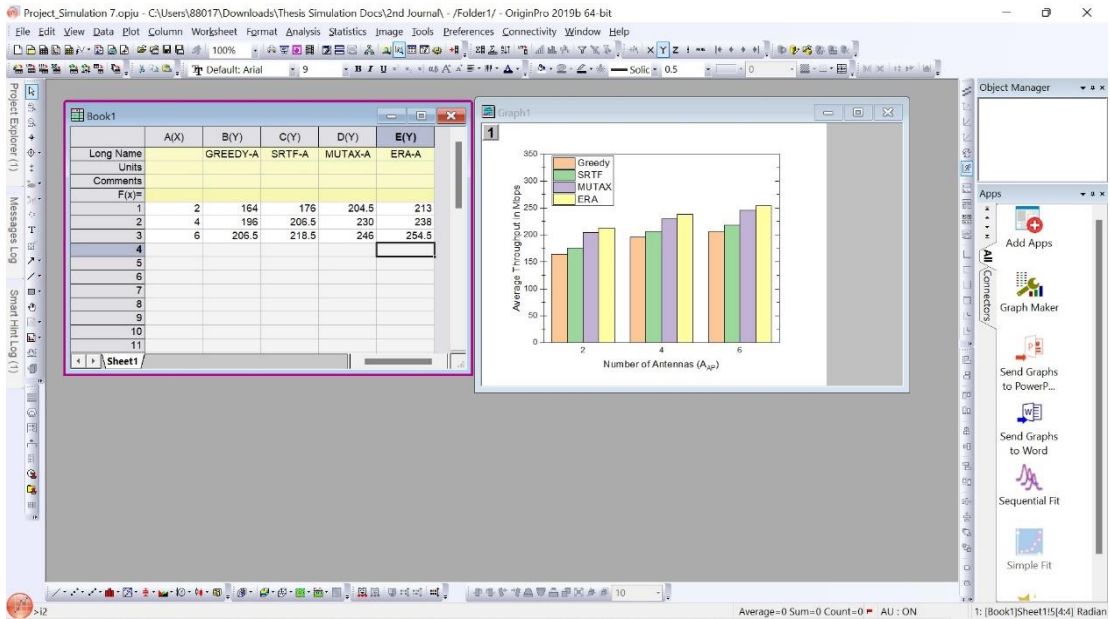

**Experiment 3: Impact of the number of antennas of the AP on the throughput**

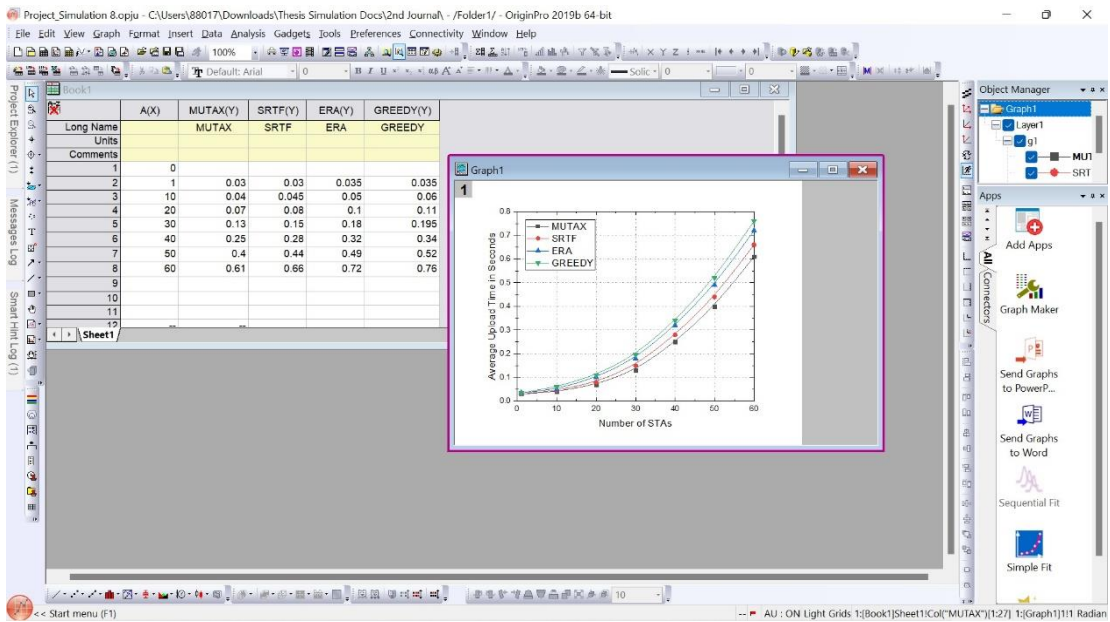

**Experiment 4: Upload time vs number of stations**

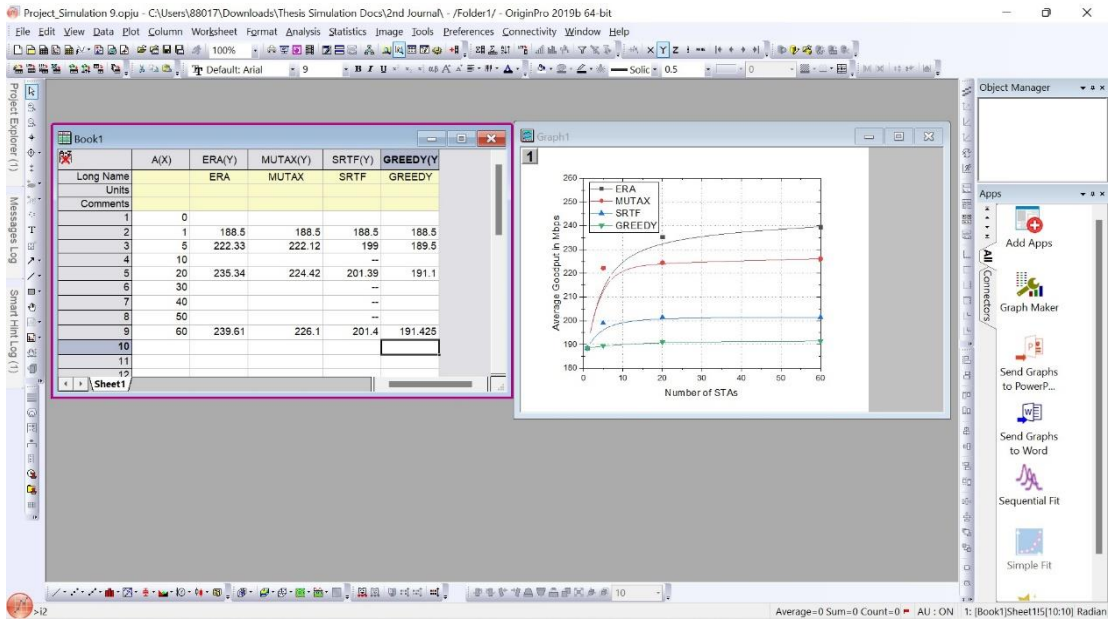

**Experiment 5: Goodput vs number of stations**

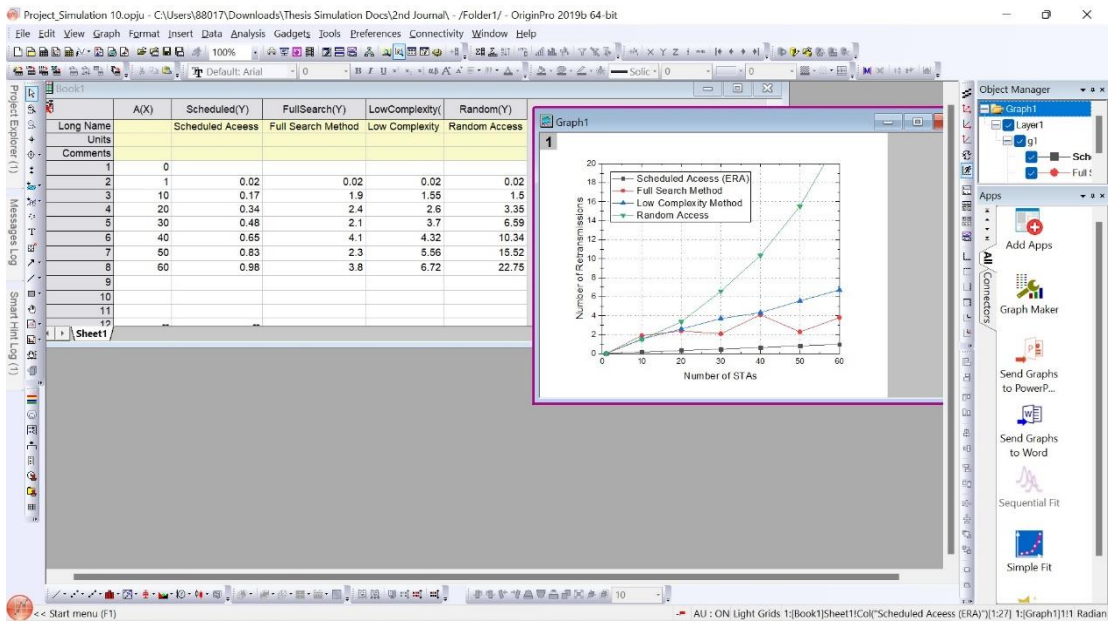

**Experiment 6: Number of retransmissions vs number of stations**

# Appendix F

# NS-3 Code for PRS Protocol

## Experiment 1: Throughput of the PRS protocol for different distribution of stations

```
#include "ns3/core-module.h"
#include "ns3/network-module.h"
#include "ns3/applications-module.h"
#include "ns3/mobility-module.h"
#include "ns3/wifi-module.h"
#include "ns3/internet-module.h"
#include "ns3/flow-monitor-helper.h"
#include "ns3/olsr-helper.h"
#include "ns3/ipv4-list-routing-helper.h"

#include <iostream>
#include <fstream>

using namespace ns3;

NS_LOG_COMPONENT_DEFINE ("multirate");

class Experiment
{
public:

  Experiment ();
  Experiment (std::string name);
  Gnuplot2dDataset Run (const WifiHelper &wifi, const YansWifiPhyHelper &wifiPhy,

  bool CommandSetup (int argc, char **argv);
  bool IsRouting () { return (enableRouting == 1) ? 1 : 0; }
  bool IsMobility () { return (enableMobility == 1) ? 1 : 0; }

  uint32_t GetScenario () { return scenario; }

  std::string GetRtsThreshold () { return rtsThreshold; }
  std::string GetOutputFileName () { return outputFileName; }
  std::string GetRateManager () { return rateManager; }

private:

  Ptr<Socket> SetupPacketReceive (Ptr<Node> node);
  NodeContainer GenerateNeighbors (NodeContainer c, uint32_t senderId);

  void ApplicationSetup (Ptr<Node> client, Ptr<Node> server, double start, double stop);
  void AssignNeighbors (NodeContainer c);
  void SelectSrcDest (NodeContainer c);
  void ReceivePacket (Ptr<Socket> socket);
  void CheckThroughput ();
  void SendMultiDestinations (Ptr<Node> sender, NodeContainer c);
```

```cpp
  Gnuplot2dDataset m_output;

  double totalTime;
  double expMean;
  double samplingPeriod;
  bool enablePcap;
  bool enableTracing;
  bool enableFlowMon;
  bool enableRouting;
  bool enableMobility;

  NodeContainer containerA, containerB, containerC, containerD;
  std::string rtsThreshold, rateManager, outputFileName;
};

Experiment::Experiment (std::string name) :
  m_output (name),
  totalTime (0.3),
  expMean (0.1), //flows being exponentially distributed
  samplingPeriod(0.1),
  bytesTotal (0),
  packetSize (2000),
  gridSize (10), //10x10 grid  for a total of 100 nodes
  nodeDistance (30),
  port (5000),
  scenario (4),
  enablePcap (false),
  enableTracing (true),
  enableFlowMon (false),
  enableRouting (false),
  enableMobility (false),
  rtsThreshold ("2200"), //0 for enabling rts/cts
  rateManager ("ns3::MinstrelWifiManager"),
  outputFileName ("minstrel")
{
  m_output.SetStyle (Gnuplot2dDataset::LINES);
}

void
Experiment::CheckThroughput ()
{
  double mbs = ((bytesTotal * 8.0) /1000000 /samplingPeriod);
  bytesTotal = 0;
  m_output.Add ((Simulator::Now ()).GetSeconds (), mbs);

  //check throughput every samplingPeriod second
  Simulator::Schedule (Seconds (samplingPeriod), &Experiment::CheckThroughput, this);
}

static inline Vector
GetPosition (Ptr<Node> node)
{
  Ptr<MobilityModel> mobility = node->GetObject<MobilityModel> ();
  return mobility->GetPosition ();
}

  //Configure the IP stack
```

```
  InternetStackHelper stack;
  stack.Install (wifiApNodes);
  stack.Install (wifiStaNodes);
  Ipv4AddressHelper address;
  address.SetBase ("10.1.1.0", "255.255.255.0");
  Ipv4InterfaceContainer i = address.Assign (wifiDevices);
  Ipv4Address sinkAddress = i.GetAddress (0);
  uint16_t port = 9;

  apps_sink.Start (Seconds (0.5));
  apps_sink.Stop (Seconds (simuTime));

static inline std::string
PrintPosition (Ptr<Node> client, Ptr<Node> server)
{
  Vector serverPos = GetPosition (server);
  Vector clientPos = GetPosition (client);

  Ptr<Ipv4> ipv4Server = server->GetObject<Ipv4>();
  Ptr<Ipv4> ipv4Client = client->GetObject<Ipv4>();

  Ipv4InterfaceAddress iaddrServer = ipv4Server->GetAddress (1,0);
  Ipv4InterfaceAddress iaddrClient = ipv4Client->GetAddress (1,0);

  Ipv4Address ipv4AddrServer = iaddrServer.GetLocal ();
  Ipv4Address ipv4AddrClient = iaddrClient.GetLocal ();

  std::ostringstream oss;
  oss << "Set up Server Device " <<  (server->GetDevice (0))->GetAddress ()
                  << " with ip " << ipv4AddrServer
                  << " position (" << serverPos.x << "," << serverPos.y << "," << serverPos.z
<< ")";

  oss << "Set up Client Device " <<  (client->GetDevice (0))->GetAddress ()
                  << " with ip " << ipv4AddrClient
                  << " position (" << clientPos.x << "," << clientPos.y << "," << clientPos.z
<< ")"
                  << "\n";
  return oss.str ();
}

  ApplicationContainer apps = onoff.Install (client);
  apps.Start (Seconds (start));
  apps.Stop (Seconds (stop));

  Ptr<Socket> sink = SetupPacketReceive (server);

}

coverage
    //you might have to change these values for other grids
    NodeContainer c1, c2, c3, c4, c5, c6, c7, c8, c9;

    c1 = GenerateNeighbors (c, 22);
    c2 = GenerateNeighbors (c, 24);;
    c3 = GenerateNeighbors (c, 26);;
```

```
      c4 = GenerateNeighbors (c, 42);;
      c5 = GenerateNeighbors (c, 44);;
      c6 = GenerateNeighbors (c, 46);;
      c7 = GenerateNeighbors (c, 62);;
      c8 = GenerateNeighbors (c, 64);;
      c9 = GenerateNeighbors (c, 66);;

      SendMultiDestinations (c.Get (22), c1);
      SendMultiDestinations (c.Get (24), c2);
      SendMultiDestinations (c.Get (26), c3);
      SendMultiDestinations (c.Get (42), c4);
      SendMultiDestinations (c.Get (44), c5);
      SendMultiDestinations (c.Get (46), c6);
      SendMultiDestinations (c.Get (62), c7);
      SendMultiDestinations (c.Get (64), c8);
      SendMultiDestinations (c.Get (66), c9);
    }

  CheckThroughput ();

  if (enablePcap)
    {
      phy.EnablePcapAll (GetOutputFileName ());
    }

  if (enableTracing)
    {
      AsciiTraceHelper ascii;
      phy.EnableAsciiAll (ascii.CreateFileStream (GetOutputFileName () + ".tr"));
    }

  FlowMonitorHelper flowmonHelper;

  if (enableFlowMon)
    {
      flowmonHelper.InstallAll ();
    }

  Simulator::Stop (Seconds (totalTime));
  Simulator::Run ();

  if (enableFlowMon)
    {
      flowmonHelper.SerializeToXmlFile ((GetOutputFileName () + ".flomon"), false, false);
    }

  Simulator::Destroy ();

  return m_output;
}

  return 0;
}
```

# Experiment 2: Number of retransmissions of the PRS protocol for different distribution of stations

```cpp
#include "ns3/core-module.h"
#include "ns3/network-module.h"
#include "ns3/applications-module.h"
#include "ns3/wifi-module.h"
#include "ns3/mobility-module.h"
#include "ns3/ipv4-global-routing-helper.h"
#include "ns3/internet-module.h"

// This example considers two hidden STAs in an 802.11ax network which supports MPDU aggregation.
// The user can specify whether RTS/CTS is used and can set the number of aggregated MPDUs.
//
// Example: ./waf --run "simple-ht-hidden-stations --enableRts=1 --nMpdus=8"
//
// Network topology:
//
//   Wifi 192.168.1.0
//
//            AP
//    *       *       *
//    |       |       |
//   STA1   STA2    STA3
//
// Packets in this simulation aren't marked with a QosTag so they are considered
// belonging to BestEffort Access Class (AC_BE).

using namespace ns3;

NS_LOG_COMPONENT_DEFINE ("SimplesHtHiddenStations");

int main (int argc, char *argv[])
{
  uint32_t payloadSize = 1472; //bytes
  uint64_t simulationTime = 10; //seconds
  uint32_t nMpdus = 1;
  uint32_t maxAmpduSize = 0;
  bool enableRts = 0;

  CommandLine cmd;
  cmd.AddValue ("nMpdus", "Number of aggregated MPDUs", nMpdus);
  cmd.AddValue ("payloadSize", "Payload size in bytes", payloadSize);
  cmd.AddValue ("enableRts", "Enable RTS/CTS", enableRts); // 1: RTS/CTS enabled; 0: RTS/CTS disabled
  cmd.AddValue ("simulationTime", "Simulation time in seconds", simulationTime);
  cmd.Parse (argc, argv);

  if (!enableRts)
    {
      Config::SetDefault ("ns3::WifiRemoteStationManager::RtsCtsThreshold", StringValue ("999999"));
    }
  else
    {
```

```
      Config::SetDefault ("ns3::WifiRemoteStationManager::RtsCtsThreshold", StringValue ("0"));
    }

  Config::SetDefault ("ns3::WifiRemoteStationManager::FragmentationThreshold", StringValue ("990000"));

  //Set the maximum size for A-MPDU with regards to the payload size
  maxAmpduSize = nMpdus * (payloadSize + 200);

  // Set the maximum wireless range to 5 meters in order to reproduce a hidden nodes scenario, i.e. the distance between hidden stations is larger than 5 meters
  Config::SetDefault ("ns3::RangePropagationLossModel::MaxRange", DoubleValue (5));

  NodeContainer wifiStaNodes;
  wifiStaNodes.Create (2);
  NodeContainer wifiApNode;
  wifiApNode.Create (1);

  YansWifiChannelHelper channel = YansWifiChannelHelper::Default ();
  channel.AddPropagationLoss ("ns3::RangePropagationLossModel"); //wireless range limited to 5 meters!

  YansWifiPhyHelper phy = YansWifiPhyHelper::Default ();
  phy.SetPcapDataLinkType (YansWifiPhyHelper::DLT_IEEE802_11_RADIO);
  phy.SetChannel (channel.Create ());

  WifiHelper wifi;
  wifi.SetStandard (WIFI_PHY_STANDARD_80211n_5GHZ);
  wifi.SetRemoteStationManager ("ns3::ConstantRateWifiManager", "DataMode", StringValue ("HtMcs7"), "ControlMode", StringValue ("HtMcs0"));
  WifiMacHelper mac;

  Ssid ssid = Ssid ("simple-mpdu-aggregation");
  mac.SetType ("ns3::StaWifiMac",
         "Ssid", SsidValue (ssid),
         "BE_MaxAmpduSize", UintegerValue (maxAmpduSize));

  NetDeviceContainer staDevices;
  staDevices = wifi.Install (phy, mac, wifiStaNodes);

  mac.SetType ("ns3::ApWifiMac",
         "Ssid", SsidValue (ssid),
         "BeaconGeneration", BooleanValue (true),
         "BE_MaxAmpduSize", UintegerValue (maxAmpduSize));

  NetDeviceContainer apDevice;
  apDevice = wifi.Install (phy, mac, wifiApNode);

  // Setting mobility model
  MobilityHelper mobility;
  Ptr<ListPositionAllocator> positionAlloc = CreateObject<ListPositionAllocator> ();

  // Access Point is between the two STAs, each STA being located at 5 meters from the Access Point.
  // The distance between the two stations is thus equal to 10 meters.
  // Since the wireless range is limited to 5 meters, the two stations are hidden from each other.
```

```cpp
  positionAlloc->Add (Vector (5.0, 0.0, 0.0));
  positionAlloc->Add (Vector (0.0, 0.0, 0.0));
  positionAlloc->Add (Vector (10.0, 0.0, 0.0));
  mobility.SetPositionAllocator (positionAlloc);

  mobility.SetMobilityModel ("ns3::ConstantPositionMobilityModel");

  mobility.Install (wifiApNode);
  mobility.Install (wifiStaNodes);

  // Internet stack
  InternetStackHelper stack;
  stack.Install (wifiApNode);
  stack.Install (wifiStaNodes);

        Ipv4AddressHelper address;

        address.SetBase ("192.168.1.0", "255.255.255.0");
        Ipv4InterfaceContainer staNodeInterface;
        Ipv4InterfaceContainer apNodeInterface;

        staNodeInterface = address.Assign (staDevice);
        apNodeInterface = address.Assign (apDevice);

          OnOffHelper onoff ("ns3::TcpSocketFactory",Ipv4Address::GetAny ());
          onoff.SetAttribute ("OnTime",  StringValue
  // Setting applications
  UdpServerHelper myServer (9);
  ApplicationContainer serverApp = myServer.Install (wifiApNode);
  serverApp.Start (Seconds (0.0));
  serverApp.Stop (Seconds (simulationTime + 1));

  UdpClientHelper myClient (ApInterface.GetAddress (0), 9);
  myClient.SetAttribute ("MaxPackets", UintegerValue (4294967295u));
  myClient.SetAttribute ("Interval", TimeValue (Time ("0.00002"))); //packets/s
  myClient.SetAttribute ("PacketSize", UintegerValue (payloadSize));

  // Saturated UDP traffic from stations to AP
  ApplicationContainer clientApp1 = myClient.Install (wifiStaNodes);
  clientApp1.Start (Seconds (1.0));
  clientApp1.Stop (Seconds (simulationTime + 1));

  phy.EnablePcap ("SimpleHtHiddenStations_Ap", apDevice.Get (0));
  phy.EnablePcap ("SimpleHtHiddenStations_Sta1", staDevices.Get (0));
  phy.EnablePcap ("SimpleHtHiddenStations_Sta2", staDevices.Get (1));

  Simulator::Stop (Seconds (simulationTime + 1));
  Simulator::Run ();
  Simulator::Destroy ();

  uint32_t totalPacketsThrough = DynamicCast<UdpServer> (serverApp.Get (0))->GetReceived ();
  double throughput = totalPacketsThrough * payloadSize * 8 / (simulationTime * 1000000.0);
  std::cout << "Throughput: " << throughput << " Mbit/s" << '\n';

  return 0;
}
```

# Experiment 3: Throughput comparison between different protocols

```
#include "ns3/core-module.h"
#include "ns3/network-module.h"
#include "ns3/applications-module.h"
#include "ns3/wifi-module.h"
#include "ns3/mobility-module.h"
#include "ns3/ipv4-global-routing-helper.h"
#include "ns3/internet-module.h"

// This is an example that estimates the throughput of 802.11ax STAa of different protocols,
namely, PRS, PF, MR, Divide & Conquer.
// It defines 4 independant Wi-Fi networks (working on different channels).
// Each network contains one access point and one station. Each station
// continously transmits data packets to its respective AP.
//
// Network topology (numbers in parentheses are the number of STAs):
//
//  Network A (35)   Network B (35)   Network C (35)   Network D (35)
//    *      *         *      *         *      *         *      *
//    |      |         |      |         |      |         |      |
//  AP A   STA A     AP B   STA B     AP C   STA C     AP D   STA D

// A-MSDU size is limited to 7935 bytes (whereas the maximum A-MPDU size is limited to
65535 bytes). When A-MSDU and A-MPDU are both enabled
// (= two-level aggregation), the throughput is slightly smaller than the first scenario since we set
a smaller maximum A-MPDU size.

using namespace ns3;

NS_LOG_COMPONENT_DEFINE ("SimpleMpduAggregation");

int main (int argc, char *argv[])
{
  uint32_t payloadSize = 1472; //bytes
  uint64_t simulationTime = 10; //seconds
  double distance = 5; //meters
  bool enablePcap = 0;

  CommandLine cmd;
  cmd.AddValue ("payloadSize", "Payload size in bytes", payloadSize);
  cmd.AddValue ("simulationTime", "Simulation time in seconds", simulationTime);
  cmd.AddValue ("distance", "Distance in meters between the station and the access point", distance);
  cmd.AddValue ("enablePcap", "Enable/disable pcap file generation", enablePcap);
  cmd.Parse (argc, argv);

  //Network A
  ssid = Ssid ("network-A");
  phy.Set ("ChannelNumber", UintegerValue(36));
  mac.SetType ("ns3::StaWifiMac",
        "Ssid", SsidValue (ssid));
  staDeviceA = wifi.Install (phy, mac, wifiStaNode.Get(0));

  mac.SetType ("ns3::ApWifiMac",
        "Ssid", SsidValue (ssid),
        "BeaconGeneration", BooleanValue (true));
```

```cpp
  apDeviceA = wifi.Install (phy, mac, wifiApNode.Get(0));

  //Network B
  ssid = Ssid ("network-B");
  phy.Set ("ChannelNumber", UintegerValue(40));
  mac.SetType ("ns3::StaWifiMac",
         "Ssid", SsidValue (ssid),
         "BE_MaxAmpduSize", UintegerValue (0)); //Disable A-MPDU

  staDeviceB = wifi.Install (phy, mac, wifiStaNode.Get(1));

  mac.SetType ("ns3::ApWifiMac",
         "Ssid", SsidValue (ssid),
         "BeaconGeneration", BooleanValue (true));
  apDeviceB = wifi.Install (phy, mac, wifiApNode.Get(1));

  //Network C
  ssid = Ssid ("network-C");
  phy.Set ("ChannelNumber", UintegerValue(44));
  mac.SetType ("ns3::StaWifiMac",
         "Ssid", SsidValue (ssid),
         "BE_MaxAmpduSize", UintegerValue (0), //Disable A-MPDU
         "BE_MaxAmsduSize", UintegerValue (7935)); //Enable A-MSDU with the highest maximum size allowed by the standard (7935 bytes)

  staDeviceC = wifi.Install (phy, mac, wifiStaNode.Get(2));

  mac.SetType ("ns3::ApWifiMac",
         "Ssid", SsidValue (ssid),
         "BeaconGeneration", BooleanValue (true));
  apDeviceC = wifi.Install (phy, mac, wifiApNode.Get(2));

  //Network D
  ssid = Ssid ("network-D");
  phy.Set ("ChannelNumber", UintegerValue(48));
  mac.SetType ("ns3::StaWifiMac",
         "Ssid", SsidValue (ssid),
         "BE_MaxAmpduSize", UintegerValue (32768), //Enable A-MPDU with a smaller size than the default one
         "BE_MaxAmsduSize", UintegerValue (3839)); //Enable A-MSDU with the smallest maximum size allowed by the standard (3839 bytes)

  staDeviceD = wifi.Install (phy, mac, wifiStaNode.Get(3));

  mac.SetType ("ns3::ApWifiMac",
         "Ssid", SsidValue (ssid),
         "BeaconGeneration", BooleanValue (true));
  apDeviceD = wifi.Install (phy, mac, wifiApNode.Get(3));

  /* Setting mobility model */
  MobilityHelper mobility;
  Ptr<ListPositionAllocator> positionAlloc = CreateObject<ListPositionAllocator> ();
  mobility.SetMobilityModel ("ns3::ConstantPositionMobilityModel");

  //Set position for APs
  positionAlloc->Add (Vector (0.0, 0.0, 0.0));
  positionAlloc->Add (Vector (10.0, 0.0, 0.0));
```

```
    positionAlloc->Add (Vector (20.0, 0.0, 0.0));
    positionAlloc->Add (Vector (30.0, 0.0, 0.0));
    //Set position for STAs
    positionAlloc->Add (Vector (distance, 0.0, 0.0));
    positionAlloc->Add (Vector (10 + distance, 0.0, 0.0));
    positionAlloc->Add (Vector (20 + distance, 0.0, 0.0));
    positionAlloc->Add (Vector (30 + distance, 0.0, 0.0));
    //Remark: while we set these positions 10 meters apart, the networks do not interact
    //and the only variable that affects transmission performance is the distance.

    mobility.SetPositionAllocator (positionAlloc);
    mobility.Install (wifiApNode);
    mobility.Install (wifiStaNode);

            /* Internet stack*/
            InternetStackHelper stack;
            stack.Install (wifiApNode);
            stack.Install (wifiStaNode);

    Ipv4AddressHelper address;

    address.SetBase ("192.168.1.0", "255.255.255.0");
    Ipv4InterfaceContainer StaInterfaceA;
    StaInterfaceA = address.Assign (staDeviceA);
    Ipv4InterfaceContainer ApInterfaceA;
    ApInterfaceA = address.Assign (apDeviceA);

    address.SetBase ("192.168.2.0", "255.255.255.0");
    Ipv4InterfaceContainer StaInterfaceB;
    StaInterfaceB = address.Assign (staDeviceB);
    Ipv4InterfaceContainer ApInterfaceB;
    ApInterfaceB = address.Assign (apDeviceB);

    /* Setting applications */
    UdpServerHelper myServerA (9);
    ApplicationContainer serverAppA = myServerA.Install (wifiStaNode.Get (0));

    if (enablePcap)
      {
        phy.EnablePcap ("AP_A", apDeviceA.Get (0));
        phy.EnablePcap ("STA_A", staDeviceA.Get (0));
        phy.EnablePcap ("AP_B", apDeviceB.Get (0));
        phy.EnablePcap ("STA_B", staDeviceB.Get (0));
        phy.EnablePcap ("AP_C", apDeviceC.Get (0));
        phy.EnablePcap ("STA_C", staDeviceC.Get (0));
        phy.EnablePcap ("AP_D", apDeviceD.Get (0));
        phy.EnablePcap ("STA_D", staDeviceD.Get (0));
      }

    Simulator::Stop (Seconds (simulationTime + 1));
    Simulator::Run ();
    Simulator::Destroy ();

    /* Show results */
    uint32_t totalPacketsThrough = DynamicCast<UdpServer> (serverAppA.Get (0))->GetReceived ();
    double throughput = totalPacketsThrough * payloadSize * 8 / (simulationTime * 1000000.0);
```

```cpp
  std::cout << "Throughput with default configuration (A-MPDU aggregation enabled, 65kB): " << throughput << " Mbit/s" << '\n';

  totalPacketsThrough = DynamicCast<UdpServer> (serverAppB.Get (0))->GetReceived ();
  throughput = totalPacketsThrough * payloadSize * 8 / (simulationTime * 1000000.0);
  std::cout << "Throughput with aggregation disabled: " << throughput << " Mbit/s" << '\n';

  totalPacketsThrough = DynamicCast<UdpServer> (serverAppC.Get (0))->GetReceived ();
  throughput = totalPacketsThrough * payloadSize * 8 / (simulationTime * 1000000.0);
  std::cout << "Throughput with A-MPDU disabled and A-MSDU enabled (8kB): " << throughput << " Mbit/s" << '\n';

  totalPacketsThrough = DynamicCast<UdpServer> (serverAppD.Get (0))->GetReceived ();
  throughput = totalPacketsThrough * payloadSize * 8 / (simulationTime * 1000000.0);
  std::cout << "Throughput with A-MPDU enabled (32kB) and A-MSDU enabled (4kB): " << throughput << " Mbit/s" << '\n';

  return 0;
}
```

## Experiment 4: Fairness comparison between different protocols

```cpp
#include "ns3/core-module.h"
#include "ns3/network-module.h"
#include "ns3/mobility-module.h"
#include "ns3/stats-module.h"
#include "ns3/wifi-module.h"
#include "ns3/internet-module.h"

#include <iostream>
#include <fstream>
#include <string>

using namespace ns3;

NS_LOG_COMPONENT_DEFINE ("Main");

class Experiment
{
public:
  Experiment ();
  Experiment (std::string name);
  uint32_t Run (const WifiHelper &wifi, const YansWifiPhyHelper &wifiPhy,
                const WifiMacHelper &wifiMac, const YansWifiChannelHelper &wifiChannel);
private:
  void ReceivePacket (Ptr<Socket> socket);
  void SetPosition (Ptr<Node> node, Vector position);
  Vector GetPosition (Ptr<Node> node);
  Ptr<Socket> SetupPacketReceive (Ptr<Node> node);
  void GenerateTraffic (Ptr<Socket> socket, uint32_t pktSize,
                       uint32_t pktCount, Time pktInterval );

  uint32_t m_pktsTotal;
  Gnuplot2dDataset m_output;
};
```

```cpp
Experiment::Experiment ()
{
}

Experiment::Experiment (std::string name)
 : m_output (name)
{
  m_output.SetStyle (Gnuplot2dDataset::LINES);
}

void
Experiment::ReceivePacket (Ptr<Socket> socket)
{
  Ptr<Packet> packet;
  while ((packet = socket->Recv ()))
    {
      m_pktsTotal++;
    }
}

Ptr<Socket>
Experiment::SetupPacketReceive (Ptr<Node> node)
{
  TypeId tid = TypeId::LookupByName ("ns3::UdpSocketFactory");
  Ptr<Socket> sink = Socket::CreateSocket (node, tid);
  InetSocketAddress local = InetSocketAddress (Ipv4Address::GetAny (), 80);
  sink->Bind (local);
  sink->SetRecvCallback (MakeCallback (&Experiment::ReceivePacket, this));
  return sink;
}

void
Experiment::GenerateTraffic (Ptr<Socket> socket, uint32_t pktSize,
                 uint32_t pktCount, Time pktInterval )
{
  if (pktCount > 0)
    {
      socket->Send (Create<Packet> (pktSize));
      Simulator::Schedule (pktInterval, &Experiment::GenerateTraffic, this,
                 socket, pktSize,pktCount-1, pktInterval);
    }
  else
    {
      socket->Close ();
    }
}

uint32_t
Experiment::Run (const WifiHelper &wifi, const YansWifiPhyHelper &wifiPhy,
          const WifiMacHelper &wifiMac, const YansWifiChannelHelper &wifiChannel)
{
  m_pktsTotal = 0;

  NodeContainer c;
  c.Create (2);
```

F-12

```cpp
  InternetStackHelper internet;
  internet.Install (c);

  YansWifiPhyHelper phy = wifiPhy;
  phy.SetChannel (wifiChannel.Create ());

  WifiMacHelper mac = wifiMac;
  NetDeviceContainer devices = wifi.Install (phy, mac, c);

  MobilityHelper mobility;

  yansplot.GenerateOutput (yansfile);
  yansfile.close ();

  nistplot.SetTerminal ("postscript eps color enh \"Times-BoldItalic\"");
  nistplot.SetLegend ("SNR(dB)", "Frame Success Rate");
  nistplot.SetExtra  ("set xrange [-5:30]\n\

  Ipv4AddressHelper ipv4;
  NS_LOG_INFO ("Assign IP Addresses.");
  ipv4.SetBase ("10.1.1.0", "255.255.255.0");
  Ipv4InterfaceContainer i = ipv4.Assign (devices);

  Ptr<Socket> recvSink = SetupPacketReceive (c.Get (0));

  TypeId tid = TypeId::LookupByName ("ns3::UdpSocketFactory");
  Ptr<Socket> source = Socket::CreateSocket (c.Get (1), tid);
  InetSocketAddress remote = InetSocketAddress (Ipv4Address ("255.255.255.255"), 80);
  source->SetAllowBroadcast (true);
  source->Connect (remote);
  uint32_t packetSize = 1014;
  uint32_t maxPacketCount = 200;
  Time interPacketInterval = Seconds (1.);
  Simulator::Schedule (Seconds (1.0), &Experiment::GenerateTraffic,
                       this, source, packetSize, maxPacketCount,interPacketInterval);
  Simulator::Run ();

  Simulator::Destroy ();

  return m_pktsTotal;
}

int main (int argc, char *argv[])
{
  std::ofstream outfile ("clear-channel.plt");
  std::vector <std::string> modes;

  modes.push_back ("DsssRate1Mbps");
  modes.push_back ("DsssRate2Mbps");
  modes.push_back ("DsssRate5_5Mbps");
  modes.push_back ("DsssRate11Mbps");
  // disable fragmentation

  Gnuplot gnuplot = Gnuplot ("clear-channel.eps");

  for (uint32_t i = 0; i < modes.size (); i++)
    {
```

```
    std::cout << modes[i] << std::endl;
    Gnuplot2dDataset dataset (modes[i]);

    for (double rss = -102.0; rss <= -80.0; rss += 0.5)
      {
       Experiment experiment;
       dataset.SetStyle (Gnuplot2dDataset::LINES);

       WifiHelper wifi;
       wifi.SetStandard (WIFI_PHY_STANDARD_80211b);
       WifiMacHelper wifiMac;
       Config::SetDefault ("ns3::WifiRemoteStationManager::NonUnicastMode",
                 StringValue (modes[i]));
       wifi.SetRemoteStationManager ("ns3::ConstantRateWifiManager",
                      "DataMode",StringValue (modes[i]),
("ns3::ConstantSpeedPropagationDelayModel");
       wifiChannel.AddPropagationLoss ("ns3::FixedRssLossModel","Rss",DoubleValue (rss));

        wifiPhy.Set ("CcaMode1Threshold", DoubleValue (-110.0) );
       wifiPhy.Set ("TxPowerStart", DoubleValue (15.0) );
       wifiPhy.Set ("TxPowerEnd", DoubleValue (15.0) );
       wifiPhy.Set ("RxGain", DoubleValue (0) );
       wifiPhy.Set ("RxNoiseFigure", DoubleValue (7) );
       uint32_t pktsRecvd = experiment.Run (wifi, wifiPhy, wifiMac, wifiChannel);
       dataset.Add (rss, pktsRecvd);
      }

   gnuplot.AddDataset (dataset);
 }
 gnuplot.SetTerminal ("postscript eps color enh \"Times-BoldItalic\"");
 gnuplot.SetLegend ("RSS(dBm)", "Number of packets received");
 gnuplot.SetExtra  ("set xrange [-102:-83]");
 gnuplot.GenerateOutput (outfile);
 outfile.close ();

 return 0;
}
```

# Experiment 5: Impact of the number of stations on the throughput

```
#include "ns3/core-module.h"
#include "ns3/network-module.h"
#include "ns3/mobility-module.h"
#include "ns3/config-store-module.h"
#include "ns3/wifi-module.h"
#include "ns3/internet-module.h"

#include <iostream>
#include <vector>
#include <string>

using namespace ns3;

NS_LOG_COMPONENT_DEFINE ("WifiSimpleAdhoc");
```

```cpp
void ReceivePacket (Ptr<Socket> socket)
{
  while (socket->Recv ())
    {
      NS_LOG_UNCOND ("Received one packet!");
    }
}

static void GenerateTraffic (Ptr<Socket> socket, uint32_t pktSize,
                 uint32_t pktCount, Time pktInterval )
static void
SetPosition (Ptr<Node> node, Vector position)
{
  Ptr<MobilityModel> mobility = node->GetObject<MobilityModel> ();
  mobility->SetPosition (position);
}

static Vector
GetPosition (Ptr<Node> node)
{
  Ptr<MobilityModel> mobility = node->GetObject<MobilityModel> ();
  return mobility->GetPosition ();
}

{
  if (pktCount > 0)
    {
      socket->Send (Create<Packet> (pktSize));
      Simulator::Schedule (pktInterval, &GenerateTraffic,
                  socket, pktSize,pktCount-1, pktInterval);
    }
  else
    {
      socket->Close ();
    }
}

int main (int argc, char *argv[])
{
  std::string phyMode ("DsssRate1Mbps");
  double rss = -80;  // -dBm
  uint32_t packetSize = 1000; // bytes
  uint32_t numPackets = 1;
  double interval = 1.0; // seconds
  bool verbose = false;

  CommandLine cmd;

  cmd.AddValue ("phyMode", "Wifi Phy mode", phyMode);
  cmd.AddValue ("rss", "received signal strength", rss);
  cmd.AddValue ("packetSize", "size of application packet sent", packetSize);
  cmd.AddValue ("numPackets", "number of packets generated", numPackets);
  cmd.AddValue ("interval", "interval (seconds) between packets", interval);
  cmd.AddValue ("verbose", "turn on all WifiNetDevice log components", verbose);

  cmd.Parse (argc, argv);
  // Convert to time object
```

```cpp
  Time interPacketInterval = Seconds (interval);

  // disable fragmentation for frames below 2200 bytes
  Config::SetDefault ("ns3::WifiRemoteStationManager::FragmentationThreshold", StringValue ("2200"));
  // turn off RTS/CTS for frames below 2200 bytes
  Config::SetDefault ("ns3::WifiRemoteStationManager::RtsCtsThreshold", StringValue ("2200"));
  // Fix non-unicast data rate to be the same as that of unicast
  Config::SetDefault ("ns3::WifiRemoteStationManager::NonUnicastMode",
                     StringValue (phyMode));

  NodeContainer c;
  c.Create (2);

  // The below set of helpers will help us to put together the wifi NICs we want
  WifiHelper wifi;
  if (verbose)
    {
      wifi.EnableLogComponents ();  // Turn on all Wifi logging
    }
  wifi.SetStandard (WIFI_PHY_STANDARD_80211b);

  YansWifiChannelHelper wifiChannel;
  wifiChannel.SetPropagationDelay ("ns3::ConstantSpeedPropagationDelayModel");
  // The below FixedRssLossModel will cause the rss to be fixed regardless
  // of the distance between the two stations, and the transmit power
  wifiChannel.AddPropagationLoss ("ns3::FixedRssLossModel","Rss",DoubleValue (rss));
  wifiPhy.SetChannel (wifiChannel.Create ());

  // Note that with FixedRssLossModel, the positions below are not
  // used for received signal strength.
  MobilityHelper mobility;
  Ptr<ListPositionAllocator> positionAlloc = CreateObject<ListPositionAllocator> ();
  positionAlloc->Add (Vector (0.0, 0.0, 0.0));
  positionAlloc->Add (Vector (5.0, 0.0, 0.0));
  mobility.SetPositionAllocator (positionAlloc);
  mobility.SetMobilityModel ("ns3::ConstantPositionMobilityModel");
  mobility.Install (c);

  InternetStackHelper internet;
  internet.Install (c);

  Ipv4AddressHelper ipv4;
  NS_LOG_INFO ("Assign IP Addresses.");
  ipv4.SetBase ("10.1.1.0", "255.255.255.0");
  Ipv4InterfaceContainer i = ipv4.Assign (devices);

  TypeId tid = TypeId::LookupByName ("ns3::UdpSocketFactory");
  Ptr<Socket> recvSink = Socket::CreateSocket (c.Get (0), tid);
  InetSocketAddress local = InetSocketAddress (Ipv4Address::GetAny (), 80);
  recvSink->Bind (local);
  recvSink->SetRecvCallback (MakeCallback (&ReceivePacket));

  Ptr<Socket> source = Socket::CreateSocket (c.Get (1), tid);
  InetSocketAddress remote = InetSocketAddress (Ipv4Address ("255.255.255.255"), 80);
  source->SetAllowBroadcast (true);
```

```cpp
  source->Connect (remote);

  // Tracing
  wifiPhy.EnablePcap ("wifi-simple-adhoc", devices);

  // Output what we are doing
  NS_LOG_UNCOND ("Testing " << numPackets  << " packets sent with receiver rss " << rss );

  Simulator::ScheduleWithContext (source->GetNode ()->GetId (),
                  Seconds (1.0), &GenerateTraffic,
                  source, packetSize, numPackets, interPacketInterval);

  Simulator::Run ();
  Simulator::Destroy ();

  return 0;
}
```

# Experiment 6: Impact of the number of antennas on the throughput

```cpp
#include "ns3/core-module.h"
#include "ns3/network-module.h"
#include "ns3/mobility-module.h"
#include "ns3/config-store-module.h"
#include "ns3/wifi-module.h"
#include "ns3/internet-module.h"
#include "ns3/ipv4-static-routing-helper.h"
#include <iostream>
#include <fstream>
#include <vector>
#include <string>
using namespace ns3;
NS_LOG_COMPONENT_DEFINE ("WifiSimpleGrid");
void ReceivePacket (Ptr<Socket> socket)
{
  while (socket->Recv ())
    {
      NS_LOG_UNCOND ("Received one packet!");
    }
}
static void GenerateTraffic (Ptr<Socket> socket, uint32_t pktSize,
                 uint32_t pktCount, Time pktInterval )
{
  if (pktCount > 0)
    {
      socket->Send (Create<Packet> (pktSize));
      Simulator::Schedule (pktInterval, &GenerateTraffic,
                socket, pktSize,pktCount-1, pktInterval);
    }
  else
    {
      socket->Close ();
    }
}
int main (int argc, char *argv[])
```

```cpp
{
  std::string phyMode ("DsssRate1Mbps");
  double distance = 500;  // m
  uint32_t packetSize = 1000; // bytes
  uint32_t numPackets = 1;
  uint32_t numNodes = 25;  // by default, 5x5
  uint32_t sinkNode = 0;
  uint32_t sourceNode = 24;
  double interval = 1.0; // seconds
  bool verbose = false;
  bool tracing = false;
  CommandLine cmd;
  cmd.AddValue ("phyMode", "Wifi Phy mode", phyMode);
  cmd.AddValue ("distance", "distance (m)", distance);
  cmd.AddValue ("packetSize", "size of application packet sent", packetSize);
  cmd.AddValue ("numPackets", "number of packets generated", numPackets);
  cmd.AddValue ("interval", "interval (seconds) between packets", interval);
  cmd.AddValue ("verbose", "turn on all WifiNetDevice log components", verbose);
  cmd.AddValue ("tracing", "turn on ascii and pcap tracing", tracing);
  cmd.AddValue ("numNodes", "number of nodes", numNodes);
  cmd.AddValue ("sinkNode", "Receiver node number", sinkNode);
  cmd.AddValue ("sourceNode", "Sender node number", sourceNode);
  cmd.Parse (argc, argv);
  // Convert to time object
  Time interPacketInterval = Seconds (interval);
  // disable fragmentation for frames below 2200 bytes
  Config::SetDefault ("ns3::WifiRemoteStationManager::FragmentationThreshold", StringValue ("2200"));
  // turn off RTS/CTS for frames below 2200 bytes
  Config::SetDefault ("ns3::WifiRemoteStationManager::RtsCtsThreshold", StringValue ("2200"));
  // Fix non-unicast data rate to be the same as that of unicast
  Config::SetDefault ("ns3::WifiRemoteStationManager::NonUnicastMode",
                     StringValue (phyMode));
  NodeContainer c;
  c.Create (numNodes);
  // The below set of helpers will help us to put together the wifi NICs we want
  WifiHelper wifi;
  if (verbose)
    {
      wifi.EnableLogComponents ();  // Turn on all Wifi logging
    }
  YansWifiPhyHelper wifiPhy =  YansWifiPhyHelper::Default ();
  // set it to zero; otherwise, gain will be added
  wifiPhy.Set ("RxGain", DoubleValue (-10) );
  // ns-3 supports RadioTap and Prism tracing extensions for 802.11b
  wifiPhy.SetPcapDataLinkType (YansWifiPhyHelper::DLT_IEEE802_11_RADIO);
  YansWifiChannelHelper wifiChannel;
  wifiChannel.SetPropagationDelay ("ns3::ConstantSpeedPropagationDelayModel");
  wifiChannel.AddPropagationLoss ("ns3::FriisPropagationLossModel");
  wifiPhy.SetChannel (wifiChannel.Create ());
  // Add an upper mac and disable rate control
  WifiMacHelper wifiMac;
  wifi.SetStandard (WIFI_PHY_STANDARD_80211b);
  wifi.SetRemoteStationManager ("ns3::ConstantRateWifiManager",
                                "DataMode",StringValue (phyMode),
                                "ControlMode",StringValue (phyMode));
```

```cpp
  // Enable OLSR
  OlsrHelper olsr;
  Ipv4StaticRoutingHelper staticRouting;
  Ipv4ListRoutingHelper list;
  list.Add (staticRouting, 0);
  list.Add (olsr, 10);
  InternetStackHelper internet;
  internet.SetRoutingHelper (list); // has effect on the next Install ()
  internet.Install (c);
  Ipv4AddressHelper ipv4;
  NS_LOG_INFO ("Assign IP Addresses.");
  ipv4.SetBase ("10.1.1.0", "255.255.255.0");
  Ipv4InterfaceContainer i = ipv4.Assign (devices);

  //Configure the STA node
    wifi.SetRemoteStationManager (staManager, "RtsCtsThreshold", UintegerValue
(rtsThreshold));

    Ssid ssid = Ssid ("AP");
    wifiMac.SetType ("ns3::StaWifiMac",
              "Ssid", SsidValue (ssid));
    wifiStaDevices.Add (wifi.Install (wifiPhy, wifiMac, wifiStaNodes.Get (0)));

    //Configure the AP node
    wifi.SetRemoteStationManager (apManager, "RtsCtsThreshold", UintegerValue
(rtsThreshold));

    ssid = Ssid ("AP");
    wifiMac.SetType ("ns3::ApWifiMac",
              "Ssid", SsidValue (ssid));
    wifiApDevices.Add (wifi.Install (wifiPhy, wifiMac, wifiApNodes.Get (0)));
  }
  if (tracing == true)
   {
    AsciiTraceHelper ascii;
    wifiPhy.EnableAsciiAll (ascii.CreateFileStream ("wifi-simple-adhoc-grid.tr"));
    wifiPhy.EnablePcap ("wifi-simple-adhoc-grid", devices);
    // Trace routing tables
    Ptr<OutputStreamWrapper> routingStream = Create<OutputStreamWrapper> ("wifi-simple-adhoc-grid.routes", std::ios::out);
    olsr.PrintRoutingTableAllEvery (Seconds (2), routingStream);
    Ptr<OutputStreamWrapper> neighborStream = Create<OutputStreamWrapper> ("wifi-simple-adhoc-grid.neighbors", std::ios::out);
    olsr.PrintNeighborCacheAllEvery (Seconds (2), neighborStream);
    // To do-- enable an IP-level trace that shows forwarding events only
   }
  // Output
  NS_LOG_UNCOND ("Testing from node " << sourceNode << " to " << sinkNode << " with grid distance " << distance);
  Simulator::Stop (Seconds (33.0));
  Simulator::Run ();
  Simulator::Destroy ();
  return 0;
}
```

# Experiment 7: Goodput vs number of stations

```cpp
#include <sstream>
#include <iomanip>
#include "ns3/core-module.h"
#include "ns3/config-store-module.h"
#include "ns3/network-module.h"
#include "ns3/wifi-module.h"
#include "ns3/mobility-module.h"
#include "ns3/spectrum-module.h"
#include "ns3/internet-module.h"
// This is a simple example of an IEEE 802.11ax Wi-Fi network.
//
// The main use case is to enable and test Goodput vs Number of STAs
// under saturation conditions (for max throughput).
using namespace ns3;
NS_LOG_COMPONENT_DEFINE ("WifiSpectrumSaturationExample");
int main (int argc, char *argv[])
{
  double distance = 1;
  double simulationTime = 10; //seconds
  uint16_t index = 256;
  uint32_t channelWidth = 0;
  std::string wifiType = "ns3::SpectrumWifiPhy";
  std::string errorModelType = "ns3::NistErrorRateModel";
  bool enablePcap = false;
  CommandLine cmd;
  cmd.AddValue ("frequency", "Whether working in the 2.4 or 5.0 GHz band (other values gets rejected)", frequency);
  cmd.AddValue ("distance", "Distance in meters between the station and the access point", distance);
  cmd.AddValue ("simulationTime", "Simulation time in seconds", simulationTime);
  cmd.AddValue ("udp", "UDP if set to 1, TCP otherwise", udp);
  cmd.Parse (argc,argv);
  uint16_t startIndex = 0;
  uint16_t stopIndex = 63;
  if (index < 64)
    {
      startIndex = index;
      stopIndex = index;
    }
  std::cout << "wifiType: " << wifiType << " distance: " << distance << "m" << std::endl;
  std::cout << std::setw (5) << "index" <<
    std::setw (6) << "MCS" <<
    std::setw (8) << "width" <<
    std::setw (12) << "Rate (Mb/s)" <<
    std::setw (12) << "Tput (Mb/s)" <<
    std::setw (10) << "Received " <<
    std::endl;
  for (uint16_t i = startIndex; i <= stopIndex; i++)
    {
      uint32_t payloadSize;
      payloadSize = 1472; // 1500 bytes IPv4
      NodeContainer wifiStaNode;
      wifiStaNode.Create (1);
      NodeContainer wifiApNode;
      wifiApNode.Create (1);
```

```cpp
      if (i <= 7)
        {
          phy.Set ("ShortGuardEnabled", BooleanValue (false));
          channelWidth = 20;
        }
      else if (i > 7 && i <= 15)
        {
          phy.Set ("ShortGuardEnabled", BooleanValue (true));
          channelWidth = 20;
        }
      else if (i > 15 && i <= 23)
        {
          phy.Set ("ShortGuardEnabled", BooleanValue (false));
          channelWidth = 40;
        }
      else if (i > 23 && i <= 31)
        {
          phy.Set ("ShortGuardEnabled", BooleanValue (true));
          channelWidth = 40;
        }
      else if (i > 31 && i <= 39)
        {
          phy.Set ("ShortGuardEnabled", BooleanValue (false));
          phy.Set ("RxAntennas", UintegerValue (2));
          phy.Set ("TxAntennas", UintegerValue (2));
          channelWidth = 20;
        }
      else if (i > 39 && i <= 47)
        {
          phy.Set ("ShortGuardEnabled", BooleanValue (true));
          phy.Set ("RxAntennas", UintegerValue (2));
          phy.Set ("TxAntennas", UintegerValue (2));
          channelWidth = 20;
        }
      else if (i > 47 && i <= 55)
        {
          phy.Set ("ShortGuardEnabled", BooleanValue (false));
          phy.Set ("RxAntennas", UintegerValue (2));
          phy.Set ("TxAntennas", UintegerValue (2));
          channelWidth = 40;
        }
      else if (i > 55 && i <= 63)
        {
          phy.Set ("ShortGuardEnabled", BooleanValue (true));
          phy.Set ("RxAntennas", UintegerValue (2));
          phy.Set ("TxAntennas", UintegerValue (2));
          channelWidth = 40;
        }
    }
  else if (wifiType == "ns3::SpectrumWifiPhy")
    {
      //Bug 2460: CcaMode1Threshold default should be set to -62 dBm
WifiHelper wifi;
wifi.SetStandard (WIFI_PHY_STANDARD_80211n_5GHZ);
WifiMacHelper mac;
Ssid ssid = Ssid ("ns380211ax");
wifi.SetRemoteStationManager ("ns3::ConstantRateWifiManager","DataMode", DataRate,
```

```cpp
                      "ControlMode", DataRate);
NetDeviceContainer staDevice;
NetDeviceContainer apDevice;
if (wifiType == "ns3::YansWifiPhy")
  {
    mac.SetType ("ns3::StaWifiMac",
            "Ssid", SsidValue (ssid),
            "ActiveProbing", BooleanValue (false));
    staDevice = wifi.Install (phy, mac, wifiStaNode);
    mac.SetType ("ns3::ApWifiMac",
            "Ssid", SsidValue (ssid));
    apDevice = wifi.Install (phy, mac, wifiApNode);
  }
else if (wifiType == "ns3::SpectrumWifiPhy")
  {
    mac.SetType ("ns3::StaWifiMac",
            "Ssid", SsidValue (ssid),
            "ActiveProbing", BooleanValue (false));
    staDevice = wifi.Install (spectrumPhy, mac, wifiStaNode);
    mac.SetType ("ns3::ApWifiMac",
            "Ssid", SsidValue (ssid));
    apDevice = wifi.Install (spectrumPhy, mac, wifiApNode);
  }
// mobility.
MobilityHelper mobility;
Ipv4AddressHelper address;
address.SetBase ("192.168.1.0", "255.255.255.0");
Ipv4InterfaceContainer staNodeInterface;
Ipv4InterfaceContainer apNodeInterface;
staNodeInterface = address.Assign (staDevice);
apNodeInterface = address.Assign (apDevice);
/* Setting applications */
ApplicationContainer serverApp, sinkApp;
//UDP flow
UdpServerHelper myServer (9);
serverApp = myServer.Install (wifiStaNode.Get (0));
serverApp.Start (Seconds (0.0));
serverApp.Stop (Seconds (simulationTime + 1));
UdpClientHelper myClient (staNodeInterface.GetAddress (0), 9);
myClient.SetAttribute ("MaxPackets", UintegerValue (4294967295u));
myClient.SetAttribute ("Interval", TimeValue (Time ("0.00002"))); //packets/s
myClient.SetAttribute ("PacketSize", UintegerValue (payloadSize));
ApplicationContainer clientApp = myClient.Install (wifiApNode.Get (0));
clientApp.Start (Seconds (1.0));
clientApp.Stop (Seconds (simulationTime + 1));
if (enablePcap)
  {
    std::stringstream ss;
    ss << "wifi-spectrum-saturation-example-" << i;
    phy.EnablePcap (ss.str (), apDevice);
  }
Simulator::Stop (Seconds (simulationTime + 1));
Simulator::Run ();
Simulator::Destroy ();
double throughput;
uint32_t totalPacketsThrough;
totalPacketsThrough = DynamicCast<UdpServer> (serverApp.Get (0))->GetReceived ();
```

```
      throughput = totalPacketsThrough * payloadSize * 8 / (simulationTime * 1000000.0);
//Mbit/s
      std::cout << std::setw (5) << i <<
        std::setw (6) << (i % 8) + 8 * (i / 32) <<
        std::setw (8) << channelWidth <<
        std::setw (10) << datarate <<
        std::setw (12) << throughput <<
        std::setw (8) << totalPacketsThrough <<
        std::endl;
    }
  return 0;
}
```

# Appendix G

# PRS Protocol Data for Graphing

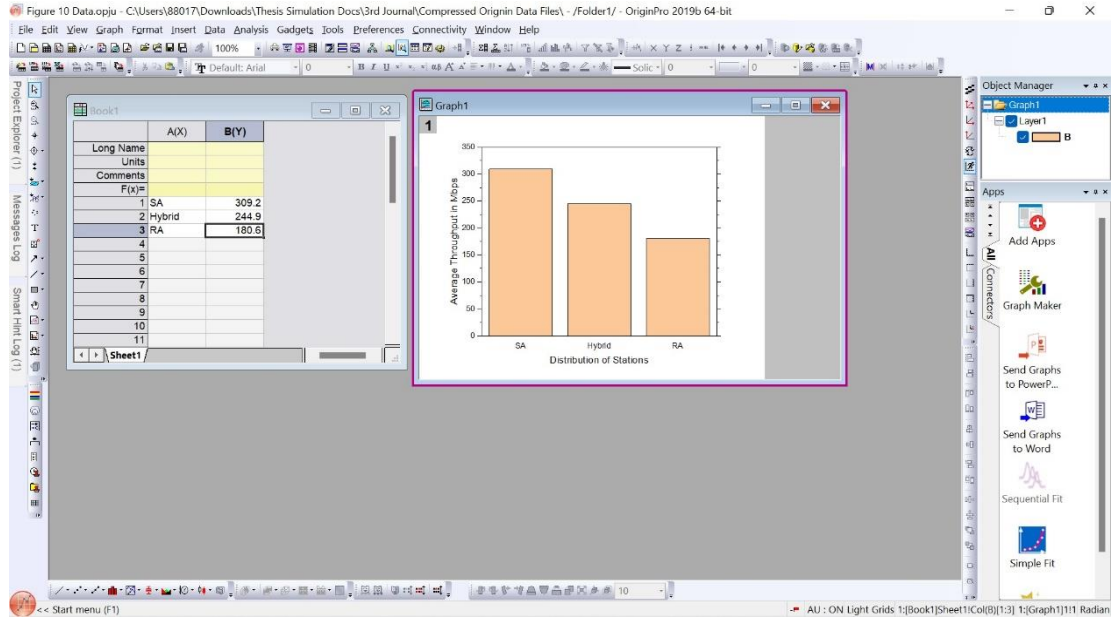

**Experiment 1: Throughput of the PRS protocol for different distribution of stations**

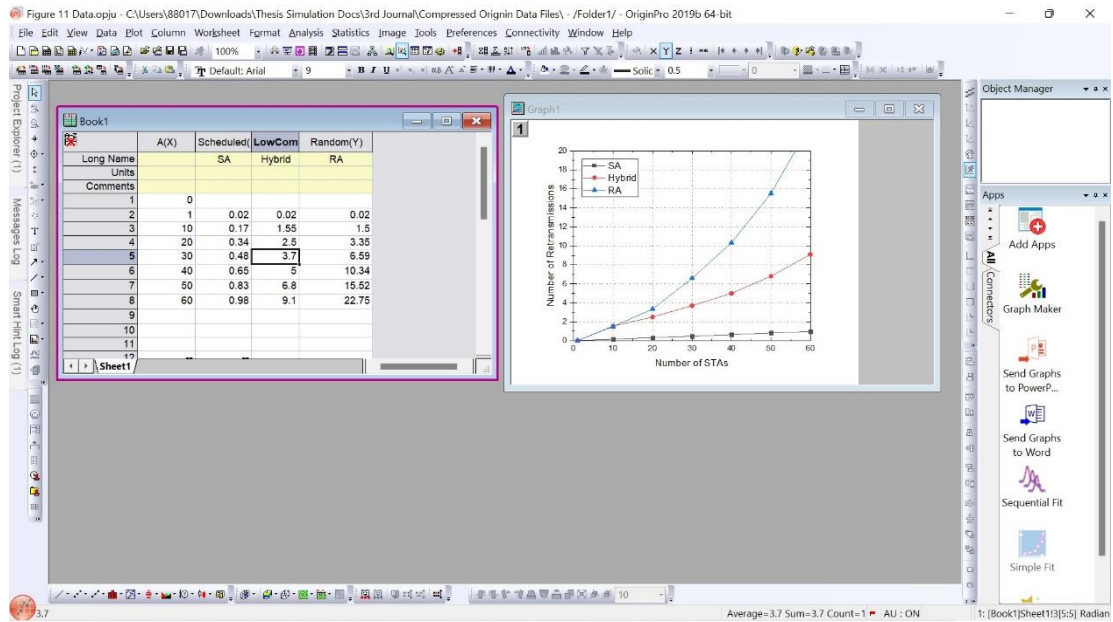

**Experiment 2: Number of retransmissions of the PRS protocol for different distribution of stations**

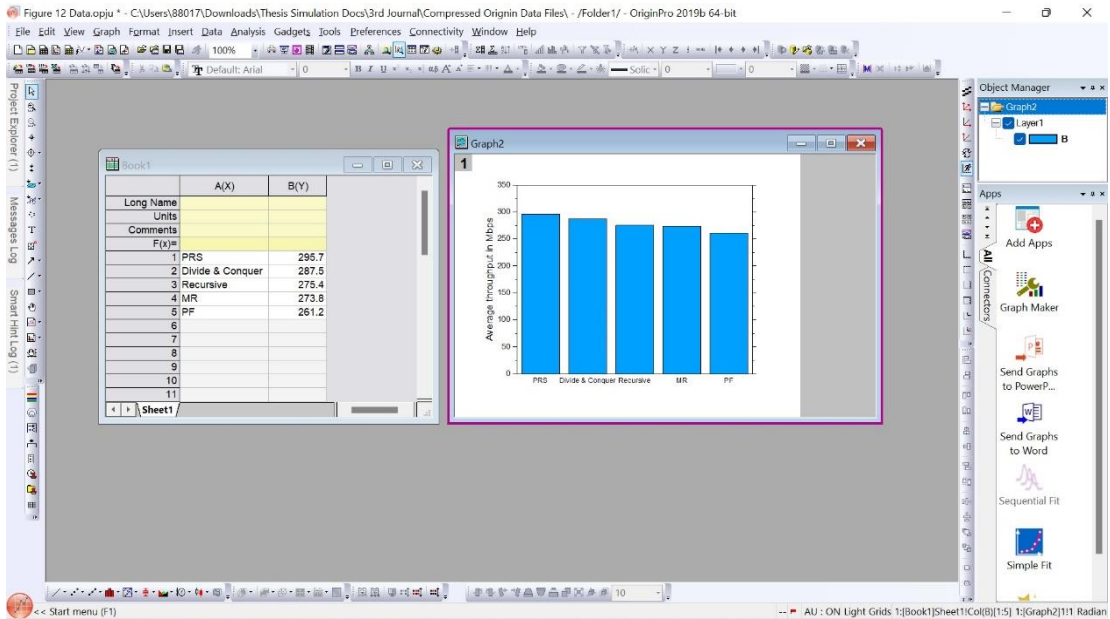

**Experiment 3: Throughput comparison between different protocols**

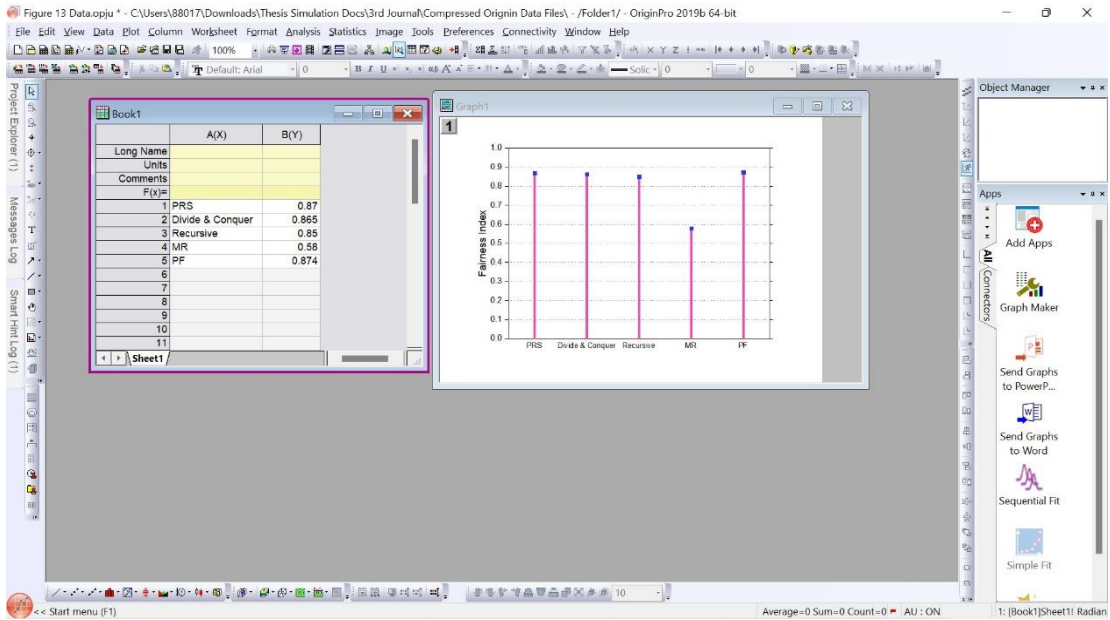

**Experiment 4: Fairness comparison between different protocols**

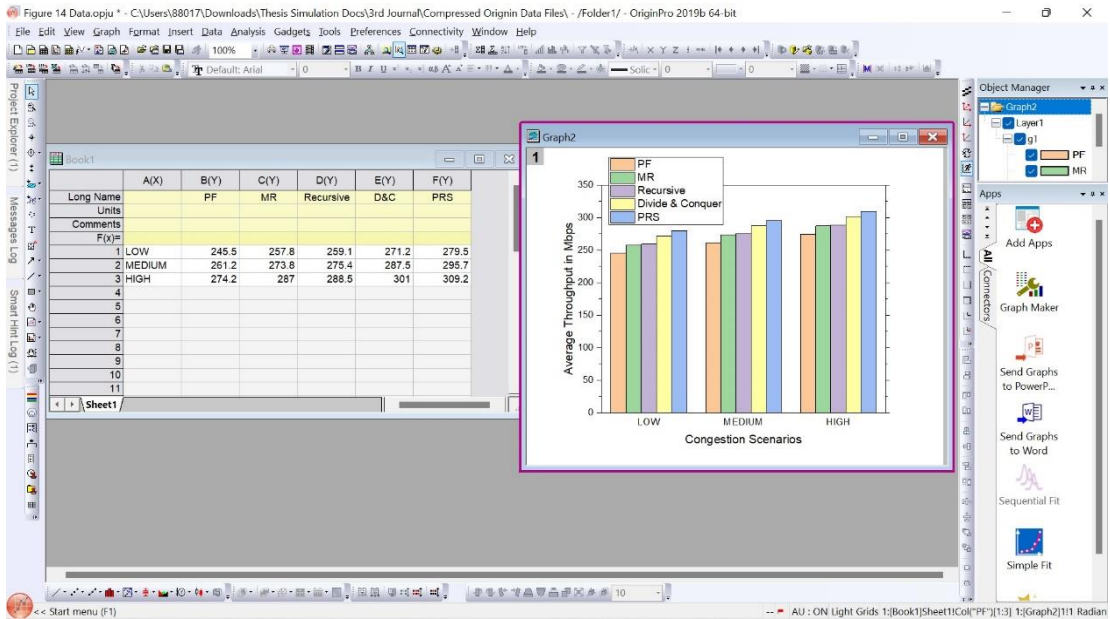

Experiment 5: Impact of the number of stations on the throughput

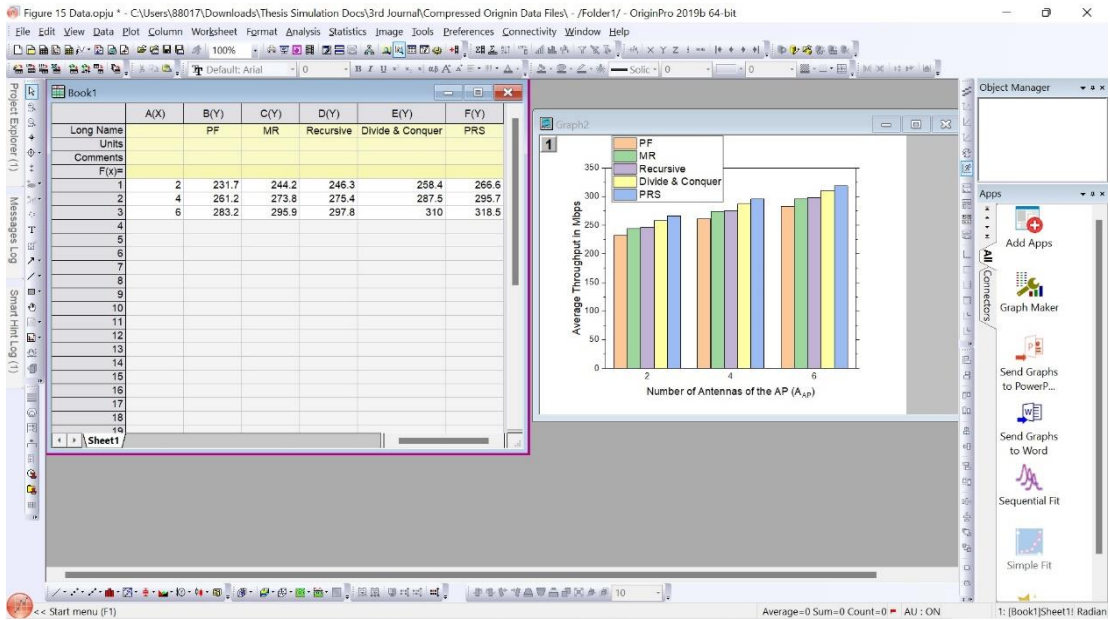

Experiment 6: Impact of the number of antennas on the throughput

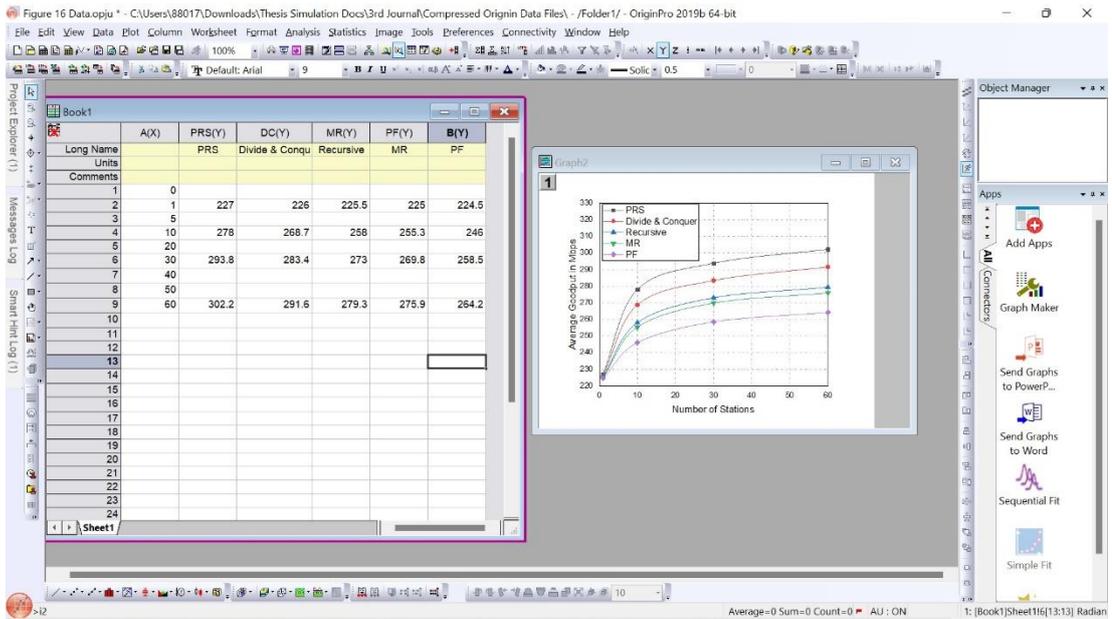

**Experiment 7: Goodput vs number of stations**